%% file: main.tex
\documentclass[12pt,openany]{Thesis}

\usepackage{cancel}
\usepackage{soul}

\input{./packages.tex}

\begin{document}

\input{./titlepage.tex}

\pagenumbering{roman}

\addcontentsline{toc}{chapter}{Summary}
\chapter*{Summary}
\thispagestyle{chapterfirstpage}
\input{./abstract.tex}

\addcontentsline{toc}{chapter}{Acknowledgements}
\chapter*{Acknowledgements}
\thispagestyle{chapterfirstpage}
\input{./acknowledgements.tex}

\pagestyle{fancy}

\addtocontents{toc}{\protect\setcounter{tocdepth}{-1}}
\tableofcontents
\addtocontents{toc}{\protect\setcounter{tocdepth}{3}}
\thispagestyle{chapterfirstpage}

\clearpage

\mainmatter 

\pagestyle{fancy} 

\chapter{Introduction} 
\thispagestyle{chapterfirstpage}

\label{ch:intro}

\input{./intro.tex}

\chapter{AdS/CFT correspondence} 
\thispagestyle{chapterfirstpage}

\label{ch:adscft}

\input{./adscft.tex}

\chapter{Line defects in AdS/CFT} 

\thispagestyle{chapterfirstpage}

\label{ch: line defects}

\input{./introdefects.tex}

\chapter{Line defects in AdS$_3$/CFT$_2$} 

\thispagestyle{chapterfirstpage}

\label{ch: line defects in AdS3CFT2} 

\input{./defectsAdS3CFT2.tex}

\chapter{Integrable line defects in ABJM} 

\thispagestyle{chapterfirstpage}

\label{ch: integrable line defects in ABJM}

\input{./integrableWLABJM.tex}

\chapter{Scattering amplitudes and positive geometries in AdS/CFT} 

\thispagestyle{chapterfirstpage}

\label{ch: scatt ampl}

\input{./introamplitudes.tex}

\chapter{Integrated negative geometries in ABJM} 

\thispagestyle{chapterfirstpage}

\label{ch: neg geoms ABJM} 

\input{./integrated_neg_geoms.tex}

\chapter{Conclusions}
\thispagestyle{chapterfirstpage}
\label{ch: conclusions} 
\input{./conclusions.tex}

\appendix 

\chapter{Scalar and spinor fields in AdS/CFT} 

\thispagestyle{chapterfirstpage}

\label{ch: scalar and spinor fields in AdS/CFT}

\input{./Scalar_and_spinor_fields_in_AdSCFT.tex}

\chapter{AdS$_3\times$S$^3_+\times$S$^3_-\times$S$^1$ Killing spinors}

\thispagestyle{chapterfirstpage}

\label{killingspinors}

\input{ads3cft2_killing_spinors.tex}

\chapter{Analysis of fluctuations: effective action and supersymmetry transformations}

\thispagestyle{chapterfirstpage}

\label{app: quadratic fluctuations}

\input{quadratic_fluctuations.tex}

\chapter{Representation theory of the $\mathfrak{psu}(1,1|2)\times \mathfrak{su}(2)_A$ algebra}

\thispagestyle{chapterfirstpage}

\label{app: psu(1,1|2)}

\input{defectalgebra.tex}

\chapter{AdS$_2$ propagators and $D$-integrals}

\thispagestyle{chapterfirstpage}

\label{app: D integrals}

\input{D_integrals.tex}

\chapter{Integrable Wilson lines in ABJM: holographic dual of spin chain's vacuum state}

\thispagestyle{chapterfirstpage}

\label{app: dual string vacuum state}

\input{vacuumapp.tex}

\chapter{Kinematics of massless particles} 

\thispagestyle{chapterfirstpage}

\label{ch: massless kinematics} 

\input{./kinematics.tex}

\chapter{Normalization of negative geometries} 

\thispagestyle{chapterfirstpage}

\label{ch: normalizations} 

\input{./app_normalizations.tex}

\chapter{Five-dimensional notation} 

\thispagestyle{chapterfirstpage}

\label{app: five-dimensional notation} 

\input{./app_five_dim_not.tex}

\chapter{Useful integrals} 

\thispagestyle{chapterfirstpage}

\label{app: useful integrals} 

\input{./app_useful_integrals.tex}

\bibliographystyle{JHEP} 
\addcontentsline{toc}{chapter}{Bibliography}
\bibliography{ref.bib} 
\thispagestyle{chapterfirstpage}
\label{Bibliography}

\end{document}

%% file: packages.tex
\usepackage[square, numbers, comma, sort&compress]{natbib} 
\usepackage[english]{babel}
\usepackage[utf8]{inputenc}
\usepackage{amsfonts,euscript,amssymb,stmaryrd,braket}
\usepackage{graphics,tikz}
\usetikzlibrary{arrows,decorations.markings,patterns}
\usepackage{slashed}
\usepackage{multirow}
\usepackage{epsfig}
\usepackage{graphicx}
\usepackage{tabularx}
\usepackage{longtable}
\usepackage{bbold}
\usepackage{amsthm}
\usepackage[dvipsnames]{xcolor}
\usepackage{sectsty}

\usetikzlibrary{matrix,arrows,snakes,shapes,decorations.fractals,decorations.pathmorphing,patterns,decorations.markings}

\tikzset{middlearrow/.style={
        decoration={markings,
            mark= at position 0.5 with {\arrow{#1}} ,
        },
        postaction={decorate}
    }
}

\pgfdeclaredecoration{cappedcurveto}{initial}{%
  \state{initial}[width=\pgfdecoratedinputsegmentlength/100]
  {
    \pgfpathlineto{\pgfpointorigin}
  }%
  \state{final}{
  }%
}%
\makeatother

    \tikzset{
        draw left/.style={
            decorate,
            decoration={
                cappedcurveto,
                raise={.5*\the\pgflinewidth}
            }
        },
        draw right/.style={
            decorate,
            decoration={
            cappedcurveto,
            raise={-.5*\the\pgflinewidth}}
        }
    }

\usepackage{colortbl}

\usepackage{array}
\newcolumntype{C}[1]{>{\centering\arraybackslash}m{#1}}

\usepackage[table]{xcolor}

\usepackage{empheq}

\usepackage{upgreek}

\usepackage{tikz}
\usetikzlibrary{arrows,shapes}
\usetikzlibrary{patterns,fadings}

\usetikzlibrary{snakes,arrows,shapes,positioning,decorations}

\tikzset{
BPSbox/.style={
       fill={red!20},
       draw={red!20},
       text width=2cm,
       inner sep=2pt,
       text centered,
       }}
\tikzset{ 
BPSbox2/.style={
       fill={green!20},
       draw={green!20}, thick,
       text width=2.2cm,
       inner sep=2pt,
       text centered,
       }
}

\tikzset{ 
BPSbox3/.style={
       fill={orange!20},
       draw={orange!20}, thick,
       text width=2.2cm,
       inner sep=2pt,
       text centered,
       }
}

\newcommand{\mc}{\mathcal}

\newcommand{\be}{\begin{equation}}
\newcommand{\ee}{\end{equation}}
\newcommand{\ba}{\begin{eqnarray}}
\newcommand{\ea}{\end{eqnarray}}
\newcommand{\nn}{{\nonumber}}

\newcommand{\beaa}{\begin{eqnarray}}
\newcommand{\eeaa}{\end{eqnarray}}

\usepackage{subcaption}

%% file: titlepage.tex
\begin{titlepage}

\centering

\includegraphics[width= 0.20\linewidth]{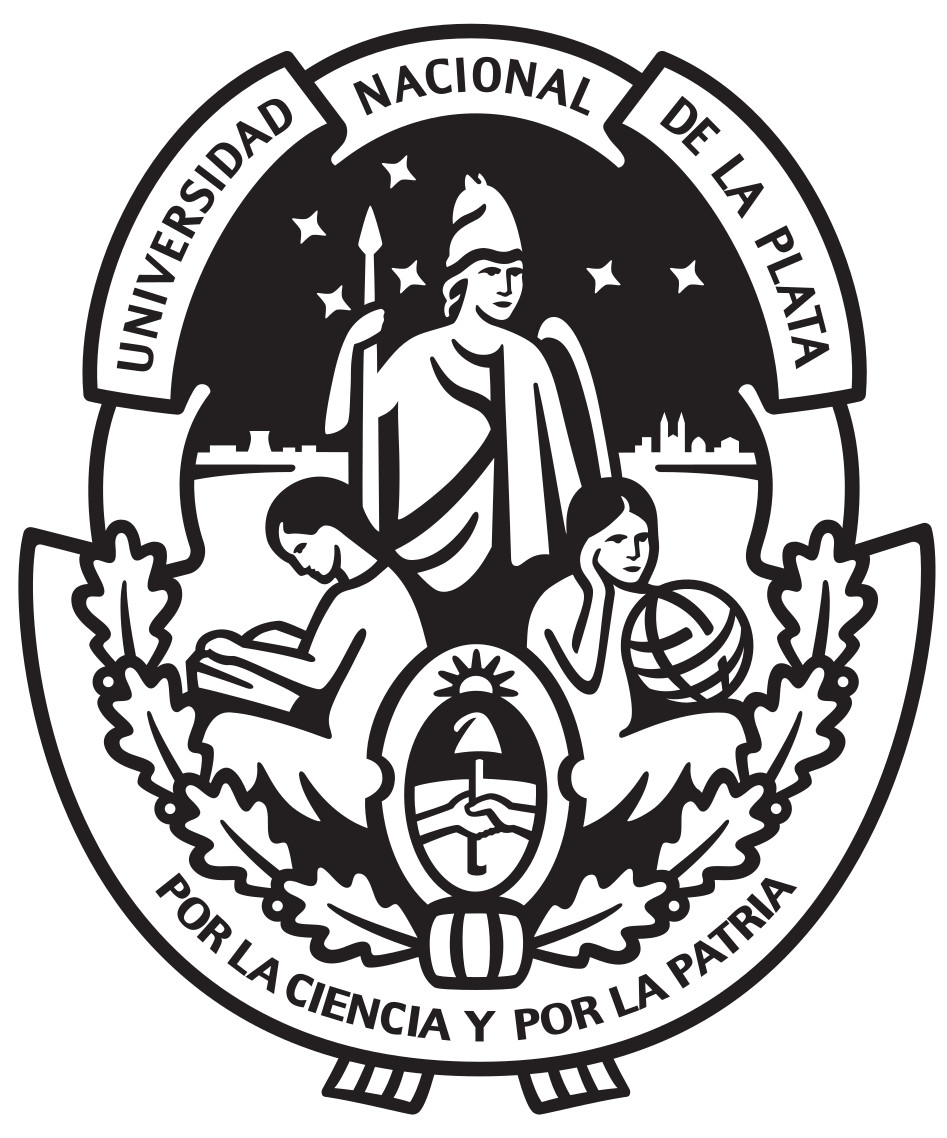}

\vspace{0.5cm}

\Large{\textbf{National University of La Plata}}

\vspace{-0.25cm}

\normalsize{\textbf{School of Exact Sciences}}

\vspace{-0.25cm}

\normalsize{\textbf{Department of Physics}}

\vspace{1cm}

\textcolor{gray}{\rule{\textwidth}{0.4pt}}

\large{ \bf{PhD Thesis:}}

\huge{\textbf{Line defects and scattering amplitudes in the context of the AdS/CFT correspondence}}
\vspace{-0.5cm}

\textcolor{gray}{\rule{\textwidth}{0.4pt}}

\vspace{0.65cm}

\vspace{1.25cm}

\Large{\textbf{Student: Martín Lagares}}

\vspace{1cm}

\large{\textbf{Supervisor: Diego Correa}} 

\vspace{2cm}

\normalsize{\textbf{Year: 2025}}

\end{titlepage}

%% file: abstract.tex
Line defects and scattering amplitudes have proven to be fruitful objects of study in the context of holographic dualities. The former serve as valuable theoretical laboratories for the development and refinement of non-perturbative methods, given the low dimensionality of the defect conformal field theories that they define and the existence of a precise holographic dual. Meanwhile, the analysis of the latter has unveiled remarkable structures that have deepened our understanding of quantum field theory, such as the positive geometry description of scattering amplitudes in theories like ${\cal N}=4$ super Yang-Mills and ABJM. 

This thesis focuses on two main goals. The first concerns the application of analytic conformal bootstrap and integrability methods to the study of superconformal line defects in AdS$_3$/CFT$_2$ and AdS$_4$/CFT$_3$ dualities. The second pertains to the analysis of infrared-finite functions, referred to as integrated negative geometries, that arise from the positive geometry description of scattering amplitudes in the ABJM theory. The topics discussed throughout this thesis are:

{\bf 1a. Analytic conformal bootstrap description of line defects in AdS$_3$/CFT$_2$}

We focus on AdS$_3$/CFT$_2$ realizations in which the supergravity background consists of an $AdS_3 \times S^3_+ \times S^3_- \times S^1$ metric with mixed Ramond-Ramond and Neveu Schwarz-Neveu Schwarz three-form fluxes. By means of a string theory analysis in the gravity side of the correspondence we find a vast family of supersymmetric line defects in the CFT$_2$, that range from 1/2 BPS to 1/8 BPS, and which are dual to strings with Dirichlet, smeared or Neumann boundary conditions. We also find a network of strings with interpolating boundary conditions that link all these defects. We then perform an analytic conformal bootstrap description of 1/2 BPS defects in the limit in which the dual metric becomes $AdS_3 \times S^3 \times T^4$. We study two-, three- and four-point functions of operators in the displacement and tilt supermultiplets defined by these defects, and the bootstrap analysis is performed up to next-to-leading order in the strong-coupling expansion. We obtain a bootstrap result that is fixed up to two coefficients and is in complete agreement with the Witten diagram expansion of the correlators, which provides a holographic interpretation of the coefficients not determined by the bootstrap procedure. Our analysis provides a successful application of the analytic conformal bootstrap program to the study of a line defect invariant under four Poincaré supercharges, thereby reducing the amount of supersymmetry compared to other cases studied in the literature.

{\bf 1b. Integrability approach to line defects in ABJM}

We study the integrability properties of the 1/2 BPS line defect of the three-dimensional ABJM theory, presenting evidence for its all-loop integrability. We propose an open spin chain that describes the anomalous dimensions of operators inserted along the defect, and we construct the corresponding all-loop reflection matrix, including the overall dressing phases. These phases, which solve a crossing symmetry equation, are shown to be consistent with weak- and strong-coupling expectations. Furthermore, we provide a $Y$-system of equations for the cusped Wilson line, which we test by reproducing the one-loop cusp anomalous dimension of the theory, finding perfect agreement with the literature. Finally, we write a set of Boundary Thermodynamic Bethe Ansatz equations compatible with the $Y$-system proposal. This analysis provides the first step towards an all-loop integrability-based computation of the bremsstrahlung function of ABJM, which would enable a derivation of the interpolating function of the theory to all orders in the coupling.

{\bf 2. Analysis of integrated negative geometries in ABJM}

We study infrared-finite functions that naturally emerge within the context of the positive geometry description of scattering amplitudes in the ABJM theory. These functions, known as integrated negative geometries, are obtained by performing $L-1$ loop integrations over the $L$ loop integrand for the logarithm of the scattering amplitude. We focus on the four-particle case, and we obtain the integrated results up to the $L=4$ order. To that end, we compute one-, two-, and three-loop integrals through either direct integration or the method of differential equations. The resulting integrated negative geometries are polylogarithmic functions of uniform transcendental weight, and follow an alternating sign pattern in the Euclidean region. We find an apparent simplicity in the leading singularities of the integrated results, provided one works in the frame in which the unintegrated loop variable goes to infinity. Finally, we propose a prescription to compute the light-like cusp anomalous dimension of the theory in terms of the integrated negative geometries, which we use to obtain the cusp anomalous dimension up to four loops. This constitutes a direct computation of the four-loop cusp anomalous dimension of ABJM, and perfectly agrees with the all-loop integrability-based prediction present in the literature.

%% file: acknowledgements.tex
As my grandfather Antonio often says, there is no \textit{I} without a \textit{You}. Without a doubt, this is no exception, and that is why I want to thank everyone who, in one way or another, was part of this journey:

To my parents, for their unconditional support. For all their effort and dedication over the years. For being an example of affection, humanity, and kindness. For constantly encourage me to enjoy every step of the way. Words fall short of doing you justice, thank you for being a true inspiration to follow.

To Diego, for being an excellent thesis advisor. Not only from an academic point of view, but above all for your human quality. For always having your door open for talks and discussions. Thank you for allowing me to work and learn a lot alongside you, and at the same time giving me the freedom and support to pursue my own projects.

To Johannes, for giving me the chance to work with you and your group during my visits to the Max Planck Institute for Physics (MPP). I learned a lot working at your side, and I truly enjoyed my time at the MPP. Thank you for your constant willingness to help and to give me opportunities to grow in my academic and professional development.

To my whole family. To my brothers Anto, Fede, and Andrés, who have been by my side since day one. I am truly lucky to have you, thank you for your unwavering support. To my nephews Nico and Sofi, who give me the best title of all, being an uncle. Without realizing it, you teach me more than I could ever teach you. I hope to be lucky enough to walk with you along whichever paths you choose. To my grandparents, and especially my grandfather Antonio, who—with tea and a slice of cake—teaches me, among many things, the value of will and the eagerness for knowledge. Thank you for your endless affection and for being a constant example to follow. To my uncles, cousins, sisters in law, and the rest of my family, thank you for always being there. And to Ramona, for your ever-present affection. Searching for letters together in the newspaper for school homework eventually paid off.

To my friends, for all the support and company. Thank you to all of you, some even since kindergarten, for always being there. Thank you for the relaxed laughs but also for the deep conversations. Thank you for every gathering, snack, dinner, and afternoon of mate.

To Gabriel, Nacho, Shun, and Victor, for collaborating with me on the projects that are part of this thesis. It was a true privilege to do science with all of you, I learned a lot from each one. Thank you for always fostering a human and relaxed work atmosphere.

To all the members of the High Energy Theoretical Physics Group from the Physics Institute of La Plata (IFLP). Thank you for creating a friendly and academically enriching work environment. Sharing the everyday with you —at each seminar, discussion, lunch, or coffee— was an amazing experience. And to the IFLP and its entire community, thank you for giving me a pleasant place to work during my PhD.

To the members of the Quantum Field Theory group at the Max Planck Institute for Physics (MPP), for warmly welcoming me during each of my visits to the MPP. It was a great pleasure to spend my time in Munich with you, both in the office and in the Biergarten. And to the MPP, for hosting me during my research stays.

To the National University of La Plata, and to the Argentine public education system in general, for allowing me to access high-quality undergraduate and graduate education free of charge. To CONICET, for granting me a scholarship to pursue my PhD. And to the DAAD, for funding my research stays in Germany through a scholarship.

%% file: intro.tex
The last century has been the scenario of outstanding progress in our knowledge of Quantum Field Theory (QFT), as a result of a synergy between theoretical and experimental developments. Canonical examples of the success of QFT as a precise model of nature are the remarkable agreement between the theoretical and experimental values for the gyromagnetic ratio of the electron \cite{Fan:2022eto}, and the prediction and subsequent observation of the Higgs boson \cite{ATLAS:2012yve}.

Nonetheless, much remains to be understood about the structure and properties of QFT. As an example, one pressing challenge concerns the analysis and development of efficient methods to probe QFTs beyond their weak-coupling regime, i.e. the study of non-perturbative techniques. Progress in this direction would provide crucial insight into the behavior of Quantum Chromodynamics (QCD) at low energies, where perturbative computations are no longer applicable. Furthermore, the extraordinary accuracy achieved by collider physics experiments calls for new and more precise ways to compute scattering amplitudes, which serve as the building blocks for the cross sections ultimately measured at the experiments. These objects often exhibit a remarkable simplicity, that is obscured in their Feynman-diagram expansions and from which much can be learned about the general properties of QFT. A notable case is the Parke-Taylor formula for $n$ gluon tree-level maximally-helicity-violating amplitudes in Yang-Mills theories \cite{Parke:1986gb}, which, despite arising from the sum of hundreds of Feynman diagrams, can be concisely written in a line.

One way to make progress towards addressing the challenging questions mentioned above is to consider appropriate toy models which, despite their simplicity compared to more realistic set ups, could help in our comprehension of the fundamental aspects of QFT. In this regard, highly symmetric theories play a crucial role, given the large number of constraints that symmetries impose on their observables. Paradigmatic examples are provided by Conformal Field Theories (CFTs), which naturally generalize Poincaré-invariant QFTs by being also invariant under dilatations and special conformal transformations. These symmetries are enlarged in the case of Superconformal Field Theories (SCFTs), which are additionally invariant under supersymmetry and thus provide an interesting framework for theoretical analyses.

Particularly useful examples of SCFTs arise in the context of holographic correspondences. These dualities, commonly referred to as AdS/CFT, conjecture the equivalence between certain CFTs in flat space and some string theories in Anti de Sitter (AdS) spacetimes. The best known example of such correspondences was proposed in a seminal paper by Maldacena \cite{Maldacena:1997re}, and is given by the duality between the four-dimensional ${\cal N}=4$ super Yang-Mills theory with gauge group $SU(N)$ and $SU(4)$ R-symmetry and type IIB string theory in $AdS_5 \times S^5$. Since then many other AdS/CFT dualities have been proposed and studied. That is the case of the conjectured equivalence between the three-dimensional ${\cal N}=6$ superconformal ABJM theory and M-theory in $AdS_4 \times S^7/\mathbb{Z}_k$ \cite{Aharony:2008ug}, which in the large $k$ limit reduces to type IIA string theory in $AdS_4 \times \mathbb{CP}^3$. Other noteworthy examples of holographic correspondences, which will be of particular interest in this thesis, are the lower-dimensional dualities known as AdS$_3$/CFT$_2$ \cite{Maldacena:1997re, Elitzur:1998mm, deBoer:1999gea, Gukov:2004ym, Tong:2014yna, Eberhardt:2017pty, Eberhardt:2018ouy, Eberhardt:2019niq, Eberhardt:2019ywk, Gaberdiel:2024dva}.

Perhaps one of the most striking features of AdS/CFT correspondences relies in the way in which they relate the coupling constants of both theories. The strong-coupling regime of the gauge theory can be accessed with perturbative computations on the string theory side of the duality (and vice versa), which has made superconformal field theories with holographic duals a unique testing ground for non-perturbative methods. As a result, numerous non-perturbative techniques have been developed and refined in the AdS/CFT framework. Over the past two decades a plethora of results have been obtained regarding the integrability of the ${\cal N}=4$ super Yang Mills and ABJM theories in their planar limit (see \cite{Beisert:2010jr} for a review), extending the analysis of integrable QFTs beyond two dimensions and enabling all-loop computations of quantities such as the cusp anomalous dimension. Moreover, many exact results have been derived using supersymmetric localization techniques (see \cite{Pestun:2007rz,Kapustin:2009kz,Drukker:2009hy, Correa:2012at,Bianchi:2018scb} for examples), and holographic SCFTs have provided a fertile ground for the development of the conformal bootstrap program (see 
for example \cite{Beem:2013qxa,Liendo:2016ymz,Beem:2016wfs, Agmon:2017xes, Liendo:2015cgi}). 

Superconformal field theories with holographic duals have also been the setting of many advances in our understanding of scattering amplitudes. The large amount of symmetries that is usually present in these SCFTs has shown to give rise to remarkable structures in their scattering amplitudes. For instance, the BCFW recursion relation that was originally proposed for tree-level amplitudes of pure Yang-Mills theories \cite{Britto:2004ap,Britto:2005fq} has been extended to the computation of all-loop integrands in the context of the ${\cal N}=4$ super Yang-Mills theory \cite{Arkani-Hamed:2010zjl}. These ideas ultimately led to the discovery of the Grassmannian structure of that theory \cite{Arkani-Hamed:2012zlh} and to the proposal of the Amplituhedron \cite{Arkani-Hamed:2013jha}. The latter offers a geometric formulation of the S-matrix, in which tree-level amplitudes and loop-order integrands are expressed in terms of the volume forms of objects known as positive geometries \cite{Arkani-Hamed:2017tmz}. This formalism, that provides a perturbative description of scattering amplitudes which is independent of the traditional Feynman diagram expansion, has been extended to many other theories. Notably, scattering amplitudes in the ABJM theory \cite{Huang:2021jlh,He:2021llb,He:2022cup,He:2023rou,Lukowski:2023nnf} and the bi-adjoint $\phi^3$ theory \cite{Arkani-Hamed:2017mur,Salvatori:2018fjp,Salvatori:2018aha} have been showed to be described by positive geometries. Interesting generalizations to the study of cosmological correlators have also been proposed \cite{Arkani-Hamed:2017fdk, Arkani-Hamed:2024jbp}.

All the considerations outlined above turn SCFTs with holographic duals into exceptional theoretical laboratories for the analysis of non-perturbative methods and the study of scattering amplitudes, which constitute the main goal of this thesis. As we will discuss in the rest of this chapter, the objectives of this thesis are twofold. On the one hand, we will explore the use of superconformal line defects as toy models for the development of non-perturbative methods. In particular, we will examine the application of analytic conformal boostrap techniques and integrability methods in the context of AdS$_3$/CFT$_2$ and AdS$_4$/CFT$_3$ realizations, respectively. On the other hand, we will study infrared-finite functions that naturally arise in the AdS/CFT context from the geometrical description of amplitudes in terms of positive geometries. 

\subsubsection*{Line defects in AdS/CFT: a tool for exploring non-perturbative methods}

Throughout this thesis, special emphasis will be placed on the application of non-perturbative methods to the description of one-dimensional conformal defects (typically denoted as dCFT$_1$) within the AdS/CFT framework. These line defects are usually defined through the insertion of operators along the contour of Wilson lines  \cite{Giombi:2017cqn}. Since the early days of AdS/CFT, Wilson lines have played a crucial role in our understanding of holographic dualities. A concrete holographic interpretation of these observables was provided by Maldacena right after the AdS/CFT proposal \cite{Maldacena:1998im}, arguing that a Wilson line is dual to a string which ends along the corresponding contour at the AdS boundary. The low dimensionality of the dCFTs defined by some Wilson lines, together with the existence of a precise and well-understood holographic dual, turn line defects in the AdS/CFT framework into remarkable models for the analysis of non-perturbative techniques. A brief discussion on the properties of Wilson lines in the context of AdS/CFT dualities is provided in Chapter \ref{ch: line defects}, along with a short review on their use for the study and development of non-perturbative techniques.

Numerous non-perturbative methods have been applied to the study of Wilson lines and their corresponding dCFTs, yielding a profusion of non-trivial tests of AdS/CFT and contributing to the understanding and development of the corresponding techniques. As an example, line defects have been the scenario of many advances in the context of the conformal bootstrap program, especially in the study of analytic bootstrap techniques. The bootstrap ideas, which aim to constrain the dynamics of the defects solely based on the use of symmetry constraints and other consistency conditions, have been successfully applied for the analytic strong-coupling computation of four-point functions along the 1/2 BPS line defects of ${\cal N}=4$ super Yang-Mills \cite{Liendo:2016ymz, Liendo:2018ukf,Ferrero:2021bsb, Ferrero:2023gnu, Ferrero:2023znz, Bonomi:2024lky} and ABJM \cite{Bianchi:2020hsz,Bliard:2023zpe}. These analyses showed that the four-point functions of the corresponding displacement supermultiplets can be completely fixed up to N$^3$LO in the strong-coupling expansion by imposing crossing symmetry, braiding symmetry, consistency of the correlator with its superconformal block expansion, and a dynamical assumption on the behavior of the anomalous dimensions of heavy operators. It is worth emphasizing that obtaining those results through explicit Witten diagram computations is currently out of reach, which represents an outstanding triumph of the bootstrap program.

An interesting extension of the above ideas will be explored in detail in Chapter \ref{ch: line defects in AdS3CFT2}, where we will study the application of analytic conformal bootstrap methods for the description of line defects in the context of AdS$_3$/CFT$_2$ dualities. Certain realizations of these correspondences provide continuous interpolations, described by a parameter $\vartheta$, between pure Ramond-Ramond (R-R) and pure Neveu Schwarz-Neveu Schwarz (NS-NS) supergravity backgrounds. The $\vartheta$ variable and the 't Hooft coupling $\lambda$ define a two-parameter bootstrap problem, which contrasts with the single-parameter analysis of the ${\cal N}=4$ sYM and ABJM cases. Furthermore, the use of bootstrap ideas is specially motivated in the AdS$_3$/CFT$_2$ framework, given the absence of a known lagrangian description of the CFT$_2$ for arbitrary values of $\vartheta$.

As we will prove in Chapter \ref{ch: line defects in AdS3CFT2}, there is a rich family of supersymmetric line defects in the AdS$_3$/CFT$_2$ context, particularly in the case in which the gravity dual is given by type IIB string theory in $AdS_3 \times S^3_+ \times S^3_- \times S^1$ with mixed Ramond-Ramond (R-R) and Neveu Schwarz-Neveu Schwarz (NS-NS) three-form fluxes. These defects will range from 1/2 BPS to 1/8 BPS, and will correspond to dual strings with either Dirichlet, smeared or Neumann boundary conditions. Interestingly, we will show that analytic conformal bootstrap ideas can be successfully applied for the description of four-point functions of 1/2 BPS defects in the limit in which the dual background becomes an $AdS_3 \times S^3 \times T^4$ space. We will center the analysis on the displacement and tilt supermultiplets that are defined by the breaking of the bulk symmetries of the CFT$_2$ because of the presence of the defect. We will prove that the bootstrap procedure fixes their four-point function at NLO in their strong-coupling expansion up to two parameters. Remarkably, we will show that the bootstrap results completely agree with the expectations coming from the dual Witten-diagram expansion of the correlators. Our results provide a successful application of analytic bootstrap methods to the description of a line defect that is invariant under only 4 Poincaré supercharges. This is a step forward in the study of the analytic conformal bootstrap program, given that the previously studied cases of the 1/2 BPS lines of ${\cal N}=4$ sYM and ABJM were respectively invariant under 8 and 6 Poincaré supercharges\footnote{Another recent example where a line invariant under four supercharges is studied with bootstrap techniques is presented in \cite{Pozzi:2024xnu}.}.

Integrability techniques constitute another instance of a non-perturbative method whose application to the study of line defects yields valuable results. As an example, insertions along the 1/2 BPS Wilson line of ${\cal N}=4$ super Yang-Mills have been proven to define an integrable dCFT \cite{Drukker:2006xg}. This has been used to derive a set of Thermodynamic Bethe Ansatz (TBA) equations for the angle-dependent cusp anomalous dimension of the theory \cite{Drukker:2012de,Correa:2012hh}, which regulates both the UV divergences of cusped Wilson lines \cite{Polyakov:1980ca,Dotsenko:1979wb,Brandt:1981kf,Korchemskaya:1992je} and the IR divergences of scattering amplitudes \cite{Alday:2009zm, Henn:2010bk}. The TBA equations for the cusped Wilson line of ${\cal N}=4$ sYM were solved to all loops in the small angle limit, allowing for an all-loop integrability derivation of the bremsstrahlung of the theory \cite{Gromov:2012eu} (i.e. the quadratic order of the cusp anomalous dimension in the small-angle limit). The notable agreement between that result and the one obtained for the same function through supersymmetric localization \cite{Correa:2012at} enabled an all-loop derivation of the interpolating function of the theory \cite{Gromov:2012eu}, which parametrizes the dispersion relation of magnons and percolates in every integrability-based result. 

A natural question that arises from the integrability analysis of line defects in ${\cal N}=4$ sYM is whether similar results can be derived in the context of the ABJM theory. An all-loop integrability computation of the bremsstrahlung function would allow for its comparison with the corresponding localization-based result \cite{Lewkowycz:2013laa,Bianchi:2014laa,Bianchi:2017svd,Bianchi:2018scb}, leading to an all-loop derivation of the ABJM interpolating function. A conjecture for such a function, which regulates the coupling dependence of every all-loop result obtained through integrability methods, was proposed in \cite{Gromov:2014eha,Cavaglia:2016ide}. A first step towards this goal will be provided in Chapter \ref{ch: integrable line defects in ABJM}, where we will study the integrability properties of the 1/2 BPS Wilson line of ABJM and we will construct a set of TBA equations for the cusped Wilson line of the theory.

The analysis of Chapter \ref{ch: integrable line defects in ABJM} will show that the anomalous dimensions of insertions along the contour of the 1/2 BPS line of ABJM are described by an open and integrable spin chain. We will compute the all-loop boundary reflection matrix for impurities that propagate over the spin chain, which we will show to be fixed by the symmetries of the problem up to an overall dressing phase. The resulting reflection matrix will satisfy the boundary Yang-Baxter equation, providing strong evidence on the integrability of the problem. We will use a crossing equation to compute the all-loop dressing phase, which we will show to be compatible with weak- and strong-coupling expectations. Furthermore, we will propose a $Y$-system of equations for the cusped Wilson line, which we will test by reproducing the one-loop cusp anomalous dimension of the theory. Finally, we will provide a set of integral TBA equations that are compatible with the $Y$-system proposal.

\subsubsection*{Infrared-finite functions from geometry}

As previously discussed, holographic dualities constitute notable frameworks for analyzing the properties of scattering amplitudes. In particular, the geometric description of amplitudes in terms of positive geometries \cite{Arkani-Hamed:2017tmz} was first introduced in the context of the ${\cal N}=4$ sYM theory \cite{Arkani-Hamed:2013jha,Arkani-Hamed:2017vfh}. This construction directly relates the tree-level amplitudes and loop integrands of the theory to the volume form of a certain manifold in momentum-twistor space, known as the Amplituhedron. A key feature of this formalism is that the $L$-loop Amplituhedron is obtained by taking $L$ copies of the corresponding one-loop geometry (which is defined by a set of inequalities involving the loop variables and the external kinematics) and then imposing positivity constraints between all the loop lines. A brief review of the geometric approach to scattering amplitudes in the ${\cal N}=4$ sYM theory, as well as its extension to the ABJM case, will be presented in Chapter \ref{ch: scatt ampl}.

As shown in \cite{Arkani-Hamed:2021iya}, the above-mentioned formulation of the S-matrix allows for a natural expansion of the logarithm of the scattering amplitude in terms of objects known as \textit{negative} geometries, which are defined by a change of sign in the constraints that relate different loop variables at loop level. Interestingly, this construction implies that performing all but one of the loop integrations over the $L$-loop integrand for the logarithm of the amplitude results in an infrared-finite object \cite{Arkani-Hamed:2021iya}, referred to as an \textit{integrated} negative geometry.

The integrated negative geometries of the ${\cal N}=4$ sYM theory have attracted considerable attention in recent years \cite{Alday:2011ga,Hernandez:2013kb,Alday:2013ip,Chicherin:2022bov,Chicherin:2022zxo,Chicherin:2024hes,Brown:2025plq,Carrolo:2025pue}. These infrared-finite objects, which are one loop integration away from the amplitude, exhibit notable properties. For instance, the $L$-loop contribution to the light-like cusp anomalous dimension $\Gamma_{\rm cusp}^{\infty}$ of the theory can be computed from the $(L-1)$-th loop order of the integrated negative geometries \cite{Alday:2013ip}, allowing one to access the $L$-loop term of $\Gamma_{\rm cusp}^{\infty}$ by performing only $L-1$ loop integrations. This approach was employed in \cite{Henn:2019swt} to compute the four-loop contribution to $\Gamma_{\rm cusp}^{\infty}$ in ${\cal N}=4$ sYM and QCD, including non-planar corrections. Furthermore, a surprising relation has been found between all-plus amplitudes in pure Yang-Mills theories and the integrated negative geometries of ${\cal N}=4$ sYM \cite{Chicherin:2022bov}. These divergence-free objects have also proven to be interesting models for developing efficient methods to obtain loop-order results without explicitly performing the loop integrations. Noteworthy examples are the recent bootstrap computation of the two-loop six-particle integrated negative geometry \cite{Carrolo:2025pue} or the all-loop summation of certain negative geometry diagrams \cite{Arkani-Hamed:2021iya}.

The above discussion naturally motivates the study of integrated negative geometries in the ABJM theory, which is the goal of Chapter \ref{ch: neg geoms ABJM}. As shown in \cite{Huang:2021jlh,He:2021llb,He:2022cup,He:2023rou}, scattering amplitudes in the ABJM theory can be described in terms of positive geometries by considering a symplectic projection of the Amplituhedron of ${\cal N}=4$ sYM. Similarly to the ${\cal N}=4$ sYM case, the logarithm of the amplitude can be expanded in terms of negative geometries \cite{He:2022cup,He:2023rou}, allowing the computation of infrared-finite integrated negative geometries. Starting from the four-point negative geometries presented in \cite{He:2022cup} up to the $L \leq 5$ loop order, we will obtain the corresponding four-point integrated negative geometries for $L \leq 4$. This will involve the computation of one-, two- and three-loop integrals, which we will perform either by direct integration or by the method of differential equations. As we will show, the integrated results will be given by polylogarithmic functions of uniform transcendentality, and will follow an interesting sign pattern. Moreover, the corresponding leading singularities will belong to the set $\{ s\sqrt{t},t\sqrt{s}, \sqrt{st (s+t)} \}$ in the limit in which the unintegrated loop variable goes to infinity. These results give useful insight for possible bootstrap computations of higher-loop integrated negative geometries. Finally, we will provide a prescription to compute the $L$-loop order of the light-like cusp anomalous dimension $\Gamma_{\rm cusp}^{\infty}$ of ABJM from the $(L-1)$-loop contribution to the integrated negative geometries. We will use this result to obtain $\Gamma_{\rm cusp}^{\infty}$ up to four loops. This constitutes the first explicit four-loop computation of $\Gamma_{\rm cusp}^{\infty}$ in ABJM, and is in perfect agreement with the all-loop integrability-based prediction given in \cite{Gromov:2008qe}.

\subsubsection*{Overview}

The rest of this thesis is organized as follows. In Chapter \ref{ch:adscft} we provide a review of the main properties of the AdS/CFT correspondence, which serves as the framework for the analysis of this thesis. Chapter \ref{ch: line defects} reviews the role of line defects in the AdS/CFT framework, with an emphasis on their application to the study and development of non-perturbative techniques. In Chapter \ref{ch: line defects in AdS3CFT2} we focus on the analysis of line defects in AdS$_3$/CFT$_2$, while in Chapter \ref{ch: integrable line defects in ABJM} we study line defects in ABJM from an integrability perspective. Chapter \ref{ch: scatt ampl} gives a brief review on the properties of scattering amplitudes in the AdS/CFT context, and provides a discussion on the positive geometry construction of scattering amplitudes in ${\cal N}=4$ super Yang-Mills and ABJM. We then move to the analysis of the integrated negative geometries of ABJM in Chapter \ref{ch: neg geoms ABJM}. Finally, we give our conclusions in Chapter \ref{ch: conclusions}. The thesis includes several appendices that complement the results of the main body of the document.

%% file: adscft.tex
As discussed in Chapter \ref{ch:intro}, holographic dualities provide exceptional testing grounds for the analysis of non-perturbative methods and for the study of scattering amplitudes. In this chapter we will introduce the main properties of the AdS/CFT correspondence. We will focus on its most well-studied realizations, which we will later study throughout this thesis. We will begin with a short review of the duality between ${\cal N}=4$ super Yang-Mills theory and type IIB string theory in $AdS_5 \times S^5$, which is the most famous example of the AdS$_5$/CFT$_4$ correspondence. Although we will not work in the AdS$_5$/CFT$_4$ framework in the context of this thesis, this duality will prove useful to introduce many key concepts.
Later we will move to a discussion about the correspondence between the ABJM theory and M-theory in $AdS_4 \times S^7/\mathbb{Z}_k$, which provides a realization of the AdS$_4$/CFT$_3$ duality. Finally, we will end with an introduction to AdS$_3$/CFT$_2$ dualities.

\section{AdS$_5$/CFT$_4$}

The best-understood example of an holographic duality is the proposed equivalence between the four-dimensional ${\cal N}=4$ super Yang-Mills (sYM) with gauge group $SU(N)$ and type IIB string theory in $AdS_5 \times S^5$. Let us begin with a description of the former. Its field content is given by a gauge field $A_{\mu}$, six real scalar fields $\phi^I$ (with $I=1, \dots, 6$) in the fundamental representation of the $SO(6)$ R-symmetry group of the theory, and a ten-dimensional Majorana-Weyl spinor $\Psi$\footnote{The ten-dimensional spinor can be decomposed into four four-dimensional Weyl spinors in the fundamental representation of the $SU(4) \simeq SO(6)$ R-symmetry.}, all of which are massless and transform in the adjoint representation of the $SU(N)$ gauge group \cite{Grimm:1977xp}. The action, which can be obtained from dimensional reduction of the ten-dimensional ${\cal N}=1$ theory \cite{Brink:1976bc,Gliozzi:1976qd,Gates:1986is}, is
\begin{align}
\label{N=4 sYM action}
S_{{\cal N}=4 \, {\rm sYM}} &= \int d^4x \, {\rm Tr} \left( -\frac{1}{4} F_{\mu \nu} F^{\mu \nu} -\frac{1}{2} D_{\mu} \Phi_I \, D^{\mu} \Phi^I + \frac{g_{sYM}^2}{4} [\Phi_I, \Phi_J][\Phi^I, \Phi^J] \right. \nonumber \\
& \qquad \qquad \quad \; \; \; \left. + \frac{i}{2} \bar{\Psi} \Gamma^{\mu} D_{\mu} \Psi + \frac{g_{sYM}}{2} \bar{\Psi} \Gamma^I [\Phi_I, \Psi] + \frac{\theta}{16 \pi^2} F_{\mu\nu} \tilde{F}^{\mu\nu} \right) \,,
\end{align}
where $g_{sYM}$ is the Yang-Mills coupling and $\theta$ is the instanton angle, $F_{\mu \nu}=\partial_{\mu} A_{\nu}-\partial_{\nu} A_{\mu}-i g_{sYM} [A_{\mu}, A_{\nu}]$, $\tilde{F}_{\mu\nu}=\frac{1}{2} \epsilon_{\mu \nu \rho \sigma} F^{\rho \sigma}$, $D_{\mu}=\partial_{\mu}-i g_{sYM} [A_{\mu}, \cdot ]$ and the $\Gamma^{I}$ are ten-dimensional Gamma matrices. 

The ${\cal N}=4$ sYM theory enjoys a $PSU(2,2|4)$ superconformal symmetry, which is preserved both at the classical and quantum levels (i.e. the theory's beta function vanishes at all loops). In addition to the $P_{\mu}$ and $M_{\mu \nu}$ generators of the four-dimensional Poincaré group (which generate the translations and Lorentz transformations, respectively), the bosonic algebra consists of the dilatation operator $D$, the $K_{\mu}$ generators of special conformal transformations, and the $R^a_b$ generators of the $SU(4)$ R-symmetry (with $a,b=1, \dots, 4$). The algebra is completed with the $Q^a_{\alpha}, \bar{Q}_a^{\dot{\alpha}}, S_a^{\alpha}$ and $\bar{S}^a_{\dot{\alpha}}$ fermionic generators, which give a total of 32 real supercharges. The (anti)commutation relations of the $\mathfrak{psu}(2,2|4)$ superalgebra can be found in \cite{Minahan:2010js}.

\subsubsection*{'t Hooft limit}

A particular limit of the theory is the \textit{'t Hooft limit}, also known as the \textit{large $N$ limit} \cite{tHooft:1973alw}. It is defined as
\begin{equation}
N \to \infty\,, \qquad g_{sYM} \to 0 \qquad \text{and} \qquad \lambda:= g_{sYM}^2 N \, \text{fixed} \,,
\end{equation}
where $\lambda$ is known as the \textit{'t Hooft coupling}. Such a limit organizes the perturbative series in powers of the rank $N$. Interestingly, all the diagrams that belong to a given order in the large $N$ expansion can be drawn over surfaces of the same topological genus $g$. More specifically, the leading large $N$ order corresponds to diagrams with genus 0, which can be drawn over the surface of a sphere. Similarly, the sub-leading large $N$ order is associated to diagrams that fit over the surface of a torus. In turn, at each order in the large $N$ expansion the diagrams can be organized in a perturbative series in the 't Hooft coupling $\lambda$. This observation made by 't Hooft in the 70s was the first clue of the existence of a duality between gauge and string theories, given the resemblance between the topological expansion of gauge theories in the large $N$ limit and the perturbative expansion of string theories with respect to the string coupling $g_s$.

\subsubsection*{The Maldacena duality}

It was until a proposal of Maldacena in the late 90s \cite{Maldacena:1997re} that the first concrete example of a duality between a gauge theory and a string theory was conjectured. His proposal was that 

\textit{The four-dimensional ${\cal N}=4$ super Yang Mills theory with gauge group $SU(N)$ and R-symmetry group $SU(4)$ is dual to type IIB string theory in $AdS_5 \times S^5$, where both the $AdS_5$ and the $S^5$ spaces have the same radius $L$ and with $\int_{S^5} F_5 = N$ five-form flux. Moreover, the couplings are related as}
\begin{equation}
\label{couplings ads cft}
g_s=g_{sYM}^2 \,, \qquad L^4=4\pi g_s N (\alpha')^2 \,, \qquad \text{and} \qquad \langle C \rangle= \theta \,,
\end{equation}
\textit{where} $\langle C \rangle$ \textit{is the vacuum expectation value of the axion field.}

Let us note that \eqref{couplings ads cft} specifies how the coupling regimes of the dual theories are related. First, from \eqref{couplings ads cft} we can read that
\begin{equation}
    g_s= \frac{\lambda}{N} \,.
\end{equation}
Therefore, the $1/N$ expansion with fixed $\lambda$ that is performed in the 't Hooft limit of the gauge theory maps to a weak-coupling expansion in the string coupling $g_s$. Furthermore, we see that
\begin{equation}
    \alpha'=\frac{L^2}{\sqrt{4 \pi \lambda}} \,,
\end{equation}
which relates the $\alpha'$ expansion of the string theory with the $\frac{1}{\sqrt{\lambda}}$ expansion in the gauge theory, i.e. the leading-order strong-coupling limit in the CFT side maps to the supergravity regime of the string theory. It is precisely this way in which the duality relates the coupling constants of both theories that makes it so useful to explore their strong coupling limits. However, this is at the same time the reason why it is so hard to test the correspondence: the perturbative regime of each of the theories must be contrasted with strong-coupling computations on the other side. Nevertheless, since its proposal the conjecture has passed many non-trivial tests, and its validity is currently widely accepted. Canonical examples of duality tests include the all-loop computation of the cusp anomalous dimension of the CFT using integrability techniques \cite{Beisert:2006ez} or the all-loop derivation of the vacuum expectation value of the 1/2 BPS circular Wilson loop using supersymmetric localization \cite{Pestun:2007rz}, both of which have matched explicit perturbative computations at both sides of the correspondence.

As expected, the duality is also manifest at the level of the symmetries of both theories. The $SO(2,4)$ conformal group of the CFT is exactly the same as the isometry group of the $AdS_5$ spacetime. Similarly, the $SO(6)$ R-symmetry of ${\cal N}=4$ sYM is mapped to the isometries of the $S^5$ sphere. As for the supersymmetries, both theories are left invariant under 32 real supercharges. Finally, both sides of the duality are invariant under an $SL(2, \mathbb{Z})$ symmetry that transforms the complex coupling
\begin{equation}
\label{tau coupling}
    \tau:= \frac{\theta}{2\pi} + \frac{4 \pi i}{g_{sYM}^2} =\langle C \rangle+ i e^{-\langle \Phi \rangle} \,,
\end{equation}
as
\begin{equation}
\tau \to \frac{a \tau + b}{c \tau + d}\,, \qquad ad-bc=1\, \qquad a,b,c,d \in \mathbb{Z} \,,
\end{equation}
where $\langle \Phi \rangle$ is the vacuum expectation value of the dilaton field\footnote{In the CFT side this symmetry is conjectural, known as the Montonen-Olive conjecture \cite{Montonen:1977sn}.  }.

\subsubsection*{GKPW prescrition}

Shortly after Maldacena's proposal, a precise prescription to relate observables at both sides of the correspondence was independently suggested by Witten \cite{Witten:1998qj} and by Gubser, Klebanov and Polyakov \cite{Gubser:1998bc}. This recipe, known as the \textit{GKPW prescription}, states that
\begin{equation}
\label{GKPW}
    Z_{\rm CFT}[J]=Z_{\rm string}[J] \,,
\end{equation}
where $Z_{\rm CFT}[J]$ is the generating function of correlation  functions of local primary operators in the CFT and $Z_{\rm string}[J]$ is the string theory partition function with $J$ playing the role of the source in the corresponding boundary conditions. In particular, in the supergravity regime of the string theory one gets
\begin{equation}
\label{GKPW 2}
    Z_{\rm CFT}[J] \sim e^{-S_{\rm SUGRA, \, on-shell}[J]} \,,
\end{equation}
where $S_{\rm SUGRA, \, on-shell}$ is the on-shell supergravity action\footnote{Note that in this last step we are performing a Wick rotation to Euclidean signature.}. Therefore, strong-coupling CFT correlators can be computed by taking functional derivatives to the exponential of the on-shell supergravity action in the string theory side. See Appendix \ref{ch: scalar and spinor fields in AdS/CFT} for a discussion on the application of the GKPW prescription to the analysis of scalar and spinor fields in the AdS/CFT framework.

\subsubsection*{The original argument}

Finally, let us end this section by giving Maldacena's original argument to motivate the conjecture \cite{Maldacena:1997re}. To that aim, let us take a set of $N$ coincident D3-branes in type IIB string theory in ten-dimensional flat space, and let us analyze the system from two different but equivalent points of view. On the one hand, we can consider the effective action for the massless modes of the string, which is obtained from integrating out all the massive degrees of freedom. Such an action can be schematically written as
\begin{equation}
\label{D3 system massless action}
    S=S_{\rm bulk}+ S_{\rm brane} + S_{\rm int} \,,
\end{equation}
where $S_{\rm bulk}$ and $S_{\rm brane}$ are the actions for the degrees of freedom that propagate on the bulk and on the brane, respectively, and $S_{\rm int}$ describes the interactions between them. At this point it is useful to take $\alpha' \to 0$ with $g_s$ and $N$ fixed, which implements a low-energy limit. At this limit the interaction terms become sub-leading, and the bulk and brane dynamics decouple. Moreover, the brane degrees of freedom become effectively described by the ${\cal N}=4$ super Yang-Mills theory, while the bulk action becomes that of ten-dimensional supergravity. Therefore,
\begin{equation}
\label{D3 system low energy massless action}
    S \xrightarrow[g_s, \, N \, {\rm fixed}]{\alpha' \to 0 } S_{\rm SUGRA, \, D=10}+ S_{{\cal N}=4 \, sYM} \,,
\end{equation}
On the other hand, we can equivalently describe the system of D3-branes in terms of a theory of closed strings propagating on a background with metric given by\footnote{We are omitting the five-form field in this analysis, for simplicity.} \cite{Horowitz:1991cd}
\begin{equation}
\label{D3 background}
ds^2= f^{-1/2} (-dt^2+dx_1^2+dx_2^2+dx_3^2)+f^{1/2}(dr^2+r^2 d\Omega_5^2) \,,
\end{equation}
with
\begin{equation}
\label{D3 background parameters}
    f=1+\frac{L^4}{r^4} \,, \qquad \text{and} \qquad L^4=4\pi g_s N (\alpha')^2 \,.
\end{equation}
In order to compare with \eqref{D3 system low energy massless action} we shall consider the low energy excitations of closed strings in the \eqref{D3 background} background, as seen from an observer at infinity. To to so, it is useful to take into account that the energy $E$ measured by an observer at infinity is related to the energy $E_r$ measured by an observer at constant $r$ by
\begin{equation}
    \label{redshift eq}
    E=f^{-1/4} E_r \,.
\end{equation}
When taking the $\alpha' \to 0$ limit (with $g_s$ and $N$ fixed) we have then two types of decoupled excitations. We have low-energy excitations propagating in the bulk, which are effectively described by ten-dimensional supergravity. But we also have the degrees of freedom that propagate in the near-horizon geometry with arbitrary $E_r$. We can see this by taking the $\alpha' \to 0$ limit in \eqref{redshift eq}, which gives
\begin{equation}
E \sim E_r \, \frac{r}{\sqrt{\alpha'}} \,.
\end{equation}
Then, for $\alpha', r \to 0$ with $r/\alpha'$ fixed we see that arbitrary excitations in the $r \to 0$ near-horizon geometry are seen as low-energy excitations from the point of view of an observer in infinity. The near-horizon geometry can be obtained from \eqref{D3 background} by taking $r << L$, giving
\begin{equation}
    \label{near-horizon geometry}
    ds^2= \frac{r^2}{L^2} \left(-dt^2 +dx_1^2 +dx_2^2 +dx_3^2 \right) + L^2 \frac{dr^2}{r^2}+L^2d\Omega_5^2 \,,
\end{equation}
which is the metric of an $AdS_5 \times S^5$ spacetime where both factors have the same radius $L$. Therefore, in the low-energy limit we get
\begin{equation}
\label{D3 low energy limit, alternative approach}
    S \xrightarrow[g_s, \, N \, {\rm fixed}]{\alpha' \to 0 } S_{\rm SUGRA, D=10} + S_{\rm Type \, IIB, \, AdS_5 \times S^5} \,.
\end{equation}
Taking into account that both in \eqref{D3 system low energy massless action} and in \eqref{D3 low energy limit, alternative approach} the first term is the action for ten-dimensional supergravity, it is natural to identify the second term in both expressions, leading to the AdS/CFT conjecture\footnote{The argument presented is not a proof of the duality, which is formally a conjecture, given that the string theory was not treated non-perturbatively. The argument was presented in the $\alpha' \to 0$ limit.}. The mapping between the coupling constants comes from \eqref{tau coupling} (which is the known relationship between the type IIB parameters and the couplings of the ${\cal N}=4$ sYM action, which describes the low-energy dynamics of the D3-branes), and from \eqref{D3 background parameters} (that parametrizes the supergravity background \eqref{D3 background}).

\section{AdS$_4$/CFT$_3$}

Another example of an holographic duality was conjectured by Aharony, Bergman, Jafferis and Maldacena in \cite{Aharony:2008ug}. In its strongest form, this correspondence relates M-theory in an $AdS_4 \times S^7/\mathbb{Z}_k$ background with a three-dimensional ${\cal N}=6$ superconformal Chern-Simmons-matter theory with $U(N) \times U(N)$ gauge group and $SU(4)$ R-symmetry, known as the \textit{ABJM theory}, providing therefore a promising framework to gain a better understanding of the properties of M-theory\footnote{See \cite{Agmon:2017xes,Chester:2018aca,Binder:2018yvd,Agmon:2019imm,Alday:2022rly,Chester:2024bij,Giombi:2023vzu,Beccaria:2023hhi, Beccaria:2023ujc,Beccaria:2023cuo,Beccaria:2024gkq,Beccaria:2025vdj,Giombi:2024itd} for recent progress in this direction.}. A weaker statement can be made by considering the large $k$ limit of the correspondence, where the gravitational dual is well approximated by type IIA string theory in $AdS_4 \times \mathbb{CP}^3$. Given the lower dimensionality (of the CFT and the AdS factor) with respect to the AdS$_5$/CFT$_4$ example discussed in the last section, this conjecture constitutes an example of the AdS$_4$/CFT$_3$ correspondence.

\begin{figure}
    \centering
    \includegraphics[width=0.75\linewidth]{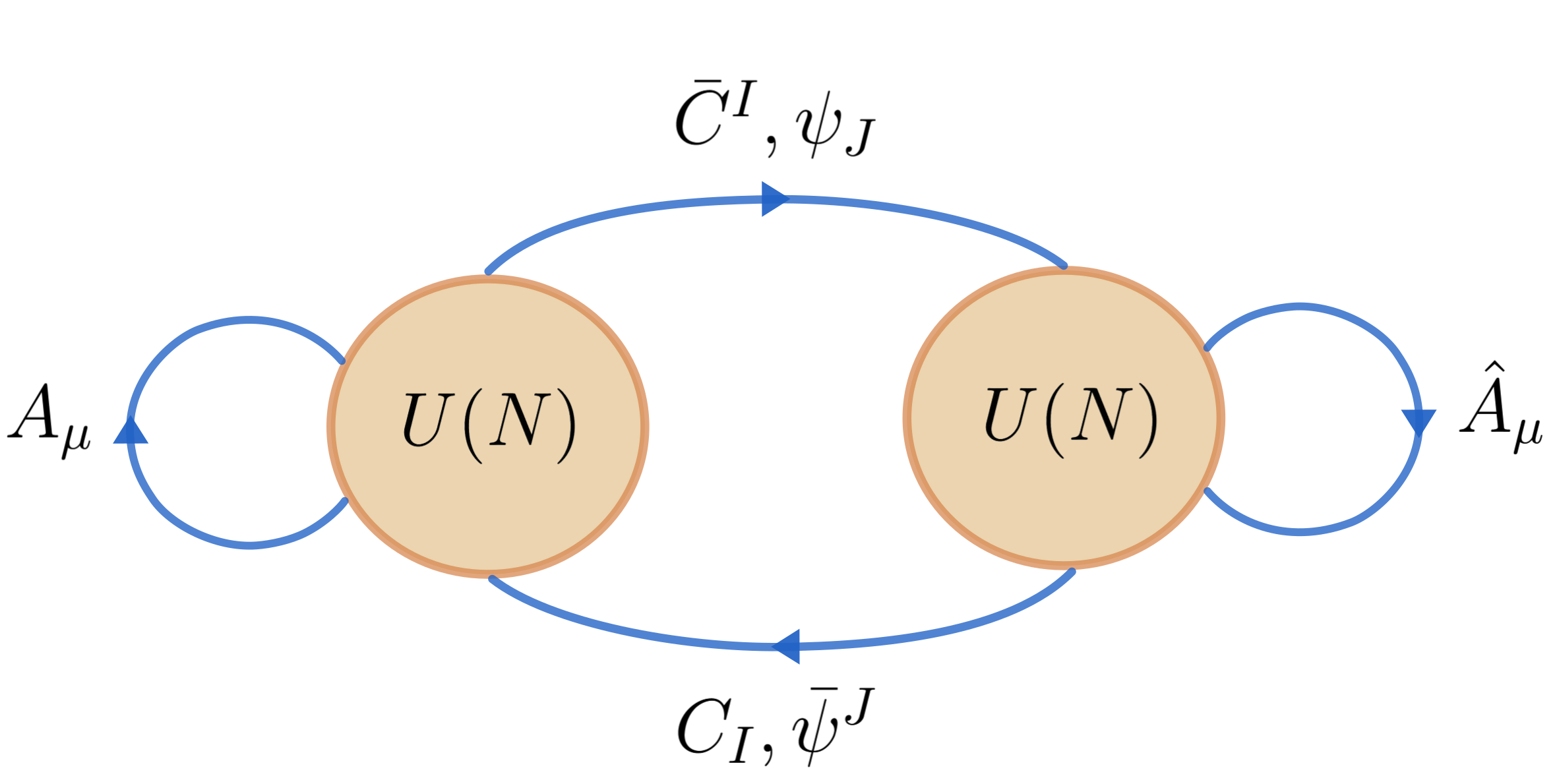}
    \caption{Quiver diagram of the ABJM theory.}
    \label{quiverABJM}
\end{figure}

\subsubsection*{The ABJM theory}

Let us begin by discussing the main properties of the ABJM theory (see \cite{Klose:2010ki} for a review). Its field content is summarized in the quiver diagram presented in Figure \ref{quiverABJM}. There are two gauge fields $\left( A_{\mu} \right)^i_j$ and $( \hat{A}_{\mu} )^{\hat{i}}_{\hat{j}}$ in the adjoint representations of the gauge groups, where we are using the notation $i,j=1, \dots, N$ and $\hat{i},\hat{j}=1, \dots, N$ for the indices of the left and right $U(N)$ factors. In addition, the theory describes the dynamics of bi-fundamental complex scalar fields $( C_I )^{\hat{i}}_j$ and $( \bar{C}^I)^{i}_{\hat{j}}$, and bi-fundamental fermions $( \psi_I )^{i}_{\hat{j}}$ and $( \bar{\psi}^I )^{\hat{i}}_{j}$, where $I=1,\dots,4$ is an $SU(4)$ R-symmetry index. The action of the theory is
\begin{equation}
\label{ABJ(M) action}
S_{\rm ABJM}=S_{\rm CS}+S_{\rm Matter}+S_{\rm int}^{\rm F}+S_{\rm int}^{\rm B} \,,
\end{equation}
with
\begin{align}
S_{\rm CS} &= - \frac{ik}{4\pi} \, \int d^3x \, \epsilon^{\mu\nu\rho} \left[ {\rm Tr} \left( A_{\mu} \partial_{\nu} A_{\rho}+ \frac{2}{3} \, A_{\mu} A_{\nu} A_{\rho} \right)- {\rm Tr} \left( \hat{A}_{\mu} \partial_{\nu} \hat{A}_{\rho}+ \frac{2}{3} \, \hat{A}_{\mu} \hat{A}_{\nu} \hat{A}_{\rho} \right) \right] \,, \nonumber \\
S_{\rm Matter} &= \frac{k}{2\pi} \int d^3x \, \left[ {\rm Tr} \left( D_{\mu} C_I D^{\mu} \bar{C}^I \right) +i {\rm Tr} \left( \bar{\psi}^I \slashed{D} \psi_I  \right)  \right] \,, \nonumber\\
S_{\rm int}^{\rm F} &= - \frac{i k}{2\pi} \,  \int d^3x \, \left[ {\rm Tr} \left( \bar{C}^I C_I \psi_J \bar{\psi}^J \right)- {\rm Tr}\left( C_I \bar{C}^I  \bar{\psi}^J \psi_J \right) + 2 {\rm Tr} \left(C_I \bar{C}^J \bar{\psi}^I \psi_J \right)  \right. \nonumber \\
& \left. \qquad \qquad \qquad \quad  -  2 {\rm Tr} \left(\bar{C}^J C_I \psi_J \bar{\psi}^I \right) - \epsilon_{IJKL} {\rm Tr} \left( \bar{C}^I \bar{\psi}^J \bar{C}^K \bar{\psi}^L \right)  \right. \nonumber \\
& \left. \qquad \qquad \qquad \quad + \epsilon^{IJKL} {\rm Tr} \left( C_I \psi_J C_K \psi_L \right)  \right] \,, \nonumber\\
S_{\rm int}^{\rm B} &= - \frac{k}{6\pi} \, \int d^3x \, \left[ {\rm Tr} \left( C_I \bar{C}^I C_J \bar{C}^J C_K \bar{C}^K  \right) + {\rm Tr} \left( \bar{C}^I C_I \bar{C}^J C_J \bar{C}^K C_K  \right) + \right. \nonumber \\
& \left. \qquad \qquad \qquad \quad + 4 {\rm Tr} \left( C_I \bar{C}^J C_K \bar{C}^I C_J \bar{C}^K  \right)-6 {\rm Tr} \left( C_I \bar{C}^J C_J \bar{C}^I C_K \bar{C}^K  \right) \right] \,, \nonumber
\end{align}
where $\epsilon^{1234}=\epsilon_{1234}=1$, $k$ is the Chern-Simons level\footnote{The Chern-Simons level takes integer values in order to preserve the gauge invariance.
}, 
and the covariant derivatives are defined as
\begin{align}
\label{cov der-1} 
D_{\mu} C_I &= \partial_{\mu} C_I +i \left( A_{\mu} C_I -C_I \hat{A}_{\mu} \right) \,, \qquad 
D_{\mu} \bar{C}^I = \partial_{\mu} \bar{C}^I -i \left( \bar{C}^I A_{\mu} -\hat{A}_{\mu} \bar{C}^I \right) \,, \\
D_{\mu} \psi_I &= \partial_{\mu} \psi_I +i \left( \hat{A}_{\mu} \psi_I -\psi_I A_{\mu} \right) \,, \qquad \;
D_{\mu} \bar{\psi}^I = \partial_{\mu} \bar{\psi}^I 
-i \left( \bar{\psi}^I \hat{A}_{\mu} -A_{\mu} \bar{\psi}^I \right) \,.
\end{align}

The theory is left invariant under an $OSP(6|4)$ supergroup, which parametrizes 24 real supercharges and whose bosonic part is given by the $SO(2,3) \times SO(6)$ product of the three-dimensional conformal group and the $SO(6)$ R-symmetries of the theory. The corresponding supersymmetry transformations of the fields are
\begin{align}
\label{susy ABJ(M) 1}
\delta A_{\mu} &= 2 i \, \bar{\Theta}^{IJ \alpha} \left( \gamma_{\mu} \right)_{\alpha}^{\beta} \left( C_I \psi_{J \beta} + \tfrac{1}{2} \epsilon_{IJKL} \bar{\psi}^K_{\beta} \bar{C}^L \right) \,, \\
\delta \hat{A}_{\mu} &= 2 i \, \bar{\Theta}^{IJ \alpha} \left( \gamma_{\mu} \right)_{\alpha}^{\beta} \left( \psi_{J \beta} C_I + \tfrac{1}{2} \epsilon_{IJKL} \bar{C}^L \bar{\psi}^K_{\beta} \right) \,, \nonumber \\
\delta C_K &= \bar{\Theta}^{IJ \alpha} \epsilon_{IJKL} \bar{\psi}^L_{\alpha} \,, \nonumber \\
\delta \bar{C}^K &= 2 \, \bar{\Theta}^{KL \alpha} \, \psi_{L \alpha} \,, \nonumber \\
\delta \psi^{\beta}_K &= -i \bar{\Theta}^{IJ \alpha} \epsilon_{IJKL} \left( \gamma_{\mu} \right)_{\alpha}^{\beta} D_{\mu} \bar{C}^L + i \,  \bar{\Theta}^{IJ\beta} \epsilon_{IJKL} \left( \bar{C}^L C_P \bar{C}^P - \bar{C}^P C_P  \bar{C}^L \right) + \nonumber \\
& \quad \, + 2i \, \bar{\Theta}^{IJ \beta} \epsilon_{IJML} \bar{C}^M C_K \bar{C}^L \,, \nonumber \\
\label{susy ABJ(M) 6}
\delta \bar{\psi}_{\beta}^K &= -2i \,\bar{\Theta}^{KL \alpha} \left( \gamma_{\mu} \right)_{\alpha \beta} D_{\mu} C_L - 2i \, \bar{\Theta}^{KL}_{\beta} \left( C_L \bar{C}^M C_M - C_M \bar{C}^M  C_L \right) - 4 i \bar{\Theta}^{IJ}_{\beta} C_I \bar{C}^K C_J \,, \nonumber
\end{align}
where the Killing spinors $\bar{\Theta}^{IJ}$ are anti-symmetric in the R-symmetry indices ($\bar{\Theta}^{IJ} =-\bar{\Theta}^{JI}$) and satisfy the reality condition
\begin{equation}
\bar{\Theta}^{IJ}=\left( \Theta_{IJ} \right)^* \,, \qquad \text{with} \qquad  \Theta_{IJ} =\frac{1}{2} \epsilon_{IJKL} \bar{\Theta}^{KL} \,.
\end{equation}

As for the ${\cal N}=4$ super Yang-Mills theory, it is also possible to define a 't Hooft limit in the ABJM theory. In this case the large $N$ limit consists of taking
\begin{equation}
\label{ABJM 't Hooft limit}
    N, k \to \infty \,, \qquad \text{with} \qquad \lambda:= \frac{N}{k} \, \text{fixed} \,.
\end{equation}

\subsubsection*{The duality between ABJM and M-theory}

Analogously to the argument used to motivate the duality between ${\cal N}=4$ sYM and type IIB strings in $AdS_5 \times S^5$, the correspondence between ABJM and M-theory in $AdS_4 \times S^7/\mathbb{Z}_k$ can be argued by considering a stack of $N$ M2-branes probing a $\mathbb{C}^4/\mathbb{Z}_k$ singularity. The statement of the conjecture is that \cite{Aharony:2008ug}

\textit{The ABJM theory with gauge group $U(N) \times U(N)$ and Chern-Simons level $k$ is dual to M-theory on $AdS_4 \times S^7/\mathbb{Z}_k$ with metric and four-form field strength given by
\begin{align}
    ds^2 &=L^2 \left( \frac{1}{4}ds^2_{AdS_4}+ds^2_{S^7/\mathbb{Z}_k} \right) \,, \\
    ds^2_{S^7/\mathbb{Z}_k} &= ds^2_{\mathbb{CP}^3}+ \left(\frac{d\xi}{4} +  A \right)^2\,, \quad \xi \equiv \xi + \frac{8\pi}{k} \,,  \\
    A&=\frac{1}{2i} (\bar{z}_a dz^a-z^a d\bar{z}_a)\,, \quad \bar{z}_a z^a=1 \,, \quad a=1,\dots,4 \,, \\
    F_4 &= dC_3= -\frac{3}{8} L^3 {\rm vol}_{AdS_4} \,,
\end{align}
and where
\begin{equation}
    \frac{L^6}{\alpha'^3}= 32 \pi^2 N \, k \,.
\end{equation}
In the 't Hooft limit \eqref{ABJM 't Hooft limit} the gravitational dual reduces to type IIA string theory in $AdS_4 \times \mathbb{CP}^3$. In units such that the string scale is set to the unity, the background fields are
\begin{align}
    ds^2 &=\frac{L^3}{k} \left( \frac{1}{4}ds^2_{AdS_4}+ds^2_{\mathbb{CP}^3} \right) \,,   \qquad     &e^{2\Phi}&= \frac{L^3}{k^3} \,, \nonumber \\
    F_4 &= \frac{3}{8} L^3 {\rm vol}_{AdS_4}\,, \qquad &F_2 &= k dA \,.
    \label{ads4cft3 other background fields}
\end{align}
}

\subsubsection*{ABJ theory}

It is also interesting to consider a generalization of ABJM, known as the \textit{ABJ theory} \cite{Aharony:2008gk}. In this case the two factors of the gauge group are not necessarily taken to have equal rank, i.e. one considers a $U(N) \times U(M)$ gauge symmetry with Chern-Simons levels $k$ and $-k$. Both the action and the supersymmetry transformations remain the same as for the $N=M$ case, and now the 't Hooft limit is defined as 
\begin{equation}
    N, M, k \to \infty \,, \qquad \text{with} \qquad \lambda:= \frac{N}{k} \, \text{and } \hat{\lambda}:= \frac{M}{k} \text{ fixed} \,.
\end{equation}
The corresponding holographic dual is only modified by the addition of a flat Kalb-Ramond field with a non-trivial holonomy along a $\mathbb{CP}^1 \subset \mathbb{CP}^3$, which is taken to be proportional to $N-M$.

\section{AdS$_3$/CFT$_2$}

The AdS/CFT correspondences discussed in the previous sections provide concrete realizations of holographic dualities in which the CFT is precisely known for all values of the couplings and parameters of the gravity dual. In this thesis we will also be interested in a rich family of lower-dimensional dualities, known as AdS$_3$/CFT$_2$ correspondences, for which the CFT side is much less understood. Despite this lesser knowledge, these dualities have attracted a lot of attention since the early days of AdS/CFT. Much can be learn about the properties of holographic correspondences from their study, as seen for example in the many integrability developments that have taken place in this framework over the past years (see \cite{Sfondrini:2014via, Seibold:2024qkh} reviews).

\subsubsection*{AdS$_3$/CFT$_2$ dualities with $AdS_3 \times S_+^3 \times S_-^3 \times S^1$ and $AdS_3 \times S^3 \times T^4$ backgrounds}

We will focus on AdS$_3$/CFT$_2$ dualities in which the gravity side of the correspondence is described by type IIB string theory in an $AdS_3 \times S_+^3 \times S_-^3 \times S^1$ background  (we use $+$ and $-$ indices to distinguish the two $S^3$ three-spheres, which are allowed to have different radius) and with mixed Ramond-Ramond (R-R) and Neveu Schwarz-Neveu Schwarz (NS-NS) three-form fluxes. The latter is one of the properties that make these dualities so appealing in contrast to other AdS/CFT realizations, given that they provide a controlled interpolation between pure R-R and pure NS-NS backgrounds. The literature on these AdS$_3$/CFT$_2$ realizations is vast, see for example \cite{Maldacena:1997re, Elitzur:1998mm, deBoer:1999gea, Gukov:2004ym, Tong:2014yna, Eberhardt:2017pty, Eberhardt:2018ouy, Eberhardt:2019niq, Eberhardt:2019ywk, Gaberdiel:2024dva,Sfondrini:2014via, Seibold:2024qkh} and references therein. More precisely, we will take the metric to be
\begin{equation}
 ds^2 =L^2 ds^2(AdS_3)+\frac{L^2}{\sin^2\Omega} ds^2(S^3_+)+ \frac{L^2}{\cos^2\Omega} ds^2(S^3_-)+l^2d\theta^2\,. 
\end{equation}
where the parameter $\Omega$, that calibrates the radii of the three-spheres, takes values in the range $0<\Omega<\frac{\pi}{2}$. This metric solves the type IIB supergravity equations of motion along with a constant dilaton $\Phi$ and the R-R and NS-NS three-form field strengths, which are given by
\begin{align}
F_{(3)} & = d C_{(2)} =  -2e^{-\Phi}L^2 \cos\vartheta
\left({\rm vol}(AdS_3)+\frac{1}{\sin^2\Omega} {\rm vol}(S^3_+)+ \frac{1}{\cos^2\Omega} {\rm vol}(S^3_-) \right)\,, \nonumber
\\
H_{(3)} & = d B_{(2)} = 2L^2 \sin\vartheta
\left({\rm vol}(AdS_3)+\frac{1}{\sin^2\Omega} {\rm vol}(S^3_+)+ \frac{1}{\cos^2\Omega} {\rm vol}(S^3_-) \right)\,.
\end{align}
The parameter $\vartheta$, taking values in the range $0\leq \vartheta \leq \frac{\pi}{2}$, interpolates between the pure R-R and the pure NS-NS backgrounds. 

The isometries of the $AdS_3 \times S_+^3 \times S_-^3 \times S^1$ background form a $D(2,1, \sin^2 \Omega)^2 \times U(1)$ supergroup. Each $D(2,1, \sin^2 \Omega)$ factor is an exceptional supergroup with 8 real supercharges (i.e. the background is 1/2 BPS, with a total of 16 real supercharges) and whose bosonic part is $SU(1,1) \times SU(2)_+ \times SU(2)_-$. 
The $D(2,1, \sin^2 \Omega)^2 \times U(1)$ supergroup of the  $AdS_3 \times S_+^3 \times S_-^3 \times S^1$ backgrounds is usually referred to as the \textit{large} two-dimensional ${\cal N}=(4,4)$ superconformal algebra.

We will also be particularly interested in the $\Omega \to \frac{\pi}{2}$ limit (or, equivalently, $\Omega \to 0$) of the above backgrounds, where the metric reduces to an $AdS_3 \times S^3 \times T^4$ geometry. The corresponding isometries become $PSU(1,1|2)^2 \times SO(4)_A \times U(1)_A$, where the $SO(4)_A \times U(1)_A$ factors are automorphisms of the supergroup. In this case, the isometry group is commonly known as the \textit{small} two-dimensional ${\cal N}=(4,4)$ superconformal algebra.

\subsubsection*{Towards the dual CFT}

As previously discussed, there is not a well understood holographic CFT for all values of the $\Omega$ and $\vartheta$ interpolating parameters. However, certain limits of the parameter space have received much attention over the past decade. Such is the case of the pure NS-NS background with $AdS_3 \times S^3 \times T^4$ metric and one unit of NS-NS three-form flux. In this limit the string theory was shown to be dual to the symmetric orbifold CFT on $T^4$, which is usually written as ${\rm Sym}^N(T^4)$ \cite{Eberhardt:2018ouy, Eberhardt:2019ywk}. It was shown that the spectrum and correlation functions of such orbifold CFT precisely match the ones of the worldsheet theory, which in this limit can be described with a Wess-Zumino-Witten (WZW) model. Similar developments have been recently achieved in the pure NS-NS case with $AdS_3 \times S^3_+ \times S^3_- \times S^1$ metric and with one unit of flux through the AdS$_3$ and two units through each of the three-spheres \cite{ Gaberdiel:2024dva}.

As in higher-dimensional dualities, the gravitational backgrounds in AdS$_3$/CFT$_2$ correspondences can be obtained as near-horizon limits of certain brane configurations, providing valuable information about the dual CFTs. To illustrate these ideas, let us consider the case of type IIB strings in $AdS_3\times S^3 \times T^4$ with pure R-R three-form flux. Such backgrounds can be obtained from the near-horizon limit of a system of intersecting D5- and D1-branes, known as the \textit{D1-D5 system} \cite{Maldacena:1997re}. The world-volume theory in the intersection of the D1-D5 system describes a theory of two-dimensional vector and hypermultiplets. Depending on which fields acquire non-zero vacuum expectation values (vev), one can fall into the theory's Higgs or Coulomb branches: in the Higgs branch only the hypermultiplets have non-vanishing vev, while the Coulomb branch is characterized by vector multiplets with non-trivial vev. Crucially, only the Higgs branch is invariant under the small ${\cal N}=(4,4)$ superconformal symmetry, which is precisely the isometry group of the $AdS_3\times S^3 \times T^4$ background. Therefore, it is conjectured that the IR limit of that Higgs branch theory is dual to type IIB string theory in $AdS_3\times S^3 \times T^4$ with pure R-R three-form flux \cite{Maldacena:1997re}.

Let us note that it is not an easy task to perform the RG flow to the IR limit of the Higgs branch. Therefore, there is currently no available lagrangian description of the corresponding IR fixed point, constraining the study of its duality with type IIB string theory in $AdS_3\times S^3 \times T^4$. The same obstacle appears in most of the known AdS$_3$/CFT$_2$ realizations, which motivates the development of bootstrap methods to study these dualities without the use of the CFT lagrangian. This is one of the objectives of this thesis, as we will discuss in Chapter \ref{ch: line defects in AdS3CFT2}.

%% file: introdefects.tex
Throughout this thesis we will be interested in the application and development of non-perturbative methods in the context of line defects, with focus on their analysis within the framework of the AdS/CFT correspondence. Therefore, in this chapter we will shortly review the main properties of Wilson lines in the context of holographic dualities, with an emphasis on the line defects that they define. We will begin with an analysis of Wilson lines in the ${\cal N}=4$ super Yang-Mills theory. We will then move on to a discussion about Wilson lines in the three-dimensional ABJM theory, and finally we will end with a brief review on the application of non-perturbative methods to the study of line defects in the AdS/CFT framework.

\section{Wilson lines in ${\cal N}=4$ super Yang-Mills}
\label{sec: WL N=4}

Given a closed contour $\mathcal{C}$\footnote{One can also consider open contours, at the cost of losing the gauge invariance of the operator.} and a gauge connection $A_{\mu}$, a Wilson loop can be defined in the standard way as the holonomy of $A_{\mu}$ along $\mathcal{C}$, i.e.
\begin{equation}
\label{standard WL}
W_{\rm stand}= \frac{1}{N} {\rm Tr} \left[ {\cal P} \exp \left( i \oint_{\cal C} A_{\mu} dx^{\mu} \right) \right] \,,
\end{equation}
where ${\cal P}$ stands for path ordering. Although gauge invariant, in the ${\cal N}=4$ super Yang-Mills the above definition is not invariant under supersymmetry. This can be overcomed by modifying the connection that it is used in the definition of the Wilson loop. As shown by Maldacena \cite{Maldacena:1998im}, one can construct supersymmetric Wilson loops by generalizing the definition \eqref{standard WL} to
\begin{equation}
\label{Wilson-Maldacena line}
W_{\rm M}= \frac{1}{N} {\rm Tr} \left[ {\cal P} \exp \left( i \oint_{\cal C} d\tau \left( A_{\mu} \dot{x}^{\mu}+ \,|\dot{x}| \,n_I (\tau) \, \phi^I \right) \right) \right] \,.
\end{equation}
where $n^I(\tau)$ has unitary norm. For generic contours and unitary $n^I$ couplings the operator defined in \eqref{Wilson-Maldacena line} is locally supersymmetric \cite{Maldacena:1998im,Zarembo:2002an}. Crucially, one can construct operators with global supersymmetry by considering certain combinations of contours and $n^I$ couplings. As an example, one obtains a 1/2 BPS operator by taking the contour to be a circumference and the $n_I$ vectors to be constant along the contour. The same result can be derived for an infinite straight line with constant coupling to the scalars. The observable introduced in \eqref{Wilson-Maldacena line} is sometimes referred to as \textit{Wilson-Maldacena loop} to distinguish it from \eqref{standard WL}. Note that the operator defined in \eqref{Wilson-Maldacena line} can be obtained by dimensional reduction of a standard Wilson line in the ten-dimensional ${\cal N}=1$ super Yang-Mills theory.

\subsubsection*{Symmetries}

Let us focus on the case in which \eqref{Wilson-Maldacena line} is defined along an infinite and straight line, with constant $n_I$ couplings. In that case the bosonic symmetries of the Wilson line are $SO(2,1) \times SO(3) \times SO(5)$, where $SO(2,1) \simeq SU(1,1)$ is the one-dimensional conformal group, $SO(3)$ stands for the transversal rotations around the line, and $SO(5)$ is the subgroup of the R-symmetries that are preserved by the operator. In addition, there are 16 real supercharges that leave invariant the Wilson line (as mentioned above, the operator is 1/2 BPS), and therefore the full set of symmetries of the line are given by an $OSp(2,2|4)$ group\footnote{This supergroup is often denoted as $OSp(4^*|4)$ in the literature.}. 

It is interesting to note that 1/4 BPS, 1/8 BPS and 1/16 BPS Wilson lines can be obtained by considering contours of different dimensionalities (e.g. 1/4 BPS loops can be constructed with contours that lie within a plane), provided that one chooses an appropiate non-constant form for the $n_I$ couplings \cite{Zarembo:2002an}.

\subsubsection*{The dual string}

\begin{figure}
     \centering
        \includegraphics[scale=0.20]{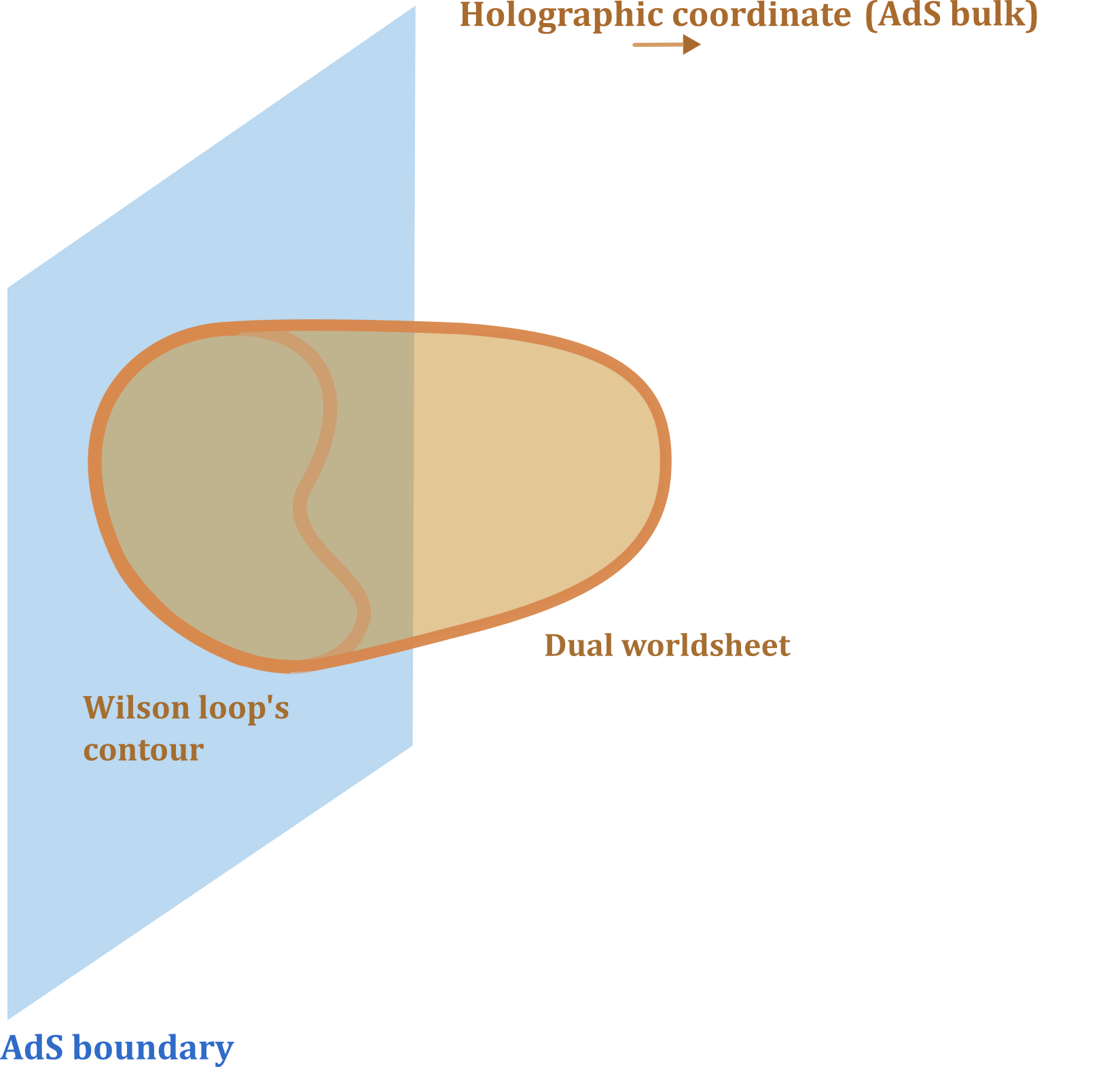}
     \caption{Schematic representation of the duality between Wilson lines and open strings in the AdS/CFT correspondence.}
     \label{fig: holographic WL}
\end{figure}

As previously mentioned, Wilson loops have a precise string dual in the AdS/CFT context. It was proposed \cite{Maldacena:1998im} that 
\begin{equation}
\label{WL duality}
    \langle W_{M} \rangle = Z_{\rm string}
\end{equation}
for a Wilson-Maldacena line with vacuum expectation value $\langle W_{M} \rangle$ and where $Z_{\rm string}$ is the partition function of a string that ends at the contour $\mathcal{C}$ at the boundary of $AdS_5$ and at the $n_I(\tau)$ curve in $S^5$. See Figure \ref{fig: holographic WL} for a schematic representation of the duality. In the supergravity limit (i.e. when $\alpha' \to 0$) the r.h.s. in \eqref{WL duality} is dominated by the minimal-area solution for a string with the aforementioned boundary conditions, i.e. 
\begin{equation}
    \langle W_{M} \rangle \sim  e^{-T A_{\rm min}}
\end{equation}
where $T$ is the string tension and $A_{\rm min}$ is the minimal string area. Beyond the leading-order supergravity limit, the r.h.s. can be studied as an $\alpha'$ expansion of $Z_{\rm string}$ in terms of quantum fluctuations around the classical minimal-area solution.

As an example, let us consider a 1/2 BPS Wilson-Maldacena line describing the straight contour $x^{\mu}=(\tau,0,0,0)$ and with constant $n_I$ couplings. We will use Poincaré coordinates
\begin{equation}
ds^2(AdS_5) =\frac{1}{z^2}\left(-dt^2+ \sum_{i=1}^3 dx_i^2 +dz^2\right)    
\end{equation}
for AdS$_5$\footnote{We are setting the AdS radius to one.} and embedding coordinates $Y^I, I=1, \dots, 6$ with $Y^I Y_I=1$ for the $S^5$ factor. The minimal area solution is then parametrized by
\begin{equation}
\label{straight 1/2 BPS WL N=4}
    t=\tau\, , \quad x^i=0 \,, \quad z=\sigma, \quad Y^I=n^I \,,
\end{equation}
with $-\infty<\tau<\infty$ and $0<\sigma<\infty$. The induced metric on the worldsheet describes an AdS$_2$ space, whose isometries are in correspondence with the conformal symmetry of the dual Wilson line.

\subsubsection*{Wilson lines as conformal defects}

Wilson lines that are conformally invariant can be used to define \textit{one-dimensional defect conformal field theories}, usually referred to as dCFT$_1$. More specifically, correlators in these defect theories are defined in terms of insertions along the contour of the Wilson line as
\begin{equation}
\langle \langle \mathcal{O}_1 (\tau_1) \dots \mathcal{O}_n (\tau_n) \rangle \rangle := \frac{ \langle {\rm Tr} \left[{\cal P} \exp \left( i \oint_{\cal C} d\tau \, {\cal L} \,  \right) \, \mathcal{O}_1 (\tau_1) \dots \, \mathcal{O}_n (\tau_n) \right] \rangle}{\langle {\rm Tr} \left[{\cal P} \exp \left( i \oint_{\cal C} d\tau \, {\cal L} \,  \right) \right] \rangle} \,,
\label{dCFTcorr}
\end{equation}
where the double-bracket is used to explicitly refer to correlators of the dCFT$_1$, and where ${\cal L}$ is the (super)connection of the Wilson line. We will henceforth denote dCFT$_1$ correlators with single brackets, omitting the double-bracket notation.

In the context of defect CFTs one can always define a protected operator that naturally arises from the breaking of the bulk spacetime symmetries because of the presence of the defect \cite{Billo:2013jda,Gaiotto:2013nva,Billo:2016cpy}. More concretely, let us consider a line defect that extends along the contour
\begin{equation}
    x^{\mu}(\tau)=(\tau, x_0^i) \,,
\end{equation}
in a $D$-dimensional theory. The conservation equation for the stress-energy tensor is then
\begin{equation}
\label{displacement def}
\partial_{\mu} T^{\mu i} (t,x^j)= \delta^{(D-1)} (x^j-x_0^j) \, \mathbb{D}^i (t) \,,
\end{equation}
where the source term defines the $\mathbb{D}^i$ \textit{displacement operator}, with $i=1, \dots, D-1$. Note that from \eqref{displacement def} one can read that $\mathbb{D}^i$ is a protected operator with scaling dimension $\Delta_{\mathbb D}=2$. In order to get explicit expressions for the $\mathbb{D}^i$ operator it is useful to consider the integrated version of the Ward identity presented in \eqref{displacement def}, which reads
\begin{equation}
\label{integrated displacement def}
\langle X\rangle_{\delta_{\epsilon} W}= \int d\tau \, \langle {\mathbb D}^i (\tau) \, X\rangle_{W} \, \epsilon + \mathcal{O}(\epsilon^2) \,,
\end{equation}
where $X$ is an arbitrary insertion on the defect, $\langle \dots \rangle_{\delta_{\epsilon} W}$ implies that the insertion is along a line slightly translated in the $i$-th direction as $x_0^i \to x_0^i+\epsilon$, and $\langle \dots \rangle_{W}$ denotes an insertion on the undeformed Wilson line. Similarly, the breaking of other bulk symmetries because of the presence of the line introduces other protected operators in the dCFT.

As an example, let us discuss the case of a Wilson-Maldacena line in ${\cal N}=4$ super Yang-Mills that extends along the time direction and which couples only to the $\Phi^6$ scalar. In that case, one can show that the displacement operator is $\mathbb{D}_i:=i F_{ti} + D_i \Phi^6$, with $i=1,2,3$ \cite{Giombi:2017cqn}. Moreover, this defect also breaks the R-symmetries of the bulk as $SO(6) \to SO(5)$, which introduces a set of five scalar protected operators $\mathbb{G}^a:=\Phi^a$ with $a=1, \dots, 5$. The latter transform in the fundamental representation of $SO(5)$ \cite{Giombi:2017cqn} and have scaling dimension $\Delta_{\mathbb G}=1$. The displacement $\mathbb{D}_i$, the $\mathbb{G}^a$ scalars, and the eight protected operators of dimension $\Delta_{\mathbb F}=3/2$ that arise from the breaking of the supersymmetries transform within a short supermultiplet of the $OSp(4^*|4)$ symmetry group of the defect, which is known as the \textit{displacement supermultiplet}. 

The fields in the displacement supermultiplet have a concrete holographic interpretation \cite{Sakaguchi:2007ba,Faraggi:2011bb,Fiol:2013iaa,Giombi:2017cqn}. To discuss it, let us again consider the 1/2 BPS straight line with coupling to the $\Phi^6$ scalar, whose dual string is described by the classical solution written in \eqref{straight 1/2 BPS WL N=4} (with $n^I=\delta^I_6$). When expanding the string action around the classical solution, one gets a spectrum of fluctuations that consists of five massless scalars in the fundamental representation of $SO(5)$, three scalars with $m^2=2$ (in units where the AdS radius is set to one) in the fundamental representation of $SO(3)$, and eight fermions with $m=1$. At this point it is useful to recall the relation
\begin{equation}
    \Delta_+= \frac{1}{2} + \sqrt{\frac{1}{4}+m_B^2} \,,
\end{equation}
between a scalar field of mass $m_B$ and regular boundary conditions in EAdS$_2$ and its dual operator of scaling dimension $\Delta_+$ (see \eqref{delta pm scalar} in Appendix \ref{ch: scalar and spinor fields in AdS/CFT}). In turn, a spinor field of mass $m_F$ with appropriate boundary conditions (see Appendix \ref{ch: scalar and spinor fields in AdS/CFT}) is dual to an operator of scaling dimension
\begin{equation}
    \Delta_+^{\psi}= \frac{1}{2} + m_F \,.
\end{equation}
The above relations imply that the quantum fluctuations map to a set of five scalar fields with $\Delta=1$, eight spinors with $\Delta=3/2$ and three scalars with $\Delta=2$. Therefore, it is natural to identify the quantum fluctuations around the classical dual string with the operators in the displacement supermultiplet \cite{Sakaguchi:2007ba,Faraggi:2011bb,Fiol:2013iaa,Giombi:2017cqn}.

\section{Wilson lines in ABJM}

Let us turn to the case of Wilson lines in the three-dimensional ABJM theory. As we will see, the construction of supersymmetric lines in this theory is much subtler than for the ${\cal N}=4$ super Yang-Mills case, and much can be learned from their study about the role of these defect operators in the AdS/CFT context. See \cite{Drukker:2019bev,Penati:2021tfj} for reviews on the study of line operators in the ABJM theory.

\subsubsection*{The 1/6 BPS and 1/2 BPS lines}

Let us define a Wilson line in the ABJM theory as
\begin{equation}
\label{general WL}
W_{\rm ABJM} = \frac{1}{2N}{\rm Tr} \left[  {\cal W}_{\rm ABJM}(\tau_1,\tau_2) \right]
 =
\frac{1}{2N} {\rm Tr} \left[
 {\cal P}\exp\left( {i \int_{\tau_1}^{\tau_2}
 {\cal L}(\tau) d\tau}\right)
\right]\,,
\end{equation}
for a given (super)connection ${\cal L}$. Following the intuition gained from the analysis of Wilson loops in the ${\cal N}=4$ super Yang-Mills theory, one could naively search for supersymmetric Wilson lines in ABJM by just imposing
\begin{equation}
\label{restrivctive susy constraint}
\delta_{\rm SUSY} {\cal L}=0 \,,
\end{equation}
for some subset of the supersymmetry transformations of the theory. Interestingly, the condition \eqref{restrivctive susy constraint} is quite restrictive, and only Wilson lines that are at most 1/6 BPS can be constructed by demanding \eqref{restrivctive susy constraint}. Indeed, by taking
\begin{equation}
\label{1/6 BPS superconnection}
{\cal L}_{1/6} =\begin{pmatrix}
A_\mu \dot{x}^\mu-i|\dot{x}|{M}^{I}_J C_I \bar{C}^J & 0
\\
0 & \hat A_\mu \dot{x}^\mu-i|\dot{x}|{M}^{I}_J \bar{C}^J C_I
\end{pmatrix}\,,
\end{equation}
with
\begin{equation}
M^I_J={\rm diag} (1,1,-1,-1) \,.
\end{equation}
one obtains a 1/6 BPS Wilson line, which preserves an $SU(1,1|1) \times SU(2) \times SU(2)$ subgroup of the $OSp(6|4)$ symmetries of the ABJM theory\footnote{A detailed discussion on the representation theories of this symmetry group is presented in \cite{Bianchi:2018scb}}. 

The construction of 1/2 BPS Wilson loops in ABJM was successfully achieved in \cite{Drukker:2009hy} by observing that the constraint \eqref{restrivctive susy constraint} could be relaxed by instead requiring
\begin{equation}
\delta_{\rm SUSY} {\cal L}= {\cal D}_{\tau} \Lambda := \partial_{\tau} \Lambda +i \{ {\cal L}, \Lambda ] \,,
\end{equation}
where $\Lambda$ is a certain supermatrix. That is, we will say that a supersymmetry transformation leaves invariant the Wilson line if under such a transformation the superconnection transforms as with a gauge transformation. With this criteria, under a finite supersymmetry transformation one gets\footnote{To construct a Wilson loop invariant under these finite transformations one has to be careful with the choice of boundary conditions. While a straight contour is compatible with taking the trace in the definition \eqref{general WL}, for a closed contour with periodic boundary conditions one should instead consider the supertrace \cite{Cardinali:2012ru}.}
\begin{equation}
\label{WL finite transformation}
{\cal W}_{1/2} (\tau_1,\tau_2) = {\cal P}\exp\left( {i \int_{\tau_1}^{\tau_2}
 {\cal L}(\tau) d\tau}\right) \to U^{-1} (\tau_1) {\cal W}_{1/2} (\tau_1,\tau_2) U (\tau_2) \,,
\end{equation}
with $U(\tau)= \exp[i \Lambda (\tau)]$. A 1/2 BPS Wilson line can be then constructed by taking
\begin{equation}
\label{1/2 BPS superconnection}
{\cal L}_{1/2} =\begin{pmatrix}
A_\mu \dot{x}^\mu-i|\dot{x}|{M}^{I}_J C_I \bar{C}^J & -i |\dot{x}| \eta^{\alpha}_{I} \bar{\psi}^{I}_{\alpha}
\\
 -i |\dot{x}| \psi^{\alpha}_{I} \bar{\eta}^{I}_{\alpha} & \hat A_\mu \dot{x}^\mu-i|\dot{x}|{M}^{I}_J \bar{C}^J C_I
\end{pmatrix}\,,
\end{equation}
where\footnote{A similar construction is obtained for a circular contour.}
\begin{equation}
x^\mu =(\tau,0,0)\,,\quad M^I_{J} = -\delta^I_J +2\delta^I_1\delta^1_J\,,\quad
\eta_I^\alpha  = \eta \, \delta_I^1\delta^\alpha_+\,,\quad
\bar\eta^I_\alpha  = \bar\eta \, \delta^I_1\delta_\alpha^+\,,
\label{WLchoice}
\end{equation}
and with
\begin{equation}
\label{eta condition}
\eta\bar\eta = -2i \,.
\end{equation}
The Wilson loop defined by \eqref{general WL}, \eqref{1/2 BPS superconnection} and \eqref{WLchoice} is invariant under an $SU(1,1|3)$ subgroup\footnote{See \cite{Bianchi:2017ozk,Bianchi:2020hsz} for a discussion on the corresponding representation theory.} of the $OSp(6|4)$ symmetries of the ABJM theory. In particular, with the choice \eqref{WLchoice} the Wilson loop \eqref{general WL} is invariant under the supersymmetry transformations \eqref{susy ABJ(M) 1} for arbitrary non-vanishing values of $\bar{\Theta}^{1J}_+ $ and $\bar{\Theta}^{IJ}_-$  with  $I,J\neq 1$. The corresponding $\Lambda$ matrix is
\begin{equation}
\label{lambda straight WL}
\Lambda = \left(
\begin{array}{cc}
0 & g_1
\\
\bar g_2 & 0
\end{array}\right)\,,
\end{equation}
with \cite{Cardinali:2012ru}
\begin{equation}
g_1 = 2 \, \eta \, \bar{\Theta}^{1I} C_I \,,
\qquad
\bar g_2 = \epsilon_{1IJK} \, \bar{\eta} \, \bar{\Theta}^{IJ} \bar{C}^K \,.
\end{equation}

\subsubsection*{Interpolating Wilson lines}

In the ABJM theory one can also construct a family of Wilson lines that continuously interpolates between the 1/6 BPS and the 1/2 BPS lines \cite{Ouyang:2015bmy, Lee:2010hk}. Interestingly, these operators are supersymmetric for all values of the interpolating parameter. More concretely, these Wilson lines can be constructed by considering the definition \eqref{general WL} with \eqref{1/2 BPS superconnection} and taking
\begin{align}
x^\mu &=(\tau,0,0)\,,\quad &M^I_{J} &= -\delta^I_J +2\delta^I_1\delta^1_J + 2 \zeta^2 \delta^I_2 \delta^2_J \,,\\
\eta_I^\alpha  &= \zeta \, \eta \, \delta_I^1\delta^\alpha_+\,,\quad
&\bar\eta^I_\alpha  &= \zeta \, \bar\eta \, \delta^I_1\delta_\alpha^+\,,
\label{interpolatingWLchoice}
\end{align}
with $0 \leq \zeta \leq 1$ and where again the condition \eqref{eta condition} has to be imposed. The resulting operators are 1/2 BPS for $\zeta=1$ and 1/6 BPS for $0\leq \zeta < 1$ \cite{Ouyang:2015bmy, Lee:2010hk}\footnote{A naive generalization of this interpolation to the ${\cal N}=4$ super Yang-Mills theory (to go from the non-supersymmetric Wilson line to the 1/2 BPS Wilson-Maldacena line) is neither supersymmetric nor conformal, as shown in \cite{Polchinski:2011im}.}. Generalizations of \eqref{interpolatingWLchoice} were studied in \cite{Castiglioni:2022yes,Castiglioni:2023uus,Castiglioni:2023tci,Castiglioni:2025iry}, were a web of interpolations between supersymmetric and non-supersymmetric lines was discussed. It is worth mentioning that framing-zero results suggest that the $\zeta$ coupling in \eqref{interpolatingWLchoice} triggers an RG flow from the 1/6 BPS Wilson loop to the 1/2 BPS loop \cite{Castiglioni:2022yes}, while framing-one arguments indicate that the interpolation preserves conformal invariance \cite{Correa:2019rdk}\footnote{See \cite{Bianchi:2013zda,Bianchi:2013rma,Bianchi:2016yzj,Bianchi:2024sod} for discussions on the choice of framing in perturbative computations in the ABJM theory.}.

\subsubsection*{Displacement supermultiplet of the 1/2 BPS line}

As discussed in previous sections, the 1/2 BPS line of the ABJM theory is invariant under an $SU(1,1|3)$ symmetry group. The breaking of ABJM the bulk symmetries by the presence of the defect gives rise to protected insertions in the corresponding dCFT, similarly to what occurs with the 1/2 BPS Wilson line in ${\cal N}=4$ sYM. The $SU(4) \to SU(3)$ R-symmetry breaking defines three complex scalars $\mathbb{O}^a$ ($a=1,2,3$) in the fundamental representation of $SU(3)$ and with $\Delta_{\mathbb O}=1$, while the displacement operator is given by a complex scalar $\mathbb{D}$ with $\Delta_{\mathbb D}=2$.  In turn, the breaking of the supersymmetries introduces the Dirac spinors $\mathbb{F}$ and $\mathbb{\Lambda}^a$, with $\Delta_{\mathbb F}=1/2$ and $\Delta_{\mathbb \Lambda}=3/2$. All these operators group into the displacement supermultiplet of the defect, whose schematic structure is
\begin{equation}
\label{displacement 1/2 BPS ABJM}
\mathbb{F} \to \mathbb{O}^a \to \mathbb{\Lambda}^a \to \mathbb{D} \,,
\end{equation}
and where the arrows indicate the application of supercharges. The expressions of the operators in the displacement supermultiplet in terms of the fundamental fields of ABJM can be found in \cite{Bianchi:2020hsz}. For a detailed analysis of the displacement supermultiplet of the 1/6 BPS line see \cite{Bianchi:2018scb}.

\subsubsection*{The holographic dual to the 1/2 BPS Wilson line}

The construction of the string dual to the 1/2 BPS Wilson line of ABJM mimics the analysis that was discussed in Section \ref{sec: WL N=4} for the 1/2 BPS line of ${\cal N}=4$ sYM \cite{Drukker:2008zx}. As before, for simplicity we will focus on the straight Wilson line, altough the analysis easily generalizes to the circular loop. Moreover, we will work in the planar limit of ABJM, at which the dual background becomes $AdS_4 \times \mathbb{CP}^3$. The dual string streches from the straight contour $\mathcal{C}$ at the boundary of $AdS_4$ into an $AdS_2$ submanifold that extends towards the bulk of $AdS_4$. Furthermore, the string has Dirichlet boundary conditions in the $\mathbb{CP}^3$, i.e. it is fixed at a point of that space. As expected, the symmetries of the dual string perfectly match those of the 1/2 BPS Wilson line. This is easily seen at the bosonic level, where $SU(1,1) \times SU(3) \subset SU(1,1|3)$ is precisely the symmetry group of a string with an $AdS_2$ induced metric and Dirichlet boundary conditions in the $\mathbb{CP}^3$.

\subsubsection*{Analysis of fluctuations}

Rich conclusions can be obtained from the analysis of the quantum fluctuations over the classical Dirichlet string described above. As shown in \cite{Forini:2012bb,Aguilera-Damia:2014bqa}, the spectrum of bosonic fluctuations over such string configuration consists of one massive complex scalar $\phi$ of mass $m^2=2$ (in units where the AdS radius is one) and three massless complex scalars $\varphi^a$ ($a=1,2,3$) in the fundamental representation of $SU(3)$. As for the fermionic fluctuations, one has a massless Dirac spinor $\psi$ and three massive spinors $\chi^a$ in the fundamental representation of $SU(3)$ and with $|m|=1$. Therefore, one obtains a spectrum of quantum fluctuations whose masses and spins precisely match with the scaling dimensions and spins of the operators in the displacement supermultiplet of the dual dCFT. 

It turns out to be fruitful to study the set of boundary conditions that can be imposed on the quantum fluctuations of the 1/2 BPS Dirichlet string. Such an analysis reveals the existence of new dCFTs, distinct from the one defined by the 1/2 BPS Wilson line. Following the discussion of Appendix \ref{ch: scalar and spinor fields in AdS/CFT}, the massive scalar $\phi$ behaves as 
\begin{equation}
\phi(z,x^{\mu}) = \alpha_{\phi}(x^{\mu}) \, z^{-1}+ \mathcal{O} (z) +\beta_{\phi}(x^{\mu}) \, z^{2} + \mathcal{O}(z^{4}) \,,
\end{equation}
near the $z=0$ boundary. As discussed in \cite{Breitenlohner:1982jf,Breitenlohner:1982bm}, these fields only admit boundary conditions that fix their $\alpha_{\phi}$ mode at the boundary. For the case of the massless $\varphi^a$ modes one gets
\begin{equation}
\varphi^a(z,x^{\mu}) = \alpha_{\varphi}^a(x^{\mu}) + \mathcal{O} (z^2) +\beta_{\varphi}^a(x^{\mu}) \, z + \mathcal{O}(z^{3}) \,.
\end{equation}
Crucially, these fluctuations admit boundary conditions that fix either $\alpha_{\varphi}^a$, $\beta_{\varphi}^a$ or a combination of both \cite{Breitenlohner:1982jf,Breitenlohner:1982bm}, i.e. one can impose Dirichlet, Neumann or mixed boundary conditions for these fields. The case where all $\varphi^a$ fluctuations are subjected to Dirichlet boundary conditions corresponds to the 1/2 BPS Dirichlet string that is dual to the 1/2 BPS Wilson line.

Interestingly, one can also study supersymmetric boundary conditions that account for delocalized strings. Following the discussion in \cite{Correa:2021sky}, one can consider boundary conditions that fix the $\dot{\alpha}_{\varphi}^a$ time derivative of a massless $\mathbb{CP}^3$ fluctuation. As we will discuss in more detail in Chapter \ref{ch: line defects in AdS3CFT2} for the case of line defects in AdS$_3$/CFT$_2$, these boundary conditions can be associated to strings that are \textit{smeared} over a given $\mathbb{CP}^1 \subset \mathbb{CP}^3$ submanifold of the compact space. To be more precise, we say that a string is smeared over a $\mathbb{CP}^1 \subset \mathbb{CP}^3$ submanifold if its partition function $Z_{\rm smeared}$ can be written as the weighted average
\begin{equation}
\label{partition function smearing}
 Z_{\rm smeared}=\int_{\mathbb{CP}^1} g_{\rm weight} \, Z_{\rm Dirichlet} \,,
\end{equation}
for some weight function $g_{\rm weight}$, where $Z_{\rm Dirichlet}$ is the partition function of a Dirichlet string that is fixed at a point of the $\mathbb{CP}^1$. We see that a \textit{uniform} smearing (i.e. with constant weight function) along a $\mathbb{CP}^1 \subset \mathbb{CP}^3$ should correspond to fixing $\alpha_{\varphi}^a= {\rm const.}$ for some $a$ and then integrating over the constant, which is equivalent to fixing $\dot{\alpha}_{\varphi}^a=0$.
These strings were studied at the classical level in \cite{Drukker:2008zx}, where the authors looked for an holographic dual to the 1/6 BPS line of ABJM. Additionally, one can also consider boundary conditions that fix the $\beta_{\varphi}^a$ mode for one of the $\varphi^a$ fluctuations, i.e. one can impose Neumann boundary conditions in a $\mathbb{CP}^1 \subset \mathbb{CP}^3$. As pointed out in \cite{Lewkowycz:2013laa,Correa:2021sky}, strings that are either smeared or have Neumann b.c. over a $\mathbb{CP}^1 \subset \mathbb{CP}^3$ have $SU(1,1|1) \times SU(2) \times SU(2)$ symmetry and are therefore 1/6 BPS. Both of these string configurations are candidates to describe the holographic dual to the 1/6 BPS Wilson line of ABJM.

One can also interpolate between different string configurations. For example, one can consider boundary conditions such as
\begin{equation}
\label{ABJM interpolating bc}
    \chi \dot{\alpha}+\beta=0 \,,
\end{equation}
which interpolate between smeared strings and strings with Neumann boundary conditions. Boundary conditions of the type \eqref{ABJM interpolating bc} where studied in \cite{Correa:2019rdk}, where they were found to describe an interpolating family of 1/6 BPS conformal defects.

\section{Line defects as toy models for the study of non-perturbative methods}

\newcolumntype{A}{ >{\centering\arraybackslash} m{4.5cm} }

\newcolumntype{B}{ >{\centering\arraybackslash} m{4.5cm} }

\newcolumntype{C}{ >{\centering\arraybackslash} m{5cm} }

\setlength{\arrayrulewidth}{1mm}
\arrayrulecolor{white}

\begin{table}
\begin{center}
\begin{tabular}{| A | B | C |}
\hline
\rowcolor{gray!20} \cellcolor{white} & \rule{0pt}{3.5ex}
    \textbf{$\mathcal{N}=4$ Super Yang Mills}
 & \rule{0pt}{3.5ex} \textbf{ABJM}  \\ [0.5cm] \hline
\rowcolor{green!20}  
& \rule{0pt}{0.8cm} Four-point correlators: \cite{Liendo:2016ymz, Liendo:2018ukf, Ferrero:2021bsb, Ferrero:2023gnu, Ferrero:2023znz,Bonomi:2024lky} 
 & \rule{0pt}{0.9cm}   Four-point correlators: \cite{Bianchi:2020hsz,Bliard:2023zpe}  \\  [0.7cm]
\rowcolor{green!20} \textbf{Analytic conformal bootstrap at strong coupling}  &  \vspace{-2cm}Higher-point functions: not developed &  \vspace{-2cm} $\,$ Higher-point functions: $\,$ not developed \\ [0.3cm]
\rowcolor{green!20}  &  \vspace{-0.25cm} Bulk-defect correlators: \cite{Barrat:2021yvp,Barrat:2022psm,Bianchi:2022ppi} & \vspace{-0.25cm} $\,$ Bulk-defect correlators: $\,$ not developed \\ [0.65cm]
\hline
\rowcolor{red!20}  \multirow{3}{4.5cm}{ \rule{0pt}{9ex} \centering \textbf{Integrability}} & \rule{0pt}{0.7cm}  One-loop analysis: \cite{Drukker:2006xg, Correa:2018fgz} & \rule{0pt}{0.7cm}  One-loop analysis: \cite{Correa:2023lsm} \\ [1.25cm]
\rowcolor{red!20} & \vspace{-0.5cm}Thermodynamic Bethe Ansatz: \cite{Correa:2012hh, Gromov:2012eu, Bajnok:2013sya} & \vspace{-0.5cm}Thermodynamic Bethe Ansatz: \cite{Correa:2023lsm} \\ [0.25cm]
\rowcolor{red!20} &  Quantum Spectral Curve: \cite{Gromov:2015dfa, Gromov:2016rrp, Cavaglia:2018lxi, Giombi:2018hsx,McGovern:2019sdd, Grabner:2020nis} & Quantum Spectral Curve: not developed \\ [0.45cm]
\hline
\rowcolor{orange!20}  
 \rule{0pt}{3.5ex} \textbf{Bootstrability} & \rule{0pt}{3.5ex}\cite{Cavaglia:2021bnz,Cavaglia:2022qpg,Niarchos:2023lot,Cavaglia:2023mmu,Cavaglia:2024dkk} & \rule{0pt}{3.5ex} Not developed \\ [0.5cm]
\hline
\rowcolor{blue!20}  
 \rule{0pt}{0.35cm} \textbf{Supersymmetric localization} & \vspace{0.25cm} Extensively studied since \cite{Pestun:2007rz}, e.g. see \cite{Correa:2012at, Fiol:2013hna, Pestun:2009nn,Giombi:2009ds}. \vspace{0.25cm}& Thoroughly developed, e.g. see \cite{Kapustin:2009kz, Drukker:2009hy, Drukker:2010nc,Lewkowycz:2013laa,Bianchi:2018scb,Bianchi:2018bke,Gorini:2020new,Penati:2021tfj} \\ [0.5cm]
\hline
\end{tabular}
\end{center}
\caption{State-of-the-art of the study of some non-perturbative methods in the context of line defects in the ${\cal N}=4$ super Yang-Mills and ABJM theories.}
\label{non pert methods table}
\end{table}

Over the past decades, line defects have been the scenario of extensive studies and developments on non-perturbative methods. 
In Chapters \ref{ch: line defects in AdS3CFT2} and \ref{ch: integrable line defects in ABJM} we will discuss the application of conformal bootstrap and integrability methods for the analysis of line defects in AdS$_3$/CFT$_2$ and AdS$_4$/CFT$_3$. Therefore, we consider useful to provide here a brief overview of the state-of-the-art of various non-perturbative methods in the context of the study of line defects in AdS/CFT. We will focus the analysis on the ${\cal N}=4$ super Yang-Mills and ABJM cases, which constitute the most famous backgrounds for the study of line defects in AdS/CFT. A summary is presented in Table \ref{non pert methods table}.

\subsubsection*{Conformal bootstrap}

Since their revival with the seminal paper of \cite{Rattazzi:2008pe}, conformal bootstrap methods have been thoroughly studied in the AdS/CFT framework. The philosophy upon which these methods are based consists on the computations of correlators by the imposition of symmetry constraints and other consistency conditions. 

Four-point functions of insertions along the 1/2 BPS Wilson line of ${\cal N}=4$ super Yang-Mills were studied using conformal bootstrap in \cite{Liendo:2016ymz, Liendo:2018ukf}. These authors provided a numerical bootstrap analysis of the defect, but also elaborated on an interesting analytical bootstrap method to study strong-coupling correlators in this framework. The latter was applied up to N$^2$LO in the strong-coupling regime in \cite{Liendo:2018ukf}, and later extended to N$^3$LO in \cite{Ferrero:2021bsb, Ferrero:2023gnu, Ferrero:2023znz}. A detailed description of this method will be provided in Chapter \ref{ch: line defects in AdS3CFT2}, when discussing its application to the study of line defects in AdS$_3$/CFT$_2$. Interestingly, the results of \cite{Liendo:2018ukf,Ferrero:2021bsb, Ferrero:2023gnu, Ferrero:2023znz} were re-derived in \cite{Bonomi:2024lky} using an alternative method based on a dispersion relation. 

The extension of the analytic bootstrap methods to the computation of higher-point correlators at strong coupling is yet to be developed. However, a weak-coupling bootstrap analysis of multipoint defect correlators in the context of the 1/2 BPS Wilson line of ${\cal N}=4$ sYM was performed in \cite{Barrat:2021tpn,Barrat:2022eim,Artico:2024wut}. Analytic bootstrap techniques have also been applied in \cite{Barrat:2021yvp,Barrat:2022psm,Bianchi:2022ppi} to the strong-coupling analysis of bulk-defect correlators, also in the context of the 1/2 BPS Wilson line of ${\cal N}=4$ sYM. The weak-coupling regime of those correlators was discussed in \cite{Barrat:2020vch, Artico:2024wnt}. The analytic conformal bootstrap ideas were also discussed in the framework of the 1/2 BPS Wilson line of ABJM in \cite{Bianchi:2020hsz,Bliard:2023zpe}, were an analysis of four-point functions at strong coupling was performed. The method was also applied in \cite{Pozzi:2024xnu} for the study of 1/2 BPS defects in three-dimensional ${\cal N}=4$ Chern-Simons-matter theories. For an extension of the conformal bootstrap ideas to the analysis of line defects in non-supersymmetric set ups see for example \cite{Gimenez-Grau:2022czc,Bianchi:2022sbz,Bianchi:2023gkk}.

\subsubsection*{Integrability}

Integrable structures were first found in the AdS/CFT framework in the seminal paper of \cite{Minahan:2002ve}, where the matrix of anomalous dimensions for single-trace operators in the planar limit of ${\cal N}=4$ sYM was perturbatively identified with the hamiltonian of an integrable and periodic spin chain. These results led firstly to the construction of \textit{Asymptotic Bethe Ansatz} (ABA) equations for the spectrum of anomalous dimensions of single-trace operators in the large-size limit \cite{Beisert:2005fw, Beisert:2005tm}. These equations were later extended to describe the full finite-size spectrum by means of a \textit{Thermodynamic Bethe Ansatz} (TBA) set of equations \cite{Gromov:2009tv,Gromov:2009bc,Arutyunov:2009ur,Bombardelli:2009ns}. We will describe this method in detail in Chapter \ref{ch: integrable line defects in ABJM}, when studying its application for the computation of the cusp anomalous dimension of the ABJM theory. The computational power of the TBA equations was later surpassed with the proposal of the \textit{Quantum Spectral Curve} (QSC) \cite{Gromov:2013pga,Gromov:2014caa}, which provided valuable access to the finite-coupling regime of the scaling dimensions of arbitrary-length single-trace operators (e.g. see \cite{Gromov:2015wca}). These developments were later extended to the AdS$_4$/CFT$_3$ and AdS$_3$/CFT$_2$ frameworks, where ultimately led to the corresponding QSC proposals \cite{Cavaglia:2014exa,Bombardelli:2017vhk,Cavaglia:2021eqr,Cavaglia:2022xld}.

The spectrum of planar gauge theories is not only composed by local single-trace operators. As previously discussed in this chapter, one may also study the insertion of operators along the contour of non-local observables such as Wilson lines. It is therefore a natural question whether the integrable structures observed for single-trace operators also extend to the dCFTs defined by certain Wilson lines. Progress in this direction was initiated in \cite{Drukker:2006xg}, where the one-loop integrability of insertions along the 1/2 BPS Wilson line of ${\cal N}=4$ sYM was studied. Those authors found that the anomalous dimensions of a certain set of insertions (known as the $SU(2)$ sector) are described by the hamiltonian of an integrable and \textit{open} spin chain. These results were extended to the case of the non-supersymmetric Wilson line in \cite{Correa:2018fgz}, where the one-loop integrability of insertions in the $SU(2|3)$ sector was showed. An interesting application of the integrability ideas was performed in \cite{Drukker:2012de, Correa:2012hh}, where a Thermodynamic Bethe Ansatz (TBA) was proposed and used to study the cusp anomalous dimension of the ${\cal N}=4$ super Yang-Mills theory. A two-loop analysis of those TBA equations was performed in \cite{Bajnok:2013sya}. In turn, an all-loop solution of the TBA equations in the small-angle limit was found in \cite{Gromov:2012eu}, where it was used to compute the all-loop bremsstrahlung function of the theory (i.e. the quadratic order of the cusp anomalous dimension in its small-angle expansion). These studies ultimately culminated in the proposal of a QSC for the dCFT defined by the 1/2 BPS Wilson line of ${\cal N}=4$ sYM \cite{Gromov:2015dfa, Gromov:2016rrp, Cavaglia:2018lxi, Giombi:2018hsx,McGovern:2019sdd, Grabner:2020nis}. 

A natural follow-up of the above findings is to study whether they extend to the analysis of Wilson lines in the ABJM theory. A first step in this direction was provided in \cite{Correa:2023lsm}, which constitutes a part of this thesis and will be discussed in detail in Chapter \ref{ch: integrable line defects in ABJM}. Evidence was found there for the integrability of insertions along the countour of the 1/2 BPS Wilson line of ABJM, and a set of TBA equations was proposed and used to reproduce the one-loop cusp anomalous dimension of the theory. It remains as an open question to see if the aforementioned TBA can be improved into a set of QSC equations.

\subsubsection*{Bootstrability}

An interesting combination of the integrability and conformal bootstrap ideas was proposed in the context of the 1/2 BPS line defect of ${\cal N}=4$ sYM by \cite{Cavaglia:2021bnz}. These ideas, known as the \textit{bootstrability} techniques, merge the numerical conformal bootstrap methods with input from the QSC formalism, which notoriously improves the precision of the bootstrap outputs. Interesting developments have been made in the analysis of four-point functions with bootstrability techniques in \cite{Cavaglia:2022qpg,Niarchos:2023lot,Cavaglia:2023mmu,Cavaglia:2024dkk}. The application of these methods to the ABJM theory is still an open problem, given the lack of a QSC formalism for the corresponding Wilson lines.

\subsubsection*{Supersymmetric localization}

Since the pioneering paper of Pestun \cite{Pestun:2007rz}, supersymmetric localization techniques have been widely used in the AdS/CFT framework to yield valuable all-loop results. These methods, which provide a way to translate infinite-dimensional path integrals into finite-dimensional integrals, allowed for the all-loop computation of the vacuum expectation value of the circular 1/2 BPS Wilson line of ${\cal N}=4$ sYM via the  analysis of a Gaussian Hermitian matrix model \cite{Pestun:2007rz}. The agreement of such result with its expected weak- and strong-coupling expansions (provided respectively by the perturbative regimes of the gauge theory and the string theory) furnished a highly non-trivial test of the duality between the ${\cal N}=4$ super Yang-Mills theory and type IIB string theory in $AdS_5 \times S^5$. Since then, many applications of supersymmetric localization methods have been developed for Wilson lines in the AdS/CFT framework, e.g. see \cite{Correa:2012at, Fiol:2013hna, Pestun:2009nn,Giombi:2009ds} for applications in ${\cal N}=4$ sYM and \cite{Kapustin:2009kz, Drukker:2009hy, Drukker:2010nc,Lewkowycz:2013laa,Bianchi:2018scb,Bianchi:2018bke,Gorini:2020new,Penati:2021tfj} for the ABJM case.

%% file: defectsAdS3CFT2.tex
As outlined in the previous chapter, line defects have been instrumental in the study of holographic dualities during the last decades, providing an ideal framework to study and develop non-perturbative methods. 
As reviewed in Chapter \ref{ch: line defects}, e.g. see Table \ref{non pert methods table}, considerable progress has been achieved in the analysis of line operators in the context of AdS$_5$/CFT$_4$ and AdS$_4$/CFT$_3$ correspondences. However, much less is known about the role of line defects in the AdS$_3$/CFT$_2$ framework, which provides a way to continuously interpolate between pure R-R and pure NS-NS supergravity backgrounds (see Chapter \ref{ch:adscft} for a discussion). The aim of this chapter is to study some properties of line defects in these lower-dimensional dualities. To be more precise, we will pursue a twofold goal throughout this chapter. On the one hand, we will study the existence of BPS line defects in the AdS$_3$/CFT$_2$ framework. As will be shown, our analysis will reveal a rich family of supersymmetric defects preserving different symmetry groups. On the other hand, we will study the application of analytic conformal bootstrap methods to the description of these defects. We will focus on the computation of two-, three- and four-point functions along the contour of 1/2 BPS line defects, showing that bootstrap techniques determine certain four-point correlators up to two coefficients.

We will begin by centering the analysis on AdS$_3$/CFT$_2$ realizations in which the gravity background is type IIB string theory in $AdS_3 \times S^3_+ \times S^3_- \times S^1$ with mixed Ramond-Ramond (R-R) and Neveu Schwarz-Neveu Schwarz (NS-NS) three-form flux. We will study the existence of BPS strings that could describe supersymmetric line defects in the holographic CFT$_2$, i.e. we will look for BPS strings ending along a straight line at the boundary of $AdS_3$. We will study both classical string configurations and the quadratic fluctuations around them, and we will show that there is a rich family of BPS line defects in the dual conformal theory. These defects will correspond to strings which are either fixed at a point in $S^3_+ \times S^3_- \times S^1$ or that are delocalized over a given submanifold of such compact space. The dual strings (and therefore the corresponding line defects) will range from 1/2 BPS to 1/8 BPS, depending on their boundary condition. Moreover, we will describe a network of interpolating strings that connects all the aforementioned defects.

As pointed out in Chapter \ref{ch:adscft}, up to date there is not a well-understood lagrangian description of the holographic CFT$_2$ for arbitrary points in the parameter space of the $AdS_3 \times S^3_+ \times S^3_- \times S^1$ background. This motivates the use of conformal bootstrap techniques for the study of the dual CFT, given that these methods do not require the precise knowledge of the lagrangian. With this in mind, we will take the $AdS_3 \times S^3 \times T^4$ limit of the gravity background and we will use analytic conformal bootstrap methods to study correlators along 1/2 BPS line defects in the dual CFT$_2$. We will center the analysis in the strong coupling regime of the CFT$_2$, where bootstrap techniques provide a way to bypass the use of Witten diagrams when computing correlators. 

Following the discussion given in Chapter \ref{ch: line defects}, the analytic conformal bootstrap program has been successfully applied in the ${\cal N}=4$ super Yang-Mills and ABJM theories to study their corresponding 1/2 BPS line defects, which are respectively invariant under eight and six Poincaré supercharges. A natural question is whether analytic conformal bootstrap methods can be applied in setups with less supersymmetry. The fact that 1/2 BPS line defects in the holographic dual to string theory in $AdS_3 \times S^3 \times T^4$ are invariant under only four supercharges provides therefore another motivation to apply the analytic conformal bootstrap program for their study. 

Let us also note that line defects in the holographic dual to string theory in $AdS_3 \times S^3 \times T^4$ are characterized by correlators which depend on two parameters, the 't Hooft coupling and the $\vartheta$ parameter that interpolates between pure R-R and pure NS-NS dual backgrounds (see Chapter \ref{ch:adscft}). Consequently, those correlators define a two-parameter bootstrap problem, which contrasts with the single-parameter problem of defects in ${\cal N}=4$ sYM and ABJM (which are characterized only by the 't Hooft coupling) and provides further reasons to study the application of bootstrap methods in the AdS$_3$/CFT$_2$ context.

The 1/2 BPS line defects that we will study break the symmetries of the bulk CFT$_2$ as $PSU(1,1|2)^2\times SO(4) \rightarrow PSU(1,1|2)\times SU(2)$, defining displacement and tilt supermultiplets.
We will focus on the two-, three- and four-point functions of these supermultiplets, which we will compute using analytic conformal bootstrap up to next-to-leading order in their strong-coupling expansion. We will obtain a bootstrap result that only depends on two OPE coefficients. Moreover, we will perform a Witten diagram check of the bootstrap result, obtaining an holographic interpretation of the two OPE coefficients that are not constrained by the bootstrap procedure.

The chapter is organized as follows. In Sections \ref{sec: classical BPS strings} and \ref{sec: BPS strings fluctuations} we will focus on $AdS_3 \times S^3_+ \times S^3_- \times S^1$ backgrounds and we will study the existence of supersymmetric strings that could describe BPS line defects in the holographic CFT$_2$. We will begin with a classical analysis in Section \ref{sec: classical BPS strings}, and in Section \ref{sec: BPS strings fluctuations} we will work at the level of the quadratic fluctuations. We will later focus the analysis on 1/2 BPS line defects in the holographic dual to string theory in $AdS_3 \times S^3 \times T^4$. In Section \ref{sec: algebra and repr} we will introduce the displacement and tilt supermultiplets of these defects. Section \ref{sec: 2 and 3 pts} will be devoted to the computation of the two- and three-point functions of fields in the displacement and tilt supermultiplets, and in Sections \ref{sec: 4 pts preliminaries} and  \ref{sec: 4 pts bootstrap} we will perform an analytic bootstrap computation of the corresponding four-point functions. Finally, in Section \ref{sec: holographic 4 pts} we will discuss an holographic check of the bootstrap computation. 

This chapter is based on \cite{Correa:2021sky} and \cite{Bliard:2024bcz}, which is work perform by the author of this thesis in collaboration with G. Bliard, D.H. Correa, V. I. Giraldo-Rivera and I. Salazar-Landea.

\section{BPS strings in AdS$_3\times$S$_+^3\times$S$_-^3\times$S$^1$: classical analysis}
\label{sec: classical BPS strings}

In this section we will study classical strings moving in an $AdS_3 \times S^3_+ \times S^3_- \times S^1$ background, and we will look for BPS strings ending on a straight contour at the $AdS_3$ boundary (i.e. strings which could be identified with a BPS line defect in the holographic CFT$_2$). As described in Chapter \ref{ch:adscft}, we will center our analysis on AdS$_3$/CFT$_2$ realizations in which the supergravity background is given by the metric
\begin{equation}
\label{AdS3xS3xS3xS1 metric}
 ds^2 =L^2 ds^2(AdS_3)+\frac{L^2}{\sin^2\Omega} ds^2(S^3_+)+ \frac{L^2}{\cos^2\Omega} ds^2(S^3_-)+l^2d\theta^2\,. 
\end{equation}
together with a constant dilaton $\Phi$ and the Ramond-Ramond (R-R) and Neveu Schwarz-Neveu Schwarz (NS-NS) three-form field strengths 
\begin{align}
F_{(3)} & = d C_{(2)} =  -2e^{-\Phi}L^2 \cos\vartheta
\left({\rm vol}(AdS_3)+\frac{1}{\sin^2\Omega} {\rm vol}(S^3_+)+ \frac{1}{\cos^2\Omega} {\rm vol}(S^3_-) \right)\,,
\nonumber \\
H_{(3)} & = d B_{(2)} = 2L^2 \sin\vartheta
\left({\rm vol}(AdS_3)+\frac{1}{\sin^2\Omega} {\rm vol}(S^3_+)+ \frac{1}{\cos^2\Omega} {\rm vol}(S^3_-) \right)\,.
\label{NS NS form}
\end{align}
The parameter $\Omega$ in \eqref{AdS3xS3xS3xS1 metric} takes values in $0<\Omega<\frac{\pi}{2}$ and measures the relative size of the two three-spheres, while the parameter $\vartheta$ in \eqref{NS NS form} takes values in $0 \leq \vartheta \leq \frac{\pi}{2}$ and interpolates between pure R-R and pure NS-NS backgrounds.

In the following we will parametrize $AdS_3$ with Poincar\'e coordinates
\begin{equation}
ds^2(AdS_3) =\frac{1}{z^2}\left(-dt^2+dx^2+dz^2\right) \,,
    \label{poincare}
\end{equation}
and the three-spheres as
\begin{equation}
ds^2(S^3_\pm) =d\beta_\pm^2 + \cos^2 \beta_\pm \left(d\gamma_{\pm}^2+\cos^2\gamma_\pm d\varphi_\pm^2\right) \,.
    \label{spherecoor}
\end{equation}

\subsubsection*{Killing spinors}

The $AdS_3 \times S^3_+ \times S^3_- \times S^1$ background described above is a solution of the type IIB supergravity equations. As described in Appendix \ref{killingspinors}, the background is invariant under sixteen real supercharges, i.e. it is 1/2 BPS. The supersymmetries close a $D(2,1, \sin^2 \Omega)^2 \times U(1)$ supergroup, and are parametrized by a set of Killing spinors $\epsilon= \eta+i\xi$ which take the form
\begin{align}
\eta & = U\epsilon_0 + \frac{\cos\vartheta}{\sin\vartheta + 1} V \epsilon_1 \,,
\label{etadef}
\\
\xi & = -\frac{\cos\vartheta}{\sin\vartheta + 1} U\epsilon_0 +  V \epsilon_1 \,,
\label{xidef}
\end{align}
with
\begin{align}
\label{U}
U & =  e^{\beta_+M_{\beta_+}}e^{\beta_-M_{\beta_-}}
e^{\gamma_+M_{\gamma_+}}e^{\gamma_-M_{\gamma_-}}e^{\varphi_+M_{\varphi_+}}e^{\varphi_-M_{\varphi_-}} e^{\log z M_z} e^{t (M_t+M_x)}e^{x (M_t+M_x)}\,,
\\
\label{V}
V & =  e^{-\beta_+M_{\beta_+}}e^{-\beta_-M_{\beta_-}}e^{-\gamma_+M_{\gamma_+}}e^{-\gamma_-M_{\gamma_-}}e^{-\varphi_+M_{\varphi_+}}e^{-\varphi_-M_{\varphi_-}}e^{-\log z M_z}e^{-t (M_t-M_x)}e^{x (M_t-M_x)}
\,,
\end{align}
and where
\begin{empheq}{alignat=9}
    M_t & = \tfrac{1}{2}\gamma^1\gamma^2 \,, &\qquad  M_x & = -\tfrac{1}{2}\gamma^0\gamma^2  \,, &\qquad & M_z & = \tfrac{1}{2}\gamma^0\gamma^1 \,, &
    \label{Mtxz}
    \\
    M_{\beta_+} & = \tfrac{1}{2}\gamma^4\gamma^5 \,, &\qquad  M_{\gamma_+} & =- \tfrac{1}{2}\gamma^3\gamma^5  \,, &\qquad & M_{\varphi_+} & = \tfrac{1}{2}\gamma^3\gamma^4  \,, &
\label{Mplus}
\\
M_{\beta_-} & = \tfrac{1}{2}\gamma^7\gamma^8 \,, &\qquad  M_{\gamma_-} & =- \tfrac{1}{2}\gamma^6\gamma^8  \,, &\qquad & M_{\varphi_-} & = \tfrac{1}{2}\gamma^6\gamma^7  \,. &
\label{Mminus}
    \end{empheq}
The constant spinors $\epsilon_0$ and $\epsilon_1$ are Majorana-Weyl spinors with the same chirality
\begin{equation}
\gamma_{11}\epsilon_0 = -\epsilon_0 \,,
\qquad
\gamma_{11}\epsilon_1 = -\epsilon_1 \,,
\end{equation}
with $\gamma_{11} = \gamma^0\gamma^1\cdots\gamma^9$. Moreover, they are further subjected to the conditions
\begin{equation}
P^{\Sigma}_- \epsilon_0 = 0\,,\qquad
P^{\Sigma}_- \epsilon_1 = 0\,,
\end{equation}
where the $P^{\Sigma}_{\pm}$ projectors are defined in \eqref{sigmaprojector}. We refer to Appendix \ref{killingspinors} for a detailed computation of the above Killing spinors.

\subsubsection*{Supersymmetric strings with Dirichlet boundary conditions}

Let us use the Green-Schwarz formalism to describe a string with coordinates $X^\mu(\tau, \sigma)$ and $\Theta(\tau, \sigma)$ in the ten-dimensional superspace. We will say that a given string configuration is supersymmetric whenever a combined transformation
$\delta:=\delta{\sf susy}+\delta_\kappa$ leaves the configuration invariant, where $\delta{\sf susy}$ is a supersymmetry transformation and $\delta_\kappa$ is a $\kappa$-symmetry transformation. As discussed in Appendix \ref{app: quadratic fluctuations} (see eq. \eqref{kappapro0}), a string is classically supersymmetric if there exists a Killing spinor $\epsilon$ which satisfies 
\begin{equation}
\Gamma\epsilon = \epsilon\,,  \qquad    \Gamma = -\frac{\partial_\tau X^\mu \partial_\sigma X^\nu\Gamma_{\mu\nu}}{\sqrt{-h}}K\,,
\label{kappapro0-ch}
\end{equation}
where $K$ indicates complex conjugation and $h$ is the determinant of the induced metric on the worldsheet. 

\begin{figure}
    \centering
    \includegraphics[width=0.5\linewidth]{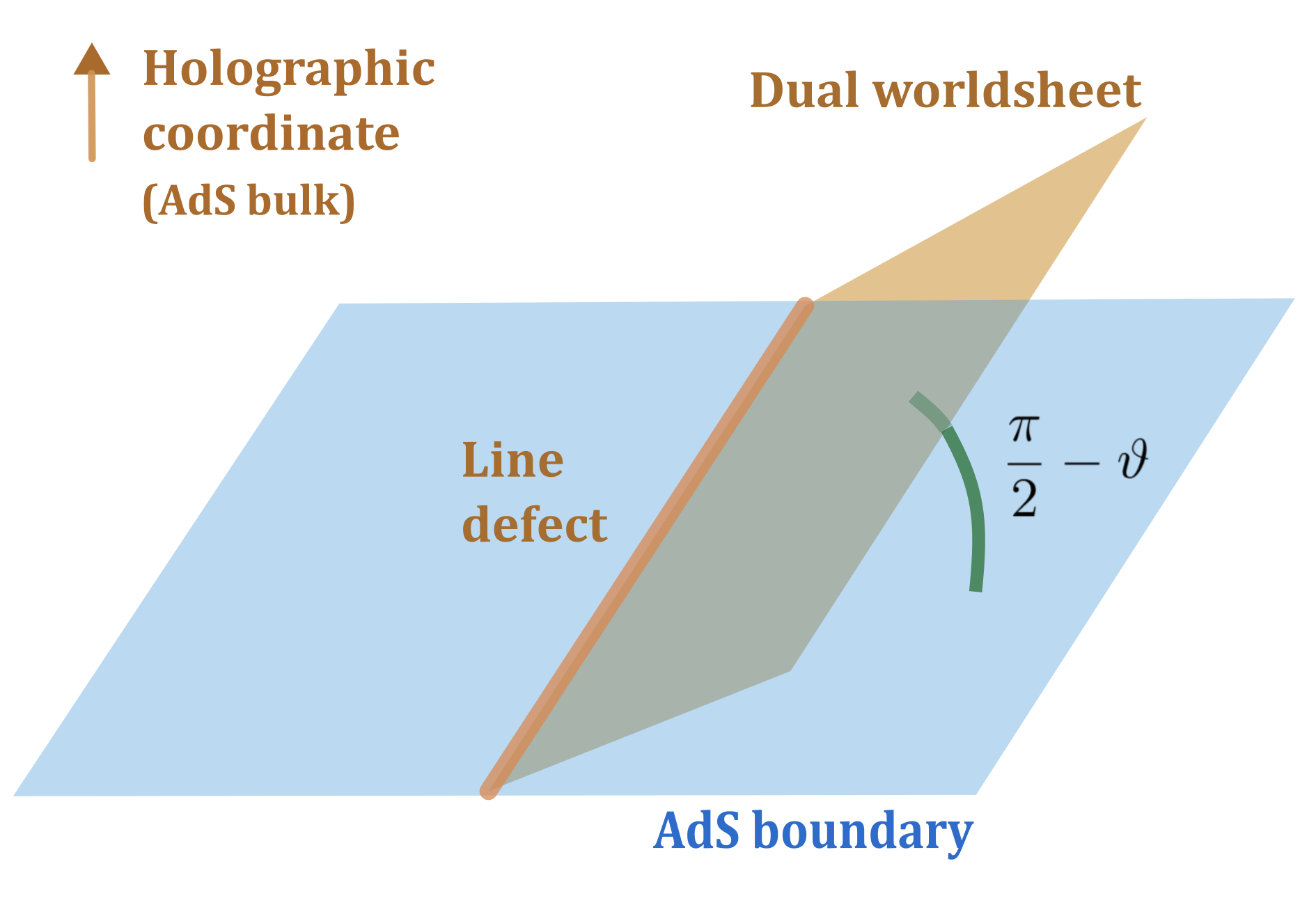}
    \caption{Schematic representation of the string configuration presented in \eqref{susyconfiguration}. The string is fixed at a point in the compact $S^3_+ \times S^3_- \times S^1$ space.}
    \label{fig: ads3string}
\end{figure}

We shall first consider strings ending at a straight line at the boundary of $AdS_3$ and sitting at fixed points in the spheres, i.e. strings with Dirichlet boundary conditions in the $S^3_+ \times S^3_- \times S^1$. An ansatz to describe them is
\begin{equation}
t = \omega\tau\,,\qquad
x = x(\sigma)\,,\qquad 
z = \sigma\,,\qquad 
\beta_\pm,\gamma_\pm,\varphi_\pm,\theta = \text{const.}
\label{ansatz}
\end{equation}
where $-\infty<\tau<\infty$ and $0<\sigma<\infty$ are the two coordinates that parametrize the string worldsheet. For the ansatz \eqref{ansatz} the supersymmetry condition \eqref{kappapro0-ch} becomes
\begin{align}
    \epsilon_0 & =
    \frac{U_7^{-1}V_7}{2\sigma\cos\vartheta}
   \left[\tfrac{2\sigma x(\sigma)+x'\left(1+x(\sigma)^2 -\sigma^2\right)}{\sqrt{1+(x')^2}}+\left(1+x(\sigma)^2 +\sigma^2\right)\sin\vartheta \right.
\nonumber\\
   & \hspace{2.3cm}
   \left.  -2\left(\tfrac{2\sigma x(\sigma)-x'\left(1-x(\sigma)^2 +\sigma^2\right)}{\sqrt{1+(x')^2}}-\left(1-x(\sigma)^2 -\sigma^2\right)\sin\vartheta\right)M_z  \right.\nonumber \\
   & \hspace{2.3cm}
   \left. -4\left(
   \tfrac{ x'x(\sigma)+\sigma}{\sqrt{1+(x')^2}}+x(\sigma)\sin\vartheta
   \right)M_x
   \right]\epsilon_1 \nonumber \\
   &\hspace{0.4cm}
   -\frac{U_7^{-1}V_7}{2\sigma\cos\vartheta}
   \left[\tfrac{x'}{\sqrt{1+(x')^2}} +\sin\vartheta\right] \left(4\omega\tau M_t+ (1-2M_z)\omega^2\tau^2\right)\epsilon_1\,. 
   \label{e0e1replaced}
     \end{align}
where $U_7$ and $V_7$ are defined in \eqref{UyV}. Since $\epsilon_0$ and $\epsilon_1$ are constant spinors, the string configuration will be supersymmetric if $x(\sigma)$ can be chosen so that all the dependence on $\sigma$ and $\tau$ goes away from the r.h.s. of \eqref{e0e1replaced}. Only the second term is $\tau$-dependent, which is cancelled provided that
\begin{equation}
    x'(\sigma) = -\tan\vartheta\,,
    \qquad
    \Leftrightarrow
    \qquad 
    x(\sigma) = x_0 - \tan\vartheta \ \sigma
    \,.
    \label{susyprofile}
\end{equation}
If this is the case, \eqref{e0e1replaced} can be rewritten as
\begin{align}
\label{susy constraint dirichlet}
    \epsilon_0 & =
    U_7^{-1}V_7\left(x_0 -2 x_0 M_z-2 M_x\right)\epsilon_1\,.
\end{align}
Therefore, the string configuration
\begin{equation}
t = \omega\tau\,,\qquad
x =  x_0 - \tan\vartheta \ \sigma\,,\qquad 
z = \sigma\,,\qquad 
\beta_\pm,\gamma_\pm,\varphi_\pm,\theta = \text{const.}
\label{susyconfiguration}
\end{equation}
is 1/2 BPS. The worldsheet ends along the line $x = x_0$ at the boundary and it is tilt with respect to such a boundary by an angle $\frac{\pi}{2}-\vartheta$. The induced geometry turns out to be that of an $AdS_2$ space. In what follows we will choose 
\begin{equation}
\label{omega}
\omega=\frac{1}{\cos \vartheta}\,,
\end{equation}
so that the induced metric becomes
\begin{equation}
ds^2 = \frac{L^2 }{\sigma^2 \cos^2 \vartheta }
\left(-d\tau^2 + d\sigma^2\right)\,,
\end{equation}
and the radius of the $AdS_2$ space is $R:=L/\cos \vartheta$. See Figure \ref{fig: ads3string} for a schematic representation of the string presented in \eqref{susyconfiguration}. As one might have expected, the configuration \eqref{susyconfiguration} solves the equations of motion for an open string that couples to the NS-NS $B$-field (see Appendix \ref{app: quadratic fluctuations} for a discussion). This coupling is precisely the reason for the tilt. The configuration \eqref{susyconfiguration} has been previously found as a classical solution in \cite{Hernandez:2019huf}.

Finally, though in the rest of this chapter we will focus on the case $0 \leq \vartheta < \frac{\pi}{2}$, it is interesting to see what happens in the pure NS-NS limit. When $\vartheta\to\frac\pi{2}$ the worldsheet becomes parallel to the boundary, so the limit $\vartheta\to\frac\pi{2}$ of \eqref{susyconfiguration} is just one particular case of a more general solution of the type
\begin{equation}
\label{ns ns string}
t = \tau\,,\qquad
x = - \sigma\,,\qquad 
z = z_0 \,,\qquad 
\beta_\pm,\gamma_\pm,\varphi_\pm,\theta = \text{const.}
\end{equation}
In this case the projection \eqref{kappapro0-ch} translates into
\begin{align}
    \label{susy condition ns ns 1 0}
    2 U^{-1} M_z U \epsilon_0 &=-\epsilon_0\,,  \\
     \label{susy condition ns ns 2 0}
    2 V^{-1} M_z V \epsilon_1 &=\epsilon_1\,.
\end{align}
Using \eqref{U} and \eqref{V} we get
\begin{align}
    \label{susy condition ns ns 1}
    2 U^{-1} M_z U &= 2 M_z [1+2(M_t+M_x) (\tau-\sigma)]\,, \\
    \label{susy condition ns ns 2}
2 V^{-1} M_z V &= 2 M_z [1-2(M_t-M_x) (\tau+\sigma)]\,.
\end{align}
Thus, in order to remove the $\tau-$dependence in \eqref{susy condition ns ns 1} and \eqref{susy condition ns ns 2} and get a BPS string,  it suffices to impose
\begin{align}
    \label{susy constraint ns ns 1}
    (M_t+M_x) \; \epsilon_0 &=0\,, \qquad \Leftrightarrow \qquad 2 M_z \epsilon_0 = - \epsilon_0\,, \\
    \label{susy constraint ns ns 2}
   (M_t-M_x) \; \epsilon_1 &=0\,, \qquad \Leftrightarrow \qquad 2 M_z \epsilon_1 =  \epsilon_1\,.
\end{align}
The strings satisfying \eqref{ns ns string} preserve 8 independent Killing spinors for all values of $z_0$, and we get therefore a one-parameter family of 1/2 BPS strings which extend parallel to the boundary of AdS$_3$. Thus, this ceases to be the dual representation of a line operator in the CFT$_2$. It is immediate to see that the induced metric for these strings is flat.

\subsubsection*{Supersymmetric smeared strings}

Generically, and because $U_7$ and $V_7$ depend on the values of $\beta_\pm,\gamma_\pm$ and $\varphi_\pm$,
string configurations of the form \eqref{susyconfiguration} that sit at different points of the internal space $S^3\times S^3$ preserve different sets of supersymmetries. We shall now observe the existence of subspaces ${\cal M}\subset S^3\times S^3\times S^1$ such that all the strings sitting in ${\cal M}$ share some fraction of supersymmetry, which enables to have supersymmetric smeared string configurations. As discussed in Chapter \ref{ch: line defects}, we say that a string is smeared over ${\cal M}$ if its partition function $Z_{\rm smeared}$ accounts for the superposition of strings with Dirichlet boundary conditions sitting at the different points of ${\cal M}$, i.e. if
\begin{equation}
\label{partition function smearing}
 Z_{\rm smeared}=\int_{\cal M} g_{\rm weight} \, Z_{\rm Dirichlet} \,,
\end{equation}
for some weight function $g_{\rm weight}$ and where $Z_{\rm Dirichlet}$ is the partition function of a Dirichlet string sitting at a point in ${\cal M}$.

One obvious possibility is to take ${\cal M}_1 = S^1$, as the Killing spinors are independent of the angle $\theta$ (see \eqref{susy constraint dirichlet}). A  configuration of strings smeared along that circle should therefore be 1/2 BPS as well. This is in stark contrast with the analogue cases in $AdS_5\times S^5$ or $AdS_4\times \mathbb{CP}^3$. Smearing open strings along compact subspaces in those backgrounds breaks the supersymmetry either fully or partially.

Less obvious examples involve smearing in the $S^3_+\times S^3_-$ factor. The angular dependence of the projection \eqref{susy constraint dirichlet} is given by the matrix
\begin{align}
U_7^{-1} V_7 &= e^{-\varphi_+ M_{\varphi_+}-\varphi_- M_{\varphi_-}} e^{-\gamma_+ M_{\gamma_+}-\gamma_- M_{\gamma_-}} e^{-2 ( \beta_+ M_{\beta_+}+ \beta_- M_{\beta_-})} \nonumber \\
& \quad\times e^{-\gamma_+ M_{\gamma_+}-\gamma_- M_{\gamma_-}} e^{-\varphi_+ M_{\varphi_+}-\varphi_- M_{\varphi_-}}\,.
\label{angular dependence matrix}
\end{align}
As a concrete example, let us consider strings sitting in two maximal circles of the three-spheres, by setting $\beta_\pm = \gamma_\pm = 0$. Therefore,
\begin{equation}
\left. U_7^{-1} V_7\right|_{\beta_\pm = \gamma_\pm = 0} =  e^{-2 ( \varphi_+ M_{\varphi_+}+ \varphi_- M_{\varphi_-})}
= e^{-(\varphi_++\varphi_-)(M_{\varphi_+}+M_{\varphi_-})}
e^{-(\varphi_+-\varphi_-)(M_{\varphi_+}-M_{\varphi_-})}
\,.
\label{UV2circles}
\end{equation}
We then note that, for Killing spinors satisfying
\begin{equation}
    \label{delocalized string constraint 0}
    (M_{\varphi_+}+M_{\varphi_-}) \; \epsilon_1=0\,, \qquad \Leftrightarrow \qquad 4 M_{\varphi_+} M_{\varphi_-} \epsilon_1= \epsilon_1\,,
\end{equation}
the matrix \eqref{UV2circles} acts trivially on $\epsilon_1$ provided that the strings additionally satisfy $\varphi_+-\varphi_-=0$. Thus, strings sitting at the submanifold ${\cal M}_2 = S^1_D\times S^1$ given by
\begin{equation}
    \label{submanifold 1}
      \beta_{\pm} =  \gamma_{\pm} = 0\,,
      \qquad
     \varphi_+-\varphi_-=0\,,
\end{equation}
will be invariant under common supersymmetry transformations parametrized by four real parameters, because the additional projection \eqref{delocalized string constraint 0} halves the number of supersymmetries. The corresponding Killing spinors satisfy
\begin{align}
    \label{submanifold 1-non zero values-1}
     & (M_{\varphi_+}+M_{\varphi_-}) \; \epsilon_1 =0\,,\\
         \label{submanifold 1-non zero values-2}
     & \epsilon_0 =
     \left(x_0 -2 x_0 M_z-2 M_x\right) \; \epsilon_1\,.
\end{align}
In analogy with the case of type IIA strings on $AdS_4\times \mathbb{CP}^3$ analyzed in \cite{Drukker:2008zx}, we conclude that a string smeared over the region $S^1_D\times S^1$ defined by \eqref{submanifold 1} is 1/4 BPS.
For different choices of the maximal circles one would encounter different projections but the smeared configurations will continue to be 1/4 BPS.

It is also possible to smear over larger submanifolds. Another example arises when we consider strings sitting in two maximal two-spheres,  setting in this case $\beta_\pm = 0$. Smearing over the ${\cal M}_3 = S^2_D\times S^1\supset {\cal M}_2$ submanifold given by
\begin{equation}
    \label{submanifold 2}
      \beta_{\pm} =  0\,,
      \qquad
     \gamma_+-\gamma_-=0\,,
      \qquad
     \varphi_+-\varphi_-=0\,,
\end{equation}
defines a supersymmetric configuration if, in addition to
\eqref{submanifold 1-non zero values-1} and \eqref{submanifold 1-non zero values-2}, one imposes that
\begin{align}
    \label{submanifold 2-non zero values-1}
     & (M_{\gamma_+}+M_{\gamma_-}) \; \epsilon_1 =0\,.
\end{align}
With this additional projection the Killing spinors contain two real parameters, which means that this smearing is 1/8 BPS.

As a final example, we can consider a string smeared over ${\cal M}_4 = S^3_D\times S^1\supset {\cal M}_3$, defined by
\begin{equation}
    \label{submanifold 3}
    \beta_+-\beta_-=0, \qquad
    \gamma_+-\gamma_-=0, \qquad 
    \varphi_+-\varphi_-=0. 
\end{equation}
In addition to \eqref{submanifold 1-non zero values-1}, \eqref{submanifold 1-non zero values-2} and \eqref{submanifold 2-non zero values-1}, Killing spinors should also be required that
\begin{align}
    \label{submanifold 3-non zero values-1}
    & (M_{\beta_+}+M_{\beta_-}) \; \epsilon_1 =0\,,
\end{align}
to have common supersymmetries in this larger submanifold. Notice, however, that 
\begin{equation}
\label{relation between delocalized string constraints}
4M_{\beta_+} M_{\beta_-} =(4 M_{\gamma_+} M_{\gamma_-} ) (4M_{\varphi_+} M_{\varphi_-})\,,
\end{equation}
which implies that  \eqref{submanifold 3-non zero values-1} is not an independent constraint with respect to \eqref{submanifold 1-non zero values-1} and \eqref{submanifold 2-non zero values-1}. Then, a string smeared over the larger ${\cal M}_4$ will also be 1/8 BPS.

\section{BPS strings in AdS$_3\times$S$_+^3\times$S$_-^3\times$S$^1$: analysis of fluctuations}
\label{sec: BPS strings fluctuations}

In the previous section we found a rich variety of classical supersymmetric strings that correspond to BPS line defects in the dual CFT$_2$. In the following, we show how this abundance is manifested when studying the quadratic fluctuations around the classical string configuration \eqref{susyconfiguration}. In particular, we will study the existence of supersymmetric boundary conditions that could be imposed on the quadratic fluctuations. Some sets of supersymmetric boundary conditions that we will find account for the Dirichlet and smeared strings that we introduced in the previous section. Interestingly, our analysis will also uncover new types of BPS strings, and it will reveal interpolations that connect all the different BPS configurations.

\subsubsection*{Mass spectrum and supersymmetry transformations}

The mass spectrum for the fluctuations around the string presented in \eqref{susyconfiguration} was calculated in \cite{Hernandez:2019huf} for the mixed-flux type IIB string theory in $AdS_3\times S^3 \times T^4$. We will discuss here the extension of those results to the $AdS_3\times S^3_+ \times S^3_- \times S^1$ case, for which we present the detail computations in Appendix \ref{app: quadratic fluctuations}. As shown there, a quadratic expansion of the action around \eqref{susyconfiguration} yields seven massless scalars  corresponding to fluctuations in the $S^{3}_+ \times S^3_- \times S^1$ directions and one massive scalar $\phi^{\sf tr}$, with mass $m_{\sf tr}^2=\tfrac{2}{R^2}$,  for the unique fluctuation transverse  to the worldsheet in $AdS_3$. The classical solution is sitting at a point in each of the three-spheres, which breaks their $SO(4)$ isommetries down to $SO(3)$.  The fluctuations must therefore accommodate into representations of these residual symmetries. Three of the scalar fluctuations $\phi^{a}_{b}$ form a  $\square\!\square_+$ of the $SO(3)_+$\footnote{The $a$,$b$ indices take values $+$ or $-$ and $\phi^a_b$ is traceless.}, another three $\phi^{\dot{a}}_{\dot{b}}$ form a $\square\!\square_-$ of the $SO(3)_-$, and the remaining $\phi^{\sf tr}$ and $\phi^9$ are in the trivial representation of both $SO(3)$.

Concerning the fermionic fluctuations, after reducing them to $AdS_2$ spinors (see Appendix \ref{app: quadratic fluctuations}) we obtain four massless $\psi^{ a\dot{a}}$ spinors while the remaining four $\chi^{a\dot{a}}$ fields have masses $m_{F}=\tfrac{1}{R}$. These two sets of spinors transform in the
$\square_+\otimes\square_-$ of $SO(3)_+\times SO(3)_-$.

The action for the quadratic fluctuations can then be written as 
\begin{align}
\label{quadratic action expansion}
S= &-\frac{1}{2} \int d^2 \sigma \; \sqrt{-h} \,  \bigg( \partial_{\alpha} \phi^{\sf tr} \partial^{\alpha} \phi^{\sf tr} + 2 \cos^2 \vartheta (\phi^{\sf tr})^2 + 
\frac{1}{2}\partial_{\alpha} \phi^{a}_{b}
\partial^{\alpha} \phi^{b}_{a} 
+
\frac{1}{2}\partial_{\alpha} \phi^{\dot{a}}_{\dot{b}}
\partial^{\alpha} \phi^{\dot{b}}_{\dot{a}}
 \nonumber\\
&
+ \partial_{\alpha} \phi^{9} \partial^{\alpha} \phi^{9} \bigg)-\frac{i}2 \int d^2 \sigma\; \sqrt{-h} \left(\bar{\psi}_{{a}\dot{a}} \ \slash\!\!\!\!{\cal D}^{(2)}  \psi^{{a}\dot{a}}  +
    \bar{\chi}_{{a}\dot{a}} \left( \slash\!\!\!\!{\cal D}^{(2)}
    -\cos \vartheta \right)\chi^{{a}\dot{a}}\right)\,,
\end{align}
where $h_{\rm \alpha \beta}$ is the induced metric on the worldsheet, evaluated on the classical solution \eqref{susyconfiguration}, and $\mathcal{D}^{(2)}$ is the covariant derivative in $AdS_2$. In the following, the two-dimensional Dirac matrices will be denoted as $\uptau^0$ and $\uptau^1$, with $\uptau^3:=\uptau^0\uptau^1$ the chirality matrix.  The supersymmetry transformations of the action \eqref{quadratic action expansion}, derived in Appendix \ref{app: quadratic fluctuations},
 are
\begin{align}
\label{susy massive fermions-1}
\delta \chi^{a \dot{a}} &= \frac{1}{2} \left( \slashed{\partial} \phi^{\sf tr} + \cos \vartheta \, \phi^{\sf tr}  \right) \uptau^3 \kappa^{a\dot{a}} \nonumber \\
&\quad -\frac{i}{2} \left[ \sin \Omega \left( \slashed{\partial} \phi^{a}_{b} - \cos \vartheta \, \phi^{a}_{b} \right) \kappa^{b\dot{a}} + \cos \Omega \left( \slashed{\partial} \phi^{\dot{a}}_{\dot{b}} - \cos \vartheta \, \phi^{\rm \dot{a}}_{\dot{b}} \right) \kappa^{a\dot{b}} \right] \,, \\
\label{susy massless fermions-1}
\delta \psi^{a \dot{a}} &= \frac{1}{2} \slashed{\partial} \phi^{9} \kappa^{a \dot{a}}+\frac{i}{2} \left( \cos \Omega \; \slashed{\partial} \phi^{a}_{b} \kappa^{b \dot{a}}-\sin \Omega \; \slashed{\partial} \phi^{\dot{a}}_{\dot{b}} \kappa^{a \dot{b}} \right) \,, \\
\label{susy s1 scalar-1}
\delta \phi^{9} &=  -\frac{1}{2} \; \bar{ \psi}_{a \dot{a}}  \kappa^{a \dot{a}} \,, \\
\label{susy tr scalar-1}
\delta \phi^{\sf tr} &=  -\frac{1}{2} \; \bar{ \chi}_{\rm a \dot{a}} \uptau^3 \kappa^{a \dot{a}} \,, \\
\label{susy s3+ scalars-1}
\delta \phi^{a}_{b} &=-\frac{i}{2} \left[ \cos \Omega \left( 2 \; \bar{\psi}_{b\dot{c}} \kappa^{a\dot{c}} - \delta^{a}_{b} \bar{\psi}_{c\dot{c}} \kappa^{c\dot{c}} \right) -\sin \Omega \left(2 \; \bar{\chi}_{b\dot{c}} \kappa^{a\dot{c}} - \delta^{a}_{b} \bar{\chi}_{c\dot{c}} \kappa^{c\dot{c}} \right) \right] \,, \\
\label{susy s3- scalars-1}
\delta \phi^{\dot{a}}_{\dot{b}} &= \frac{i}{2} \left[ \sin \Omega \left( 2 \; \bar{\psi}_{c\dot{b}} \kappa^{c\dot{a}} - \delta^{\dot{a}}_{\dot{b}} \bar{\psi}_{c\dot{c}} \kappa^{c\dot{c}} \right) +\cos \Omega  \left(2 \; \bar{\chi}_{c\dot{b}} \kappa^{c\dot{a}} - \delta^{\dot{a}}_{\dot{b}} \bar{\chi}_{c\dot{c}} \kappa^{c\dot{c}} \right) \right] \,,
\end{align}
where $\kappa^{a \dot{a}}$ are $AdS_2$ Killing spinors that satisfy
\begin{equation}
 \kappa^{a\dot{a}} =  \sigma^{-1/2} \varepsilon^{a\dot{a}} 
 +\sigma^{1/2} \uptau^0 \dot{\varepsilon}^{a\dot{a}} \,,
\end{equation}
with $\varepsilon(\tau)$ a spinor such that 
\begin{equation}
\uptau^1 \varepsilon^{{a}\dot{a}} =- \varepsilon^{{a}\dot{a}}\,, \qquad \quad \ddot{\varepsilon}^{{a}\dot{a}}=0 \,.
\end{equation}
These constraints, in addition to the property $\left( \varepsilon \right)^*_{a \dot{a}}= \epsilon_{a b} \epsilon_{\dot{a} \dot{b}} \; \varepsilon^{b \dot{b}}$ (where $\epsilon_{\dot{a} \dot{b}}$ and $\epsilon_{\dot{a} \dot{b}}$ are Levi-Civita tensors), imply that $\varepsilon^{a\dot a}$ parametrizes eight real degrees of freedom, as expected.

\subsubsection*{Asymptotic behavior}

In order to search for supersymmetric boundary conditions appropriate to describe BPS strings we shall now derive the asymptotic expansion of the supersymmetry transformations \eqref{susy massive fermions-1} to \eqref{susy s3- scalars-1}. To that aim we shall perform a Wick rotation to Euclidean signature, i.e. we will work in EAdS$_2$\footnote{We perform the Wick rotation in order to get a unique solution to the equations of motion after fixing the boundary conditions at the boundary of the spacetime.}. In Appendix \ref{ch: scalar and spinor fields in AdS/CFT} we provide an analysis of the asymptotic behavior of scalar and spinor fields in EAdS$_{D+1}$, from which we obtain that the  $\sigma \to 0$ expansion of the fluctuations introduced above is
\begin{align}
\label{chi expansion}
\chi^{a\dot{a}}(\tau,\sigma) &= \sigma^{-1/2} \left( \alpha_{\chi}^{a\dot{a}}(\tau) -\sigma \uptau^3 \dot{\alpha}_{\chi}^{a\dot{a}}(\tau) + \dots \right) +  \sigma^{3/2} \left( \beta_{\chi}^{a\dot{a}}(\tau) + \tfrac{1}{3}\sigma  \uptau^3\dot{\beta}_{\chi}^{a\dot{a}}(\tau) + \dots \right) \,, \\
\label{psi expansion}
\psi^{a\dot{a}} (\tau,\sigma) &= \sigma^{1/2} \left( \alpha_{\psi}^{a\dot{a}} (\tau) +\sigma \uptau^3 \dot{\alpha}_{\psi}^{a\dot{a}} (\tau) + \dots \right)+ \sigma^{1/2} \left( \beta_{\psi}^{a\dot{a}} (\tau) + \sigma \uptau^3 \dot{\beta}_{\psi}^{a\dot{a}} (\tau) + \dots \right) \,,  \\
\label{theta expansion}
\phi^{9} (\tau,\sigma) & = \left( \alpha_{9} (\tau)+ \dots \right) + \sigma \left( \beta_{9} (\tau) + \dots \right) \,, \\
\label{tr expansion}
\phi^{\sf tr} (\tau,\sigma) &= \sigma^{-1} \left( \alpha_{\sf tr} (\tau) + \dots \right) + \sigma^2 \left( \beta_{\sf tr} (\tau) + \dots \right) \,, \\
\label{phi s3+ expansion}
\phi^{a}_{b} (\tau,\sigma) &= \left( \alpha^{a}_{b} (\tau)+ \dots \right)+ \sigma \left( \beta^{a}_{b} (\tau) + \dots \right) \,, \\
\label{phi s3- expansion}
\phi^{\dot{a}}_{\dot{b}} (\tau,\sigma) &= \left( \alpha^{\dot{a}}_{\dot{b}} (\tau)+ \dots \right) + \sigma \left( \beta^{\dot{a}}_{\dot{b}} (\tau) + \dots \right) \,,
\end{align}
where
\begin{align}
\uptau^{1} \alpha_{\chi}^{a\dot{a}}=- \alpha_{\chi}^{a\dot{a}} \,, \qquad \uptau^{1} \beta_{\chi}^{a\dot{a}}= \beta_{\chi}^{a\dot{a}} \,, \\
\uptau^{1} \alpha_{\psi}^{a\dot{a}}=- \alpha_{\psi}^{a\dot{a}}\,,  \qquad \uptau^{1} \beta_{\psi}^{a\dot{a}}= \beta_{\psi}^{a\dot{a}} \,.
\end{align}

Due to the Breitenlohner-Freedman bound\footnote{The Breitenlohner-Freedman bound establishes that only regular boundary conditions (i.e. $\alpha$-fixing conditions) can be imposed for scalar fields with $\frac{D^2}{4}+m_B^2 \geq 1$, where $D$ is the dimension of the boundary of EAdS$_{D+1}$ and $m_B$ is the mass of the field. However, more general boundary conditions that mix the $\alpha$ and $\beta$ modes of the field can be imposed if $0<\frac{D^2}{4}+m_B^2 \leq  1$. For the case of spinor fields one obtains that mixed boundary conditions can be imposed only if $|m_F| <\frac{1}{2}$, where $m_F$ is the mass of the field.}\cite{Breitenlohner:1982jf,Breitenlohner:1982bm} we must always take
\begin{align}
\label{bf bc 1}
\alpha_{\chi}^{a\dot{a}}=0 \quad \forall a,\dot{a} \,; 
\qquad
\alpha_{\sf tr}=0 \,,
\end{align}
as boundary conditions for the $\chi^{a\dot{a}}$ and $\phi^{\sf tr}$ fields. However, we have some freedom to impose more general boundary conditions on the other fluctuations. In the following, conditions such as $\alpha_{\chi}^{a\dot{a}}=0 \; \, \forall a,\dot{a}$ will be written simply as $\alpha_{\chi}^{a\dot{a}}=0$, without explicitly specifying that indices without contraction take all possible values.

Then, taking \eqref{bf bc 1} into consideration  
and inserting \eqref{chi expansion}-\eqref{phi s3- expansion} into the transformations \eqref{susy massive fermions-1}-\eqref{susy s3- scalars-1}, the transformations relevant to the analysis of supersymmetric boundary conditions become 
\begin{align}
\label{susy alpha chi}
\delta \alpha_{\chi}^{a\dot{a}} &= \frac{i \cos \vartheta}{2} \left( \sin \Omega \; \alpha^a_b \varepsilon^{b\dot{a}}+ \cos \Omega \; \alpha^{\dot{a}}_{\dot{b}} \varepsilon^{a\dot{b}} \right) \,,\hspace{8cm}\\
\label{susy alpha psi}
\delta \alpha_{\psi}^{a\dot{a}} &= - \frac{\cos \vartheta}{2} \beta_{9} \varepsilon^{a\dot{a}}  - \frac{i\cos \vartheta}{2} \left( \cos \Omega \; \beta^{a}_b \varepsilon^{b\dot{a}} - \sin \Omega \; \beta^{\dot{a}}_{\dot{b}} \varepsilon^{a\dot{b}} \right) \,, \\
\label{susy beta psi}
\delta \beta_{\psi}^{a\dot{a}} &= - \frac{\uptau^3\cos \vartheta}{2} \dot{\alpha}_{9} \varepsilon^{a\dot{a}} - \frac{i\cos \vartheta}{2} \uptau^3 \left( \cos \Omega \; \dot{\alpha}^{a}_b \varepsilon^{b\dot{a}} - \sin \Omega \; \dot{\alpha}^{\dot{a}}_{\dot{b}} \varepsilon^{a\dot{b}} \right) \,, \\
\label{susy alpha 9}
\delta \alpha_{9} & = -\frac{1}{2} \overline{\beta}_{\psi,a\dot{a}} \varepsilon^{a\dot{a}} \,,\\
\label{susy beta 9}
\delta \beta_{9} & = \frac{1}{2} \left( \dot{\overline{\alpha}}_{\psi,a\dot{a}} \uptau^3 \varepsilon^{a\dot{a}} + \overline{\alpha}_{\psi,a\dot{a}} \uptau^3 \dot{\varepsilon}^{a\dot{a}} \right) \,, \\
\label{susy alpha tr}
\delta \alpha_{\sf tr} &= 0 \,, \\
\label{susy alpha su2+}
\delta \alpha^a_b &= -\frac{i \cos \Omega}{2} \left( 2 \overline{\beta}_{\psi,b\dot{c}} \varepsilon^{a\dot{c}} -\delta^a_b \overline{\beta}_{\psi,c\dot{c}} \varepsilon^{c\dot{c}} \right) \,, \\
\label{susy beta su2+}
\delta \beta^a_b &= -\frac{i \cos \Omega}{2} \left[ -2 \left(  \dot{\overline{\alpha}}_{\psi,b\dot{c}} \uptau^3 \varepsilon^{a\dot{c}} + \overline{\alpha}_{\psi,b\dot{c}} \uptau^3 \dot{\varepsilon}^{a\dot{c}} \right) +\delta^a_b \left(  \dot{\overline{\alpha}}_{\psi,c\dot{c}} \uptau^3 \varepsilon^{c\dot{c}} + \overline{\alpha}_{\psi,c\dot{c}} \uptau^3 \dot{\varepsilon}^{c\dot{c}} \right) \right] \nonumber \\
& \hspace{0.40cm} +\frac{i \sin \Omega}{2} \left( 2 \overline{\beta}_{\chi,b\dot{c}} \varepsilon^{a\dot{c}}-\delta^a_b \overline{\beta}_{\chi,c\dot{c}} \varepsilon^{c\dot{c}} \right) \,,\\
\label{susy alpha su2-}
\delta \alpha^{\dot{a}}_{\dot{b}} &= \frac{i \sin \Omega}{2} \left( 2 \overline{\beta}_{\psi,c\dot{b}} \varepsilon^{c\dot{a}} -\delta^{\dot{a}}_{\dot{b}} \overline{\beta}_{\psi,c\dot{c}} \varepsilon^{c\dot{c}} \right) \,, \\
\label{susy beta su2-}
\delta \beta^{\dot{a}}_{\dot{b}} &=  \frac{i \sin \Omega}{2} \left[ -2 \left(  \dot{\overline{\alpha}}_{\psi,c\dot{b}} \uptau^3 \varepsilon^{c\dot{a}} + \overline{\alpha}_{\psi,c\dot{b}} \uptau^3 \dot{\varepsilon}^{c\dot{a}} \right) +\delta^{\dot{a}}_{\dot{b}} \left(  \dot{\overline{\alpha}}_{\psi,c\dot{c}} \uptau^3 \varepsilon^{c\dot{c}} + \overline{\alpha}_{\psi,c\dot{c}} \uptau^3 \dot{\varepsilon}^{c\dot{c}} \right) \right] \nonumber \\
& \hspace{0.40cm}  + \frac{i \cos \Omega}{2} \left( 2 \overline{\beta}_{\chi,c\dot{b}} \varepsilon^{c\dot{a}}-\delta^{\dot{a}}_{\dot{b}} \overline{\beta}_{\chi,c\dot{c}} \varepsilon^{c\dot{c}} \right) \,.
\end{align}

\subsubsection*{Dirichlet boundary conditions}

We will say that a set of boundary conditions is supersymmetric when the variation of them under \eqref{susy alpha chi}-\eqref{susy beta su2-} is vanishing for some  non-trivial choice of parameters $\varepsilon^{a\dot a}$.
It is straightforward to see from these expressions that Dirichlet boundary conditions
\begin{equation}
\alpha_{\chi}^{a\dot{a}}=  0\,, \qquad
\beta_{\psi}^{a\dot{a}} =  0\,, \qquad 
{\alpha}_{9} = 0  \,,\qquad
\alpha^{\dot{a}}_{\dot{b}} =  0\,, \qquad 
\alpha^a_b =  0\,,  \qquad  \alpha_{\sf tr} = 0 \,,
\end{equation}
preserve the full set of supersymmetries of the action (\ref{quadratic action expansion}), and thus matches the fact that the string given by (\ref{susyconfiguration}) is 1/2 BPS\footnote{All the supersymmetries of the action for quadratic fluctuations are one half of the supersymmetries of the $AdS_3 \times S^3_+ \times S^3_- \times S^1 $ background.}.  

\subsubsection*{Smeared boundary conditions}

Let us now study delocalized supersymmetric string configurations. In particular, we will first focus on the smeared strings which were introduced at the classical level in Section \ref{sec: classical BPS strings}, and we will look for ways to describe them in terms boundary conditions for the fluctuations.

As discussed in Section \ref{sec: classical BPS strings}, a smeared string is defined as a string whose partition function $Z_{\rm smeared}$ can be obtained from a weighted average of the partition functions $Z_{\rm Dirichlet}$ of Dirichlet strings that sit over a given manifold ${\cal M}$ of the background spacetime, see \eqref{partition function smearing}. Let us focus first on the case of strings delocalized over the submanifold $\mathcal{M}_1=S^1$. In terms of the fluctuations, imposing $\dot\alpha_9 = 0$ corresponds to a  \emph{uniform} smearing (that is, with a constant weight function in \eqref{partition function smearing}) over $\mathcal{M}_1$, given that it is equivalent to imposing $\alpha_9 = \text{const.}$ and integrating over the constant. Therefore, we propose that the boundary conditions for a string uniformly smeared over ${\cal M}_1$ should be
\begin{equation}
\label{smeared bc s1}
\alpha_{\chi}^{a\dot{a}}=  0\,, \qquad
\beta_{\psi}^{a\dot{a}} =  0\,, \qquad 
\dot{\alpha}_{9} = 0  \,,\qquad
\alpha^{\dot{a}}_{\dot{b}} =  0\,, \qquad 
\alpha^a_b =  0\,,  \qquad  \alpha_{\sf tr} = 0 \,.
\end{equation}
By looking at the transformations \eqref{susy alpha chi}-\eqref{susy beta su2-} we see that these boundary conditions are invariant under the action of the eight supercharges parametrized by the $\varepsilon^{a \dot{a}}$
spinors, in agreement with the results obtained in Section \ref{sec: classical BPS strings} when analyzing the classical limit of this delocalized string.

On the other hand, for the string smeared over the submanifold $\mathcal{M}_2 = S^1_D\times S^1$ given at \eqref{submanifold 1} we must impose now
\begin{equation}
\label{smeared bc 1-zeta fluctuations}
\zeta^{\beta_{\pm}}=0 \,, \qquad
\zeta^{\gamma_{\pm}}=0 \,, \qquad
\zeta^{\varphi_+}-\zeta^{\varphi_-}=0 \,, \qquad
\dot{\zeta}^{\theta}=0 \,,
\end{equation}
on the $\zeta^{\mu}$ fluctuations defined in Appendix \ref{app: quadratic fluctuations}. Using the $\phi^a_b$ and $\phi^{\dot{a}}_{\dot{b}}$ fields (see \eqref{definition phi su2+} and \eqref{definition phi su2-}) these boundary conditions can be expressed as
\begin{equation}
\label{smeared bc 1-scalar fluctuations}
\alpha^{1}_{2}=\alpha^{2}_{1}=0 \,, \qquad 
\alpha^{\dot{1}}_{\dot{2}}=\alpha^{\dot{2}}_{\dot{1}}=0 \,, \qquad 
\sin \Omega \; \alpha^{1}_{1}-\cos \Omega \; \alpha^{\dot{1}}_{\dot{1}}=0 \,, \qquad
\dot{\alpha}_9=0 \,,
\end{equation}
where the $\sin \Omega$ and $\cos \Omega$ factors come from the vielbein that relates the $\zeta^{\mu}$ fields with the scalar fluctuations (we have assumed the classical string to be sitting at the origin of the compact space). Conditions \eqref{smeared bc 1-scalar fluctuations} have to be supplemented with a condition that accounts for the smearing along the diagonal $S^1_D$. This condition is simply written as 
$\cos \Omega \; \dot{\alpha}^{1}_{1}+\sin \Omega \; \dot{\alpha}^{\dot{1}}_{\dot{1}}=0$.

Looking again at the transformations \eqref{susy alpha chi}-\eqref{susy beta su2-} we see that 
\begin{equation}
\label{smeared bc 1-complete}
\begin{aligned}
 \alpha_{\chi}^{a\dot{a}} &=0\,, \quad \; \,
 & \beta_{\psi}^{a\dot{a}}&=0\,, \quad & \alpha_{\sf tr}&=0\,, \\   \dot{\alpha}_{9}&=0 \,, \quad 
&\alpha^{1}_{2}=\alpha^{2}_{1} &=0 \,,  \qquad  &
\alpha^{\dot{1}}_{\dot{2}}=\alpha^{\dot{2}}_{\dot{1}}&=0 \,, \qquad \\
\sin \Omega \; \alpha^{1}_{1}-\cos \Omega \; \alpha^{\dot{1}}_{\dot{1}}&=0 \,, \qquad &
\cos \Omega \; \dot{\alpha}^{1}_{1}+\sin \Omega \; \dot{\alpha}^{\dot{1}}_{\dot{1}}&=0 \,,
\end{aligned}
\end{equation}
define a  set of boundary conditions which are invariant only under the supersymmetries parametrized by the spinors $\varepsilon^{1 \dot{2}}$ and $\varepsilon^{2 \dot{1}}$. This becomes clear when demanding that $\delta \alpha_{\chi}^{a \dot{a}}$ must be 0. 
Thus, the boundary conditions \eqref{smeared bc 1-complete} preserve 4 real supercharges, in agreement with the results presented in Section \ref{sec: classical BPS strings} for the 1/4 BPS classical string smeared over $\mathcal{M}_2$.

As for the string smeared over the submanifold $\mathcal{M}_3 = S^2_D\times S^1$ given at \eqref{submanifold 2}, in this case we must impose
\begin{equation}
\label{smeared bc 2-zeta fluctuations}
\begin{aligned}
& \zeta^{\beta_{\pm}}=0 \,, \qquad
\zeta^{\gamma_+}-\zeta^{\gamma_-}=0 \,, \qquad
\zeta^{\varphi_+}-\zeta^{\varphi_-}=0 \,, \qquad
\dot{\zeta}^{\theta}=0 \,,
\end{aligned}
\end{equation}
or, equivalently,
\begin{align}
\label{smeared bc 2-scalar fluctuations}
\alpha^{1}_{2}+ \alpha^{2}_{1}&=0 \,, \qquad 
&\alpha^{\dot{1}}_{\dot{2}}+ \alpha^{\dot{2}}_{\dot{1}}&=0 \,, \qquad &\dot{\alpha}_9&=0 \,, \nonumber \\
\sin \Omega \; \alpha^{1}_{2}-\cos \Omega \; \alpha^{\dot{1}}_{\dot{2}}&=0 \,, \quad
&\sin \Omega \; \alpha^{1}_{1}-\cos \Omega \; \alpha^{\dot{1}}_{\dot{1}}&=0 \,. & &
\end{align}
The conditions associated to the smearing are in this case $\cos \Omega \; \dot{\alpha}^{1}_{2}+\sin \Omega \; \dot{\alpha}^{\dot{1}}_{\dot{2}}=0$ and $\cos \Omega \; \dot{\alpha}^{1}_{1}+\sin \Omega \; \dot{\alpha}^{\dot{1}}_{\dot{1}}=0$. Then, the complete set of boundary conditions in this case is
\begin{equation}
\label{smeared bc 2-complete}
\begin{aligned}
\alpha_{\chi}^{a\dot{a}} &=0\,, \quad \; \,
 & \beta_{\psi}^{a\dot{a}}&=0\,, \quad & \alpha_{\sf tr}=0\,,   \qquad  \dot{\alpha}_{9}&=0 \,, 
 \\
\alpha^{1}_{2}+ \alpha^{2}_{1} &=0 \,, \qquad &
\sin \Omega \; \alpha^{1}_{2}-\cos \Omega \; \alpha^{\dot{1}}_{\dot{2}} &=0 \,, \qquad &
\cos \Omega \; \dot{\alpha}^{1}_{2}+\sin \Omega \; \dot{\alpha}^{\dot{1}}_{\dot{2}} &=0 \,, \\
\alpha^{\dot{1}}_{\dot{2}}+ \alpha^{\dot{2}}_{\dot{1}} &=0 \,,&
\sin \Omega \; \alpha^{1}_{1}-\cos \Omega \; \alpha^{\dot{1}}_{\dot{1}}&=0 \,, \quad &
\cos \Omega \; \dot{\alpha}^{1}_{1}+\sin \Omega \; \dot{\alpha}^{\dot{1}}_{\dot{1}}&=0 \,,
\end{aligned}
\end{equation}
and they are invariant under \eqref{susy alpha chi}-\eqref{susy beta su2-} for transformations
 which satisfy
\begin{equation}
\label{constraints susy 1/8 BPS}
\varepsilon^{1 \dot{1}}=\varepsilon^{2 \dot{2}}=0 \,, \qquad \varepsilon^{1 \dot{2}}=-\varepsilon^{2 \dot{1}} \;. 
\end{equation}
These constraints are solved by two real supercharges, and therefore the conditions \eqref{smeared bc 2-complete} match the results obtained previously for the classical string smeared over $\mathcal{M}_3$.

The boundary conditions \eqref{smeared bc 2-complete} can be slightly modified in order to describe the string smeared over the submanifold $\mathcal{M}_4 = S^3_D\times S^1$. In this case it suffices to impose
\begin{equation}
\label{smeared bc 3-complete}
\begin{aligned}
\alpha_{\chi}^{a\dot{a}} =0\,, \qquad \beta_{\psi}^{a\dot{a}}&=0\,, \quad & \alpha_{\sf tr}=0\,,   \qquad  \dot{\alpha}_{9}&=0 \,, 
\\
\sin \Omega \; \alpha^{1}_{2}-\cos \Omega \; \alpha^{\dot{1}}_{\dot{2}} &=0 \,, \qquad &
\cos \Omega \; \dot{\alpha}^{1}_{2}+\sin \Omega \; \dot{\alpha}^{\dot{1}}_{\dot{2}} &=0 \,, \\
\sin \Omega \; \alpha^{1}_{1}-\cos \Omega \; \alpha^{\dot{1}}_{\dot{1}}&=0 \,, \quad &
\cos \Omega \; \dot{\alpha}^{1}_{1}+\sin \Omega \; \dot{\alpha}^{\dot{1}}_{\dot{1}}&=0 \,,
\end{aligned}
\end{equation}
These conditions are again preserved only by the supersymmetries which solve \eqref{constraints susy 1/8 BPS}, in agreement with the results expected for the 1/8 BPS string smeared over $\mathcal{M}_4$.

\subsubsection*{Neumann boundary conditions}
\label{neumann bc}

The analysis of fluctuations not only appears to be consistent with the results obtained when studying classical strings, but also enables us to characterize some other delocalized supersymmetric strings not accounted in Section \ref{sec: classical BPS strings}. Instead of smearing over regions ${\cal M}\subset S^3\times S^3\times S^1$ we can also impose Neumann boundary conditions on the coordinates spanning them. In this section we will study in which cases those boundary conditions are supersymmetric.

We can start by considering the imposition of a Neumann boundary condition for the fluctuation along ${\cal M}_1$, with Dirichlet conditions on the rest of the directions. We get
\begin{equation}
\label{neumann bc s1}
  \beta_{9} = 0  \,,
   \qquad  \alpha^a_b =  0\,,  \qquad  \alpha^{\dot{a}}_{\dot{b}}=  0\,, 
 \qquad  \alpha_{\sf tr} = 0 \,.
\end{equation}
A quick inspection of the transformations \eqref{susy alpha chi} to \eqref{susy beta su2-} reveals that 
\eqref{neumann bc s1} cannot preserve all the supersymmetries. In order to have $\delta\beta_9 = 0$ with the most general parameters the vanishing of all  $\alpha_\psi^{a\dot{a}}$ would be needed. 
The variation of the latter depend on $\beta^a_b$ and
$\beta^{\dot a}_{\dot b}$, which are not necessarily vanishing when imposing Dirichlet boundary conditions on $\phi^a_b$ and $\phi^{\dot a}_{\dot b}$. 

Nonetheless, \eqref{neumann bc s1} can preserve some fraction of the supersymmetry by demanding
\begin{equation}
\label{neumann bc s1fer}
\alpha_{\chi}^{a\dot{a}} = 0\,,\qquad
\alpha_{\psi}^{1\dot{2}}-\alpha_{\psi}^{2\dot{1}}=0\,,
\qquad
\beta_{\psi}^{1\dot{2}}+\beta_{\psi}^{2\dot{1}}=0\,,
\qquad 
\beta_{\psi}^{1\dot{1}}=\beta_{\psi}^{2\dot{2}}=0\,.
\end{equation}
for the fermionic fluctuations. In particular, \eqref{neumann bc s1} and \eqref{neumann bc s1fer} are invariant under transformations with $\varepsilon^{1 \dot{1}}=\varepsilon^{2 \dot{2}}=0$  and $ \varepsilon^{1 \dot{2}}=-\varepsilon^{2 \dot{1}}$. This example represents a notorious difference between the $AdS_3 \times S^3_+ \times S^3_- \times S^1$ and the $AdS_4 \times \mathbb{CP}^3$ cases. We see that in this case imposing $\dot{\alpha}=0$ or $\beta=0$ on the fluctuations of a given submanifold does not preserve the same amount of supersymmetry. This reinforces the proposal which states that the delocalized configurations presented in Section \ref{sec: classical BPS strings} are accounted in terms of boundary conditions that fix $\dot{\alpha}$ for the corresponding scalar fluctuations.

Imposing Neumann boundary conditions for the fluctuation along ${\cal M}_2$ can lead to supersymmetry invariance if some fermionic boundary conditions are accordingly changed. More precisely, from \eqref{susy alpha chi} to \eqref{susy beta su2-} we find that the boundary conditions
\begin{equation}
\label{extra bc 1-complete}
\begin{aligned}
  \alpha_{\sf tr}&=0\,,   \quad & \beta_{9}&=0 \,,
  & \alpha_{\chi}^{a\dot{a}} &=0\,, 
 \\
\alpha^{1}_{2}=\alpha^{2}_{1} &=0 \,,  \quad  &
\alpha^{\dot{1}}_{\dot{2}}=\alpha^{\dot{2}}_{\dot{1}}&=0 \,, \quad &\beta_{\psi}^{1\dot{1}}=\beta_{\psi}^{2\dot{2}}=\alpha_{\psi}^{1\dot{2}}=\alpha_{\psi}^{2\dot{1}}&=0\,, \\ 
\sin \Omega \; \alpha^{1}_{1}-\cos \Omega \; \alpha^{\dot{1}}_{\dot{1}}&=0 \,,  &
\cos \Omega \; \beta^{1}_{1}+\sin \Omega \; \beta^{\dot{1}}_{\dot{1}}&=0 \,,
\end{aligned}
\end{equation}
are invariant under the supersymmetries parametrized by $\varepsilon^{1\dot{2}}$ and $\varepsilon^{2\dot{1}}$, and thus describe a 1/4 BPS string. In contrast to the previous example, in this case the supersymmetries preserved are exactly the same as 
the ones preserved in the smearing over ${\cal M}_2$.

When turning to Neumann boundary conditions in ${\cal M}_3$, we can impose
\begin{equation}
\label{extra bc M3}
\begin{aligned}
\alpha_{\chi}^{a\dot{a}}&=0 \,, \qquad &\alpha_{\psi}^{1\dot{2}}=\alpha_{\psi}^{2\dot{1}}&=0 \,,\\
\alpha_{\psi}^{1\dot{1}}+\alpha_{\psi}^{2\dot{2}}&=0 \,, \qquad &\beta_{\psi}^{1\dot{1}}-\beta_{\psi}^{2\dot{2}}&=0 \,,\\
\alpha^1_2+\alpha^2_1=\alpha^{\dot{1}}_{\dot{2}}+\alpha^{\dot{2}}_{\dot{1}}&=0 \,, \qquad
&\alpha_{\sf tr}=0 \,, \quad \beta_9&=0 \,,\\
\sin \Omega  \left( \alpha^{1}_{2}- \alpha^{2}_{1} \right) - \cos \Omega \left( \alpha^{\dot{1}}_{\dot{2}} - \alpha^{\dot{2}}_{\dot{1}} \right) &=0 \,, \qquad
&\sin \Omega \; \alpha^{1}_{1} - \cos \Omega \; \alpha^{\dot{1}}_{\dot{1}}&=0 \,,\\
\cos \Omega  \left( \beta^{1}_{2}- \beta^{2}_{1} \right) + \sin \Omega \left( \beta^{\dot{1}}_{\dot{2}} - \beta^{\dot{2}}_{\dot{1}} \right) &=0 \,, \qquad
&\cos \Omega \; \beta^{1}_{1} + \sin \Omega \; \beta^{\dot{1}}_{\dot{1}}&=0 \,.
\end{aligned}
\end{equation}
These conditions are preserved by the supersymmetries which satisfy \eqref{constraints susy 1/8 BPS}. Therefore, they describe a 1/8 BPS string, as in the case presented in \eqref{smeared bc 3-complete} for the smearing over $\mathcal{M}_3$.

Our final example is the case of Neumann boundary conditions for the fluctuations along ${\cal M}_4$. In that case, the conditions
\begin{equation}
\label{extra bc 2-complete}
\begin{aligned}
\alpha_{\chi}^{a\dot{a}} =0\,, \quad \; \,
 \alpha_{\psi}^{a\dot{a}}&=0\,, \quad & \alpha_{\sf tr}=0\,,   \qquad  \beta_{9}&=0 \,, 
\\
\sin \Omega \; \alpha^{1}_{2}-\cos \Omega \; \alpha^{\dot{1}}_{\dot{2}} &=0\,, \qquad &
\sin \Omega \; \alpha^{1}_{1}-\cos \Omega \; \alpha^{\dot{1}}_{\dot{1}} &=0\,, \qquad &
\\
\cos \Omega \; \beta^{1}_{2}+\sin \Omega \; \beta^{\dot{1}}_{\dot{2}} &=0 \,, &
\cos \Omega \; \beta^{1}_{1}+\sin \Omega \; \beta^{\dot{1}}_{\dot{1}} &=0 \,, &
\end{aligned}
\end{equation}
are preserved only by the supersymmetries which satisfy \eqref{constraints susy 1/8 BPS}, and thus correspond to a 1/8 BPS string.

\subsubsection*{Interpolating strings}

In the previous sections we have found diverse sets of 1/2 BPS, 1/4 BPS and 1/8 BPS strings which are described either by Dirichlet, smeared or Neumann boundary conditions. As we shall see in the following, there is a network of supersymmetric boundary conditions that allows us to interpolate between all those BPS strings.

Let us begin our analysis by studying the interpolation between the Dirichlet and smeared strings discussed in the previous sections. For this it is useful to consider first the case of the string smeared over $\mathcal{M}_1=S^1$, and to note that a Dirichlet string can be thought as a limiting case of such delocalized configuration. For the uniformly smeared string the partition function can be obtained by taking Dirichlet strings with boundary condition $\phi^9=\text{const.}$ and then integrating over the values of such constant (see eq. \eqref{partition function smearing}).  We could in principle perform this integration with an arbitrary weight function and the configuration would continue to be supersymmetric. We can then think of  the Dirichlet string as the limit in which the region of support of the weight function collapses to a point. Thus, by smoothly deforming the domain of support we can interpolate between the Dirichlet string and the string uniformly smeared over $\mathcal{M}_1$. This idea can be generalized to interpolate between any of the strings considered in Section \ref{sec: classical BPS strings}. These configurations preserve the same amount of supersymmetry as the least supersymmetric endpoint of the interpolation.

We can also interpolate between smeared and Neumann strings. This is similar to the mixed boundary conditions presented in \cite{Correa:2019rdk} for the $AdS_4 \times \mathbb{CP}^3$ case. For example, we can interpolate between the smeared and Neumann boundary conditions over $\mathcal{M}_1$ if we impose
\begin{equation}
\label{interpolating bc M1}
\begin{aligned}
\alpha_{\chi}^{a\dot{a}}&=  0\,, \qquad &\beta_{\psi}^{1\dot{1}}=\beta_{\psi}^{2\dot{2}}&=0\,, \\
\beta_{\psi}^{1\dot{2}}+\beta_{\psi}^{2\dot{1}}&=0\,, \qquad &\left( \Lambda \, \alpha_{\psi}^{1\dot{2}} +\uptau^3 \beta_{\psi}^{1\dot{2}} \right)-\left( \Lambda \, \alpha_{\psi}^{2\dot{1}} +\uptau^3 \beta_{\psi}^{2\dot{1}} \right)&=0\,, \\
\alpha^a_b =\alpha^{\dot{a}}_{\dot{b}} =  0\,, \quad \alpha_{\sf tr} &= 0 \,, \qquad &\Lambda \, \beta_{9} + \dot{\alpha}_{9} &= 0  \,,
\end{aligned}
\end{equation}
where $\Lambda \in \mathbb{R}$. As expected, this set of conditions is invariant under two of the supersymmetries of the action \eqref{quadratic action expansion} for $\Lambda>0$, and in the limit $\Lambda=0$ preserves all the supersymmetries.

\begin{figure}
\begin{center}
\small{
\begin{tikzpicture}
\node [BPSbox] at (1.05,0) {\small Dirichlet string \\ {\bf 1/2 BPS}};
\node [BPSbox2] at (3.5,1.2) {\small Smearing over ${\cal M}_1$ \\ {\bf 1/2 BPS}};
\node [BPSbox2] at (6.9,1.2) {\small Smearing over ${\cal M}_2$ \\ {\bf 1/4 BPS}};
\node [BPSbox2] at (10.3,1.2) {\small Smearing over ${\cal M}_3$ \\ {\bf 1/8 BPS}};
\node [BPSbox2] at (13.7,1.2) {\small Smearing over ${\cal M}_4$ \\ {\bf 1/8 BPS}};
\node [BPSbox3] at (3.5,-1.2) {\small Neumann on ${\cal M}_1$ \\ {\bf 1/8 BPS}};
\node [BPSbox3] at (6.9,-1.2) {\small Neumann on ${\cal M}_2$ \\ {\bf 1/4 BPS}};
\node [BPSbox3] at (10.3,-1.2) {\small Neumann on ${\cal M}_3$ \\ {\bf 1/8 BPS}};
\node [BPSbox3] at (13.7,-1.2) {\small Neumann on ${\cal M}_4$ \\ {\bf 1/8 BPS}};
\draw[<->,thick] (4.8,1.2)--(5.6,1.2);
\draw[<->,thick] (8.2,1.2)--(9,1.2);
\draw[<->,thick] (11.6,1.2)--(12.4,1.2);
\draw[<->,thick] (4.8,-1.2)--(5.6,-1.2);
\draw[<->,thick] (8.2,-1.2)--(9,-1.2);
\draw[<->,thick] (11.6,-1.2)--(12.4,-1.2);
\draw[<->,thick] (1.05,0.90)--(2.2,1.25);
\draw[<->,thick] (1.05,-0.90)--(2.2,-1.25);
\draw[<->,thick] (3.5,0.35)--(3.5,-0.35);
\draw[<->,thick] (6.9,0.35)--(6.9,-0.35);
\draw[<->,thick] (10.3,0.35)--(10.3,-0.35);
\draw[<->,thick] (13.7,0.35)--(13.7,-0.35);
\end{tikzpicture}
}
\end{center}
\normalsize{
\caption{Examples of BPS string configurations. Arrows indicate BPS interpolations between them. The submanifolds are ${\cal M}_1 = S^1$, ${\cal M}_2 = S^1_D\times S^1$, ${\cal M}_3 = S^2_D\times S^1$ and ${\cal M}_4 = S^3_D\times S^1$.}
\label{resumefig}
}
\end{figure}
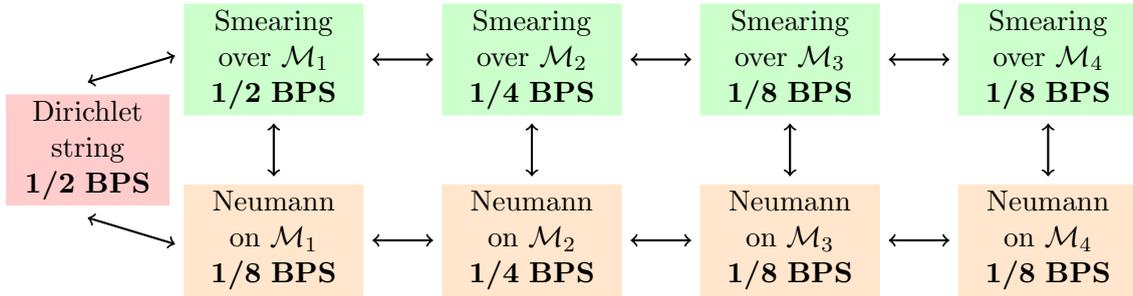

In a similar way we can interpolate between all the smeared and Neumann boundary conditions presented in the previous sections. All those interpolating conditions preserve the same amount of supersymmetry as the least supersymmetric end of the interpolation. 
We summarize the different BPS boundary conditions and the corresponding interpolations in Figure \ref{resumefig}.

Let us conclude this section by noticing that the possibility of imposing boundary conditions other than Dirichlet is intimately related to the presence of massless fermionic degrees of freedom within the spectrum of fluctuations. The same result was obtained for the fluctuations over the 1/2 BPS string of ABJM \cite{Correa:2019rdk}. A key difference between that case and the one studied here lies in the larger number of massless fermionic modes that is associated to the string \eqref{susyconfiguration}, which unveils a richer family of supersymmetric boundary conditions.

\section{1/2 BPS defects in the AdS$_3 \times$S$^3\times$T$^4$ limit}
\label{sec: algebra and repr}

In the previous sections we revealed that type IIB supergravity backgrounds with $AdS_3 \times S^3_+ \times S^3_- \times S^1$ metric and mixed three-form flux admit an abundant variety of BPS string configurations ending along a line at the boundary of the $AdS_3$ space (see Fig.  \ref{resumefig}). These strings are in correspondence with supersymmetric line defects in the holographic CFT$_2$, which preserve different subgroups of the $D(2,1, \sin^2 \Omega)^2 \times U(1)$ symmetry group of the bulk theory. It is natural to expect that these preserved symmetries impose constraints on the correlators of the corresponding dCFTs. In the rest of this chapter we will study those constraints in more detail, and to that aim we will focus the discussion on the 1/2 BPS Dirichlet string found in the previous sections (see Fig.  \ref{resumefig}). To simplify the analysis we will take the $\Omega \to \frac{\pi}{2}$ limit, in which the supergravity background is described by an $AdS_3 \times S^3 \times T^4$ metric, mixed three-form flux, and a constant dilaton. In this limit the symmetry group of the 1/2 BPS line defects reduces to $PSU(1,1|2)\times SU(2)_A$, where $SU(2)_A$ is an automorphism of the superconformal $PSU(1,1|2)$ factor.

We will perform an analytic conformal bootstrap computation of two-, three- and four-point functions of insertions along the contour of the aforementioned 1/2 BPS defects. As discussed at the beginning of this chapter, bootstrap methods are particularly useful in the context of the AdS$_3$/CFT$_2$ dualities that we are studying, given the absence of a known lagrangian description of the bulk CFT$_2$ for arbitrary values of the $\vartheta$ parameter. Moreover, the 1/2 BPS defects that we will study provide interesting toy models to further develop the analytic conformal bootstrap techniques, given that they are invariant under only four supercharges (while previously studied examples in ${\cal N}=4$ sYM and ABJM are invariant under eight and six supercharges, respectively). Furthermore, these defects define a two-parameter bootstrap problem, given that the theory is characterized both by the 't Hooft coupling and by the $\vartheta$ parameter, in contrast to the single-parameter cases observed for line defects in ${\cal N}=4$ sYM and ABJM.

In this section we will introduce and describe the displacement and tilt supermultiplets of the 1/2 BPS defects, which are the multiplets that we will study with the bootstrap techniques. We refer to Appendix \ref{app: psu(1,1|2)} for a detailed analysis of the $\mathfrak{psu}(1,1|2)\times \mathfrak{su}(2)_A$ symmetry algebra, together with a description of its corresponding representation theory and a derivation of the existence of a topological sector for correlators of short supermultiplets.

\subsubsection*{Displacement and tilt supermultiplets}

In order to apply the analytic conformal bootstrap program we need to first specify which operators we will insert in the correlation functions that we will study. To that aim we shall consider how the symmetries of the bulk CFT$_2$ break as a consequence of the presence of the line defect. More precisely, we will be interested in the operators with protected scaling dimensions that are associated to those broken CFT$_2$ symmetries. Following the discussion of Chapter \ref{ch: line defects}, this will allow us to define \textit{displacement} and \textit{tilt} supermultiplets.

The most obvious effect of the presence of a line defect is the breaking of translation invariance, which is described in the Ward identity \cite{McAvity:1993ue}
\begin{equation}
\label{ward translations}
    \partial_\mu T^{\mu x}(t,x)= \delta(x) \rho(t) \,,
\end{equation}
where $\mu=t,x$, $T^{\mu\nu}$ is the stress-energy tensor of the bulk CFT$_2$ and $\rho$ is the \textit{displacement operator}. In the following it will prove useful to use the notation introduced in Appendix \ref{app: psu(1,1|2)} for the basis of states in the Hilbert space of the dCFT$_1$: we will write them as $|\Delta,r,\ell\rangle$, where $\Delta$ is the scaling dimension and $r$ and $\ell$ are the R-symmetry and automorphism charges, respectively. With this notation we see that the displacement operator $\rho$ can be identified with a $|2,0,0 \rangle$ state. Moreover, since $\rho$ must have protected quantum numbers, it must fit within a short representation. These multiplets are presented in eq. \eqref{short multiplets} of Appendix \ref{app: psu(1,1|2)}, from which we see that the only short multiplet that can account for a $|2,0,0 \rangle$ state is 
\begin{equation}
\mc{B}^{1}_{0}: [1,2,0]\rightarrow [\frac 32,1,1]\rightarrow[2,0,0] \,.
\end{equation}
The latter can therefore be defined as the \textit{displacement supermultiplet}. One can relate the remaining operators in this supermultiplet with those associated to the breaking of supersymmetry and R-symmetry. Indeed, the breaking $SO(4) \rightarrow SU(2)_R$  of the R-symmetry implies
\begin{equation}
\label{ward R symm}
    \partial_\mu J^{\mu ab}(t,x)= \delta(x) \, \upsilon^{ab}(t) \,,
\end{equation}
where $J^{\mu ab}$ is the broken R-symmetry current and $\upsilon^{ab}$ is an operator of dimension $\Delta=1$, which we can identify with the superprimary $|1,2,0 \rangle$ of the displacement supermultiplet. Let us emphasize that $\upsilon^{ab}$ is symmetric in its indices. As for the $\chi^{a\dot{a}}$ fermion described by the $|\tfrac{3}{2},1,1\rangle$ state, it can be similarly associated to the breaking of supersymmetry. Therefore, we see that the displacement supermultiplet is related to the breaking $PSU(1,1|2)^2\to PSU(1,1|2) $. The supersymmetry transformations of the fields are
\begin{align}
    Q_{a\dot{a}} \upsilon_{bc}&= \frac{1}{2} \epsilon_{ab} \chi_{c\dot{a}} +\frac{1}{2} \epsilon_{ac} \chi_{b\dot{a}} \,, \nonumber\\
    Q_{a\dot{a}} \chi_{b\dot{b}}&=  -\epsilon_{\dot{a}\dot{b}} \partial_t\upsilon_{ab} + \epsilon_{ab}\epsilon_{\dot{a}\dot{b}} \rho \,, \nonumber\\
    Q_{a\dot{a}} \rho&= \frac{1}{2} \partial_t \chi_{a \dot{a}} \,.
\end{align}

On the other hand, the Ward identity that describes the breaking $SO(4)\times U(1) \rightarrow SU(2)_A$ of the automorphism
\begin{equation}
\label{ward automorphism}
    \partial_\mu W^{\mu \dot{a}\dot{b}}(t,x)= \delta(x) \varphi^{\dot{a}\dot{b}}(t) \,,
\end{equation}
allows us to define a $\varphi^{\dot{a}\dot{b}}$ operator, which is identified with a $|1,0,2 \rangle \oplus |1,0,0 \rangle$ state. Above we are introducing the automorphism currrent $W^{\mu \dot{a} \dot{b}}$. Let us note that $\varphi^{\dot{a}\dot{b}}$ is not symmetric, which comes from the fact that we have a broken $U(2)$ subgroup of the automorphism. We can associate this operator with a descendant of a $\mc{B}^{1/2}_{1}$ supermultiplet
\begin{align}
    \mc{B}^{1/2}_{1}:\qquad {[\frac 12,1,1]}\rightarrow   [1,0, 2] \oplus [1,0, 0] \,,
\end{align}
which we will refer to as the \textit{tilt supermultiplet}. The remaining states in this supermultiplet can be combined into a $\psi_{a\dot{a}}$ spinor. The corresponding supersymmetry transformations are
\begin{align}
    Q_{a\dot{a}} \psi_{b\dot{b}}&=  \epsilon_{ab} \varphi_{\dot{a}\dot{b}} \,, \nonumber\\
     Q_{a\dot{a}} \varphi_{\dot{b}\dot{c}}&= -\, \epsilon_{\dot{a}\dot{b}} \partial_t\psi_{a\dot{c}} \,.
\end{align}

Finally, let us notice that the displacement and tilt supermultiplets combine into one bigger supermultiplet in the more general $\mathfrak{d}(2,1;\sin^2{\Omega})$ algebra, which for $\Omega \rightarrow 0$ or $\Omega \rightarrow \frac{\pi}{2}$ reduces to the $\mathfrak{psu}(1,1|2)\times \mathfrak{su}(2)_A$ algebra of our defect.

\section{Two- and three-point functions}
\label{sec: 2 and 3 pts}

Correlation functions are strongly constrained by superconformal symmetry. For two- and three-point functions, conformal symmetry completely fixes the time dependence of correlators up to overall constants that depend on the operators included in the correlation function. On top of this, for correlators of fields belonging to the same supermultiplet those constants are further related by supersymmetry.
In this section we will apply supersymmetry to relate the two- and three-point functions of operators in the tilt and displacement supermultiplets.

\subsubsection*{Tilt supermultiplet}

Let us begin with the tilt supermultiplet. The most general ansatz that one can in principle write for the two-point functions is 
\begin{align}
     \langle \psi_{a\dot{a}}(t_1) \psi_{b\dot{b}}(t_2)\rangle &=\frac{\epsilon_{ab}\epsilon_{\dot{a}\dot{b}}}{t_{12}}  \,, \nonumber \\
     \langle \varphi_{\dot{a}\dot{b}}(t_1)\varphi_{\dot{c} \dot{d}}(t_2) \rangle &=  \frac{\kappa_1 \, \epsilon_{\dot{a}\dot{c}}\epsilon_{\dot{b}\dot{d}}+\kappa_2 \,\epsilon_{\dot{a}\dot{d}}\epsilon_{\dot{b}\dot{c}}+\kappa_3 \,\epsilon_{\dot{a}\dot{b}}\epsilon_{\dot{c}\dot{d}}}{t_{12}^2} \,,
\end{align}
where we have arbitrarily chosen a unit normalization for the superprimary. We can fix the unknown constants in the above ansatz by noticing that
\begin{equation}
0= Q_{a \dot{a}} \langle \psi_{b\dot{b}}(t_1) \phi_{\dot{c}\dot{d}}(t_2)\rangle=\epsilon_{a b} \langle \varphi_{\dot{a}\dot{b}}(t_1) \varphi_{\dot{c}\dot{d}}(t_2)\rangle + \epsilon_{\dot{a}\dot{c}}  \langle \psi_{b\dot{b}}(t_1) \psi_{a\dot{d}}(t_2)\rangle \,,
\end{equation}
which gives us
\begin{align}
\label{tilt2pt}
     \langle \psi_{a\dot{a}}(t_1) \psi_{b\dot{b}}(t_2)\rangle &=\frac{\epsilon_{ab}\epsilon_{\dot{a}\dot{b}}}{t_{12}}  \,,  \nonumber \\
     \langle \varphi_{\dot{a}\dot{b}}(t_1)\varphi_{\dot{c} \dot{d}}(t_2) \rangle &=  \frac{\epsilon_{\dot{a}\dot{c}}\epsilon_{\dot{b}\dot{d}}}{t_{12}^2} \,.
\end{align}

Applying the same procedure to determine relations among the three-point functions of operators in the tilt supermultiplet one obtains that all the three-point functions of the this supermultiplet are vanishing. This is a natural consequence of the vanishing of the three-point function
of the $\psi_{a\dot{a}}$ superprimary.

\subsubsection*{Displacement supermultiplet}

Let us now turn to the two- and three-point functions of the displacement multiplet. By imposing supersymmetry we can constrain the corresponding two-point functions to be 
\begin{align}
    \langle \upsilon_{ab}(t_1)\upsilon_{cd}(t_2) \rangle &=  -\frac{1}{2} \frac{\epsilon_{ac}\epsilon_{bd}+\epsilon_{ad}\epsilon_{bc}}{t_{12}^2}  \,,  \nonumber \\
     \langle \chi_{a\dot{a}}(t_1) \chi_{b\dot{b}}(t_2)\rangle &=2\frac{\epsilon_{ab}\epsilon_{\dot{a}\dot{b}}}{t_{12}^3}  \,,  \nonumber \\
     \langle \rho(t_1)\rho(t_2) \rangle &= \frac{3}{t_{12}^4}  \,,
     \label{rhonorm}
\end{align}
where again we have arbitrarily fixed the normalization of the superprimary. As for the 
three-point functions, we get
\begin{align}
    \langle \upsilon_{ab}(t_1)\upsilon_{cd}(t_2) \upsilon_{ef}(t_3) \rangle &=  \frac{\sigma \,(\epsilon_{ad}\epsilon_{cf}\epsilon_{eb}-\epsilon_{af}\epsilon_{cb}\epsilon_{ed})}{t_{12} t_{23} t_{13}} \,, \nonumber \\
     \langle \rho(t_1)\rho(t_2) \rho(t_3) \rangle &=\frac{6\, \sigma}{t_{12}^2t_{23}^2t_{13}^2} \,, \nonumber \\
     \langle \upsilon_{ab}(t_1)\upsilon_{cd}(t_2) \rho(t_3) \rangle &= \frac{-\frac{\sigma}{2} (\epsilon_{ac}\epsilon_{bd}+\epsilon_{bc}\epsilon_{ad})}{t_{23}^2t_{13}^2} \,, \nonumber \\
      \langle \chi_{a\dot{a}}(t_1) \chi_{b\dot{b}}(t_2) \upsilon_{cd}(t_3) \rangle &= -\frac{\sigma \, (\epsilon_{ac} \epsilon_{bd}\epsilon_{\dot{a}\dot{b}}+\epsilon_{ad} \epsilon_{bc}\epsilon_{\dot{a}\dot{b}})}{t_{12}^2t_{23}t_{13}} \,, \nonumber \\
        \langle \chi_{a\dot{a}}(t_1) \chi_{b\dot{b}}(t_2) \rho(t_3) \rangle &= \frac{3 \sigma \, \epsilon_{ab}\epsilon_{\dot{a}\dot{b}}}{t_{12}t_{23}^2t_{13}^2} \,,
        \label{3ptdispl}
\end{align}
where $\sigma$ is a constant that cannot be fixed with superconformal symmetry.

\section{Four-point functions: preliminaries}
\label{sec: 4 pts preliminaries}

In the following we will turn to the computation of the four-point functions of fields within the displacement and tilt supermultiplets. As it is well known, four-point functions in one-dimensional CFTs are constrained by conformal symmetry up to functions of a conformal cross ratio. In turn, supersymmetry imposes non-trivial constraints between correlators of fields that belong to a same supermultiplet. In particular, in this section we will show that we can use supersymmetry to express every correlator of insertions within the displacement supermultiplet in terms of two unknown cross-ratio functions and a constant. Similarly, correlators of the tilt supermultiplet will be completely determined by two cross-ratio functions and two constants. Moreover, we will discuss the OPE channels associated to each of the four-point correlators, and the superconformal block expansions that can be derived from them. The results presented in this section will be crucial in the bootstrap analysis that we will discuss in the next section.

\subsubsection*{Constraints from supersymmetry}

Let us begin by studying the constraints that supersymmetry imposes within four-point functions of fields that belong to a same supermultiplet. In particular, we will begin our analysis by focusing on the case of the displacement supermultiplet.

To write down the most general four-point correlators of components of the displacement multiplet it is useful to define the following tensor structures 
\begin{align}
S^{(2)}_{ab,cd} &\ := \epsilon_{ac}\epsilon_{bd}+\epsilon_{ad}\epsilon_{bc}\,,
\nonumber\\
S^{(3)}_{ab,cd,ef} &\ := \epsilon_{ad}\epsilon_{cf}\epsilon_{eb}+\epsilon_{af}\epsilon_{bc}\epsilon_{ed}\,.
\end{align}
In terms of them, we get that R-symmetry implies, for example
\begin{align}
\langle \upsilon_{a_1b_1}(t_1)\upsilon_{a_2b_2}(t_2)\upsilon_{a_3b_3}(t_3)\upsilon_{a_4b_4}(t_4)\rangle
= & \ \frac{1}{t_{12}^2t_{34}^2}\left( 
S^{(2)}_{a_1b_1,a_2b_2}\, S^{(2)}_{a_3b_3,a_4b_4} \; d_1(\chi)\right.
\nonumber\\
& \hspace{-4cm}
\left. + S^{(2)}_{a_1b_1,a_3b_3}\, S^{(2)}_{a_2b_2,a_4b_4} \; d_2(\chi)
 +       S^{(2)}_{a_1b_1,a_4b_4}\, S^{(2)}_{a_2b_2,a_3b_3} \; d_3(\chi)\right) \,,
 \label{phi4}
\\
\langle \upsilon_{a_1b_1}(t_1)\upsilon_{a_2b_2}(t_2)\chi_{a_3\dot a_3}(t_3)\chi_{a_4\dot a_4}(x_4)\rangle
= & \ \frac{\epsilon_{\dot a_3\dot a_4}}{t_{12}^2t_{34}^3}\left(S^{(2)}_{a_1b_1,a_2b_2}\,\epsilon_{a_3 a_4} \; d_4(\chi)
\right.
\nonumber\\
& \left. 
\hspace{-4cm} +(S^{(2)}_{a_1b_1,a_2a_3}\,\epsilon_{b_2 a_4} + S^{(2)}_{a_1b_1,b_2a_3}\,\epsilon_{a_2 a_4} ) \; d_5(\chi) \right) \,,
\nonumber\label{phi2chi2}
\\
\nonumber \langle \upsilon_{a_1b_1}(t_1)\upsilon_{a_2b_2}(t_2)\upsilon_{a_3b_3}(t_3)\rho(t_4)\rangle
= & \ \frac{1}{t_{12}t_{14}t_{24}t_{34}^2}
S^{(3)}_{a_1b_2,a_2b_2,a_3b_3} d_6(\chi) \,,
\end{align}
where 
\begin{equation}
\chi= \frac{t_{12}t_{34}}{t_{13}t_{24}} \,,
\end{equation}
is a spacetime cross ratio. Similarly, we can use R-symmetry to constrain all the remaining correlators of components of the displacement supermultiplet. We get that they can be expressed in terms of a total of 14 functions of the cross ratio $\chi$. As we have done for two- and three-point functions, we can find relations between these functions by considering the supersymmetry transformations of trivial four-point functions. For example, by studying the behaviour under supersymmetry  of $\langle \upsilon_{a_1b_1}(t_1)\upsilon_{a_2b_2}(t_2)\upsilon_{a_3 b_3}(t_3)\chi_{a_4\dot a_4}(x_4)\rangle$ we get
\begin{equation}
\left(d_1(\chi)+ \frac{1}{\chi^2} d_2(\chi)+  \frac{(1 - \chi)^2}{\chi^2} d_3(\chi) 
\right)'=0 \,,
\label{susyconditiondisp}
\end{equation}
which allows us to easily express $d_3(\chi)$ in terms of $d_1(\chi)$, $d_2(\chi)$ and an integration constant $D_0$
\begin{equation}
d_3(\chi)=\frac{D_0 \chi^2 - 4 \chi^2 d_1(\chi) - 4 d_2(\chi)}{4 (1 - \chi)^2} \,.
\end{equation}
Moreover, we obtain that
\begin{align}
d_4(\chi)= &\ 
-4 d_1(\chi) + \frac{4 d_2(\chi)}{1 - \chi} + \frac{(2 - 4 \chi) \chi d_1'(\chi)}{1 - \chi} - \frac{2 \chi d_2'(\chi)}{1 - \chi} \,,
\nonumber\\
d_5(\chi)= &\
-\frac{2 (2 - \chi) d_2(\chi)}{1 - \chi} + \frac{\chi^2 d_1'(\chi)}{1 - \chi} + \frac{(2 - \chi) \chi d_2'(\chi)}{1 - \chi} \,,
\nonumber\\
d_6(\chi)= &\
-\frac{4 d_2(\chi)}{\chi} + 2\chi d_1'(\chi) + 2d_2'(\chi)\,.
\end{align}
Proceeding in the same way with other trivial four-point correlators we can fix the remaining functions of the cross ratio in terms of $d_1(\chi)$, $d_2(\chi)$ and the integration constant $D_0$. In particular, for correlator of four displacement operators we get
\begin{equation}
\langle \rho(t_1)\rho(t_2)\rho(t_3)\rho(t_4)
\rangle =     
\frac{d_{14}(\chi)}{t_{12}^4t_{34}^4} \,,
\label{rho4pt}
\end{equation}
with
\begin{align}
d_{14}(\chi) & =
\frac{9 D_0 \chi^4}{(-1 + \chi)^4} 
- \frac{36 (-1 + 4 \chi - 6 \chi^2 + 4 \chi^3) d_1(\chi)}{(-1 + \chi)^4} 
\nonumber \\
&+ \frac{4 \chi (-14 + 43 \chi - 46 \chi^2 + 8 \chi^3) d_2(\chi)}{(-1 + \chi)^4} 
\nonumber
\\
& - \frac{2 \chi (16 - 65 \chi + 86 \chi^2 - 13 \chi^3 - 36 \chi^4 + 12 \chi^5) d_1'(\chi)}{(-1 + \chi)^4} 
\nonumber
\\
&
 + \frac{2 \chi^2 (16 - 50 \chi + 33 \chi^2 + 6 \chi^3 - 7 \chi^4 + 2 \chi^5) d_2'(\chi)}{(-1 + \chi)^4} \nonumber \\
&- \frac{2 \chi^2 (-7 + 15 \chi - 3 \chi^2 + 2 \chi^3) d_1''(\chi)}{(-1 + \chi)^2} 
\nonumber
\\
&
+ \frac{2 \chi^3 (-2 + 3 \chi - 15 \chi^2 + 7 \chi^3) d_2''(\chi)}{(-1 + \chi)^2} 
+ \frac{4 \chi^3 (1 - \chi + \chi^2) d_1'''(\chi)}{(-1 + \chi)} \nonumber \\
&+ \frac{4 \chi^4 (3 - 4 \chi + 2 \chi^2) d_2'''(\chi)}{(-1 + \chi)} 
\nonumber
\\
&
- \chi^4 (-1 + 2 \chi) d_1''''(\chi) 
+ \frac{(-2 \chi^5 + 9 \chi^6 - 16 \chi^7 + 14 \chi^8 - 6 \chi^9 + \chi^{10}) d_2''''(\chi)}{(-1 + \chi)^4} \,.
\label{d14}
\end{align}

In Section \ref{sec: 4 pts bootstrap} we will perform a bootstrap analysis that will allow us to compute the unknown functions $d_1(\chi)$ and $d_2(\chi)$. In order to do so, we will find useful to work with correlators that are invariant under R-symmetry and the automorphism. As in Appendix \ref{app: psu(1,1|2)}, we can easily do this by introducing two auxiliary variables $Y^a$ and $W^{\dot{a}}$ in the fundamental representation of the R-symmetry and automorphism groups, respectively. For example, by defining
\begin{equation}
\Phi(t,Y):= Y^a Y^b \phi_{ab}(t) \,,
\end{equation}
we obtain
\begin{align}
\label{4-pt displ primary}
\langle \Phi(t_1,Y_1) \Phi(t_2,Y_2) \Phi(t_3,Y_3) \Phi(t_4,Y_4) \rangle &= \frac{Y_{12}^2 Y_{34}^2}{t_{12}^2 t_{34}^2} \, \mathbf{f}(\chi,\rho_1) \,,
\end{align}
where 
\begin{equation}
\rho_1= \frac{Y_{12}Y_{34}}{Y_{13}Y_{24}} \,,
\end{equation}
and with
\begin{equation}
\label{4-pt displ Ward}
\mathbf{f}(\chi,\rho_1)= D_0 + \left( 1 - \frac{\chi}{\rho_1}\right) f_1(\chi) + \left( 1 - \frac{\chi}{\rho_1}\right)^2 f_2(\chi) \,,
\end{equation}
where
\begin{align}
f_1(\chi)&= \frac{2 D_0}{\chi -1}-\frac{8 \, d_1(\chi )}{\chi -1}-\frac{8 \, d_2(\chi )}{(\chi -1) \chi } \,,
\nonumber
\\
f_2(\chi)&= \frac{D_0}{(\chi -1)^2}-\frac{4 d_1(\chi )}{(\chi -1)^2}+ \frac{d_2(\chi)}{(\chi -1)^2} \left( 4-\frac{8}{\chi }\right) \,.
\label{dstof2}
\end{align}

Finally, let us turn to the constraints imposed by supersymmetry on the correlators of the tilt supermultiplet. Proceeding in a similar way as we have done for the displacement supermultiplet, we obtain that all the correlators of the tilt supermultiplet can be expressed in terms of two constants $T_1$ and $T_2$ and two functions $h_1(\chi)$ and $h_2(\chi)$ of the cross ratio $\chi$. In particular, for the superprimary $\psi_{a \dot{a}}$ we get
\begin{align}
\label{4 pt tilt primary}
\langle \Psi(t_1,Y_1,W_1) \Psi(t_2,Y_2,W_2) &\Psi(t_3,Y_3,W_3) \Psi(t_4,Y_4,W_4) \rangle = \frac{Y_{12} Y_{34} W_{12}W_{34}}{t_{12} t_{34}} \, \mathbf{h}(\chi,\rho_1,\rho_2) \,,
\end{align}
where 
\begin{equation}
\rho_2= \frac{W_{12}W_{34}}{W_{13}W_{24}}\,,
\end{equation}
and with
\begin{equation}
\label{4-pt tilt Ward}
\mathbf{h}(\chi,\rho_1,\rho_2)= T_1 + \left( 1 - \frac{\chi}{\rho_1}\right) h_1(\chi) + \frac{1}{\rho_2} \left(T_2+ \left( 1 - \frac{\chi}{\rho_1}\right) h_2(\chi) \right) \,,
\end{equation}
and
\begin{equation}
\Psi(t,Y,W):= Y^a W^{\dot{a}} \psi_{a\dot{a}}(t) \,.
\end{equation}

To compare with holographic results we will be interested in the scalar four-point function
\begin{equation}
\langle \hat{\varphi} (t_1) \hat{\varphi} (t_2) \hat{\varphi} (t_3) \hat{\varphi} (t_4) \rangle= \frac{h_3 (\chi)}{t_{12}^2 t_{34}^2} \,,
\label{varphi4pt}
\end{equation}
defined by the contraction
\begin{equation}
\hat{\varphi}(t)= \epsilon^{\dot{a}\dot{b}} \varphi_{\dot{a}\dot{b}} (t) \,.
\label{varphidef}
\end{equation}
Supersymmetry relates $h_3$ with the functions appearing in the super primary correlator through 
\begin{align}
    h_3(\chi)&= \left(2 \chi ^4-6 \chi ^3+4 \chi ^2\right) h_1''(\chi )+\left(6 \chi ^3-4 \chi ^2-4 \chi \right) h_1'(\chi )+\left(2 \chi ^2+4\right) h_1(\chi
   ) \nonumber\\
   &\quad +\left(4 \chi ^4-6 \chi ^3+2 \chi ^2\right) h_2''(\chi )+\left(12 \chi ^3-8 \chi ^2-2 \chi \right) h_2'(\chi )+\left(4 \chi ^2+2\right)
   h_2(\chi ) \nonumber \\
   &\quad +2 T_1+4 T_2 \,.
   \label{h3}
   \end{align}

It is worth noticing that the expressions \eqref{4-pt displ Ward} and \eqref{4-pt tilt Ward}, which were derived from supersymmetry constraints, verify the condition \eqref{topfin} of Appendix \ref{app: psu(1,1|2)}, which was derived from the existence of a topological sector. We would also like to point out that the structure imposed by supersymmetry in \eqref{4-pt displ Ward} and \eqref{4-pt tilt Ward} can also be found from the constraints of appropriate superconformal Ward identities \cite{Dolan:2004mu}.

\subsubsection*{Operator product expansion and superconformal blocks}

In the following we will perform a bootstrap computation of the \eqref{4-pt displ Ward} and \eqref{4-pt tilt Ward} four-point functions. In order to do so, it will prove crucial to expand those correlators in terms of exchanged operators using the operator product expansion (OPE). 

In particular, for the bootstrap analysis of \eqref{4-pt displ Ward} and \eqref{4-pt tilt Ward} we will be interested in the $\mathcal{B}^{1/2}_{1} \times \mathcal{B}^{1/2}_{1}$ and $\mathcal{B}^{1}_{0} \times \mathcal{B}^{1}_{0}$ OPEs. Using the selection rules for the R-symmetry and automorphism $SU(2)$'s along with the operator content of the theory described in Appendix \ref{app: psu(1,1|2)} we get
 \begin{align}
 \nonumber
     &\mathcal{B}^{1/2}_{1} \times \mathcal{B}^{1/2}_{1} \subset \mc{I}+\mathcal{B}^{1}_{0}+\mathcal{B}^{1}_{2}+\sum_{\Delta>0} \left(  \mathcal{L}_{0,0}^\Delta+\mathcal{L}_{0,2}^\Delta+\mathcal{L}_{2,2}^\Delta+\mathcal{L}_{2,0}^\Delta \right) \,, \\
     \label{displ ope}
     &\mathcal{B}^{1}_{0} \times \mathcal{B}^{1}_{0} \subset \mc{I}+\mathcal{B}^{1}_{0}  +\mathcal{B}^{2}_{0} + \sum_{\Delta>0} \left(  \mathcal{L}_{0,0}^\Delta+\mathcal{L}_{2,0}^\Delta+\mathcal{L}_{4,0}^\Delta \right) \,.
\end{align}
where the long and short supermultiplets ${\cal L}^{\Delta}_{r,\ell}$ and ${\cal B}^{\Delta}_{\ell}$ are introduced in Appendix \ref{app: psu(1,1|2)}, and where ${\cal I}$ is the identity operator. Translating the above OPEs into superconformal block expansions for the $\mathbf{h}$ and $\mathbf{f}$ functions defined in \eqref{4-pt displ Ward} and \eqref{4-pt tilt Ward} we obtain
\begin{align}
\nonumber
\mathbf{h}&= b_{\cal I}+ b_{1,0} \, \mathcal{G}_{\mc{B}^{1}_{0}}+  b_{1,2} \, \mathcal{G}_{\mc{B}^{1}_{2}}+ \sum_{\Delta>0} \left(  c_{\Delta,0,0} \, \mathcal{G}_{\mc{L}^{\Delta}_{0,0}}+ c_{\Delta,0,2} \, \mathcal{G}_{\mc{L}^{\Delta}_{0,2}}+ c_{\Delta,2,2} \, \mathcal{G}_{\mc{L}^{\Delta}_{2,2}}+ c_{\Delta,2,0} \, \mathcal{G}_{\mc{L}^{\Delta}_{2,0}}  \right) \,, \\
\label{f ope 1}
\mathbf{f}&=  \tilde{b}_{\cal I}+ \tilde{b}_{1,0} \, \mathcal{G}_{\mc{B}^{1}_{0}} + \tilde{b}_{2,0}  \, \mathcal{G}_{\mc{B}^{2}_{0}}+ \sum_{\Delta>0} \left( \tilde{c}_{\Delta,0,0} \,  \mathcal{G}_{\mc{L}^{\Delta}_{0,0}} + \tilde{c}_{\Delta, 2,0} \, \mathcal{G}_{\mc{L}^{\Delta}_{2,0}} + \tilde{c}_{\Delta,4,0} \, \mathcal{G}_{\mc{L}^{\Delta}_{4,0}} \right) \,,
\end{align}
where $\mathcal{G}_{\mc{B}^{\Delta}_{\ell}}$ and $\mathcal{G}_{\mc{L}^{\Delta}_{r,\ell}}$ are the superconformal blocks associated to short and long supermultiplets, respectively, and $b_{\Delta, \ell}, \, \tilde{b}_{\Delta, \ell}, \, c_{\Delta, r, \ell}, \, \tilde{c}_{\Delta,r, \ell}$ are their corresponding OPE coefficients. 
Generically, we can expand the superblocks as
\begin{align}
\label{generic exp 1}
\mathcal{G}_{\mc{B}^{\Delta}_{\ell}} 
   & = \sum_i \alpha^{\Delta,i}_{\ell} G^{\Delta_i}_{r_i,\ell_i}(\rho_1,\rho_2,\chi) \,,
   \\
   \label{generic exp 2}
   \mathcal{G}_{\mc{L}^{\Delta}_{r,\ell}}
   & = \sum_i \beta^{\Delta,i}_{r,\ell} G^{\Delta_i}_{r_i,\ell_i}(\rho_1,\rho_2,\chi) \,,
\end{align}
for some set of coefficients $\alpha^{\Delta,i}_{\ell}$ and $\beta^{\Delta,i}_{r,\ell}$ and where the sum runs over all conformal primaries in the respective supermultiplet, with
\begin{equation}
\label{conf block def}
G^\Delta_{r,\ell} = g_{-\frac r 2}(\rho_1)g_{-\frac \ell 2}(\rho_2) g_\Delta(\chi) \,, \quad g_{k}(x)=x^{k} \, _2F_1(k ,k ;2 k ;x) \,.
\end{equation}
In order to constrain \eqref{generic exp 1} and \eqref{generic exp 2} we should first note that odd-level descendants should give a vanishing contribution, since they correspond to three-point functions with an odd number of fermions. For example, for the $\mc{B}^{\Delta}_{\ell}$ short supermultiplets we only need to consider the contributions from the superprimary and the level-two superdescendant, i.e. from the red multiplets in 
\begin{align}
    \mc{B}^{\Delta}_{\ell}:\qquad {\color{red}{[\Delta,2\Delta,\ell]}}\rightarrow [\Delta+\frac 12,2\Delta-1,\ell\pm 1]\rightarrow {\color{red}{[\Delta+ 1,2\Delta-2,\ell]}}
\end{align}
Similarly, for long operators we have the contributions of the superprimary and the superdescendants at level two and four, i.e.
\begin{align}
   \mc{L}^\Delta_{r,\ell}:\hspace{5cm}\color{red}[\Delta,&\color{red}r,\ell]\nonumber \\
    [\Delta+\frac 12,r+1,\ell+1] \; \; [\Delta+\frac 12,r+1,\ell-1] &\; \; [\Delta+\frac 12,r-1,\ell+1] \; \; [\Delta+\frac 12,r-1,\ell-1] \nonumber\\
    \color{red}[\Delta+1,r\pm2,\ell]\quad 2 \, [\Delta+1,&\color{red}r,\ell]\quad[\Delta+1,r,\ell\pm2]\\
   [\Delta+\frac 32,r+1,\ell+1] \; \; [\Delta+\frac 32,r+1,\ell-1] &\; \; [\Delta+\frac 32,r-1,\ell+1] \; \; [\Delta+\frac 32,r-1,\ell-1] \nonumber\\
    \color{red}[\Delta+2,&\color{red}r,\ell] \nonumber
\end{align}

At this point it proves useful to impose the existence of a topological sector for correlators of superprimaries of the $\mc{B}^{\Delta}_{\ell}$ supermultiplets, as discussed in Appendix \ref{app: psu(1,1|2)}, by requiring 
\begin{align}
\nonumber 
    \partial_{\chi} \mathbf{h} |_{\rho_1=\chi}&=0 \,, \\
\label{top sector f}
    \partial_{\chi} \mathbf{f} |_{\rho_1=\chi}&=0 \,.
\end{align}
Equivalently, one can use the superconformal Ward identities presented in \cite{Baume:2019aid, Bliard:2024und} or the constraints \eqref{susyconditiondisp} derived above. Let us note that \eqref{top sector f}   
are symmetry constraints that do not depend on the dynamics of the theory. Therefore, imposing \eqref{top sector f} on \eqref{f ope 1} we see that each superconformal block contributing to the \eqref{f ope 1} expansions should independently satisfy the topological sector constraints, regardless of the values of the OPE coefficients. Furthermore, those constraints should be valid for four-point functions of arbitrary $\mc{B}^{\Delta}_{\ell}$ superprimaries, and therefore we obtain
\begin{align}
\nonumber 
\mc{G}_{\mc{B}^{\Delta}_{\ell}}&=G^{\Delta}{}_{2\Delta,\ell}-\frac{\Delta^2}{4(4\Delta^2-1)} G^{\Delta+1}{}_{2\Delta-2,\ell} \,, \\
\label{L blocks}
\mc{G}_{\mc{L}^\Delta_{r,\ell}}&=G^{\Delta }{}_{r,\ell}-G^{\Delta +1}{}_{r+2,\ell}-\frac{r^2}{16(r^2-1)} G^{\Delta +1}{}_{r-2,\ell}+\frac{(\Delta +1)^2 G^{\Delta +2}{}_{r,\ell}}{4 (2 \Delta +1) (2 \Delta +3)} \,,
\end{align}
for generic $\mc{B}^{\Delta}_{\ell}$ and $\mc{L}^\Delta_{r,\ell}$ supermultiplets. Since the $SU(2)_A$ automorphism acts as a spectator in these superconformal blocks, we recover the structure derived in \cite{Baume:2019aid}.

Let us return now to \eqref{f ope 1}. From 
\eqref{L blocks} we see that we can use the $SU(2)_R$ and $SU(2)_A$ selection rules to further constrain the $\mathbf{h}$ and $\mathbf{f}$ OPEs. To be more specific, we note from \eqref{L blocks} that each $\mc{L}^\Delta_{r,\ell}$ multiplet gives a non-vanishing contribution from a conformal primary of spin $r+2$. Therefore, the $\mc{L}^\Delta_{2,2}$ and $\mc{L}^\Delta_{2,0}$ supermultiplets can not be part of the \eqref{f ope 1} expansion, as they would contribute with a spin four operator, in contradiction with the $SU(2)$ selection rules for the $\mathcal{B}^{1/2}_{1} \times \mathcal{B}^{1/2}_{1}$ OPE. Similarly, we can rule out the $\mc{L}^\Delta_{4,0}$ from the $\mathcal{B}^{1}_{0} \times \mathcal{B}^{1}_{0}$ OPE. Therefore, we arrive at
 \begin{align}
\nonumber 
&\mathcal{B}^{1/2}_{1} \times \mathcal{B}^{1/2}_{1}  \sim \mc{I}+\mathcal{B}^{1}_{0}+\mathcal{B}^{1}_{2}+ \sum_{\Delta>0} \left(  \mathcal{L}_{0,0}^\Delta+\mathcal{L}_{0,2}^\Delta \right) \,, \\
\label{B1 OPE}
     &\mathcal{B}^{1}_{0} \times \mathcal{B}^{1}_{0} \sim \mc{I}+\mathcal{B}^{1}_{0}  +\mathcal{B}^{2}_{0} + \sum_{\Delta>0} \left(  \mathcal{L}_{0,0}^\Delta+\mathcal{L}_{2,0}^\Delta \right) \,,
\end{align}
and
\begin{align}
\nonumber 
\mathbf{h}&= b_{\cal I}+ b_{1,0} \, \mathcal{G}_{\mc{B}^{1}_{0}}+ b_{1,2} \, \mathcal{G}_{\mc{B}^{1}_{2}}+ \sum_{\Delta>0} \left(  c_{\Delta,0,0} \, \mathcal{G}_{\mc{L}^{\Delta}_{0,0}}+ c_{\Delta, 0,2} \, \mathcal{G}_{\mc{L}^{\Delta}_{0,2}} \right) \,, \\
\label{f ope 2}
\mathbf{f}&=  \tilde{b}_{\cal I}+ \tilde{b}_{1,0} \, \mathcal{G}_{\mc{B}^{1}_{0}} + \tilde{b}_{2,0} \, \mathcal{G}_{\mc{B}^{2}_{0}}+ \sum_{\Delta>0} \left(  \tilde{c}_{\Delta,0,0} \, \mathcal{G}_{\mc{L}^{\Delta}_{0,0}} + \tilde{c}_{\Delta,2,0} \, \mathcal{G}_{\mc{L}^{\Delta}_{2,0}} \right) \,.
\end{align}

\section{Four-point functions: analytic conformal bootstrap}
\label{sec: 4 pts bootstrap}

In the spirit of \cite{Liendo:2018ukf,Ferrero:2021bsb,Ferrero:2023znz,Ferrero:2023gnu,Bianchi:2020hsz,Pozzi:2024xnu}, in this section we will follow an analytic conformal bootstrap approach for the computation of \eqref{4-pt displ primary} and \eqref{4 pt tilt primary} at strong coupling. This method consists of deriving the correlators by imposing superconformal symmetry and other consistency conditions. For reviews applied to the cases of one-dimensional superconformal defects in $\mc{N}=4$ SYM and ABJM see \cite{Ferrero:2023gnu,Bliard:2023zpe}. We will obtain that, up to next-to-leading order in the strong coupling expansion, the four-point functions of the displacement and tilt supermultiplets can be completely fixed in terms of two parameters, for which we will provide an holographic interpretation in Section \ref{sec: holographic 4 pts}.

To be more specific, we will compute the ${\bf f}(\chi,\rho_1)$ and ${\bf h}(\chi,\rho_1,\rho_2)$ functions that give the cross-ratio dependence of \eqref{4-pt displ primary} and \eqref{4 pt tilt primary}. We will focus in their strong coupling limit, and therefore we will study the expansions
\begin{align}
{\bf h}(\chi,\rho_1,\rho_2) &={\bf h}^{(0)}(\chi,\rho_1,\rho_2)+ \epsilon \, {\bf h}^{(1)}(\chi,\rho_1,\rho_2)+ \epsilon^2 \,  {\bf h}^{(2)}(\chi,\rho_1,\rho_2) +\dots,\nonumber\\
{\bf f}(\chi,\rho_1) &={\bf f}^{(0)}(\chi,\rho_1)+ \epsilon\, {\bf f}^{(1)}(\chi,\rho_1)+ \epsilon^2 \,  {\bf f}^{(2)}(\chi,\rho_1) +\dots \,,
\end{align}
where $\epsilon$ is a coupling constant proportional to the inverse of the string tension. When comparing with the OPE given in \eqref{B1 OPE}, we get 
\begin{align}
b_{\Delta,\ell} &= b_{\Delta,\ell}^{(0)} + \epsilon \, b_{\Delta,\ell}^{(1)} + \dots \,, \nonumber \\
\tilde{b}_{\Delta,\ell} &= \tilde{b}_{\Delta,\ell}^{(0)} + \epsilon \, \tilde{b}_{\Delta,\ell}^{(1)} + \dots \,, \nonumber\\
c_{\Delta,r,\ell} &= c_{\Delta,r,\ell}^{(0)} + \epsilon \, c_{\Delta,r,\ell}^{(1)}+ \dots \,, \nonumber\\
\tilde{c}_{\Delta,r,\ell} &= \tilde{c}_{\Delta,r,\ell}^{(0)} + \epsilon \, \tilde{c}_{\Delta,r,\ell}^{(1)}  + \dots \,,
\end{align}
for the OPE coefficients. For the identity we will consider
$b_{\cal I}=\tilde{b}_{\cal I}=1$, which amounts to
arbitrarily setting a unit two-point function coefficient of the superprimaries at all orders in $\epsilon$. In particular, this 
is convenient to 
relate the OPE coefficients to the square of the constants appearing in three-point functions.
Similarly,
\begin{align}
\Delta_{r,\ell}(\epsilon) &= \Delta_{r,\ell}^{(0)} + \epsilon \, \gamma_{\Delta,r,\ell}^{(1)}+ \dots \nonumber \,, \\
\tilde{\Delta}_{r,\ell}(\epsilon) &= \tilde{\Delta}_{r,\ell}^{(0)} + \epsilon \, \tilde{\gamma}_{\Delta,r,\ell}^{(1)} + \dots \,,
\end{align}
where $\Delta_{r,\ell}(\epsilon)$ and $\tilde{\Delta}_{r,\ell}(\epsilon)$ are the scaling dimensions of the ${\cal L}^{\Delta}_{r,\ell}$ operators that contribute to \eqref{B1 OPE}. 

As discussed in previous sections, one can use supersymmetry to fully constrain the R-symmetry dependence of ${\bf f}$ and ${\bf h}$. In the end we are left with three constants $D_0,\, T_1$ and $T_2$ and four functions $f_1(\chi), f_2(\chi), h_1(\chi)$ and $h_2(\chi)$ that completely determine ${\bf f}$ and ${\bf h}$. 
All of them will be expanded in powers of the coupling constant $\epsilon$ as
\begin{align}
\nonumber
T_i &=T_i^{(0)} + \epsilon \, T_i^{(1)} + \dots \,, \\
\nonumber
D_0 &= D_0^{(0)} + \epsilon \, D_0^{(1)} + \dots \,, \\
\nonumber
h_i(\chi)&=h_i^{(0)}(\chi)+ \epsilon\,  h_i^{(1)}(\chi) +\dots \,, \\
\label{strong coupling fi}
f_i(\chi)&=f_i^{(0)}(\chi)+ \epsilon\,  f_i^{(1)}(\chi) +\dots,
\end{align}
for $i=1,2$.

\subsubsection*{Leading order}

From AdS/CFT we know that the strong coupling expansion of the \eqref{4-pt displ primary} and \eqref{4 pt tilt primary} correlators can be recast as an expansion in terms of Witten diagrams. Therefore, from Wick contractions we get
\begin{align}
\nonumber 
\mathbf{h}^{(0)}(\chi,\rho_1,\rho_2) &=  
1-\frac{\chi}{\rho_1\rho_2}+\frac{\chi(1-\rho_1)(1-\rho_2)}{(1-\chi)\rho_1\rho_2} \,, \\
\label{LO f}
\mathbf{f}^{(0)} (\chi,\rho_1)&=1+\frac{\chi ^2}{\rho_1^2}+\frac{(1-\rho_1)^2 \chi ^2}{\rho_1^2 (\chi -1)^2} \,,
\end{align}
at leading order. This fixes
\begin{align}
\nonumber
 D_0^{(0)} &= 3\,, &T_1^{(0)}& = 0\,, &T_2^{(0)}&=0 \,,
\\
\nonumber
h_1^{(0)} (\chi)  &= \frac{1}{1-\chi}\,, 
&h_2^{(0)} (\chi)&= -\frac{\chi}{1-\chi}  \,,& &\\
\label{LO fs}
f_1^{(0)}(\chi) &= -\frac{2(\chi-2)}{\chi-1}
\,, &f_2^{(0)}(\chi) &= 1 + \frac{1}{(\chi-1)^2} \,. & &
\end{align}
Then, from the above results we can read that the leading-order CFT data is
\begin{align}
b_{1,0}^{(0)}&=-1 ,  & & & b_{1,2}^{(0)}&=0, & & \nonumber\\
\tilde{b}_{1,0}^{(0)}&=0, & & & \tilde{b}_{2,0}^{(0)}&=2, & &\nonumber  \\
c_{\Delta,0,0}^{(0)}&= 
\frac{\Gamma (\Delta) \Gamma (\Delta +1)}{2\Gamma \left(2\Delta\right)}
&\text{for even } \Delta,&    & c_{\Delta,0,0}^{(0)} &= 0 &\text{for odd } \Delta,& \nonumber
\\
c_{\Delta,0,2}^{(0)}&=
-\frac{\Gamma (\Delta) \Gamma (\Delta +1)}{\Gamma \left(2\Delta\right)} 
&\text{for odd } \Delta,&
 &c_{\Delta,0,2}^{(0)}& = 0 &\text{for even } \Delta,&\nonumber\\
\tilde{c}_{\Delta,0,0}^{(0)}&=
\frac{(\Delta -1) \Delta  (\Delta +2) \Gamma (\Delta )^2}{2 \Gamma (2 \Delta )} 
&\text{for even } \Delta ,&   &\tilde{c}_{\Delta,0,0}^{(0)} &= 0 &\text{for odd } \Delta,& \nonumber
\\
\tilde{c}_{\Delta,2,0}^{(0)}&=-\frac{(\Delta +1) 
\Gamma (\Delta +1)^2}{\Gamma (2 \Delta )}
&\text{for odd } \Delta\geq 3,&
&\tilde{c}_{\Delta,2,0}^{(0)} &= 0 &\text{for even } \Delta,& \nonumber\\
\label{LO CFT data-7}
\tilde{c}_{1,2,0}^{(0)}&=0. & & & & & & 
\end{align}

\subsubsection*{Next-to-leading order: the ansatz}

For the next-to-leading order (NLO) in the strong coupling expansion, the starting point of the analytic bootstrap process consists in writing a suitable ansatz for the correlators \cite{Ferrero:2019luz}. In our case, we will propose an ansatz for the $f_i^{(1)}(\chi)$ and $h_i^{(1)}(\chi)$ functions that were introduced in 
\eqref{strong coupling fi}. When comparing to higher-dimensional setups, here is where the first simplification occurs. Crucially, in one-dimensional CFTs there is only one independent conformal cross ratio that can be constructed out of the four time coordinates of the insertions, as opposed to the two independent cross ratios that one has to consider in higher dimensions.

In order to formulate an ansatz for $f_i(\chi)$ and $h_i(\chi)$ it is convenient to use the intuition coming from the AdS/CFT correspondence. As presented in \eqref{LO f}, at leading order $f_i^{(0)}(\chi)$ and $h_i^{(0)}(\chi)$ are rational functions, given that there is no vertex integral involved in their computations. When moving to subleading corrections we expect transcendental functions to appear in the ansatz. Borrowing intuition from the analysis of 1/2 BPS line defects in the ${\cal N}=4$ super Yang-Mills and ABJM theories \cite{Liendo:2018ukf,Ferrero:2021bsb,Ferrero:2023znz,Ferrero:2023gnu,Bianchi:2020hsz,Pozzi:2024xnu}, we expect the N$^{L}$LO correction to $f_i(\chi)$ and $h_i(\chi)$ to be expressed in terms of harmonic polylogarithms \cite{Remiddi:1999ew, Alday:2015eya, Ferrero:2019luz} of transcendental weight less or equal to $L$.
In particular, for the NLO term we will make the ansatz
\begin{align}
\nonumber
    h_i^{(1)}(\chi)&=s_{i0}(\chi)+s_{i1}(\chi)\log(\chi)+s_{i2}(\chi)\log(1-\chi)\,, \\
\label{NLO fi ansatz}    
    f_i^{(1)}(\chi)&=\tilde{s}_{i0}(\chi)+\tilde{s}_{i1}(\chi)\log(\chi)+\tilde{s}_{i2}(\chi)\log(1-\chi) \,,
\end{align}
where $s_{ij}$ and $\tilde{s}_{ij}$ are rational functions.

\subsubsection*{NLO constraints: crossing and braiding symmetry}

The ansatz proposed above for the NLO term of the correlators is further constrained by crossing symmetry, which relates correlators with points 1 and 3 exchanged. More precisely, crossing symmetry implies
\begin{align}
\nonumber
    {\bf h}(\chi,\rho_1,\rho_2) &= \left(\frac{\chi}{1-\chi}\right)\left(\frac{\rho_1}{1-\rho1}\right)^{-1} \left(\frac{\rho_2}{1-\rho_2}\right)^{-1} {\bf h}(1-\chi,1-\rho_1,1-\rho_2) \,, \\
\label{crossing f}
    {\bf f}(\chi,\rho_1) &= \left(\frac{\chi}{1-\chi}\right)^{2}\left(\frac{\rho_1}{1-\rho1}\right)^{-2}  {\bf f}(1-\chi,1-\rho_1) \,.
\end{align}
Equivalently, using the expansions \eqref{4-pt displ Ward} and \eqref{4-pt tilt Ward}, we get
\begin{align}
\nonumber
T_1&=0 \,, \\
\nonumber\chi \, h_1(1-\chi) - (1-\chi) \, h_1(\chi) &=0 \,, \\
\nonumber
(1-\chi) \, h_2 (\chi)+ \chi (h_1(1-\chi)+h_2(1-\chi))+T_2&=0 \,, \\
\nonumber
(1-\chi) \, f_1 (\chi) + \chi f_1 (1-\chi) + 2 D_0 &=0 \,, \\
\label{crossing-5}
(1-\chi)^2 f_2 (\chi)-\chi f_1 (1-\chi)- \chi^2 f_2 (1-\chi) +D_0 &=0 \,.
\end{align}

Additionally, there is another symmetry, known as \textit{braiding} symmetry \cite{Liendo:2018ukf}, that is inherited from the identity
\begin{align}
\label{braiding blocks}
    \left(\frac{\chi}{\chi-1}\right)^\Delta {}_2F\left(\Delta,\Delta,2\Delta,\frac{\chi}{\chi-1}\right)=\left(-\chi\right)^\Delta {}_2F(\Delta,\Delta,2\Delta,\chi) \,,
\end{align}
satisfied by the conformal blocks. Let us focus for the moment on the leading-order term of ${\bf f}$, which was presented in \eqref{LO f}. It is interesting to note from \eqref{LO CFT data-7} that every block $G^\Delta_{r,\ell}$ (see eq. \eqref{conf block def}) that contributes to the leading-order OPE of ${\bf f}$ has $\ell=0$ and satisfies that the quantum numbers $\{ \Delta, \frac{r}{2}\}$ are either both even or both odd. Therefore, using \eqref{braiding blocks} we see that each of those blocks satisfies
\begin{equation}
G^\Delta_{r,\ell} \left( \frac{\chi}{\chi-1}, \frac{\rho_1}{\rho_1-1}, \frac{\rho_2}{\rho_2-2} \right) =G^\Delta_{r,\ell} \left( \chi, \rho_1, \rho_2 \right) \,,
\end{equation}
at leading order, and consequently
\begin{equation}
{\bf f}^{(0)} \left( \frac{\chi}{\chi-1}, \frac{\rho_1}{\rho_1-1} \right) ={\bf f}^{(0)} \left( \chi, \rho_1 \right) \,.
\label{brading f0}
\end{equation}
When moving to the next-to-leading order we need to take into account the perturbative corrections to the scaling dimensions of the unprotected multiplets. Assuming an all-order vanishing contribution of the OPE coefficients that are zero at leading order we get, using \eqref{braiding blocks}, that\footnote{A similar argument, used in \cite{Liendo:2018ukf,Ferrero:2021bsb}, was justified in \cite{Cavaglia:2023mmu} by looking at parity-odd operators and finding them to be absent at strong coupling.}
\begin{equation}
{\bf f}^{(1)} \left( \frac{\chi}{\chi-1}, \frac{\rho_1}{\rho_1-1} \right) \bigg|_{\log(-\chi) \to \log(\chi)}={\bf f}^{(1)} \left( \chi, \rho_1 \right) \,.
\end{equation}
Similarly, for ${\bf h}$ we have
\begin{align}
\label{brading h0}
{\bf h}^{(0)} \left( \frac{\chi}{\chi-1}, \frac{\rho_1}{\rho_1-1} \right) &={\bf h}^{(0)} \left( \chi, \rho_1 \right) \,, \\
\label{bradingh1}
{\bf h}^{(1)} \left( \frac{\chi}{\chi-1}, \frac{\rho_1}{\rho_1-1} \right) \bigg|_{\log(-\chi) \to \log(\chi)} &={\bf h}^{(1)} \left( \chi, \rho_1 \right) \,.
\end{align}
It is easy to check that the braiding symmetry conditions \eqref{brading f0} and \eqref{brading h0}, as well as crossing symmetry conditions
\eqref{crossing f}, are satisfied by the leading order expressions \eqref{LO f}.

In what follows, we will study in detail how crossing and braiding constrain the functions in our NLO ansatz  \eqref{NLO fi ansatz}.
Let us begin with the ${\bf h}$ function,  which describes the tilt four-point correlator. By inserting \eqref{4-pt tilt Ward} in \eqref{crossing f}, \eqref{brading h0} and \eqref{bradingh1} we obtain
\begin{align}
    T_i^{(0)}=T_i^{(1)}=0 \,,
\end{align}
for $i=1,2$, which is consistent with the LO result found in \eqref{LO f}. 

With a little hindsight we find convenient to reparametrize the rational functions of the NLO ansatz  \eqref{NLO fi ansatz} as
\begin{align}
    s_{10}(\chi) & =\frac{(2-\chi ) \, p_1(\chi)+\frac{\chi}{2} \, p_2(\chi)}{1-\chi } \,, 
    \quad &&
    s_{11}(\chi)=\frac{\chi}{1-\chi } (r_1(\chi) -\tfrac{1}{2}r_2(\chi)) \,,
   \nonumber\\ 
    s_{12}(\chi) & = r_1  (1-\chi ) - \frac{1}{2} r_2(1-\chi)\,, && \nonumber\\
    s_{20}(\chi) & =
    \frac{\chi}{\chi-1}
    p_2(\chi)\,, \quad && \ s_{21}(\chi) = \frac{\chi }{1-\chi } r_2(\chi) \,,
\nonumber\\
 s_{22}(\chi) & =-r_1(1-\chi )-\tfrac{1}{2} r_2(1-\chi )\,. &&
\end{align}
The advantage of this parametrization is the simplification of most of the braiding constraints, which become 
\begin{align}
\label{brading cond pi ri-1}
    p_1(\chi) - p_1 \left(\frac {\chi}{1-\chi}\right) = 0  \qquad \qquad p_2(\chi) - p_2 \left( \frac {\chi}{1-\chi}\right) & = 0  \\
    \label{brading cond pi ri-2}
   r_1(\chi) + r_1\left( \frac {\chi}{1-\chi}\right)=0  \qquad  \qquad r_2(\chi )- r_2\left(\frac{\chi }{\chi -1}\right) & =0 
   \\
   2 (\chi -1) \, r_1(1-\chi )-2 \chi  \, r_1(\chi )+ 2
\,r_2\left(\frac{1}{1-\chi }\right)-(\chi -1) \, r_2(1-\chi )+\chi  \, r_2(\chi )&=0 
\label{braideingleft1} \\
\!\!\!\!r_1\left(\frac{1}{1-\chi }\right)+(\chi -1) \, r_1(1-\chi )+ \frac12 r_2\left(\frac{1}{1-\chi }\right)+\frac12 (\chi -1) \, r_2(1-\chi )+\chi  \, r_2(\chi ) &=0 
\label{braideingleft2}
\end{align}
while the crossing conditions  impose that
\begin{align}
\label{crossingleft1}
-2 (\chi +1) \, p_1(1-\chi )-2 (\chi -2) \, p_1(\chi )+(\chi -1) \, p_2(1-\chi )+\chi  \, p_2(\chi ) &=0 \,, \\
\label{crossingleft2}
 2 (\chi +1) \, p_1(1-\chi )+(\chi -1) \, p_2(1-\chi )-2 \chi  \, p_2(\chi ) &=0 \,.
\end{align}

The four conditions in \eqref{brading cond pi ri-1} and \eqref{brading cond pi ri-2} imply that $p_1(\chi)$, $p_2(\chi)$ and $r_2(\chi)$ are braiding-symmetric and that $r_1(\chi)$ is braiding-antisymmetric. Therefore, we propose the expansion
\begin{align}
\nonumber 
    p_i(\chi)&= \sum_{k=M_{p_i}}^{N_{p_i}} p_{i,k} \, \left(\frac{\chi^2}{\chi-1}\right)^k \quad \text{for} \quad i=1,2, \\
  \nonumber %
  r_2(\chi) &= \sum_{k=M_{r_2}}^{N_{r_2}} r_{2,k}\left(\frac{\chi^2}{\chi-1}\right)^k , \\
    \label{brading symm ansatz h1-3}
    r_1(\chi) &= \left(\chi-\frac{\chi}{\chi-1}\right)\sum_{k=M_{r_1}}^{N_{r_1}} r_{1,k}\left(\frac{\chi^2}{\chi-1}\right)^k,
\end{align}
for some finite integers $M_{p_i},N_{p_i},M_{r_i}$ and $N_{r_i}$\footnote{The construction of the ansatz as an expansion in powers of $\frac{\chi^2}{\chi-1}$ will be crucial to obtain finite sums.}. The next step in the analytic bootstrap will involve applying the remaining braiding and crossing constraints and further consistency conditions to establish the upper and lower bounds for these sums, and to relate the coefficients of \eqref{brading symm ansatz h1-3} among them.

We now turn to the ${\bf f}$ function, which specifies the displacement four-point correlators. We find it convenient to use the following parametrization
\begin{align} 
    \tilde{s}_{10}(\chi) & =-\chi (1-2\chi)\, \tilde{p}_1(\chi)-2D_0, \quad && \tilde{s}_{11}(\chi )=\frac{\chi^3}{(1-\chi)^2}\, \tilde{r}_1(\chi),
    \nonumber
    \\ 
   \nonumber \tilde{s}_{12}(\chi ) & = -\frac{(1-\chi)^2}{\chi} \, \tilde{r}_1(1-\chi),
\\
    \tilde{s}_{20}(\chi ) & = \chi (1-2\chi) \, \tilde{p}_1(\chi)+\chi^2 \, \tilde{p}_2(\chi)+D_0, \quad && 
    \tilde{s}_{21}(\chi )=  - \frac{(2-\chi) \chi^3}{2(1-\chi)^3} \tilde{r}_1 (\chi)+ \frac{\tilde{r}_2(\chi)}{\chi^2}, 
    \nonumber
    \\
    \tilde{s}_{22}(\chi ) & = \frac{(1-\chi)^2}{2\chi} \, \tilde{r}_1(1-\chi) + \frac{\chi^2}{(1-\chi)^4} \, \tilde{r}_2(1-\chi) .
\end{align}
Then, crossing and braiding imply
\begin{align}
\label{brading cond tpi tri-1}
    \tilde{p}_1(\chi)&= \tilde{p}_1 \left( 1-\chi\right) \,, \; \; \quad \qquad \tilde{p}_2(\chi) = \tilde{p}_2 \left( 1-\chi\right) \,, \\
    \label{brading cond tpi tri-2}
   \tilde{r}_1(\chi)&=-\tilde{r}_1\left( \frac {\chi}{1-\chi}\right) \,, \qquad \tilde{r}_2(\chi )  =\tilde{r}_2\left(\frac{\chi }{\chi -1}\right) \,,
   \\
\label{br displ extra 1}
2 \, D_0 \,(\chi -1)^2&=   (2 \chi -1) (\chi -1)^3 \, \tilde{p}_1(\chi )+(\chi +1) \, \tilde{p}_1\left(\frac{\chi }{\chi -1}\right) \,,  \\
   \chi ^4 \, \tilde{r}_1(\chi ) &=(\chi -1)^4 \tilde{r}_1(1-\chi )-\tilde{r}_1\left(\frac{1}{1-\chi }\right) \,,  \\
(1-2 \chi) \, \tilde{p}_1(\chi )&=-\chi  \, \tilde{p}_2(\chi )+\frac{\chi}{(\chi-1)^4}  \, \tilde{p}_2\left(\frac{\chi }{\chi
   -1}\right)-\frac{D_0 (\chi -2)}{(\chi-1)^2} \nonumber \\
   &\quad -\frac{(\chi +1)}{(\chi-1)^4} \, \tilde{p}_1\left(\frac{\chi }{\chi -1}\right)  \,, \\
   \label{br displ extra 4}
2 \frac{(\chi -1)^4}{\chi} \, \tilde{r}_2(\chi )&=-(\chi -1)^6
   \, \tilde{r}_1(1-\chi )-(\chi -1) \, \tilde{r}_1\left(\frac{1}{1-\chi }\right) +\chi^4  \left(\chi ^2-3 \chi +2\right) \, \tilde{r}_1(\chi )\nonumber \\
   &\quad -2\chi ^3  \left[(\chi -1)^4 \, \tilde{r}_2\left(\frac{1}{1-\chi }\right)+\, \tilde{r}_2(1-\chi )\right] \,. 
\end{align}
The conditions \eqref{brading cond tpi tri-1} and \eqref{brading cond tpi tri-2} imply that $\tilde p_i$ are crossing-symmetric, $\tilde r_1$ is braiding-antisymmetric and $\tilde r_2$ is braiding-symmetric, which naturally guide us to propose
\begin{align}
\nonumber
    \tilde{p}_i(\chi)&= \sum_{k=M_{\tilde{p}_i}}^{N_{\tilde{p}_i}} \tilde{p}_{i,k} \, \left[\chi (1-\chi) \right]^k \quad \text{for} \quad i=1,2, \\
   \nonumber  \tilde{r}_2(\chi) &= \sum_{k=M_{\tilde{r}_2}}^{N_{\tilde{r}_2}} \tilde r_{2,k}\left(\frac{\chi^2}{\chi-1}\right)^k , \\
    \label{brading symm ansatz f1-3}
    \tilde{r}_1(\chi) &= \left(\chi-\frac{\chi}{\chi-1}\right)\sum_{k=M_{\tilde{r}_1}}^{N_{\tilde{r}_1}} \tilde{r}_{1,k}\left(\frac{\chi^2}{\chi-1}\right)^k.
\end{align}
We will later impose the remaining conditions \eqref{br displ extra 1}-\eqref{br displ extra 4} in order to constrain the above sums. But first we will find useful to discuss further consistency conditions to be satisfied by the ansatz. These will imply the consistency of the correlators with their OPE and a mild growth of the anomalous dimensions at large $\Delta$.

\subsubsection*{NLO constraints: consistency with OPE and behavior of anomalous dimensions}

The consistency of the CFT requires the ability to write a convergent OPE. This will be constraining in several ways. First of all, the OPE controls the boundary behavior of the functions. In particular, the fact that the conformal blocks introduced in \eqref{conf block def} are finite for $\chi \to 0$ implies
\begin{align}
\nonumber
    h_i (\chi) & \sim \mathcal{O}(1) \quad \text{for} \quad \chi \to 0,\\
    f_i (\chi) & \sim \mathcal{O}(1) \quad \text{for} \quad \chi \to 0,
\end{align}
for $i=1,2$. These constraints, combined with the crossing conditions presented in \eqref{crossing-5}, give
\begin{align}
    h_i (\chi) & \sim \mathcal{O}[(\chi-1)^{-1}] \quad \text{for} \quad i=1,2, \nonumber\\
    f_1 (\chi) & \sim \mathcal{O}[(\chi-1)^{-1}],\nonumber \\
    f_2 (\chi) & \sim \mathcal{O}[(\chi-1)^{-2}],
\end{align}
for $\chi \to 1$.

Now we will introduce a dynamical constraint on the anomalous dimensions of exchanged long operators, and we will do it by requiring them to have a mild behaviour at large $\Delta$. Let us start by considering the N$^L$LO contribution $\gamma_{\Delta,r,\ell}^{(L)}$ to the anomalous dimensions of the long multiplets that are exchanged in \eqref{B1 OPE}. It has been suggested in the literature that the behaviour of the anomalous dimensions $\gamma_{\Delta,r,\ell}^{(L)}$ in the limit $\Delta \to \infty$ is related to the lagrangian of the AdS$_2$ dual theory \cite{Heemskerk:2009pn,Fitzpatrick:2010zm}. In particular, the more irrelevant the interactions that contribute to the N$^L$LO order term of the four-point function, the bigger the growth of $\gamma_{\Delta,r,\ell}^{(L)}$ for large $\Delta$. To be more specific, it has been found that for the 1/2 BPS line defects of ${\cal N}=4$ SYM and ABJM \cite{Liendo:2018ukf,Bianchi:2020hsz,Ferrero:2023gnu}
\begin{align}
\gamma_{\Delta,r,\ell}^{(L)} \sim \Delta^{L+1}
   \qquad\text{for\ } {\Delta\rightarrow\infty} \,.
\end{align}
As we will discuss in Section \ref{sec: holographic 4 pts}, the interaction terms that contribute to the NLO of the four point functions \eqref{4-pt displ primary} and \eqref{4 pt tilt primary} are at most quartic terms with four derivatives. Taking into account that this behaviour of the interaction terms is the same as for the 1/2 BPS line defects of ${\cal N}=4$ SYM and ABJM \cite{Heemskerk:2009pn,Fitzpatrick:2010zm}, we will therefore impose
\begin{align}
\gamma_{\Delta,r,\ell}^{(1)} \sim \Delta^{2}
   \qquad\text{for\ } {\Delta\rightarrow\infty} \,,
\end{align}
for our NLO ansatz.

The advantage of imposing consistency with the OPE and a mild behaviour for the anomalous dimensions is that they impose strong constraints on the upper and lower bounds of the \eqref{brading symm ansatz h1-3} and \eqref{brading symm ansatz f1-3} expansions. These, combined with the remaining crossing and braiding conditions discussed in \eqref{braideingleft1}-\eqref{crossingleft2}, \eqref{br displ extra 1} and \eqref{br displ extra 4}, leave only four unconstrained coefficients in the ansatz, as we will now present.

\subsubsection*{NLO constrained result}

When applying the above constraints over ${\bf h}$ and ${\bf f}$ and then matching the result with the OPE introduced in \eqref{f ope 2} we obtain 
\small{
\begin{align}
   h_1^{(1)}(\chi) = & 3 \, c_{2,0,0}^{(1)} \left(\frac{1}{\chi-1}-\frac{\chi \log (\chi )}{(\chi -1)^2}-\frac{\log (1-\chi )}{\chi }\right) \,, \nonumber\\
   h_2^{(1)}(\chi) = &    {3 \, c_{2,0,0}^{(1)}} \left(-\frac{\chi}{\chi-1} +\frac{\chi\left(\chi ^2-2 \chi +2\right)}{(\chi-1)^2} \log (\chi ) - \chi \log (1-\chi )\right) \,,
   \nonumber\\
f_1^{(1)}(\chi) = & 
   -\frac{2}{3}\frac{(\chi -2)}{(\chi-1)}
  \left( \tilde{b}^{(1)}_{1,0}+ \tilde{b}^{(1)}_{2,0}\right)
-\frac{1}{3}\frac{\chi ^2(\chi -2)}{(\chi-1)^2} \left(\tilde{b}^{(1)}_{1,0}-2 \tilde{b}^{(1)}_{2,0}\right) 
\log (\chi)\nonumber
\\
& +\frac{1}{3} \frac{\left(\chi ^2-1\right) }{\chi}\left(\tilde{b}^{(1)}_{1,0}-2 \tilde{b}^{(1)}_{2,0}\right) \log (1-\chi )\,,
   \nonumber\\
f_2^{(1)}(\chi) = & 
\frac{\left(\chi ^2-\chi +1\right)}{5(\chi -1)^2} 
\left(12 \tilde{c}^{(1)}_{2,0,0}+5 \tilde{b}^{(1)}_{1,0}-4 \tilde{b}^{(1)}_{2,0}\right)
- \frac{1}{3(\chi -1)} \left(\tilde{b}^{(1)}_{1,0}+\tilde{b}^{(1)}_{2,0}\right) \\
  - &
 \frac{ \chi^2\log (\chi )}{30 (\chi-1)^3}\left(
3\left(12 \tilde{c}^{(1)}_{2,0,0}+5 \tilde{b}^{(1)}_{1,0}-4 \tilde{b}^{(1)}_{2,0}\right)\left(2 \chi ^2-5\chi +5\right)+5\left(\tilde{b}^{(1)}_{1,0}-2 \tilde{b}^{(1)}_{2,0}\right) (\chi-1)
\right)
 \nonumber\\
   + &
  \frac{ \log (1-\chi )}{30 \chi}\left(3\left(12 \tilde{c}^{(1)}_{2,0,0}+5 \tilde{b}^{(1)}_{1,0}-4 \tilde{b}^{(1)}_{2,0}\right)\left(2 \chi ^2+\chi +2\right)+5\left(\tilde{b}^{(1)}_{1,0}-2 \tilde{b}^{(1)}_{2,0}\right) (\chi+2)\right) \,.\nonumber
   \end{align}
}\normalsize{Moreover, for the topological constants we get}
\begin{align}
   T_1^{(1)}&=T_2^{(1)}=0 \,,\qquad
\label{top const 2}
D_0=\tilde{b}^{(1)}_{1,0}+\tilde{b}^{(1)}_{2,0}\,.
\end{align}
Let us note that the parameters unfixed by the bootstrap can be related to the normalization of the three-point functions discussed in \eqref{3ptdispl} as
\begin{equation}
\tilde{b}_{1,0}^{(1)} = - 4\sigma^2 \,.
\end{equation}
So far, we see that we have obtained a solution parametrized by four in principle independent OPE coefficients. However, at this point it is useful to consider the anomalous dimensions $\gamma^{(1)}_{\Delta,0,0}$ and $\tilde{\gamma}^{(1)}_{\Delta,0,0}$ of the exchanged ${\cal L}^{\Delta}_{0,0}$ supermultiplets.
\begin{align}
\nonumber
\gamma^{(1)}_{\Delta,0,0} &= -3 \Delta (\Delta+1) \, c^{(1)}_{2,0,0} \,, \\
\tilde{\gamma}^{(1)}_{\Delta,0,0} &=\frac{\Delta  (\Delta +1) \left(\frac{1}{5} \left(\Delta ^2+\Delta +4\right) \left(12  \tilde{c}^{(1)}_{2,0,0}+5 \tilde{b}^{(1)}_{1,0}-4 \tilde{b}^{(1)}_{2,0}\right)+2 \tilde{b}^{(1)}_{1,0}-4 \tilde{b}^{(1)}_{2,0}\right)}{4
   \left(\Delta ^2+\Delta -2\right)} \,.
\end{align}
Let us recall that the bulk CFT$_2$, which is dual to type IIB string theory in $AdS_3\times S^3\times T^4$, can be obtained as a limit of the CFT$_2$ which is dual to string theory in $AdS_3\times S^3_+\times S^3_-\times S^1$. As previously discussed, in the latter case 1/2 BPS line defects are invariant under the $D(2,1;\sin^2 \Omega)$ supergroup. One can recover the $PSU(1,1|2) \times SU(2)_A$  invariant line defects associated to the $AdS_3\times S^3\times T^4$ case by taking the limit $\Omega\to 0$ or $\Omega\to \frac{\pi}{2}$. Interestingly, in this limit the displacement supermultiplet of the $D(2,1;\sin^2 \Omega)$ invariant defect splits and gives rise to the displacement and tilt supermultiplets of the $PSU(1,1|2)\times SU(2)_A$ invariant defect. Therefore, it is reasonable to assume that the ${\cal L}^{\Delta}_{0,0}$ operators that are exchanged in both the tilt and displacement OPE (see \eqref{B1 OPE}) have the same anomalous dimensions, i.e.\footnote{The same conclusion could be reached by assuming that the anomalous dimensions of long operators are proportional to the eigenvalue of the quadratic Casimir, as proved for the case of the 1/2 BPS line of ${\cal N}=4$ SYM in \cite{Ferrero:2023gnu}.}
\begin{equation}
\label{anom dim assumption}
\gamma^{(1)}_{\Delta,0,0}= \tilde{\gamma}^{(1)}_{\Delta,0,0} \,.
\end{equation}
With this last assumption we get
\begin{align}
\nonumber
c_{2,0,0}^{(1)} &= \frac{1}{36} \left(\tilde{b}_{1,0}^{(1)}-2\tilde{b}_{2,0}^{(1)}\right) \,, \\
\tilde{c}_{2,0,0}^{(1)} &= \frac{1}{18} \left(-10 \tilde{b}_{1,0}^{(1)}+11 \tilde{b}_{2,0}^{(1)}\right) \,,
\end{align}
and therefore
\begin{align}
\nonumber 
 h_1^{(1)}(\chi)&= 
 \frac{ \left(\tilde{b}_{1,0}^{(1)}-2\tilde{b}_{2,0}^{(1)}\right)}{12}
 \left(\frac{1}{\chi-1}-\frac{\chi \log (\chi )}{(\chi -1)^2}-\frac{\log (1-\chi )}{\chi }\right) \,, \nonumber\\
   h_2^{(1)}(\chi) &= 
    \frac{ \left(\tilde{b}_{1,0}^{(1)}-2\tilde{b}_{2,0}^{(1)}\right)}{12}
   \left(-\frac{\chi}{\chi-1} +\frac{\chi\left(\chi ^2-2 \chi +2\right)}{(\chi-1)^2} \log (\chi ) - \chi \log (1-\chi )\right) 
   \,, \nonumber\\
   f_1^{(1)}(\chi) &= 
   -\frac{2}{3}\frac{(\chi -2)}{(\chi-1)}
  \left( \tilde{b}^{(1)}_{1,0}+ \tilde{b}^{(1)}_{2,0}\right)
-\frac{1}{3}\frac{\chi ^2(\chi -2)}{(\chi-1)^2} \left(\tilde{b}^{(1)}_{1,0}-2 \tilde{b}^{(1)}_{2,0}\right) 
\log (\chi)\nonumber
\\
& +\frac{1}{3} \frac{\left(\chi ^2-1\right) }{\chi}\left(\tilde{b}^{(1)}_{1,0}-2 \tilde{b}^{(1)}_{2,0}\right) \log (1-\chi )  \nonumber  \\
   \label{NLO final 4}
f_2^{(1)}(\chi)&= -\frac{\chi ^2 \left(\tilde{b}_{1,0}^{(1)}-2 \tilde{b}_{2,0}^{(1)}\right)+3 \tilde{b}_{2,0}^{(1)}\chi -3 \tilde{b}_{2,0}^{(1)}}{3 (\chi -1)^2}  
+\frac{\chi ^2 \left(\chi ^2-3 \chi +3\right) \left(\tilde{b}_{1,0}^{(1)}-2
   \tilde{b}_{2,0}^{(1)}\right) \log (\chi )}{3 (\chi -1)^3}\nonumber
\\
&\quad -\frac{\chi}{3}  \left(\tilde{b}_{1,0}^{(1)}-2 \tilde{b}_{2,0}^{(1)}\right) \log (1-\chi ) \,.
\end{align}

Let us remark that the final result from the analytic bootstrap depends only on two OPE coefficients, $\tilde{b}^{(1)}_{1,0}$ and $\tilde{b}^{(1)}_{2,0}$. We would like to emphasize that this statement strongly relies on the assumption \eqref{anom dim assumption}. We will later check this hypothesis with an holographic computation, providing an holographic interpretation for the coefficients that are not fixed by the bootstrap process. To facilitate the comparison we also present the bootstrap result for the four-point functions \eqref{rho4pt} and \eqref{varphi4pt} at NLO. Using the bootstrap solution in \eqref{d14} and \eqref{h3} we
get
\small{
\begin{align}
\langle \rho(t_1)\rho(t_2)&\rho(t_3)\rho(t_4)
\rangle^{(1)} =     \nonumber
\\
&   
\frac{2\tilde{b}^{(1)}_{2,0}}{\tau_{12}^4\tau_{34}^4} \frac{(1 - \chi + \chi^2)}{(1 - \chi)^4} (12-36\chi+25\chi^2+10\chi^3+25\chi^4-36\chi^5+12 \chi^6)\nonumber
\\&   
-\frac{4\tilde{b}^{(1)}_{1,0}}{\tau_{12}^4\tau_{34}^4} \frac{(1 - \chi + \chi^2)}{(1 - \chi)^4} (3-9\chi+7\chi^2+\chi^3+7\chi^4-9\chi^5+3 \chi^6)\nonumber
\\&   
-\frac{6\left(\tilde{b}^{(1)}_{1,0}-2\tilde{b}^{(1)}_{2,0}\right)}{\tau_{12}^4\tau_{34}^4}  
\frac{\chi^4}{(1-\chi)^5}
(2 - 6 \chi + 20 \chi^2 - 30 \chi^3 + 25 \chi^4 - 
 11 \chi^5 + 2 \chi^6)\log(\chi)\nonumber
 \\&   
-\frac{6\left(\tilde{b}^{(1)}_{1,0}-2\tilde{b}^{(1)}_{2,0}\right)}{\tau_{12}^4\tau_{34}^4}  
\frac{1}{\chi}
(2 - \chi  - 
 \chi^5 + 2 \chi^6)\log(1-\chi) \,,
\label{rho4ptbootstrap}
\end{align}
}\normalsize{and}
\small{
\begin{align}
\langle \varphi(t_1)\varphi(t_2)\varphi(t_3)\varphi(t_4)
\rangle^{(1)} = & -\frac{2\left(\tilde{b}^{(1)}_{1,0}-2\tilde{b}^{(1)}_{2,0}\right)}{3\tau_{12}^2\tau_{34}^2}\left(2\frac{(1-\chi+\chi^2)^2}{(1-\chi)^2} \right. \nonumber \\
& + \left.\frac{2-\chi-\chi^3+2\chi^4}{\chi}\log(1-\chi) \right. \nonumber \\
& + \left.\frac{\chi^2(2 - 4 \chi + 9 \chi^2 - 7 \chi^3 + 2 \chi^4)}{(1-\chi)^3}
\log(\chi)\right) \,.
\label{varphi4ptbootstrap}
\end{align}
}
\normalsize{}

\subsubsection*{Extracting the NLO CFT data}

By comparing the NLO solution \eqref{top const 2} and \eqref{NLO final 4} with the corresponding superconformal block expansions \eqref{f ope 2} we obtain that the NLO CFT data is
\begin{align}
\nonumber
\gamma^{(1)}_{\Delta,0,0}&=-\frac{\tilde{b}_{1,0}^{(1)}-2\tilde{b}_{2,0}^{(1)}}{12}\Delta (\Delta+1)\,, &\gamma^{(1)}_{\Delta,0,2}&=-\frac{\tilde{b}_{1,0}^{(1)}-2\tilde{b}_{2,0}^{(1)}}{12}\Delta (\Delta+1) \,, \\
\nonumber
 b_{1,0}^{(1)}&=0 \,, &b_{1,2}^{(1)}&=0 \,, \\
c_{\Delta,0,0}^{(1)}&=\partial_\Delta \left(\gamma^{(1)}_{\Delta,0,0}\,  c_{\Delta,0,0}^{(0)}\right) \,, & c_{\Delta,0,2}^{(1)}&=\partial_\Delta \left(\gamma^{(1)}_{\Delta,0,2} \, c_{\Delta,0,2}^{(0)}\right) \,,
\end{align}
and
\begin{align} \nonumber
\tilde{\gamma}^{(1)}_{\Delta,0,0}&=-\frac{1}{12} \Delta  (\Delta +1) \left(\tilde{b}_{1,0}^{(1)}-2 \tilde{b}_{2,0}^{(1)}\right)\,, \\
\nonumber
\tilde{\gamma}^{(1)}_{\Delta,2,0}&=-\frac{1}{12} (\Delta -1) (\Delta +2) \left(\tilde{b}_{1,0}^{(1)}-2 \tilde{b}_{2,0}^{(1)}\right) \,, \\ \nonumber
\tilde{c}_{\Delta,2,0}^{(1)}&= \partial_\Delta \left(\tilde{c}_{\Delta,2,0}^{(0)} \tilde{\gamma}_{\Delta,2,0}^{(1)}\right) +\frac{\tilde{b}_{1,0}^{(1)} \, \tilde{c}_{\Delta, 2,0}^{(0)}}{2 \Delta  (\Delta +1)}\,,
\\
\tilde{c}_{\Delta,0,0}^{(1)}&= \partial_\Delta \left(\tilde{c}_{\Delta,0,0}^{(0)} \tilde{\gamma}_{\Delta,0,0}^{(1)} \right)-\frac{3 \tilde{b}_{1,0}^{(1)} \, \tilde{c}_{\Delta,0,0}^{(0)}}{2 (\Delta -1) (\Delta +2)} \,.
\end{align}
Let us highlight two interesting limits in the above formulas by doing an informal presentation of what we might expect from the string theory perspective. Firstly, for $\tilde{b}_{1,0}^{(1)} \to 0$ the OPE coefficients take an expression analogous to the one presented in \cite{Bianchi:2020hsz, Giombi:2017cqn, Liendo:2018ukf},
suggesting that this limit corresponds with the pure R-R flux case, where the dual worldsheet is perpendicular to the boundary. On the other hand, when $\tilde{b}_{1,0}^{(1)} \to 2 \tilde{b}_{2,0}^{(1)}$ all the anomalous dimensions vanish. There is a particular limit where we might expect such a behavior. This limit will correspond with  the pure NS-NS flux case, where the dual worldsheet becomes parallel to the boundary. In the next section we will present an holographic description that supports the interpretation of these two limits.

\section{Holographic check of bootstrap results}
\label{sec: holographic 4 pts}

In the previous section we obtained an analytic bootstrap result for the four-point functions of operators in the displacement and tilt supermultiplets. We will know perform an equivalent Witten diagram computation of those correlators, which will provide an holographic check of the bootstrap result. Moreover, it will give us an holographic interpretation of the two parameters that are not fixed by the bootstrap procedure.

\subsubsection*{Effective action for bosonic fluctuations}

Let us begin by computing the effective action for the bosonic fluctuations around the classical string presented in \eqref{susyconfiguration}. We will choose to work in euclidean signature, i.e. we will perform a Wick rotation to EAdS. Expanding the euclidean string action in terms of the normal coordinates introduced in eq. \eqref{geodesic eq solution}, \eqref{rho coord riemann} and \eqref{scalar phi} we obtain\footnote{We are omitting some boundary terms which are not relevant in our analysis.}
\begin{equation}
\label{bosonic action expansion-main text}
S_B= g S^{(0)}+S^{(2)}+\frac{1}{\sqrt g}S^{(3)}+ \frac{1}{g} S^{(4)}_{\phi_{\sf tr}}+ \frac{1}{g} S^{(4)}_{\phi^9}+\frac{1}{g} S^{(4)}_{S^3_+}+ \frac{1}{g} S^{(4)}_{S^3_-}+ \frac{1}{g} S^{(4)}_{\sf mix}
+{\cal O}(g^{-3/2})\,,
\end{equation}
where $S^{(0)}$ and $S^{(2)}$ are the classical and quadratic action discussed in Appendix \ref{app: quadratic fluctuations} (up to a Wick rotation) and with
\begin{align}
S^{(3)} &= -\frac{2\sin\vartheta}{3} \!\!\int\! \! d^2 \sigma  \sqrt{h} \left[  \cos^2\vartheta\ \phi_{\sf tr}^3 + \tfrac{ \sin\Omega}{2} \epsilon^{\alpha\beta} \, \phi^a_b \partial_{\alpha} \phi^b_c \partial_{\beta}\phi^c_a + \tfrac{\cos\Omega}{2} \epsilon^{\alpha\beta} \,\phi^{\dot{a}}_{\dot{b}} \partial_{\alpha} \phi^{\dot{b}}_{\dot{c}} \partial_{\beta} \phi^{\dot{c}}_{\dot{a}} \right]
\label{qubicterms}
\\
S^{(4)}_{\phi^{\sf tr}} &=  \!\int\! \! d^2 \sigma  \sqrt{h} \left[  \tfrac{\cos^2\vartheta}{3} \phi_{\sf tr}^4 - \tfrac{1}{8} \left( \partial_{\alpha} \phi_{\sf tr} \partial^{\alpha} \phi_{\sf tr} \right)^2 \right] \\
S^{(4)}_{\phi^{9}} &= -\frac{1}{8} \!\int\! \! d^2 \sigma  \sqrt{h}  \left( \partial_{\alpha} \phi^{9} \partial^{\alpha} \phi^{9} \right)^2 \label{S4phi9}\,, \\
S^{(4)}_{S^3_+} &= \int\! \! d^2 \sigma  \sqrt{h} \bigg[ \tfrac{1}{32} \left( \partial_{\alpha} \phi^{a}_b \partial^{\alpha} \phi^{b}_a \right)^2  - \tfrac{1}{16} \partial_{\alpha} \phi^{a}_b \partial_{\beta} \phi^{b}_a \partial^{\alpha} \phi^{c}_d \partial^{\beta} \phi^{d}_c \nonumber \\
&\qquad \qquad \quad \qquad \, \, -\tfrac{\sin^2 \Omega}{24} (  \phi^{a}_b  \phi^{b}_a \partial_{\alpha} \phi^{c}_d \partial^{\alpha} \phi^{d}_c - \phi^{a}_b \partial_{\alpha} \phi^{b}_a  \phi^{c}_d \partial^{\alpha} \phi^{d}_c ) \bigg]  \,, \\
S^{(4)}_{S^3_-} &= \!\int\! \! d^2 \sigma  \sqrt{h}\left[\tfrac{1}{32} \left( \partial_{\alpha} \phi^{\dot{a}}_{\dot{b}} \partial^{\alpha} \phi^{\dot{b}}_{\dot{a}} \right)^2  - \tfrac{1}{16} \partial_{\alpha} \phi^{\dot{a}}_{\dot{b}} \partial_{\beta} \phi^{\dot{b}}_{\dot{a}} \partial^{\alpha} \phi^{\dot{c}}_{\dot{d}} \partial^{\beta} \phi^{\dot{d}}_{\dot{c}} \right. \nonumber \\
&\qquad \qquad \quad \qquad \, \, - \left.\tfrac{\cos^2 \Omega}{24} (  \phi^{\dot{a}}_{\dot{b}}  \phi^{\dot{b}}_{\dot{a}} \partial_{\alpha} \phi^{\dot{c}}_{\dot{d}} \partial^{\alpha} \phi^{\dot{d}}_{\dot{c}} - \phi^{\dot{a}}_{\dot{b}} \partial_{\alpha} \phi^{\dot{b}}_{\dot{a}}  \phi^{\dot{c}}_{\dot{d}} \partial^{\alpha} \phi^{\dot{d}}_{\dot{c}} ) \right]  \,, \\
\label{bosonic action expansion-9}
S^{(4)}_{\sf mix} &= \!\int\! \! d^2 \sigma  \sqrt{h} \bigg[ \tfrac{1}{8} \left( \partial_{\alpha} \phi^{\sf tr} \partial^{\alpha} \phi^{\sf tr} \partial_{\beta} \phi^a_b \partial^{\beta} \phi^b_a + \partial_{\alpha} \phi^{\sf tr} \partial^{\alpha} \phi^{\sf tr} \partial_{\beta} \phi^{\dot{a}}_{\dot{b}} \partial^{\beta} \phi^{\dot{b}}_{\dot{a}} + 2 \, \partial_{\alpha} \phi^{\sf tr} \partial^{\alpha} \phi^{\sf tr} \partial_{\beta} \phi^9 \partial^{\beta} \phi^9 \right) 
\nonumber
\\
& \qquad \qquad  - \tfrac{1}{4} \left( \partial_{\alpha} \phi^{\sf tr} \partial_{\beta} \phi^{\sf tr} \partial^{\alpha} \phi^a_b \partial^{\beta} \phi^b_a + \partial_{\alpha} \phi^{\sf tr} \partial_{\beta} \phi^{\sf tr} \partial^{\alpha} \phi^{\dot{a}}_{\dot{b}} \partial^{\beta} \phi^{\dot{b}}_{\dot{a}} + 2 \, \partial_{\alpha} \phi^{\sf tr} \partial_{\beta} \phi^{\sf tr} \partial^{\alpha} \phi^9 \partial^{\beta} \phi^9 \right)   \nonumber \\
&\qquad \qquad  +\tfrac{1}{8} \left( \partial_{\alpha} \phi^{9} \partial^{\alpha} \phi^{9} \partial_{\beta} \phi^a_b \partial^{\beta} \phi^b_a + \partial_{\alpha} \phi^{9} \partial^{\alpha} \phi^{9} \partial_{\beta} \phi^{\dot{a}}_{\dot{b}} \partial^{\beta} \phi^{\dot{b}}_{\dot{a}} \right)  \nonumber \\
& \qquad \qquad - \tfrac{1}{4} \left( \partial_{\alpha} \phi^{9} \partial_{\beta} \phi^{9} \partial^{\alpha} \phi^a_b \partial^{\beta} \phi^b_a + \partial_{\alpha} \phi^{9} \partial_{\beta} \phi^{9} \partial^{\alpha} \phi^{\dot{a}}_{\dot{b}} \partial^{\beta} \phi^{\dot{b}}_{\dot{a}} \right)  \nonumber \\
&\qquad \qquad +\tfrac{1}{16} \partial_{\alpha} \phi^a_b \partial^{\alpha} \phi^b_a \partial_{\beta} \phi^{\dot{a}}_{\dot{b}} \partial^{\beta} \phi^{\dot{b}}_{\dot{a}} - \frac{1}{8} \partial_{\alpha} \phi^a_b \partial_{\beta} \phi^b_a \partial^{\alpha} \phi^{\dot{a}}_{\dot{b}} \partial^{\beta} \phi^{\dot{b}}_{\dot{a}} \bigg]
\,,
\end{align}

It is worth noting that the coupling with the NS-NS $B$-field gives rise to cubic interaction terms. This is novel compared to the case of open string fluctuations in $AdS_5\times S^5$, which does not exhibit cubic terms \cite{Drukker:2000ep,Giombi:2017cqn}. This is also different to the case of open string fluctuations in 
$AdS _4\times \mathbb{CP}^3$, where the coupling with the NS-NS $B$-field only leads to boundary terms \cite{Bianchi:2020hsz,Correa:2023thy}.

\subsubsection*{Correlator from $\phi_{\rm tr}$ fluctuations}

From the massive scalar field $\phi_{\sf tr}$ we can compute correlation functions of the displacement operator $\rho$, the component of the displacement supermultiplet whose scale dimension is $\Delta = 2$ (see Section \ref{sec: algebra and repr}). The corresponding bulk-to-boundary propagator is, in this case,
\begin{equation}
K_2(z,\tau,\tau') = {\cal C}_2 \left(\frac{z}{z^2+(\tau-\tau')^2}\right)^2\,,
\qquad
{\cal C}_2 := \frac{2}{3\pi}\,.
\end{equation}
The 2-point correlation function is simply
\begin{equation}
\langle {\cal O}_{\sf tr}(\tau_1){\cal O}_{\sf tr}(\tau_2)\rangle
= \frac{{\cal C}_2}{\tau_{12}^4}\,,\qquad\text{with\ } \tau_{ij}:= \tau_i-\tau_j
\end{equation}
where ${\cal O}_{\sf tr}$ is the operator dual to the $\phi_{\rm tr}$ fluctuation. To have unit normalized two-point functions one can define $\tilde{\cal O}_{\sf tr}={\cal O}_{\sf tr}/\sqrt{{\cal C}_2}$.

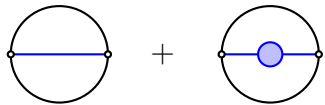
\begin{figure}[h!]
    \centering
    \begin{tikzpicture}[scale=0.8]
    \draw[thick] (0,0) circle [radius=0.8cm];
\draw[thick,cap=round,blue]  (-0.8,0) -- (0.8,0);
    \draw[thick,fill=white] (0.8,0) circle[radius=0.05cm];
    \draw[thick,fill=white] (-0.8,0) circle[radius=0.05cm];
     \node at (1.7,0) {+};
    \end{tikzpicture}
    \hspace{0.15cm}
    \begin{tikzpicture}[scale=0.8]
    \draw[thick] (0,0) circle [radius=0.8cm];
\draw[blue,thick,cap=round]  (-0.8,0) -- (0.8,0);    \draw[blue,fill=blue!25,thick] (0,0) circle [radius=0.2cm];
    \draw[thick,fill=white] (0.8,0) circle[radius=0.05cm];
    \draw[thick,fill=white] (-0.8,0) circle[radius=0.05cm];
    \end{tikzpicture}
    \hspace{0.5cm}
     \caption{Diagrammatic expansion of the 2-point function.}
    \label{fig:2pt}
\end{figure}

The two-point function will also receive $\tfrac{1}{g}$ corrections from loop diagrams, as schematically 
depicted in Fig. \ref{fig:2pt}. These corrections are not explicitly computed, as they are absorbed into the normalization of the dual operator under the assumption that
\begin{equation}
\langle \tilde{\cal O}_{\sf tr}(\tau_1)\tilde{\cal O}_{\sf tr}(\tau_2)\rangle
= \frac{1}{\tau_{12}^4}\,,
\end{equation}
remains valid at all-loop order. In what follows we will omit the symbol $\ \tilde{}\ $ when referring to the unit normalized operators.

The three-point correlator is computed from the corresponding cubic vertex in \eqref{qubicterms}
\begin{align}
\langle {\cal O}_{\sf tr}(\tau_1){\cal O}_{\sf tr}(\tau_2){\cal O}_{\sf tr}(\tau_3)\rangle
= &\ 3!\frac{{\cal C}_2^{-3/2}}{\sqrt g} \frac{2}{3}\sin\vartheta \int d^2 \sigma 
\frac{1}{z^2} K_2(z,\tau,\tau_1)K_2(z,\tau,\tau_2)K_2(z,\tau,\tau_3)
\nonumber
\\
= &\ \frac{4}{\sqrt g}\sin\vartheta\ {\cal C}_2^{3/2} D_{2,2,1,1}(\tau_1,\tau_2,\tau_3,\tau_3)
= \frac{\frac{1}{\sqrt g}\sqrt{\frac{2}{3\pi}}\sin\vartheta}{\tau_{12}^2\tau_{13}^2\tau_{23}^2}\,.
\end{align}
The conventions we use for the $D$-integrals are detailed in Appendix \ref{app: D integrals}. From this result, we read that the corresponding OPE coefficient is
\begin{equation}
C_{{\cal O}_{\sf tr}{\cal O}_{\sf tr}{\cal O}_{\sf tr}} = \frac{1}{\sqrt g}\sqrt{\frac{2}{3\pi}}\sin\vartheta\,.
\label{Crho3}
\end{equation}

Up to a relative normalization (see \eqref{rhonorm}), ${\cal O}_{\sf tr}$ should be identified with the component $\rho$ of the displacement multiplet. This allow us to fix the value of the constant $\sigma$ in \eqref{3ptdispl}, which can be related to the OPE coefficient 
$\tilde{b}_{1,0}^{(1)}$  
\begin{equation}
    \sigma = \frac{1}{\sqrt g}\frac{1}{\sqrt{2\pi}}\sin\vartheta
    \quad\Rightarrow\quad
    \tilde{b}_{1,0}^{(1)} = - 4\sigma^2 = -
    \frac{1}{g}\frac{2}{\pi}\sin^2\vartheta \,.
\end{equation}
The leading order contribution to the four-point correlator is straightforwardly computed in terms of boundary-to-boundary propagators
\begin{align}
\langle {\cal O}_{\sf tr}(\tau_1){\cal O}_{\sf tr}(\tau_2) {\cal O}_{\sf tr}(\tau_3) {\cal O}_{\sf tr}(\tau_4)\rangle^{(0)}
= \frac{1}{\tau_{12}^4\tau_{34}^4}\left(1+\chi^4+\frac{\chi^4}{(1-\chi)^4}\right)\,.
\label{LOrho4}
\end{align}
\begin{figure}[h!]
    \centering
    \begin{tikzpicture}[scale=0.8]
    \draw[thick] (0,0) circle [radius=0.8cm];
     \draw[blue,thick] (0.8*0.707,0.8*0.707)
     --(-0.8*0.707,0.8*0.707);
    \draw[blue,thick] (0.8*0.707,-0.8*0.707)
     --(-0.8*0.707,-0.8*0.707);
    \draw[thick,fill=white] (0.8*0.707,0.8*0.707) circle[radius=0.05cm];
    \draw[thick,fill=white] (0.8*0.707,-0.8*0.707) circle[radius=0.05cm];
    \draw[thick,fill=white] (-0.8*0.707,0.8*0.707) circle[radius=0.05cm];
    \draw[thick,fill=white] (-0.8*0.707,-0.8*0.707) circle[radius=0.05cm];
    \node at (1.7,0) {+};
    \end{tikzpicture}
    \hspace{0.15cm}
    \begin{tikzpicture}[scale=0.8]
    \draw[thick] (0,0) circle [radius=0.8cm];
     \draw[blue,thick] (0.8*0.707,0.8*0.707)
     --(0.8*0.707,-0.8*0.707);
     \draw[blue,thick] (-0.8*0.707, 0.8*0.707)
     --(-0.8*0.707,-0.8*0.707);
    \draw[thick,fill=white] (0.8*0.707,-0.8*0.707) circle[radius=0.05cm];
    \draw[thick,fill=white] (0.8*0.707,0.8*0.707) circle[radius=0.05cm];
    \draw[thick,fill=white] (-0.8*0.707,-0.8*0.707) circle[radius=0.05cm];
    \draw[thick,fill=white] (-0.8*0.707,0.8*0.707) circle[radius=0.05cm];
    \node at (1.7,0) {+};
    \end{tikzpicture}
\hspace{0.15cm}
 \begin{tikzpicture}[scale=0.8]
    \draw[thick] (0,0) circle [radius=0.8cm];
     \draw[blue,thick] (-0.8*0.707,0.8*0.707)
     --(0.8*0.707,-0.8*0.707);
     \draw[white,fill=white] (0,0) circle [radius=0.1cm];
     \draw[blue,thick] (-0.8*0.707, -0.8*0.707)
     --(0.8*0.707,0.8*0.707);
    \draw[thick,fill=white] (0.8*0.707,-0.8*0.707) circle[radius=0.05cm];
    \draw[thick,fill=white] (0.8*0.707,0.8*0.707) circle[radius=0.05cm];
    \draw[thick,fill=white] (-0.8*0.707,-0.8*0.707) circle[radius=0.05cm];
    \draw[thick,fill=white] (-0.8*0.707,0.8*0.707) circle[radius=0.05cm];
    \end{tikzpicture}
    \caption{Leading order contributions to the four-point function}
    \label{fig:4-ptLO}
 \end{figure}
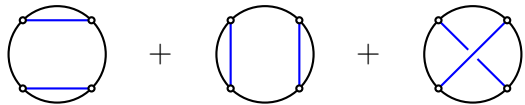
 
\begin{figure}[h!]
\centering
\begin{tikzpicture}[scale=0.8]
\draw[thick] (0,0) circle [radius=0.8cm];
\draw[blue,thick] (-0.8*0.707,0.8*0.707) --(0.8*0.707,-0.8*0.707);
\draw[blue,thick] (-0.8*0.707, -0.8*0.707)--(0.8*0.707,0.8*0.707);
\draw[thick,fill=white] (0.8*0.707,-0.8*0.707) circle[radius=0.05cm];
\draw[thick,fill=white] (0.8*0.707,0.8*0.707) circle[radius=0.05cm];
\draw[thick,fill=white] (-0.8*0.707,-0.8*0.707) circle[radius=0.05cm];
\draw[thick,fill=white] (-0.8*0.707,0.8*0.707) circle[radius=0.05cm];
\draw[thick,black,fill=black] (0,0) circle[radius=0.05cm];
\node at (1.7,0) {+};
\end{tikzpicture}
\hspace{0.15cm}
\begin{tikzpicture}[scale=0.8]
\draw[thick] (0,0) circle [radius=0.8cm];
\draw[blue,thick] (0.8*0.707,0.8*0.707) -- (0.3,0);
\draw[blue,thick] (0.8*0.707,-0.8*0.707) -- (0.3,0);
\draw[blue,thick] (-0.8*0.707,0.8*0.707) -- (-0.3,0);
\draw[blue,thick] (-0.8*0.707,-0.8*0.707) -- (-0.3,0);
\draw[blue,thick] (0.3,0) -- (-0.3,0);
\draw[thick,fill=white] (0.8*0.707,0.8*0.707) circle[radius=0.05cm];
\draw[thick,fill=white] (0.8*0.707,-0.8*0.707) circle[radius=0.05cm];
\draw[thick,fill=white] (-0.8*0.707,0.8*0.707) circle[radius=0.05cm];
\draw[thick,fill=white] (-0.8*0.707,-0.8*0.707) circle[radius=0.05cm];
\draw[thick,fill=black] (0.3,0) circle[radius=0.05cm];
\draw[thick,fill=black] (-0.3,0) circle[radius=0.05cm];
\node at (1.7,0) {+};
\end{tikzpicture}
\hspace{0.15cm}
\begin{tikzpicture}[scale=0.8]
\draw[thick] (0,0) circle [radius=0.8cm];
\draw[thick,blue] (0.8*0.707,0.8*0.707) -- (0,0.3);
\draw[thick,blue] (-0.8*0.707,0.8*0.707) -- (0,0.3);
\draw[thick,blue] (0.8*0.707,-0.8*0.707) -- (0,-0.3);
\draw[thick,blue] (-0.8*0.707,-0.8*0.707) -- (0,-0.3);
\draw[thick,blue] (0,0.3) -- (0,-0.3);
\draw[thick,fill=white] (0.8*0.707,0.8*0.707) circle[radius=0.05cm];
\draw[thick,fill=white] (0.8*0.707,-0.8*0.707) circle[radius=0.05cm];
\draw[thick,fill=white] (-0.8*0.707,0.8*0.707) circle[radius=0.05cm];
\draw[thick,fill=white] (-0.8*0.707,-0.8*0.707) circle[radius=0.05cm];
\draw[thick,fill=black] (0,0.3) circle[radius=0.05cm];
\draw[thick,fill=black] (0,-0.3) circle[radius=0.05cm];
\node at (1.7,0) {+};
\end{tikzpicture}
\hspace{0.15cm}
\begin{tikzpicture}[scale=0.8]
\draw[thick] (0,0) circle [radius=0.8cm];
\draw[blue,thick] (0.8*0.707,0.8*0.707) -- (-0.3,0);
\draw[fill=white,white] (0,0.15) circle[radius=0.12cm];    
\draw[thick,blue] (0.8*0.707,-0.8*0.707) -- (0.3,0);
\draw[thick,blue] (-0.8*0.707,0.8*0.707) -- (0.3,0);
\draw[thick,blue] (-0.8*0.707,-0.8*0.707) -- (-0.3,0);
\draw[thick,blue] (0.3,0) -- (-0.3,0);
\draw[thick,fill=white] (0.8*0.707,0.8*0.707) circle[radius=0.05cm];
\draw[thick,fill=white] (0.8*0.707,-0.8*0.707) circle[radius=0.05cm];
\draw[thick,fill=white] (-0.8*0.707,0.8*0.707) circle[radius=0.05cm];
\draw[thick,fill=white] (-0.8*0.707,-0.8*0.707) circle[radius=0.05cm];
\draw[thick,fill=black] (0.3,0) circle[radius=0.05cm];
\draw[thick,fill=black] (-0.3,0) circle[radius=0.05cm];
\end{tikzpicture}
\caption{Connected next-to-leading order contributions to the four-point function}
\label{fig:4-ptNLO}
\end{figure}
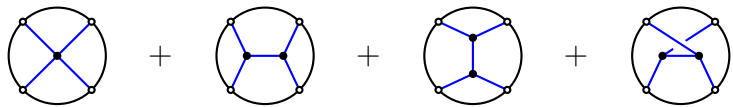

At the next-to-leading order, we only need to consider connected diagrams. Diagrams like the ones that are shown in Fig. \ref{fig:4-ptLO}, where one of the propagators is corrected by a one-loop contribution, are absorbed into the normalization of the operator. The connected diagrams that contribute to the next-to-leading order include those with a single quartic vertex and exchange diagrams involving two cubic vertices in three different channels (see Fig. \ref{fig:4-ptNLO}),
\begin{align}
\langle {\cal O}_{\sf tr}(\tau_1){\cal O}_{\sf tr}(\tau_2) {\cal O}_{\sf tr}(\tau_3){\cal O}_{\sf tr}(\tau_4)\rangle^{(1)}
= {\sf q}_1 + {\sf q}_2 + {\sf ex} \,.
\label{NLOrho4}
\end{align}
where ${\sf q}_1$ and ${\sf q}_2$ are the two types diagrams coming from quartic vertices, while ${\sf ex}$ comprises the sum of all the exchange diagrams.

For the quartic vertices without derivatives we have
\begin{align}
{\sf q}_1 \ &= - 4! \frac{1}{g}\frac{{\cal C}_2^{-2}}{3} \int d^2 \sigma 
\frac{1}{z^2} K_2(z,\tau,\tau_1)K_2(z,\tau,\tau_2)K_2(z,\tau,\tau_3)K_2(z,\tau,\tau_4)\\
\ & = -\frac{8}{g} {\cal C}_2^2 D_{2,2,2,2}(\tau_1,\tau_2,\tau_3,\tau_4) \,.
\end{align}
\noindent
The quartic vertex with derivatives gives
\begin{align}
{\sf q}_2 \ &= \frac{8 {\cal C}_2^{-2}}{g}\frac{\cos^2\vartheta}{8} \left(\delta^{\alpha\beta}\delta^{\gamma\rho}+\delta^{\alpha\gamma}\delta^{\beta\rho}+\delta^{\alpha\rho}\delta^{\beta\gamma}\right)
 \nonumber \\
& \hspace{1cm} \int d^2 \sigma \, z^2 \, \partial_\alpha K_2(z,\tau,\tau_1)\partial_\beta K_2(z,\tau,\tau_2) \partial_\gamma K_2(z,\tau,\tau_3) \partial_\rho K_2(z,\tau,\tau_4)\,,
\end{align}
which, using the identity \eqref{dKdK} in Appendix \ref{app: D integrals}, becomes
\begin{align}
{\sf q}_2  =\ & \frac{\cos^2\vartheta}{g}16{\cal C}_2^2(D_{2,2,2,2}-2\tau_{12}^2 D_{3,3,2,2}-2\tau_{34}^2 D_{2,2,3,3}+4\tau_{12}^2\tau_{34}^2 D_{3,3,3,3}) 
 \nonumber \\
& +\frac{\cos^2\vartheta}{g}16{\cal C}_2^2(D_{2,2,2,2}-2\tau_{13}^2 D_{3,2,3,2}-2\tau_{24}^2 D_{2,3,2,3}+4\tau_{13}^2\tau_{24}^2 D_{3,3,3,3}) 
\nonumber\\
& +\frac{\cos^2\vartheta}{g}16{\cal C}_2^2(D_{2,2,2,2}-2\tau_{14}^2 D_{3,2,2,3}-2\tau_{23}^2 D_{2,3,3,2}+4\tau_{14}^2\tau_{23}^2 D_{3,3,3,3}) \,.
 \end{align}
The exchange term ${\sf ex}$ in \eqref{NLOrho4} admits the expansion
\begin{align}
{\sf ex}  =\ & \frac{16\sin^2\vartheta}{g} {\cal C}_2^2
\, \frac{\chi^4}{\tau_{12}^4\tau_{34}^4}\, \left(f_{\sf ex}^{(12)}+f_{\sf ex}^{(13)}+f_{\sf ex}^{(14)}\right)
\,,
 \end{align}
where each term can be related to simpler contact integrals through a differential operator \cite{DHoker:1999mqo,DHoker:1999kzh} (see \cite{Zhou:2018sfz,Bliard:2022xsm} for recent reviews). More precisely, one has
\begin{equation}
(C_{(2)}^{ij} - 2)f^{(ij)}_{\sf ex}=  D_{2,2,2,2}\tau_{13}^4\tau_{24}^4 \,,
\end{equation}
where $C_{(2)}^{ij}$ is the quadratic Casimir acting on the two boundary points $i$ and $j$.
Some of the integration constants in $f^{(ij)}$ are fixed imposing crossing and braiding symmetry
\begin{align}
f^{(12)}_{\sf ex}(\chi) =&\ f^{(14)}_{\sf ex}(1-\chi)\,,
\\
f^{(13)}_{\sf ex}(\chi) =&\ f^{(13)}_{\sf ex}(1-\chi)\,,
\\
f^{(13)}_{\sf ex}(\chi) =&\ (1-\chi)^{-4}f^{(14)}_{\sf ex}\left(\tfrac{\chi}{\chi-1}\right)\,.
\end{align} 
Then, we obtain
\begin{align}
{\sf ex}  =\ & \frac{1}{g}\frac{2\sin^2\vartheta}{9 \pi\tau_{12}^4\tau_{34}^4}\frac{(1 - \chi + \chi^2)}{(1 - \chi)^4}
 (36 - 108 \chi + 73 \chi^2 + 34 \chi^3 + 73\chi^4 - 
    108 \chi^5 + 36 \chi^6)\nonumber\\
& + \frac{k}{g}\frac{\sin^2\vartheta}{6\pi \tau_{12}^4\tau_{34}^4}\frac{(1 - \chi + \chi^2)}{(1-\chi)^4}
 \left( 1 - 3 \chi + 2 \chi^2 + \chi^3 + 2 \chi^4 - 3 \chi^5 + \chi^6\right)\nonumber
 \\
 & + \frac{1}{g}\frac{2\sin^2\vartheta}{9 \pi\tau_{12}^4\tau_{34}^4}\frac{\chi^4}{(1 - \chi)^5}
 (31 - 93 \chi + 343 \chi^2 -531 \chi^3 + 448\chi^4 - 
    198 \chi^5 + 36 \chi^6)\log(\chi)\nonumber
 \\
 & + \frac{1}{g}\frac{3 k \sin^2\vartheta}{ \pi\tau_{12}^4\tau_{34}^4}\frac{\chi^4}{(1 - \chi)^5}
 (2 - 6 \chi + 20 \chi^2 -30 \chi^3 + 25\chi^4 - 
    11 \chi^5 + 2 \chi^6)\log(\chi)\nonumber
    \\
    & +  \frac{1}{g}\frac{\sin^2\vartheta}{ 9\pi\tau_{12}^4\tau_{34}^4}\frac{1}{\chi}
    (72 - 36 \chi -4 \chi^2 -2 \chi^3 -4\chi^4 - 
    36 \chi^5 + 72 \chi^6)\log(1-\chi)\nonumber
    \\
 & + \frac{1}{g}\frac{3 k \sin^2\vartheta}{ \pi\tau_{12}^4\tau_{34}^4}\frac{1}{\chi}(2-\chi-\chi^5+2\chi^6)
\log(1-\chi)\,,
    \end{align}
where $k$ is the remaining integration constant, which can be fixed by taking the small $\chi$ limit. Collecting all the contributions to \eqref{LOrho4} and \eqref{NLOrho4},
we have
\begin{equation}
    \langle{\cal O}_{\sf tr}(\tau_1){\cal O}_{\sf tr}(\tau_2){\cal O}_{\sf tr}(\tau_3){\cal O}_{\sf tr}(\tau_4)
\rangle = 
\frac{1}{\tau_{12}^4\tau_{34}^4}
\left(1-\frac{1}{g} \frac{k\sin^2\vartheta}{2\pi} \chi^2+ {\cal O}(\chi^3)\right) \,.
\end{equation}
On the other hand, taking the same limit in the
conformal block expansion we obtain
\begin{align}
\langle{\cal O}_{\sf tr}(\tau_1){\cal O}_{\sf tr}(\tau_2){\cal O}_{\sf tr}(\tau_3){\cal O}_{\sf tr}(\tau_4)
\rangle = &\  
\frac{1}{\tau_{12}^4\tau_{34}^4} \sum_{\Delta}
 C_{{\cal O}_{\sf tr}{\cal O}_{\sf tr}{\cal O}_{\Delta}}^2 
 g_\Delta(\chi)
\nonumber
\\
= & \  \frac{1}{\tau_{12}^4\tau_{34}^4}(1+\chi^2 C_{{\cal O}_{\sf tr}{\cal O}_{\sf tr}{\cal O}_{\sf tr}}^2+ {\cal O}(\chi^3)) \,.
\end{align}
Comparing the two expansion and using \eqref{Crho3} we determine that $k=-\frac{4}{3}$.

The final result for the next-to-leading order correlator is
\begin{align}
\langle {\cal O}_{\sf tr}(\tau_1)&{\cal O}_{\sf tr}(\tau_2) {\cal O}_{\sf tr}(\tau_3){\cal O}_{\sf tr}(\tau_4)\rangle^{(1)}
= \nonumber \\
&   
-\frac{1}{g}\frac{1}{3 \pi\tau_{12}^4\tau_{34}^4} \frac{(1 - \chi + \chi^2)}{(1 - \chi)^4} (12-36\chi+25\chi^2+10\chi^3+25\chi^4-36\chi^5+12 \chi^6)\nonumber
\\&   
+\frac{1}{g}\frac{\sin^2\vartheta}{ \pi\tau_{12}^4\tau_{34}^4} \frac{(1 - \chi + \chi^2)}{(1 - \chi)^4} (4-12\chi+9\chi^2+2\chi^3+9\chi^4-12\chi^5+4 \chi^6)\nonumber
\\&   
-\frac{2}{g}\frac{\cos^2\vartheta}{\pi\tau_{12}^4\tau_{34}^4}
\frac{\chi^4}{(1-\chi)^5}
(2 - 6 \chi + 20 \chi^2 - 30 \chi^3 + 25 \chi^4 - 
 11 \chi^5 + 2 \chi^6)\log(\chi)\nonumber
 \\&   
-\frac{2}{g}\frac{\cos^2\vartheta}{\pi\tau_{12}^4\tau_{34}^4}  
\frac{1}{\chi}
(2 - \chi  - 
 \chi^5 + 2 \chi^6)\log(1-\chi) \,.
\label{NLOrho4final}
\end{align}

Multiplying by 9, to take into account the different normalization we used for the displacement operator $\rho$, \eqref{NLOrho4final} matches exactly the expression \eqref{rho4ptbootstrap} obtained with bootstrap, provided that
\begin{equation}
\tilde{b}_{1,0}^{(1)}  =      
-\frac{1}{g}\frac{2}{\pi}\sin^2\vartheta\,,
\qquad
\tilde{b}_{2,0}^{(1)} = -\frac{1}{g}\frac{1}{2\pi}\left(3-\sin^2\vartheta\right)\,.
\label{coeffidentification}
\end{equation}
These equations provide an holographic interpretation for the two parameters $\tilde{b}_{1,0}^{(1)}$ and $\tilde{b}_{2,0}^{(1)}$ that were not fixed by the bootstrap procedure of Section \ref{sec: 4 pts bootstrap}. As we have anticipated, the $\tilde{b}_{1,0}^{(1)}\to 0$ limit corresponds to the case in which the flux of the background is pure R-R. The other critical limit we had identified in the bootstrap analysis
was $\tilde{b}_{1,0}^{(1)}\to 2\tilde{b}_{2,0}$, in which all the anomalous dimensions of long multiplets become vanishing. We can see now that this is not a physically interesting limit in our setup, given that $\tilde{b}_{1,0}^{(1)}-
2\tilde{b}_{2,0}^{(1)}\propto \cos^2\vartheta$ becomes vanishing for $\vartheta\to\frac{\pi}{2}$, i.e. when the flux of the background is pure NS-NS. In this limit the worldsheet becomes parallel to the boundary of AdS$_3$ and no longer describes a line defect in a CFT$_2$.

\subsubsection*{Correlators from $\phi^9$ fluctuations}

The massless scalar fluctuation $\phi^9$ is in correspondence with the descendant of the tilt multiplet defined in \eqref{varphidef}, whose scale
dimension is $\Delta = 1$. The bulk-to-boundary propagator in this case is
\begin{equation}
K_1(z,\tau,\tau') = {\cal C}_1 \frac{z}{z^2+(\tau-\tau')^2}\,,
\qquad
{\cal C}_1 = \frac{1}{\pi}\,.
\end{equation}
Taking into account the vertex \eqref{S4phi9} and using the identity \eqref{dKdK} the connected four-point correlator (for unit normalized operators) becomes
\begin{align}
\langle {\cal O}_{9}(\tau_1){\cal O}_{9}(\tau_2) {\cal O}_{9}(\tau_3) {\cal O}_{9}(\tau_4)\rangle
 &=  \frac{\cos^2\vartheta}{g\pi^2}(D_{1,1,1,1}-2\tau_{12}^2 D_{2,2,1,1}-2\tau_{34}^2 D_{1,1,2,2}+4\tau_{12}^2\tau_{34}^2 D_{2,2,2,2}) 
 \nonumber \\
& +\frac{\cos^2\vartheta}{g\pi^2}(D_{1,1,1,1}-2\tau_{13}^2 D_{2,1,2,1}-2\tau_{24}^2 D_{1,2,1,2}+4\tau_{13}^2\tau_{24}^2 D_{2,2,2,2}) 
\nonumber\\
& +\frac{\cos^2\vartheta}{g\pi^2}(D_{1,1,1,1}-2\tau_{14}^2 D_{2,1,1,2}-2\tau_{23}^2 D_{1,2,2,1}+4\tau_{14}^2\tau_{23}^2 D_{2,2,2,2}) 
\nonumber\\
\ &= \frac{\cos^2\vartheta}{\pi g \tau_{12}^2\tau_{34}^2} 
\left(2\frac{(1-\chi+\chi^2)^2}{(1-\chi)^2}
+\frac{2-\chi-\chi^3+2\chi^4}{\chi}\log(1-\chi)\right.
\nonumber\\
\ & \hspace{1.85cm}+ \left.
\frac{\chi^2(2 - 4 \chi + 9 \chi^2 - 7 \chi^3 + 2 \chi^4)}{(1-\chi)^3}
\log(\chi)\right) 
\label{O94ptwitten}
\end{align}

To compare with the bootstrap result \eqref{varphi4ptbootstrap}, we must multiply \eqref{O94ptwitten} by 4, as the operator $\hat\varphi$ has a norm $\sqrt2$. With the holographic identification of coefficients given in \eqref{coeffidentification}, the agreement with the bootstrap result provided in \eqref{varphi4ptbootstrap} is exact.

%% file: integrableWLABJM.tex
A problem that naturally arises when studying a fixed point of the renormalization group is the description of its spectrum of scaling dimensions. This task acquires an additional complexity when the fixed point admits exactly marginal deformations, in which case one can further consider the dependence of the scaling dimensions with respect to the couplings of the deformations. This is precisely the case of theories like ${\cal N}=4$ super Yang-Mills and ABJM, in which scaling dimensions depend on the gauge coupling of the theory. Interestingly, the coupling dependence of the scaling dimensions (i.e. the anomalous dimensions) can be studied in those theories by means of integrability methods. This has provided not only an application of integrability tools beyond two-dimensional QFTs but also a way to access the non-perturbative regime of the spectrum.

In the AdS/CFT context integrability techniques were first used for the description of the spectrum of anomalous dimensions of single-trace operators in the planar limit of ${\cal N}=4$ super Yang-Mills (for a detailed review see \cite{Beisert:2010jr}). These results were later extended in that theory to the analysis of insertions within the contour of Wilson lines, as first studied in \cite{Drukker:2006xg}. A key outcome of the integrability analysis of line defects was the construction of a set of \textit{Thermodynamic Bethe Ansatz} (TBA) equations for the computation of the angle-dependent cusp anomalous dimension of ${\cal N}=4$ sYM \cite{Drukker:2012de,Correa:2012hh}. This constitutes an ubiquitous quantity of the theory, and describes both the UV divergences of cusped Wilson lines and the IR divergences of scattering amplitudes. The aforementioned TBA equations were exactly solved in the small-angle limit, providing an all-loop computation for the  \textit{bremsstrahlung} function \cite{Gromov:2012eu} (i.e. the quadratic term in the small-angle expansion of the cusp anomalous dimension). Crucially, the same function was also computed using supersymmetric localization \cite{Correa:2012at}, and the corresponding comparison between both results led to an all-loop derivation of the \textit{interpolating function} of the theory \cite{Gromov:2012eu}, which governs the coupling dependence of every all-loop integrability-based result.

Integrable theories constitute useful laboratories for probing QFTs beyond their weak-coupling regime. However, the use of integrability methods is currently mostly restricted to the analysis of QFTs in Minkowski space, and their extension to the study of QFTs in curved backgrounds remains to be understood. As discussed in Chapter \ref{ch:adscft}, the AdS/CFT dictionary identifies conformal defects defined by Wilson lines with open strings characterized by an $AdS_2$ induced metric on their worldsheet. In particular, the expansion of the string action in terms of fluctuations defines an effective QFT in $AdS_2$. Integrable line defects in the AdS/CFT framework consequently provide valuable toy models for understanding the application of integrability methods to the study of QFTs in curved spacetime\footnote{For a recent analysis in this direction see \cite{Antunes:2025iaw}.}.  This observation serves then as a motivation to explore the integrability properties of line defects in the AdS/CFT correspondence.

The purpose of this chapter is to provide a first step towards an integrability description of insertions within Wilson lines in the ABJM theory. We will provide evidence indicating that the 1/2 BPS Wilson line of ABJM defines an integrable theory, both at weak and strong coupling. Moreover, we will construct a set of TBA equations for the description of the cusp anomalous dimension of the theory, which we will test at the one-loop order. Our results carve a way for an integrability-based computation of the bremsstrahlung function of ABJM, which is known to all loops from localization \cite{Lewkowycz:2013laa,Bianchi:2014laa,Bianchi:2017svd,Bianchi:2018scb}. A comparison of both results would give an all-loop derivation of the interpolating function of the theory, for which an all-loop conjecture is known \cite{Gromov:2014eha,Cavaglia:2016ide}. As in the ${\cal N}=4$ sYM theory, the interpolating function percolates in every integrability-based computation, and therefore its precise all-loop derivation constitutes a fundamental problem in the integrability analysis of the theory.

The chapter is structured as follows. In Section \ref{sec: review integrability N=4} we provide a review of how integrability methods can be used in the ${\cal N}=4$ super Yang-Mills theory to compute the angle-dependent cusp anomalous dimension of the theory, while Section \ref{sec: review spin chains ABJM} is dedicated to a review on the use of integrability tools for the description of single-trace operators in the ABJM theory. In Section \ref{sec: open spin chain} we discuss how one can compute the anomalous dimensions of insertions along the contour of the 1/2 BPS Wilson line of ABJM by means of an open spin chain. We show that the reflection matrix that describes the interaction of magnons with the boundary of the spin chain can be fixed by the $SU(1|2)$ symmetry of the problem up to an overall dressing phase, for which we provide an all-loop proposal in Section \ref{sec: crossing}. We check such a proposal both at weak and strong coupling, and we show that it solves a crossing equation. In Section \ref{sec: Y-system} we propose a $Y$-system of equations for the cusp Wilson line of ABJM, whose asymptotic solution we use in Section \ref{sec: gamma cusp from BTBA} to reproduce the one-loop cusp anomalous dimension of the theory. Finally, in Section \ref{sec: btba eqs} we provide the integral Boundary Thermodynamic Bethe Ansatz (BTBA) equations for the cusped Wilson line of ABJM.

This chapter is based on \cite{Correa:2023lsm}, which is work perform by the author of this thesis in collaboration with D.H. Correa and V. I. Giraldo-Rivera.

\section{From spin chains to the cusp anomalous dimension}
\label{sec: review integrability N=4}

Before discussing the application of integrability ideas to the ABJM theory, let us begin this chapter by providing a brief review on how integrable methods can be applied in ${\cal N}=4$ super Yang-Mills, in particular to the computation of the cusped anomalous dimension of the theory. A more detailed review on the use of integrability tools in the AdS/CFT     context can be found in \cite{Beisert:2010jr}.

\subsubsection*{Spin-chain description of single-trace operators}

Let us focus on the ${\cal N}=4$ super Yang-Mills theory, and let us consider a single-trace operator
\begin{equation}
\label{so(6) sector}
{\rm Tr} (\Phi^{I_1}\Phi^{I_2} \dots \Phi^{I_L}) \,,
\end{equation}
constructed with the scalar fields of the theory (see Chapter \ref{ch:adscft}). Using the notation ${\cal O}^A$ for operators of the type depicted in \eqref{so(6) sector}, we can express the renormalized operators ${\cal O}_{\rm ren}^A$ as
\begin{equation}
{\cal O}^{A}_{\rm ren}:= Z^{A}_{B} {\cal O}^{B} \,,
\end{equation}
where $Z$ is a renormalization matrix.
This allows to define a \textit{matrix of anomalous dimensions} as
\begin{equation}
    \Gamma= Z^{-1} \frac{\partial Z}{\partial \log \mu} \,,
\end{equation}
where $\mu$ is the renormalization scale (otherwise stated we will always work in dimensional regularization). Eigenvectors ${\cal O}_*$ of $\Gamma$ behave as
\begin{equation}
\langle {\cal O}_* (x) {\cal O}_* (y) \rangle \propto \frac{1}{|x-y|^{2(L+\gamma_*)}} \,.
\end{equation}
where $\gamma_*$ is the eigenvalue of ${\cal O}_*$, known as the \textit{anomalous dimension} of the ${\cal O}_*$ operator.

The operators introduced in \eqref{so(6) sector} describe what is usually referred to as the $SO(6)$ sector of the theory, given that the R-symmetry indices in \eqref{so(6) sector} account for all possible $SO(6)$ polarizations. It is convenient to think of the $\Gamma$ matrix as acting on the tensor product space $\otimes_{i=1}^L \mathbb{R}^6$, i.e. as an operator acting on an $SO(6)$ spin chain with $L$ sites. Moreover, given the presence of the trace in \eqref{so(6) sector} one should further take the spin chain to be periodic. In a groundbreaking paper \cite{Minahan:2002ve}, Minahan and Zarembo showed that the $\Gamma$ matrix for this sector is
\begin{equation}
\label{one loop so(6) Gamma}
\Gamma = \frac{\lambda}{16 \pi^2} \sum_{i=1}^{L} ({\bf K}_{i,i+1}+2-2 {\bf P}_{i,i+1}) + \mathcal{O}(\lambda^2) \,,
\end{equation}
where the permutation operator ${\bf P}_{i,i+1}$ acts on the neighboring sites $i$ and $i+1$ as ${\bf P}_{I_i I_{i+1}}^{J_i J_{i+1}}= \delta_I^{J_{i+1}}  \delta_{I_{i+1}}^{J_{i}}$ while the trace operator ${\bf K}_{i,i+1}$ satisfies ${\bf K}_{I_i I_{i+1}}^{J_i J_{i+1}}= \delta_{I_i,I_{i+1}} \delta^{J,J_{i+1}}$.
Crucially, the matrix \eqref{one loop so(6) Gamma} coincides with the hamiltonian of an integrable model, and therefore opens the door to the application of integrable methods for the study of the spectrum of anomalous dimensions in ${\cal N}=4$ super Yang-Mills. 

The above results can be extended to the full $PSU(2,2|4)$ sector of the theory, as shown in \cite{Beisert:2003jj,Beisert:2003yb}. The vacuum state of the $PSU(2,2|4)$ spin chain can be chosen to be the protected ${\rm Tr}(Z^L)$ operator, where $Z=\frac{1}{{\sqrt{2}}}(\Phi^1+i\Phi^2)$. It should be noted that the vacuum state is invariant under an $SU(2|2)^2$ subgroup of the $PSU(2,2|4)$ symmetries of the theory. Excited states (i.e. operators with a non-vanishing anomalous dimension) are described by impurities propagating as plane waves over the ${\rm Tr}(Z^L)$ vacuum, and are commonly known as \textit{magnons}. These excitations accommodate into representations of the $SU(2|2)^2$ symmetry group of the vacuum: single-particle excitations transform in the tensor product of two fundamental representations of $SU(2|2)$, while bound-state magnons accommodate into products that include symmetric representations of $SU(2|2)$. Bound states involving $Q$ particles are usually referred to as $Q$-magnons.

\subsubsection*{Asymptotic spectrum}

For simplicity, let us focus for the moment on a spin chain with an $SU(2|2)$ invariant vacuum, as the corresponding results can be easily extended to the $SU(2|2)^2$ invariant case. Moreover, let us start by discussing the asymptotic regime, i.e. the limit in which all impurities that propagate along the spin chain are well separated. The total energy of a state with $n$ $Q$-magnons can be therefore accurately approximated as
\begin{equation}
    E_{\rm total}(\{ Q_1,p_1\}, \dots, \{ Q_n,p_n\})= \sum_{i=1}^n E_{Q_i} (p_i) \,,
    \label{total asymptotic energy}
\end{equation}
where
\begin{equation}
\label{magnon energy N=4}
E_Q(p)=\sqrt{Q^2+16 \, h^2_{{\cal N}=4}(\lambda)\sin^2(p/2)}\,,
\end{equation}
is the energy of a freely propagating $Q$-magnon with momentum $p$, and $h_{{\cal N}=4}(\lambda)$ is
\begin{equation}
\label{h N=4}
    h_{{\cal N}=4}(\lambda)= \frac{\sqrt{\lambda}}{4 \pi} \,,
\end{equation}
to all loops \cite{Gromov:2012eu}. The dispersion relation \eqref{magnon energy N=4} is obtained as a corollary of the $SU(2|2)$ symmetry of the problem, as discussed in \cite{Beisert:2005tm, Beisert:2006qh}. As we will see later, a key difference between ${\cal N}=4$ SYM and ABJM lies in the fact that the function $h_{{\cal N}=4}(\lambda)$ is significantly simpler than its ABJM counterpart.

\begin{figure}
    \centering
    \includegraphics[width=0.75\linewidth]{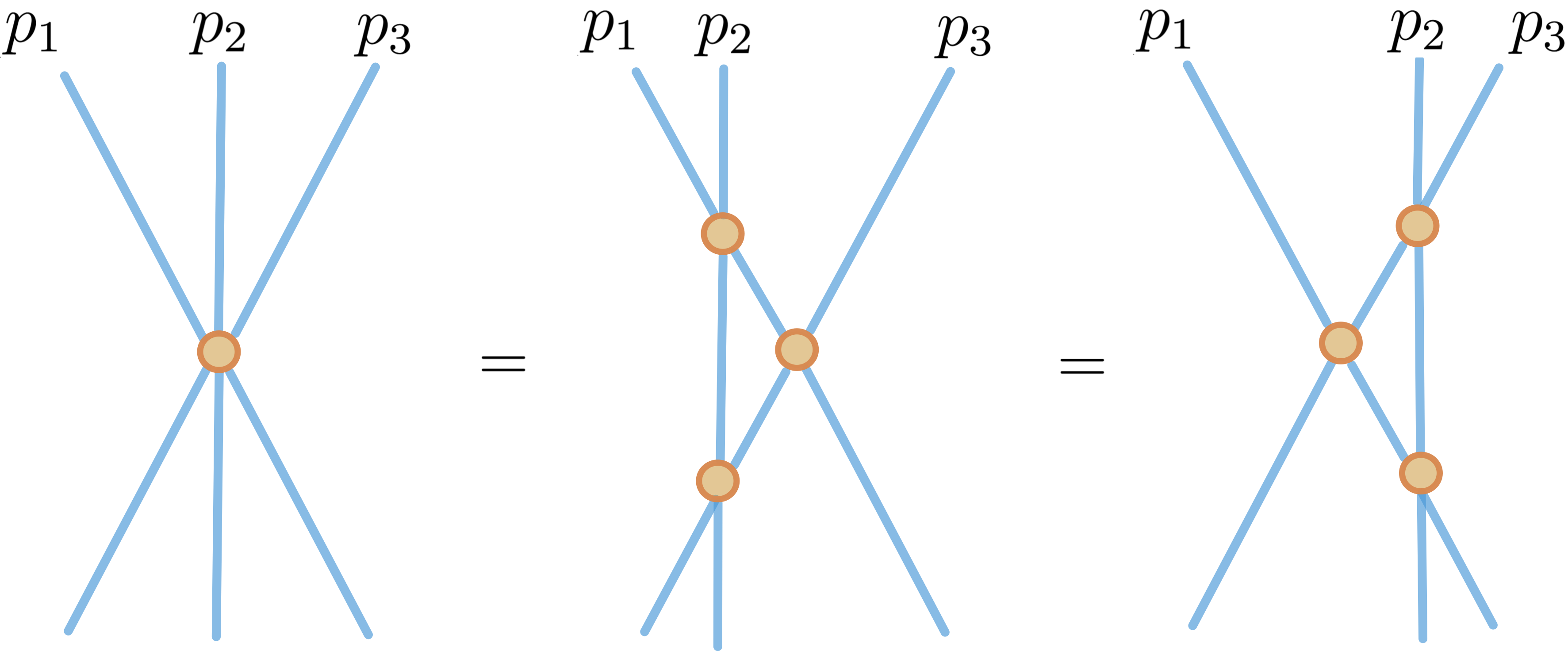}
    \caption{Factorization of a three-particle scattering process into to two-particle scatterings, which leads to the Yang-Baxter equation.}
    \label{fig:Yang-Baxter}
\end{figure}

Eq. \eqref{magnon energy N=4} implies that the problem of solving the energy spectrum of $n$-particle states in the asymptotic limit is reduced to finding the set of allowed momenta $p_i$ ($i=1, \dots, n$) of the corresponding particles. This is achieved via a set of equations known as the \textit{Asymptotic Bethe Ansatz} (ABA), which impose the periodicity constraint on the momenta and expressed them in terms of the scattering matrix of spin-chain magnons. This is were the integrable nature of the problem becomes essential: in an integrable theory the scattering matrix of an $n$-particle process can be factorized into a product of $2 \rightarrow 2$ scattering matrices. In particular, this last result implies the \textit{Yang-Baxter equation}
\begin{equation}
S(p_1,p_2) S(p_1,p_3) S(p_2,p_3)=S(p_2,p_3) S(p_1,p_3) S(p_1,p_2) \,,
\end{equation}
which can be obtained from equating the different ways into which a three-particle scattering process can be factorized (see Figure \ref{fig:Yang-Baxter}). Therefore, knowing the all-loop $2 \rightarrow 2$ S-matrix allows to formally solve the all-loop asymptotic spectrum of the theory. In a seminal paper \cite{Beisert:2005tm}, Beisert showed that the $2 \rightarrow 2$ S-matrix can be fixed up to an overall factor, known as the \textit{dressing phase}, by imposing the $SU(2|2)$ symmetry. Crucially, the resulting matrix satisfies the Yang-Baxter equation. The dressing phase was first guessed to all loops in \cite{Beisert:2006ez} and then derived from a \textit{crossing equation} \cite{Janik:2006dc,Arutyunov:2006yd} in \cite{Volin:2009uv}. The corresponding ABA equations were first conjectured in \cite{Beisert:2005fw} and then derived in \cite{Beisert:2005tm,deLeeuw:2007akd}. The precise matching of the all-loop dressing phase with its corresponding weak- and strong-coupling expectations (see \cite{Vieira:2010kb} for a review) provided one of the strongest non-trivial checks of the AdS/CFT correspondence.

\subsubsection*{Integrability in the string theory side of the duality}

The application and development of integrability methods in the AdS/CFT framework has also flourished in the string theory side of the correspondence. At the classical level the integrability of the worldsheet theory was proven in \cite{Bena:2003wd}, were the authors found a Lax connection for the non-linear sigma model in the $\frac{PSU(2,2|4)}{SO(4,1)\times SO(5)}$ coset space that describes free strings propagating in $AdS_5 \times S^5$ \cite{Metsaev:1998it}. 

To compute the anomalous dimensions in the gauge theory from a string theory perspective it is necessary to identify those quantities with the corresponding string charges. The holographic dictionary associates the scaling dimension $\Delta$ of the operators in the gauge theory with the energy $E$ of the dual string \cite{Witten:1998qj,Gubser:1998bc} \footnote{More precisely, the scaling dimension $\Delta$ is identified with the charge $E$ that is conserved in the string theory side under time shifts in global $AdS_5$ coordinates.}, while the length $L$ of the operator corresponds to the spin $J$ of the string due to its rotation around the equator of $S^5$ \cite{Berenstein:2002jq}. Therefore, anomalous dimensions are in correspondence with the $E-J$ difference. 

When studying the string spectrum at the quantum level the standard approach is to use the Green-Schwarz formalism in the light-cone gauge \cite{Berenstein:2002jq,Callan:2003xr,Callan:2004uv,Frolov:2006cc,Hentschel:2007xn}. The leading order hamiltonian in the limit in which $J \to \infty$ and $E-J= {\rm fixed}$ is that of a free theory of massive particles, and the vacuum state describes the dual of the protected ${\rm Tr}(Z^L)$ operator \cite{Berenstein:2002jq}. Corrections due to subleading interactions can be studied in the near-decompactification limit, in which the size of the string is taken to be very large and the theory can be safely considered to live on the plane: assuming that integrability holds also at the quantum order one can study the effect of interactions in an $n$ particle process by just computing the $2 \to 2$ scattering matrix. As in the gauge theory, the asymptotic spectrum can therefore be studied with the ABA equations \cite{Beisert:2005fw,Beisert:2005tm,deLeeuw:2007akd}, using that the S-matrix can be fixed by imposing $SU(2|2)^2$ invariance and a crossing equation for the overall dressing phase \cite{Beisert:2005tm,Janik:2006dc,Arutyunov:2006yd,Volin:2009uv}. For reviews and references on the integrability description of the string theory dual of ${\cal N}=4$ sYM see for example \cite{Arutyunov:2009ga,McLoughlin:2010jw,Magro:2010jx}.

\subsubsection*{Finite-size corrections and Thermodynamic Bethe Ansatz}

The one-loop hamiltonian presented in \eqref{one loop so(6) Gamma} accounts for nearest-neighbor interactions. However, this is not the case as we go to higher loops: the range of the interaction grows proportionally to the loop order \cite{Beisert:2004ry}. Generically, at $\mathcal{O}(\lambda^{\ell})$ order in perturbation theory interactions relate particles at most at $\ell$ sites of difference, e.g. two-loop corrections are described by next-to-nearest-neighbor interactions. An important consequence of this is the appearance of \textit{wrapping corrections} \cite{Ambjorn:2005wa} when considering finite-length chains, which account for interactions that can wrap around the spin chain. As an example, one could expect the ABA spectrum to fail when studying the spin chain with $L=4$ sites at the fourth loop order or beyond. The energies $E(L)$ at a given length $L$ can therefore be written as
\begin{equation}
    E(L)= E(\infty)+\Delta E(L) \,,
\end{equation}
where $E(\infty)$ is the energy \eqref{total asymptotic energy} evaluated at a solution of the ABA equations and $\Delta E(L)$ is the finite-length correction.

\begin{figure}
    \centering
    \includegraphics[width=0.75\linewidth]{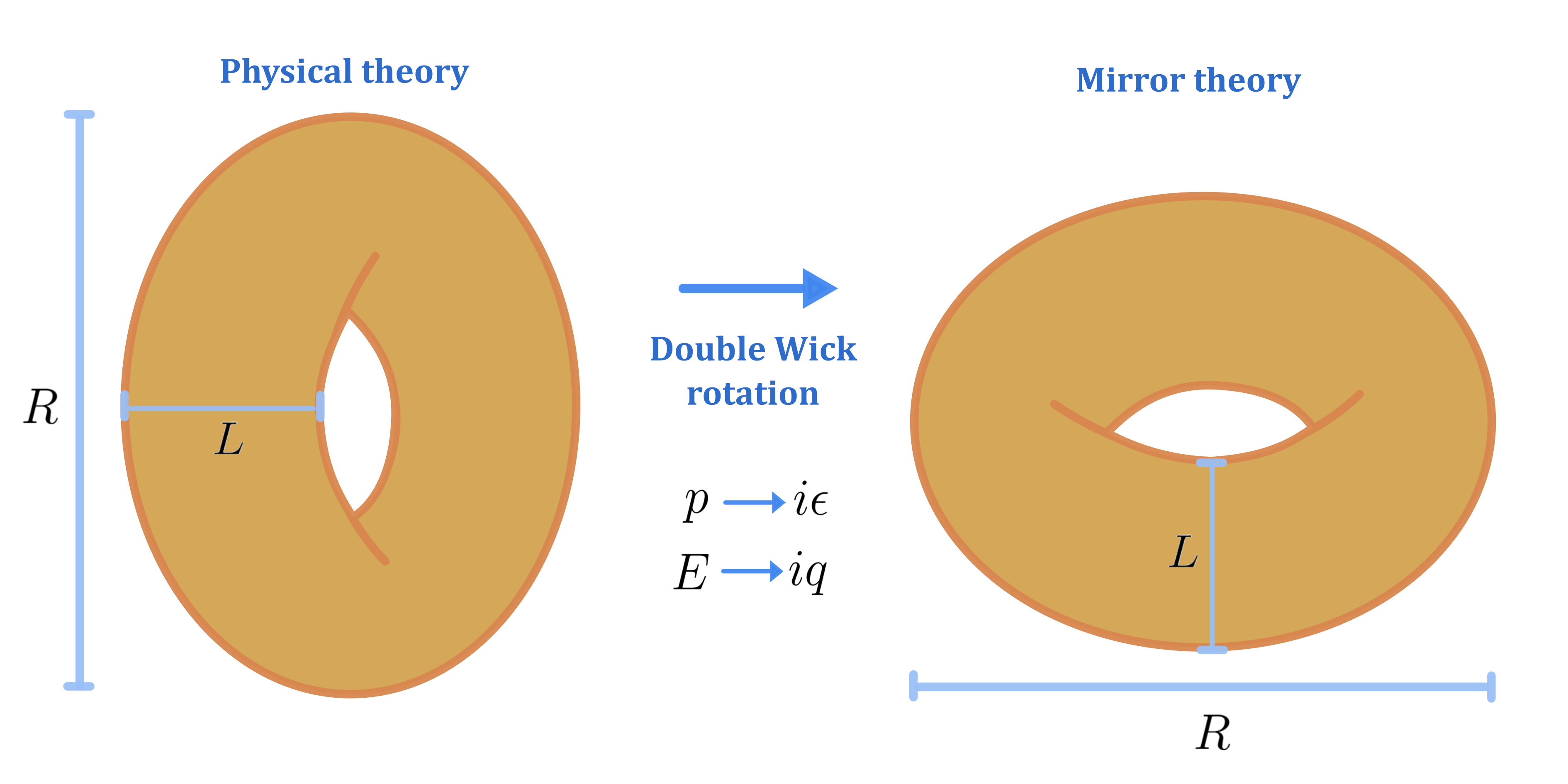}
    \caption{Schematic representation of a double Wick rotation for a system at zero temperature and with periodic boundary conditions.}
    \label{fig:doublewick}
\end{figure}

A prescription to study integrable models at finite volume was developed by Zamolodchikov \cite{Zamolodchikov:1989cf} and is known as the \textit{Thermodynamic Bethe Ansatz} (TBA). Let us consider an integrable theory in a cylinder of radius $L$ and focus on its vacuum energy $E_0(L)$. We can obtain $E_0(L)$ by considering the zero temperature limit of the system, i.e. by putting the theory on a torus of infinite radius $R$ in the time direction. In this way we get
\begin{equation}
    E_0(L) \approx -\frac{1}{R} \log Z(L,R) \,,
\end{equation}
for large $R$, where $Z$ is the partition function of the theory. As usual, $Z(L,R)$ can be obtained by performing a Wick rotation in the time direction, which puts both spacetime coordinates in the same footing at the level of the metric. As suggested by Zamolodchikov, one could further take another Wick rotation in the spatial direction, obtaining a theory in which time and space are interchanged with respect to the original set up. The energy $\epsilon$ and momentum $q$ of this new theory, known as the \textit{mirror theory}, are related to the energy $E$ and momentum $p$ of the \textit{physical theory} (i.e. the original one) via
\begin{equation}
\label{double wick rotation}
p= i \epsilon\,, \qquad \text{and} \qquad E=iq \,.
\end{equation}
The double Wick rotation that relates the physical and mirror theory is illustrated in Figure \ref{fig:doublewick}. We now have
\begin{equation}
    E_0(L) \approx -\frac{1}{R} \log \tilde{Z}(R,L) \,,
\end{equation}
where $\tilde{Z}(R,L)$ is the partition function of the mirror theory. The advantage of performing the double Wick rotation comes from the fact that the mirror theory lives in a torus with infinite radius in the spatial direction, which in integrable models allows for the use of the Asymptotic Bethe Ansatz to study the spectrum. Using the ABA equations to obtain the density of states and holes in the mirror theory one can explicitly compute the $\tilde{Z}(R,L)$ partition function at a finite $1/L$ temperature, getting \cite{Zamolodchikov:1989cf}
\begin{equation}
\label{general TBA}
E_0(L)-E_0(\infty)= - \frac{1}{2\pi} \sum_{A} \int_0^{\infty} dq \, \log[1+Y_A(q)]\,,
\end{equation}
where the $Y_A$ functions, kwown as the \textit{Y-functions}, are the solutions to a set of integral \textit{Thermodynamic Bethe Ansatz (TBA) equations}\footnote{For reviews on the derivation of the TBA equations see for example \cite{Bajnok:2010ke,vanTongeren:2016hhc}.}. 

The TBA equations for the single-trace operators of the ${\cal N}=4$ sYM theory were derived in \cite{Bombardelli:2009ns,Arutyunov:2009ur,Gromov:2009bc}. Interestingly, they can be reformulated as a set of functional equations known as the $Y$\textit{-system}, which takes the form
\begin{equation}
\label{y system N=4}
Y_{a,s}^+Y_{a,s}^- = \frac{(1+Y_{a,s+1})(1+Y_{a,s-1})}{(1+1/Y_{a+1,s})(1+1/Y_{a-1,s})} \,,
\end{equation}
where $Y_{a,s}$ are the $Y$-functions of the theory. In \eqref{y system N=4} we are using the notation $f^{[\pm a]}=f(u \pm i a/2)$, and we are writing the $Y$-functions as functions of the spectral parameter $u$, defined by
\begin{equation}
    \label{spectral parameter}
    x(u)+\frac{1}{x(u)}= \frac{u}{h_{{\cal N}=4} (\lambda)}\,,
\end{equation}
with
\begin{equation}
\frac{x^+}{x^-}=e^{ip} \,.    
\end{equation}
The set of non-vanishing nor infinite $Y$-functions satisfies
\begin{equation}
\label{Y functions N=4}
(a,s) \in \mathbb{Z} \times \mathbb{Z} \cap (\{ a=1, s\in \mathbb{Z} \} \cup \{ a\geq 1, -1 \leq s \leq 1 \} \cup \{ a=2, s= \pm2 \})\,,    
\end{equation}
in terms of which the TBA formula \eqref{general TBA} is expressed as
\begin{equation}
\label{general TBA N=4}
E_0(L)-E_0(\infty)= - \frac{1}{2\pi} \sum_{a=1} \int_0^{\infty} dq \, \log[1+Y_{a,0}(q)]\,.
\end{equation}
In addition, the $Y$-system is supplemented with set of conditions that constrain the boundary and analytical behavior of the $Y$-functions. Remarkably, changing these conditions allows the study of the spectrum of excited states of the theory \cite{Gromov:2009bc}. Let us note that, although all the $Y$-functions satisfying \eqref{Y functions N=4} have to be taken into account in the $Y$-system equations, only the $Y_{a,0}$ functions contribute to the vacuum energy in the TBA formula \eqref{general TBA N=4}.

The $Y$-system of \eqref{y system N=4}, first conjectured in \cite{Gromov:2009tv}, can alternatively be written as set of Hirota equations by making the change of variables \cite{Gromov:2009tv}
\begin{equation}
    Y_{a,s} = \frac{T_{a,s+1}T_{a,s-1}}{T_{a+1,s}T_{a-1,s}}\,.
\end{equation}
The resulting equations, known as the \textit{T-system}, are
\begin{equation}
T^+_{a,s} T^-_{a,s} = T_{a+1,s}T_{a-1,s}+T_{a,s+1}T_{a,s-1} \,,
\end{equation}
where the non-vanishing nor infinite $T$-functions have
\begin{equation}
(a,s) \in \mathbb{Z} \times \mathbb{Z} \cap (\{ 0 \leq a \leq 2, s\in \mathbb{Z} \} \cup \{ a\geq 3, -2 \leq s \leq 2 \} )\,.    
\end{equation}

Solving the TBA equations for arbitrary values of the length $L$ is usually an unfeasible task, given the infinite number of $Y$-functions. This problem was overcome by the proposal of a \textit{Quantum Spectral Curve} (QSC) \cite{Gromov:2013pga,Gromov:2014caa}, which gave a better access to a numerical solution of the spectrum by allowing to express the finite-length energies in terms of a finite number of functions.

\subsubsection*{Wilson lines and open spin chains}

The spectrum of operators of ${\cal N}=4$ super Yang-Mills includes not only local operators but also non-local operators such as Wilson lines. Moreover, as discussed in Chapter \ref{ch: line defects}, one can consider the insertion of operators along the contour of Wilson lines, which in some cases define one-dimensional defect Conformal Field Theories (dCFTs). It is therefore a natural to ask whether the integrability properties of single-trace operators extend to the case of the dCFTs defined by the standard and 1/2 BPS Wilson lines introduced in \eqref{standard WL} and \eqref{Wilson-Maldacena line}.

The perturbative integrability of insertions along the 1/2 BPS Wilson line of ${\cal N}=4$ sYM was studied in \cite{Drukker:2006xg}, where the one-loop integrability of the excitations in the $SU(2)$ sector was proven. A key difference with the case of single-trace operators relies in the fact that the 1/2 BPS Wilson line with insertions is associated to an \textit{open} spin chain, i.e. a spin chain with boundaries\footnote{Open spin chains have also appeared in other contexts within the framework of the AdS/CFT correspondence, e.g. see \cite{Berenstein:2005vf,Hofman:2007xp} for applications to the analysis of determinant operators.}. Feynman diagrams with propagators that connect the insertions to the Wilson line contribute to the \textit{boundary} hamiltonian, while diagrams without contributions coming from the Wilson line define the \textit{bulk} hamiltonian. Interestingly, the authors of \cite{Drukker:2006xg} found that for the 1/2 BPS Wilson line the complete one-loop hamiltonian for $SU(2)$ excitations is the same as the one found in \cite{Minahan:2002ve} when studying single-trace operators. The results of \cite{Drukker:2006xg} were later extended to the $SO(6)$ sector in \cite{Correa:2018fgz}. Moreover, the authors of \cite{Correa:2018fgz} found that integrability holds at the level of the $SO(6)$ and the $SU(2|3)$ sectors in the case of the standard non-supersymmetric Wilson line of the theory (see \eqref{standard WL}), but it is broken when considering the interpolation proposed in \cite{Polchinski:2011im} between the standard and 1/2 BPS Wilson lines.

\begin{figure}
    \centering
    \includegraphics[width=0.75\linewidth]{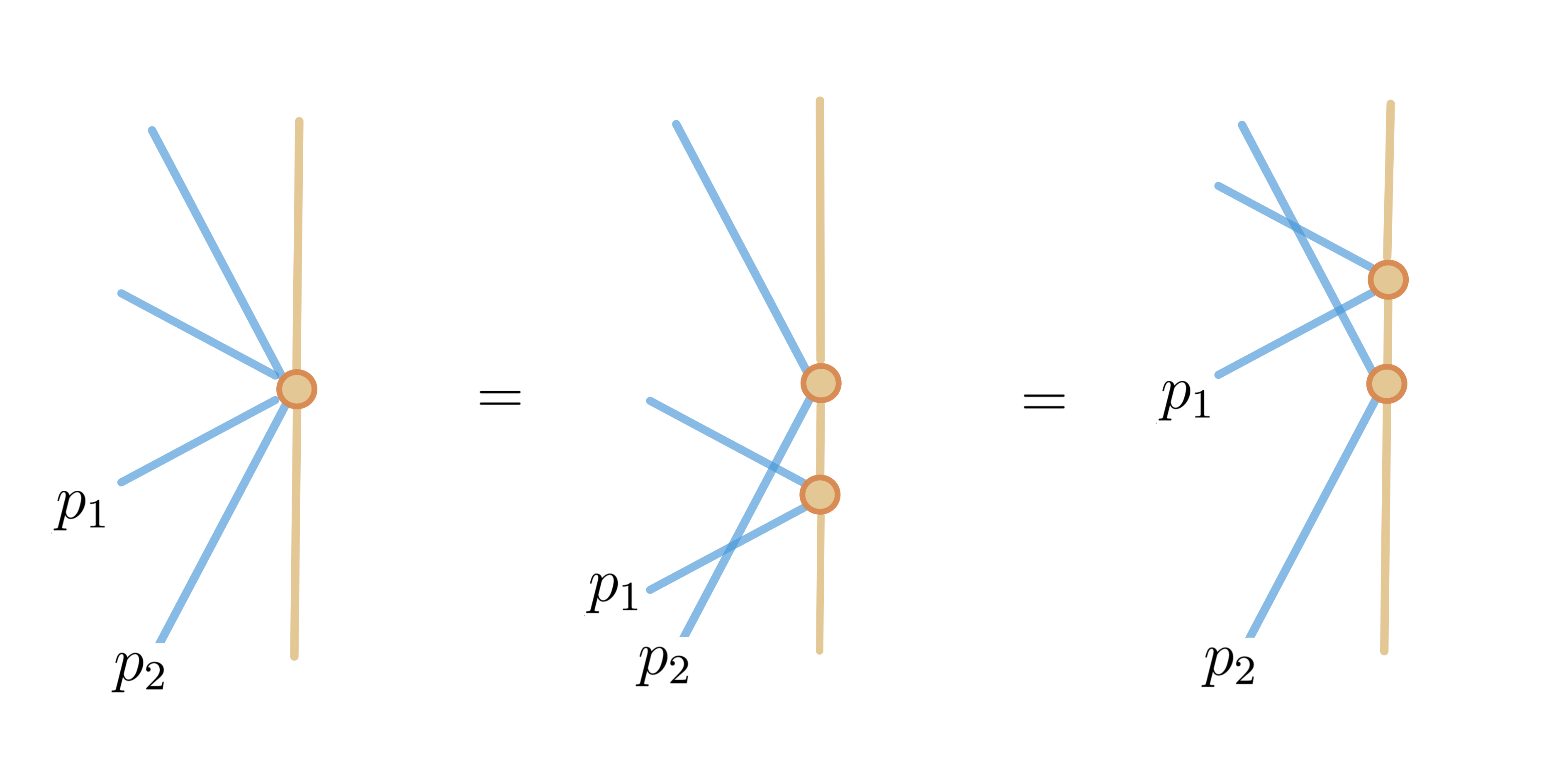}
    \caption{Factorization of a two-particle reflection process, as encoded in the boundary Yang-Baxter equation.}
    \label{fig:bybe}
\end{figure}

When dealing with open spin chains, one must consider not only the bulk scattering between propagating excitations but also the reflection of impurities at the boundaries of the spin chain, which is described by a \textit{reflection matrix}. As for the bulk scattering, integrability imposes factorization of $n$-point reflection processes into two-particle scatterings and one-particle reflections. In particular, for a two-particle reflection one gets
\begin{equation}
S(p_1,p_2)R(p_1)S(p_2,-p_1)R(p_2)=R(p_2)S(p_1,-p_2)R(p_1)S(-p_2,-p_1) \,,
\label{BYBE}
\end{equation}
where $R$ is the boundary reflection matrix. Eq. \eqref{BYBE} is know as the \textit{boundary Yang-Baxter equation}, for which a schematically is presented in Figure \ref{fig:bybe}.

As studied in \cite{Drukker:2012de,Correa:2012hh}, in the case of the 1/2 BPS line the reflection of magnons preserves only a diagonal $SU(2|2)$ subgroup of the $SU(2|2)^2$ symmetry of the bulk scattering. This residual symmetry is enough to completely fix the \textit{reflection matrix} up to an overall dressing phase, which as for the S-matrix can be computed using a crossing equation. The knowledge of the full all-loop scattering and reflection matrices drives the derivation of the \textit{open} Asymptotic Bethe Ansatz equations, that give access to the non-perturbative asymptotic spectrum of anomalous dimensions and which for the case of the 1/2 BPS line of ${\cal N}=4$ sYM were derived in \cite{Correa:2012hh}.

\subsubsection*{Boundary Thermodynamic Bethe Ansatz for the cusp anomalous dimension}

An important application of integrability tools comes when considering a Wilson line with a cusp. To that end, let us take a Wilson-Maldacena line (see \eqref{Wilson-Maldacena line}), for which one can introduce a geometrical cusp $\theta$ in the contour that the operator describes in spacetime. Moreover, at both sides of the cusp one could consider different couplings $n^I$ and $\tilde{n}^I$ to the scalar fields of the of the theory, which allows to define an extra cusp angle $\varphi$ such that $\cos \varphi= \delta_{IJ} \, n^I \tilde{n}^J$. As is well known from the renormalization theory of Wilson loops \cite{Polyakov:1980ca,Dotsenko:1979wb,Brandt:1981kf,Korchemskaya:1992je}, the presence of a cusp in the contour of a Wilson line introduces divergences that can not be absorbed with a redefinition of the couplings of the theory. More precisely, when regularizing such divergences one gets
\begin{equation}
\langle W^{\rm ren}(\theta, \varphi) \rangle= Z_{\rm cusp} (\theta, \varphi) \, \langle W(\theta, \varphi) \rangle \,,
\end{equation}
where $\langle W(\theta, \varphi) \rangle$ and $\langle W^{\rm ren}(\theta, \varphi) \rangle$ are respectively the unrenormalized and renormalized v.e.v. of the Wilson loop and $Z_{\rm cusp} (\theta, \varphi)$ is the corresponding renormalization factor. Therefore, one can naturally define a \textit{cusp anomalous dimension} as
\begin{equation}
\Gamma_{\rm cusp} (\theta, \varphi) =  \frac{\partial Z_{\rm cusp} (\theta, \varphi)}{\partial \log \mu} \,,
\end{equation}
where $\mu$ is the renormalization scale of the theory. The cusp anomalous dimension is a central object of the theory, given that it does not only controls the UV divergencies of cusped Wilson lines but also governs the IR divergences of scattering amplitudes\footnote{This can be seen as a consequence of the duality between scattering amplitudes and polygonal Wilson lines that we will discuss in Chapter \ref{ch: scatt ampl}.}.

\begin{figure}
    \centering
    \includegraphics[width=0.75\linewidth]{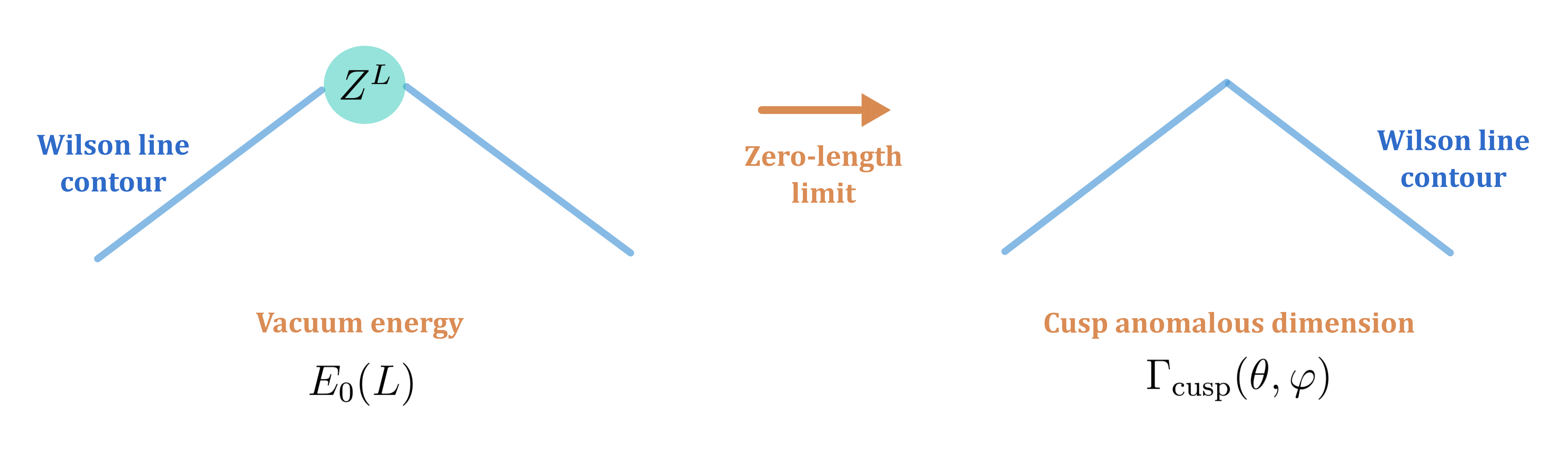}
    \caption{Zero-length limit of a cusped Wilson line with a vacuum operator $Z^L$ inserted at the position of the cusp, which allows to compute the cusp anomalous dimension as discussed in eq. \eqref{gammacuspidea}.}
    \label{fig:gammacusp}
\end{figure}

\begin{figure}
    \centering
    \includegraphics[width=0.75\linewidth]{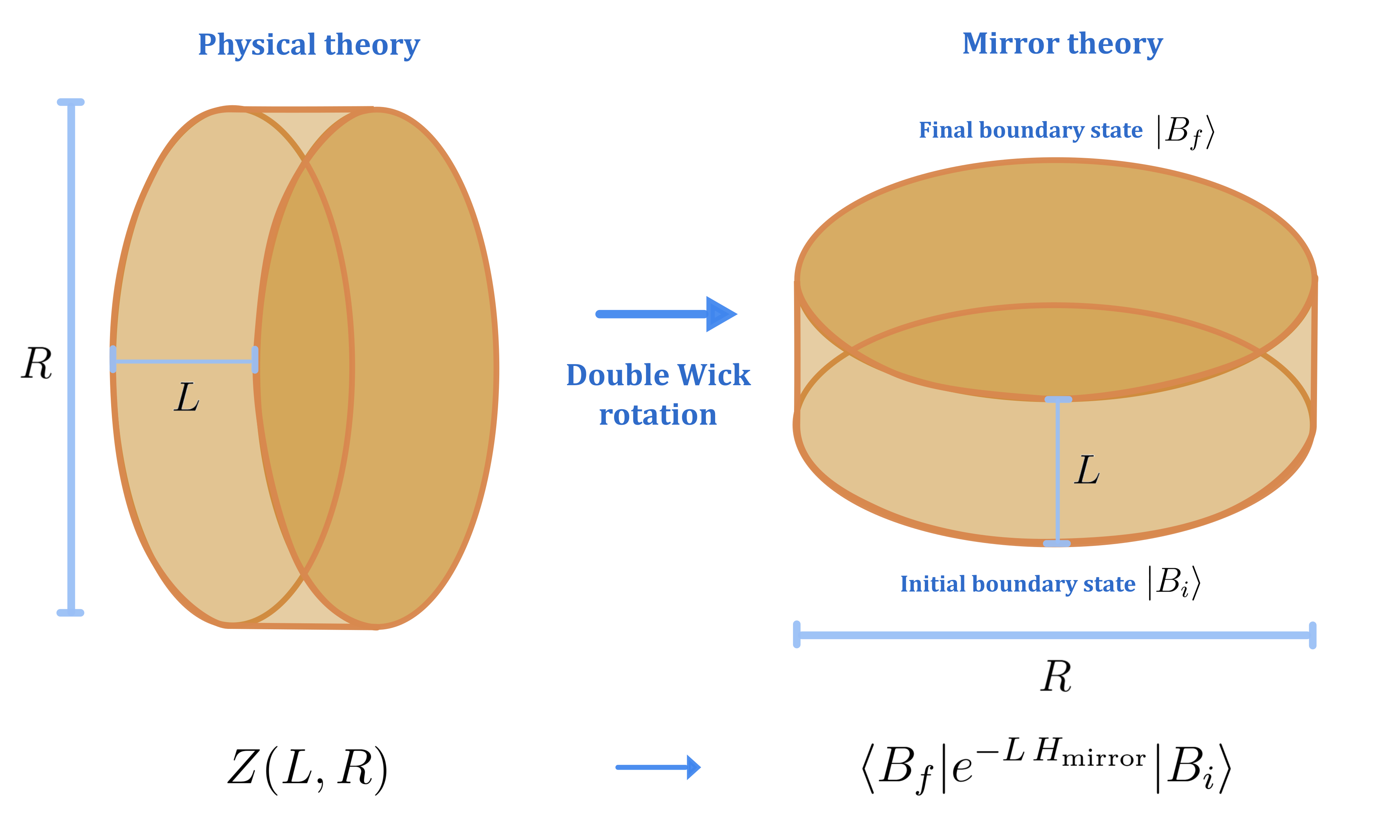}
    \caption{Schematic representation of a double Wick rotation for a system at zero temperature and with boundaries.}
    \label{fig:BTBA}
\end{figure}

A method to compute $\Gamma_{\rm cusp} (\theta, \varphi)$ by means of integrability techniques was proposed in \cite{Drukker:2012de,Correa:2012hh}. In order to introduce their idea, let us consider a cusped Wilson line with the insertion of a vacuum operator $Z^L$ at the position of the cusp. Such a state has an vacuum energy $E_0(L)$, which in the limit $L \to 0$ must reduce to the cusp anomalous dimension, i.e.
\begin{equation}
    E_0(L) \xrightarrow[L \to 0]{}\Gamma_{\rm cusp} (\theta, \varphi) \,.
    \label{gammacuspidea}
\end{equation}
Therefore, computing $E_0(L)$ for arbitrary finite values of $L$ and then taking the $L \to 0$ limit constitutes a method to obtain the value of $\Gamma_{\rm cusp} (\theta, \varphi)$. See Figure \ref{fig:gammacusp} for a schematic representation of this idea. In order to do so one can again use the double Wick rotation \eqref{double wick rotation}, which in the case of open spin chains translates the partition function of the physical theory into a transition amplitude between two boundary states $|B_i \rangle$ and $|B_f \rangle$ \cite{Ghoshal:1993tm, LeClair:1995uf} in the mirror theory, i.e.
\begin{equation}
Z(L,R) \xrightarrow[\small \begin{array}{c}
     {\rm Double \, Wick}  \\
      {\rm rotation}
\end{array}]{} \langle B_f|e^{-L \, H_{\rm mirror}}|B_i\rangle
\end{equation}\normalsize{where} $H_{\rm mirror}$ is the hamiltonian of the mirror theory, which still has periodic boundary conditions. Figure \ref{fig:BTBA} provides a schematic illustration of the double Wick rotation in the case of systems with boundaries. The $|B_i \rangle$ and $|B_f \rangle$ states are characterized by the boundaries of the theory in the physical set up, and can be expanded into operators that create and annihilate particles at the boundaries \cite{Ghoshal:1993tm, LeClair:1995uf}. Such an expansion allows to decompose the vacuum energy $E_0(L)$ into a sum of diagrams that account for the emission of virtual particles from one boundary and the later annihilation of those particles at the other boundary. In the end one obtains that the whole series sums up to an expression of the type \eqref{general TBA}, where the $Y$-functions are the solutions to a set of integral \textit{Boundary Thermodynamic Bethe Ansatz} (BTBA) equations \cite{LeClair:1995uf}. As we will discuss in more detail in the following sections for the case of the ABJM theory, the $Y$-system equations (or, equivalently, the integral BTBA equations) should not change when including integrable boundaries in a system: typically the only change should be in the asymptotic and analytic properties of the $Y$-functions \cite{Drukker:2012de,Correa:2012hh,Behrend:1995zj,OttoChui:2001xx,Gromov:2010dy,Ahn:2010ws,Ahn:2011xq,vanTongeren:2013gva,Bajnok:2012xc}. This is precisely the case observed in \cite{Drukker:2012de,Correa:2012hh} for the cusped Wilson line of the ${\cal N}=4$ super Yang-Mills theory, where the BTBA equations could be obtained by taking the TBA equations of the periodic system and then subtracting the result of evaluating them in the leading-order finite-size solution  \cite{Correa:2012hh}. 

\begin{figure}
    \centering
    \begin{subfigure}[b]{0.3\textwidth}
    \includegraphics[width=\textwidth]{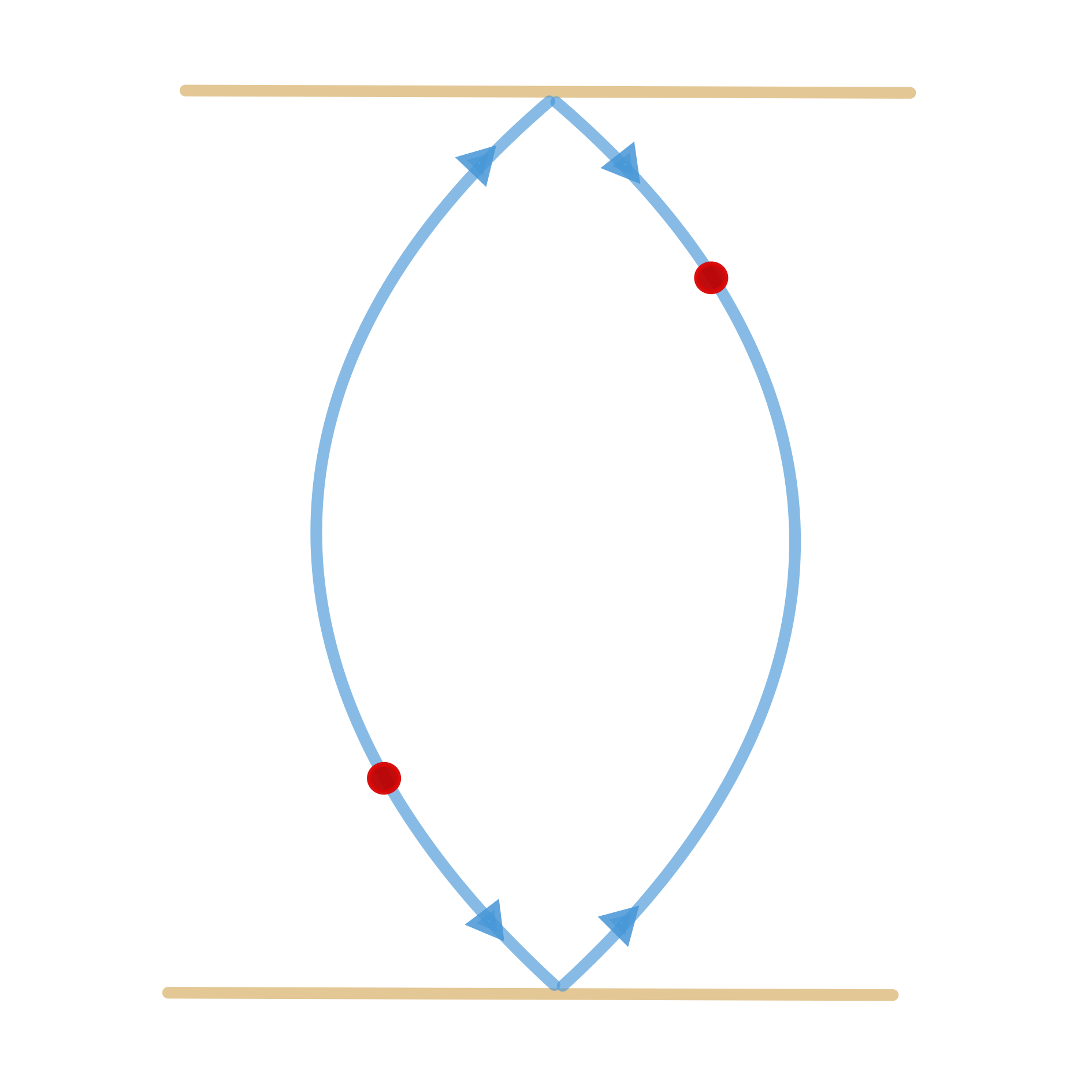}
    \caption{}
    \label{fig: luscher2}
    \end{subfigure}
    \begin{subfigure}[b]{0.3\textwidth}
    \includegraphics[width=\textwidth]{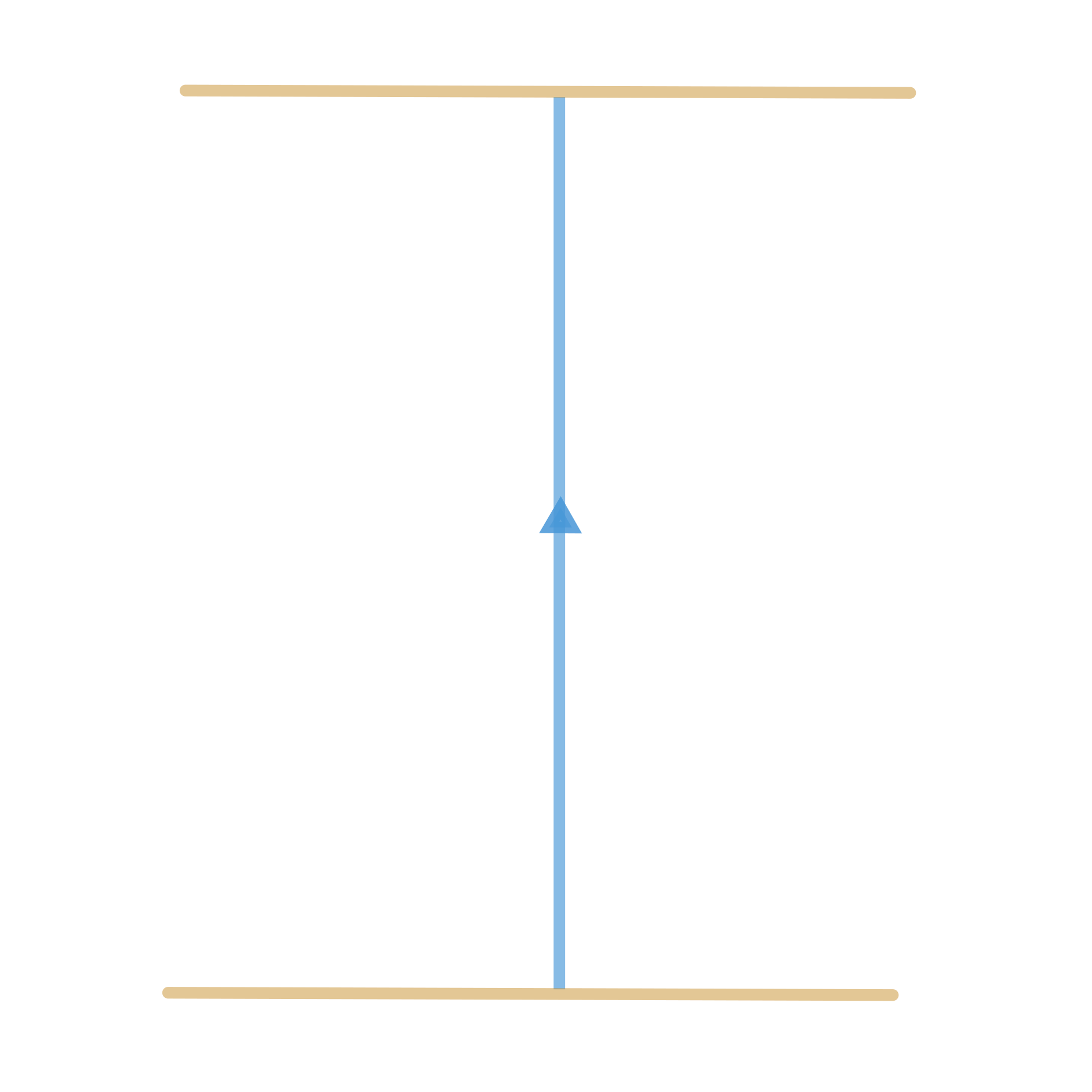}
    \caption{}
    \label{fig: luscher1}
    \end{subfigure}
    \caption{One- and two-particle Lüscher diagrams. Figure \ref{fig: luscher2} presents the two-particle process (red dots indicate charge conjugation), while Figure \ref{fig: luscher1} corresponds to the one-particle diagram.}
    \label{fig:luscher}
\end{figure}

The leading finite-size correction to the vacuum energy can be obtained via a \textit{Lüscher formula} \cite{Luscher:1986pf,Bajnok:2008bm}, which for the case of the cusped Wilson line of ${\cal N}=4$ super Yang-Mills takes the form
\begin{equation}
\label{luscher N=4}
Y_{Q,0} (q) \sim e^{-2L \, \epsilon_Q(q)} \, {\rm Tr} \left[ R_Q(q) \, C \, R_{Q,\theta, \varphi} \, (-\bar{q}) \, C^{-1} \right] \,,    
\end{equation}
where $R(q)$ is the reflection matrix for $Q$-magnons with mirror momentum $q$ and mirror energy $\epsilon_Q(q)$, $C$ is the charge-conjugation matrix, and $R_{Q,\theta, \varphi}$ is a reflection matrix obtained from $R$ via a rotation in a geometrical angle $\theta$ and in an internal angle $\varphi$. Moreover, in \eqref{luscher N=4} we have introduced the notation $\bar{q}$ for the crossing transformation
\begin{equation}
\label{crossing transformation-intro}
q \to -q  \quad \text{and} \quad E \to -E\,.
\end{equation}
Eq. \eqref{luscher N=4} gives the contribution to the transition amplitude $\langle B_f|e^{-L \, H_{\rm mirror}}|B_i\rangle$ of a process in which two particles are exchanged between the two boundaries. Such a process is represented in Figure \ref{fig: luscher2}. Nonetheless, we should also stress that for the case of the cusped Wilson line the leading correction \eqref{luscher N=4} also accounts for single-particle exchanges of virtual particles between the boundaries, whose contribution comes from double poles in the $Y_{Q,0}$ functions. This process is depicted in Figure \ref{fig: luscher1}. These poles become crucial when computing the cusp anomalous dimension from the BTBA formula.

The BTBA equations for the cusped Wilson line of ${\cal N}=4$ super Yang-Mills were used in \cite{Correa:2012hh} to obtain the first three non-trivial orders in the coupling expansion of the bremsstrahlung function of the theory, which is defined as the quadratic correction to the cusp anomalous dimension in the $\theta-\varphi$ expansion. Interestingly, it was shown in \cite{Gromov:2012eu} that the BTBA equations can be exactly solved in this small-angle limit, allowing to obtain an all-loop integrability-based expression for the bremsstrahlung function. The comparison of this result with the obtained for the same quantity using supersymmetric localization provided an all-loop derivation of the $h_{{\cal N}=4}(\lambda)$ function introduced in \eqref{magnon energy N=4}, which appears in every integrability-based result. A generalization of this idea to the ABJM theory, where an all-loop derivation of the analogous interpolating function is still lacking, constitutes one of the main motivation for deriving a set of BTBA equations for the cusped Wilson line in ABJM. This will be one of the main goals of this chapter.

Finally, let us mention that, as for the case of single-trace operators, the numerical evaluation of the finite-size spectrum of insertions along the 1/2 BPS Wilson line of ${\cal N}=4$ sYM can be improved by using the Quantum Spectral Curve formalism. For references on this topic see for example \cite{Gromov:2015dfa, Gromov:2016rrp, Cavaglia:2018lxi, Giombi:2018hsx,McGovern:2019sdd, Grabner:2020nis}. An interesting merging of the QSC and numerical conformal bootstrap ideas has been developed in the context of Wilson lines in \cite{Cavaglia:2021bnz,Cavaglia:2022qpg,Niarchos:2023lot,Cavaglia:2023mmu,Cavaglia:2024dkk} (see Chapter \ref{ch: line defects} for a brief discussion on this topic).

\section{Integrable spin chains in ABJM}
\label{sec: review spin chains ABJM}

\begin{figure}
    \centering
    \includegraphics[width=0.75\linewidth]{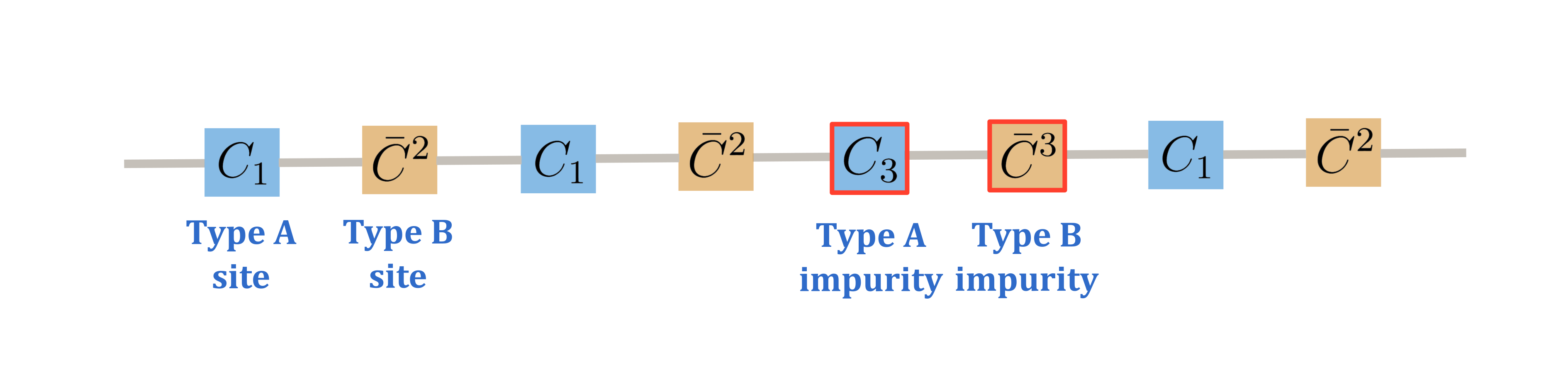}
    \caption{Schematic representation of the alternating spin chain of ABJM}
    \label{fig: alt spin chain}
\end{figure}

Before turning to the study of the integrability properties of Wilson lines in the ABJM theory, we consider instructive to review the main results regarding the integrability analysis of single-trace operators in that theory. Similar to the case discussed in the previous section for the case of ${\cal N}=4$ super Yang-Mills, in the ABJM theory the matrix of anomalous dimensions of single-trace operators has also been identified with the hamiltonian of an integrable periodic spin chain \cite{Minahan:2008hf,Minahan:2009te}. In this context, the vacuum state of the spin chain picture corresponds to the protected operator 
\begin{equation}
    \label{periodic vacuum d=3}
    {\rm Tr} \left[ \left( C_1 \bar{C}^2 \right)^{\ell} \right]\,,
\end{equation}
with $\ell \in \mathbb{N}$ (see Chapter \ref{ch:adscft} for details on the ABJM theory and its field content). When considering excited states a novel feature appears in the ABJM theory with respect to the ${\cal N}=4$ sYM case. More specifically,  in the ABJM picture one can construct an excited state either by replacing a $C_1$ or a $\bar{C}^2$ field with an impurity. Therefore, excitation waves (i.e. magnons) can be of two types,
\begin{align}
    \text{type\ } A \text{\ magnons:} & \qquad (C_3,C_4|\bar\psi^2_+,\bar\psi^2_-)\,,
    \\
     \text{type\ } B \text{\ magnons:} & \qquad
     (\bar C^3,\bar C^4|\psi_1^+,\psi_1^-)\,.
\end{align}
From the above, we say that the periodic spin chain of ABJM is \textit{alternating}, as there are two distinct sites along which impurities can propagate. See Figure \ref{fig: alt spin chain} for a schematic representation of the alternating spin chain of ABJM.

Taking into account the symmetries of the spin-chain system is crucial, as they are used to determine the scattering properties of the magnons. While in ${\cal N}=4$ sYM the vacuum state has $SU(2|2)^2$ invariance, in ABJM the corresponding symmetry is reduced to just one copy of $SU(2|2)$. In this case, each $Q$-magnon state 
accommodates in  a 
representation of the $SU(2|2)$  symmetry group \cite{Gaiotto:2008cg}, 
and each of those representations is labelled by four coefficients $({\sf a},{\sf b},{\sf c}, {\sf d})$ that characterize the action of the fermionic generators over the corresponding states. As an example, for $Q=1$ (i.e. the fundamental representation) one has
\begin{equation}
\begin{array}{ll}
   Q_{a}^{\alpha}|\phi^b\rangle
 =  {\sf a}\,\delta_{a}^{b}|\psi^\alpha\rangle,  &  Q_{a}^{\alpha}|\psi^{\beta}\rangle = {\sf b}\, \epsilon^{\alpha\beta}\epsilon_{ab}|{\phi}^{b}\rangle , \\
  S_{\alpha}^{a}|\phi^{b}\rangle
 = {\sf c}\,\epsilon_{\alpha\beta}\epsilon^{ab}|\psi^{\beta}\rangle,    & 
  S_{{\alpha}}^a|\psi^{\beta}\rangle
= {\sf d}\, \delta_{\alpha}^{\beta}|\phi^{a}\rangle ,
\end{array}
\label{QSonfundamental}
\end{equation}
with $a=1,2$ and $\alpha=3,4$ and where $Q_{a}^{\alpha}$ and $S_{{\alpha}}^a$ are the supercharges of the $SU(2|2)$ symmetry group. In the general case, the labels depend on the magnon momentum and on an unknown interpolating function $h(\lambda)$ as\footnote{We shall focus on the ABJM case, in which both gauge groups have equal ranks.}
\begin{equation}
{\sf a}=\sqrt{\frac{h(\lambda)}{Q}}\eta,\quad
{\sf b}=\sqrt{\frac{h(\lambda)}{Q}}\frac{i\zeta}{\eta}(\tfrac{x^{+}}{x^{-}}-1),\quad
{\sf c}=-\sqrt{\frac{h(\lambda)}{Q}}\frac{\eta}{\zeta x^{+}},\quad
{\sf d}=-\sqrt{\frac{h(\lambda)}{Q}}\frac{x^{+}}{i\eta}(\tfrac{x^{-}}{x^{+}}-1)\,,
\label{14:abcd}
\end{equation}
with 
\begin{equation}
x^{+}+\frac{1}{x^{+}}-x^{-}-\frac{1}{x^{-}}=\frac{iQ}{h(\lambda)},
\qquad \frac{x^+}{x^-}= e^{ip},\qquad
\eta(p,\zeta) = \zeta^{\frac 12} e^{\frac{i p}4}\sqrt{i(x^--x^+)}\,,
\label{14:mass-shell_1}
\end{equation}
and where $p$ is the magnon momentum and $\zeta$ is a phase. The unknown function is also present in the dispersion relation of magnons,
\begin{equation}
\label{magnon energy}
E_Q(p)=\frac{1}2\sqrt{Q^2+16 h^2(\lambda)\sin^2(p/2)}\,.
\end{equation}
An all-loop conjecture was made in \cite{Gromov:2014eha,Cavaglia:2016ide} for $h(\lambda)$, giving
\begin{equation}
\label{h conjecture ABJM}
\lambda= \frac{\sinh(2 \pi h)}{2\pi} \, _3F_2 \left( \frac{1}{2},\frac{1}{2},\frac{1}{2} ;1, \frac{3}{2}; - \sinh^2 (2\pi h) \right) \,.
\end{equation}
Let us note that the non-trivial functional form of $h(\lambda)$ in the ABJM theory is the main difference between \eqref{magnon energy} and the dispersion relation presented in \eqref{magnon energy N=4} for ${\cal N}=4$ sYM, for which the interpolating function was written in \eqref{h N=4}.

Being an integrable system, one can completely determine the full S-matrix of the spin chain by fixing the $2\to2$ scattering matrix. Such a matrix can be computed using the $SU(2|2)$ symmetry of the vacuum state, which fixes the scattering matrices of type $A$ and type $B$ magnons to be \cite{Ahn:2008aa}
\begin{equation}
\label{S matrix ABJM}
\begin{array}{c}
    S^{AA}(x_1,x_2) = S^{BB}(x_1,x_2) = S_0(x_1,x_2)\hat S(x_1,x_2)\,,
    \\
    S^{AB}(x_1,x_2) = S^{BA}(x_1,x_2) = \tilde S_0(x_1,x_2)\hat S(x_1,x_2)\,,
\end{array}    
\end{equation}
 where $\hat S(x_1,x_2)$ is the $SU(2|2)$-invariant matrix\footnote{We use $\hat S^{11}_{11}(x_1,x_2)=1$.} given in \cite{Arutyunov:2008zt} while $S_0(x_1,x_2)$ and $\tilde S_0(x_1,x_2)$ are the dressing phases. The scalar factors can be fixed by demanding crossing symmetry to be \cite{Ahn:2008aa,Chen:2019igg}
\begin{equation}
\label{S dressing phases}
\tilde S_0(x_1,x_2) = \sqrt{\frac{x_1^-}
{x_1^+}} \, \sqrt{\frac{x_2^+}
{x_2^-}} \, \sigma(x_1,x_2) \,, \qquad
S_0(x_1,x_2) = \frac{x_1^+-x_2^-}{x_1^--x_2^+} \,
\frac{1-\frac{1}{x_1^+ x_2^-}}{1-\frac{1}{x_1^- x_2^+}} \, \tilde S_0(x_1,x_2)  \,,
\quad 
\end{equation}
where $\sigma(x_1,x_2)$ is the BES dressing factor \cite{Beisert:2006ez}. The asymptotic Bethe ansatz equations for the periodic spin chain of ABJM where first proposed in \cite{Gromov:2008qe} and then derived in \cite{Ahn:2008aa}.

\section{Wilson lines's open spin chain}
\label{sec: open spin chain}

In order to compute the cusp anomalous dimension from an integrability approach we should first study the description of insertions within Wilson loops in terms of open spin chains. We will devote this section to such goal.

To begin with, we should identify which insertion could serve as the vacuum state of the Wilson loop spin-chain system. Following the insight obtained from the ${\cal N}=4$ sYM case \cite{Drukker:2006xg}, in ABJM one expects that a vacuum state with large R-symmetry charge should be dual to a BPS string ending on the Wilson loop's contour at the boundary of $AdS_4$ and with large angular momentum in the coordinates of the $\mathbb{CP}^3$ compact space. As shown in the Appendix \ref{app: dual string vacuum state}, one can construct a 1/6 BPS string with those properties which is invariant under a $SU(1|2)$ supersymmetry. We will therefore search for a vacuum state with the same supersymmetry.

Naively, one might expect that 
\be
{\cal D} := \left(
\begin{array}{cc}
(C_1 \bar C^2)^\ell & 0
\\
0 &  (\bar C^2 C_1)^\ell
\end{array}\right),
\label{vacuum12}
\ee
could be the vacuum state we are looking for,
as it shares some supersymmetry with the 1/2 BPS Wilson loop. However, 
the total operator, i.e. the Wilson loop with the operator \eqref{vacuum12} inserted at a position $\tau$, is not supersymmetric. Because the path-ordered exponential ${\cal W}(\tau_1,\tau_2)$ (see \eqref{general WL}) is covariant rather than invariant under supersymmetry, studying the supersymmetry transformations of insertions within Wilson loops is a bit more subtle \cite{Bianchi:2020hsz,Gorini:2022jws}.
To be more specific, let us define  ${\cal O}_W$ as the insertion of a generic operator ${\cal O}$ at the point $\tau$, i.e.
\begin{equation}
\label{WL with general insertion}
{\cal O}_{W}(\tau) := \frac{1}{2N} {\rm Tr} \left[{\cal P} {\cal W} (-\infty,\tau) \, \mathcal{O} (\tau) \, {\cal W} (\tau,\infty) \right]\,.
\end{equation}
In order to consider the transformation \eqref{WL finite transformation} of the complete operator ${\cal O}_{W}$ under supersymmetry it is instructive to introduce a \textit{covariant} supersymmetry transformation \cite{Bianchi:2020hsz,Gorini:2022jws} as
\begin{equation}
\label{covariant susy transformation}
\delta^{\rm cov} {\cal O} :=\delta {\cal O} - i [{\cal O}, \Lambda] \,.
\end{equation}
In this context, we will say that an insertion is supersymmetric if
\begin{equation}
\label{BPS condition for O}
\delta^{\rm cov} {\cal O} =0\,.
\end{equation}
Therefore, we see that despite satisfying $\delta{\cal D} =0$ the operator ${\cal D}_W$ is not supersymmetric, because the insertion does not commute with the $\Lambda$ matrix given in \eqref{lambda straight WL}. Instead, it is straightforward to verify that for
\be 
{\cal T_+} = 
\left(
\begin{array}{cc}
C_1 \bar C^2& -\frac{\eta}{2}\bar\psi^2_+
\\
0 &  \bar C^2 C_1
\end{array}\right)\,,
\label{susyinsertion}
\ee
the condition \eqref{BPS condition for O} is met. An arbitrary power of this operator will be equally BPS and provides an insertion with a large amount of the corresponding R-charge
\be 
{\cal T}_+^{\ell} = 
\left(
\begin{array}{cc}
(C_1 \bar C^2)^\ell& -\frac{\eta}{2}
\sum_k (C_1 \bar C^2)^k\bar\psi^2_+
(\bar C^2 C_1)^{\ell-k-1}
\\
0 &  (\bar C^2 C_1)^\ell
\end{array}\right)\,.
\label{susyinsertionlarge}
\ee

Although protected, we shall not consider \eqref{susyinsertionlarge} as the reference state to formulate a Bethe ansatz. Its off-diagonal block looks more like a one-impurity state (with zero momentum) than a vacuum state. Moreover, as we shall see next, the operator \eqref{susyinsertionlarge} can be regarded as a descendant when considering the covariant action of the supersymmetry transformations on a certain insertion within the Wilson line.

A more appropriate alternative to play the role of a Bethe ansatz reference state turns out to be the off-diagonal insertion
\be 
{\cal V}_\ell = 
\left(
\begin{array}{cc}
 0 & (C_1 \bar C^2)^\ell C_1 
\\
0 &  \bar 0
\end{array}\right)\,.
\label{susyinsertion0}
\ee
This operator is invariant under the supersymmetries generated by
$\bar{\Theta}^{13}_+$ and $\bar{\Theta}^{14}_+$, for which the $\Lambda$ matrix is given by
\be 
g_1 = 2{\eta}\left(\bar{\Theta}^{13}_+ C_3 + \bar{\Theta}^{14}_+ C_4\right)\,,
\qquad
\bar g_2 = 0\,.
\ee
Therefore, the insertion \eqref{susyinsertion0} breaks the $SU(1,1|3)$ symmetry of the Wilson loop to $SU(1|2)$, as expected in view of the results coming from the string theory side of the AdS$_4$/CFT$_3$ duality. Moreover, when acting with the supersymmetry transformation generated by $\bar{\Theta}^{34}_-$ on ${\cal V}_\ell$ one precisely obtains ${\cal T}_+^{\ell+1}$. Consequently, in what follows we shall consider \eqref{susyinsertion0} as our Bethe ansatz reference state.

Having identified a suitable vacuum state, we can now turn to the analysis of the impurities that can propagate along the spin chain. From the inspection of the operator \eqref{susyinsertion0}, one can see that the excited states that  propagates over such vacuum are a straightforward generalization of the type $A$ and type $B$ magnons of the periodic spin chain. Moreover, the S-matrix is the same $SU(2|2)$-invariant matrix that governs the scattering of magnons in the periodic case, as the presence of the boundary (i.e. the Wilson loop) does not affect the bulk interactions.

Let us focus now on the reflection of magnons against the boundary, which is characterized by the reflection matrix. As discussed above, there is a $SU(1|2)$ residual symmetry preserved by the Wilson loop
\eqref{WLchoice} with the insertion \eqref{susyinsertion0}. Following \cite{Hofman:2007xp,Correa:2008av}, one can use this symmetry to constrain the boundary reflection matrix of magnon excitations. The action of the right reflection matrix over the quantum numbers of a fundamental representation can be taken such that 
$(p,\zeta) \mapsto (-p,\zeta)$.
The most general reflection matrix ${\mathbf R}$ would in principle allow for the mixing of magnons of type $A$ and $B$,
\begin{equation}
{\mathbf R} = \left(
\begin{array}{cc}
R_{A}     &  \tilde{R}_{A} \\
\tilde{R}_{B}     &  R_{B}
\end{array}
\right)\,,
\label{general rotation matrix}
\end{equation}
where $R_{A/B}$ indicates the reflection of a type $A/B$ magnon into a type $A/B$ magnon. On the contrary, $\tilde{R}_{A/B}$ indicates  the reflection of a type $A/B$ magnon into a type $B/A$ magnon. The $SU(1|2)$ residual symmetry constrains the form of each of the blocks to be \cite{Drukker:2019bev}
\begin{align}
R_{A} & = R_{A}^0 \, {\rm diag}(1,1,e^{-ip/2},-e^{ip/2}) \,,
\label{RAA}
\qquad
\tilde{R}_{A}  =  \tilde{R}_{A}^0 \, {\rm diag}(1,1,e^{-ip/2},-e^{ip/2}) \,,
\\
R_{B} &  = R_{B}^0 \, {\rm diag}(1,1,e^{-ip/2},-e^{ip/2}) \,,
\qquad
\tilde{R}_{B} =  \tilde{R}_{B}^0 \, {\rm diag}(1,1,e^{-ip/2},-e^{ip/2}) \,,
\end{align}
where $R_{A}^0,R_{B}^0,  \tilde{R}_{A}^0,  \tilde{R}_{B}^0$ are dressing factors that can not be fixed with symmetry arguments.  With a reflection matrix of this form, the boundary Yang-Baxter equation is not satisfied unless $R_{A}^0 = R_{B}^0 = 0$ or $ \tilde{R}_{A}^0 = \tilde{R}_{B}^0 = 0$. The weak-coupling analysis we will present in the next section shows that, at least perturbatively, the reflection at the boundaries does not mix type $A$ and type $B$ magnons. In the following we will consider the validity of $ \tilde{R}_{A}^0 = \tilde{R}_{B}^0 = 0$
at all loops as a working assumption.

\section{Crossing symmetry and boundary dressing factors}
\label{sec: crossing}

Even in the case of  no mixing between different types of magnons at the boundary, the reflection matrix is only known up to two boundary dressing factors ${R}_{A}^0(p)$ and ${R}_{B}^0(p)$. In this section we will focus on their computation, using the standard constraints coming from boundary crossing-unitary conditions. Among the many solutions to the crossing equations we shall single the ones that are consistent with explicit weak- and strong-coupling computations.

\subsubsection*{Crossing equation}
\label{sec: crossing eq}

We will follow the ideas of \cite{Hofman:2007xp} to derive the boundary crossing equation. More specifically, we will consider the reflection of a singlet state against the boundary, and we will obtain the boundary crossing equation by demanding that such reflection must be trivial.

Let us start with the construction of the singlet state, whose defining property is its trivial interaction with any other particles. Taking this into account, one should look for a state whose quantum numbers coincide with those of the vacuum. In ABJM the spin chain has a $U(1)_{\rm extra}$ symmetry under which the fields $\bar{C}^2,C_3,C_4$ and $\bar{\psi}^2_{\pm}$ have charge +1 and the fields $C_1,\bar{C}^3, \bar{C}^4$ and $\psi_{1}^{\pm}$ have charge -1 \cite{Klose:2010ki}. Therefore, 
we should search for a singlet state with the same $U(1)_{\rm extra}$ 
charge  as the vacuum. With this in mind, we will consider
\begin{equation}
\label{singlet}
|1_{AB} \rangle (p, \bar{p}) = \epsilon_{ab} |\phi^a_A (p) \phi^b_B (\bar{p}) \rangle + \kappa \, \epsilon_{\alpha \beta} |\psi^{\alpha}_A (p) \psi^{\beta}_B (\bar{p}) \rangle \,.
\end{equation}
where $\{\phi_A^a,\psi_A^{\alpha}\}$ and $\{\phi_B^a,\psi_B^{\alpha}\}$ are respectively type A and type B impurities in the fundamental representation of $SU(2|2)$, as defined in \eqref{QSonfundamental}. In \eqref{singlet} we have $\kappa \in \mathbb{R}$ and the crossing transformation $\bar{p}$ is defined such that
\begin{equation}
\label{crossing transformation}
p \to -p  \quad \text{and} \quad E \to -E \qquad \Leftrightarrow \qquad x^{\pm} \to \frac{1}{x^{\pm}} \quad \text{and} \quad \bar{\zeta} \to \zeta \frac{x^+}{x^-} \,.
\end{equation}
For \eqref{singlet} to be a singlet state we have to further demand its invariance under all the $SU(1|2)$ generators, which implies
\begin{equation}
\label{kappa}
\kappa = - \frac{i x^-}{(x^- - x^+) \, \zeta } \, \eta(x^+,x^-,\zeta) \, \eta \left( \tfrac{1}{x^+},\tfrac{1}{x^-},\zeta \tfrac{x^+}{x^-} \right) \,,
\end{equation}
where $\eta$ was defined in \eqref{14:mass-shell_1}.

\begin{figure}
    \centering
   \includegraphics[scale=0.1]{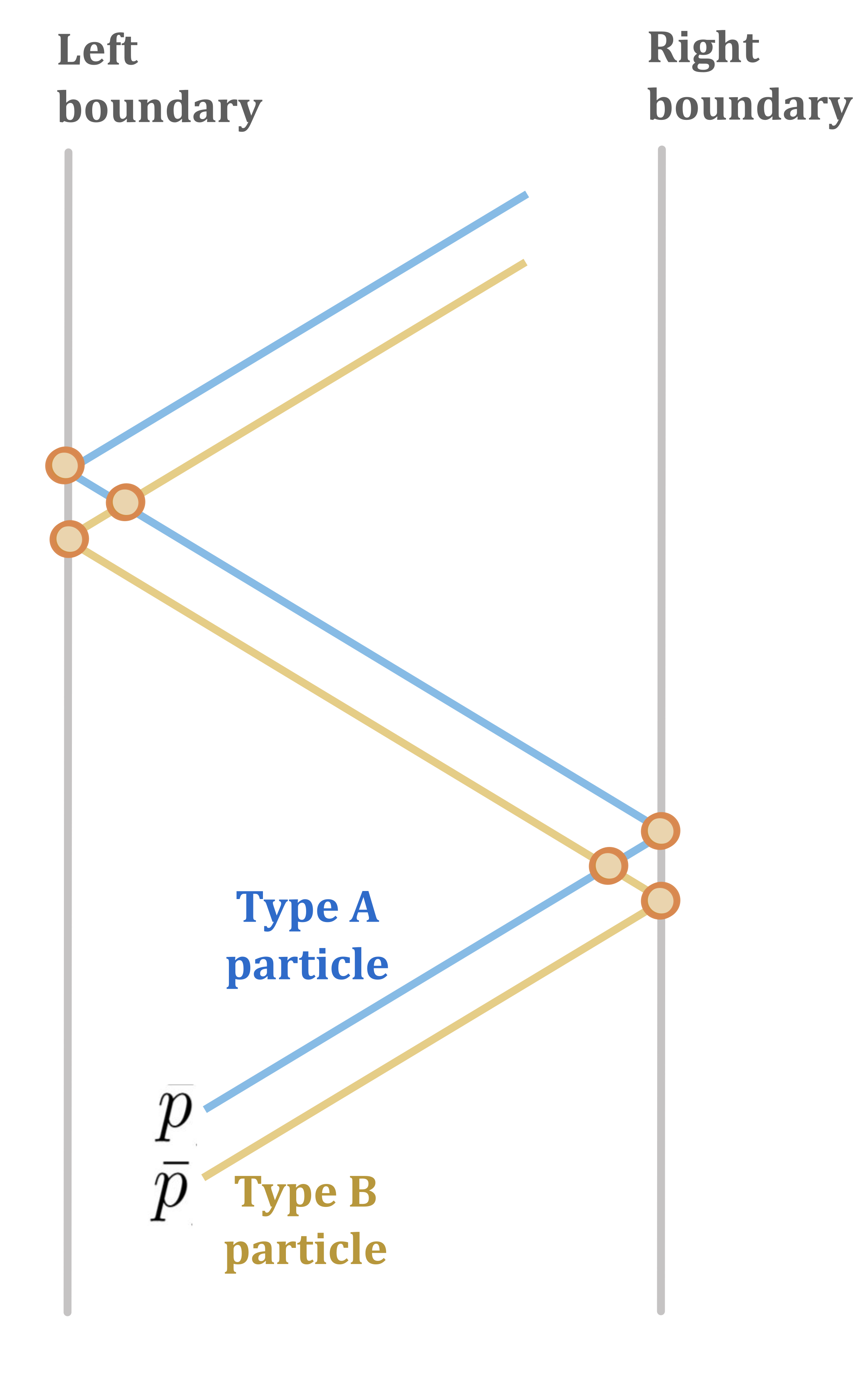}
    \caption{Sequence of right and left reflections under which the singlet state \eqref{singlet} remains invariant.}
    \label{fig: crossing}
\end{figure}

We will obtain the crossing equation by demanding that, under a sequence of right and left reflections, the singlet state \eqref{singlet} remains invariant, i.e. 
\begin{equation}
\label{right reflection singlet-invariance condition}
R_{A}^L (-p) S^{AB} (-p, \bar{p}) R_{B}^L (-\bar{p}) R_{A}^R (p) S^{AB} (p, -\bar{p}) R_{B}^R (\bar{p}) |1_{AB} \rangle (p, \bar{p}) = |1_{AB} \rangle (p, \bar{p}) \,,
\end{equation}
where we have introduced the labels $R$ and $L$ to refer to the right- and left- reflection matrices, respectively. See Figure \ref{fig: crossing} for a representation of the right and left reflections under which the singlet state remains invariant. Recalling that the action of a reflection is such that
\begin{equation}
\label{ref transformation}
p \to -p \qquad  \text{and} \qquad E \to E \qquad \Leftrightarrow \qquad x^{\pm} \to -x^{\mp} \,,
\end{equation}
we get that for the right reflection
\begin{equation}
\label{crossing condition}
R_{A}^R (p) S^{AB} (p, -\bar{p}) R_{B}^R (\bar{p}) |1_{AB} \rangle (p, \bar{p}) = r(p) |1_{AB} \rangle (-p, -\bar{p}) \,,
\end{equation}
where the reflection phase is
\begin{equation}
\label{reflection phase singlet}
r(p) = \frac{\frac{1}{x^-}+x^-}{\frac{1}{x^+}+x^+} \sigma \left( p, -\bar{p} \right)   R_{A}^0 (p) R_{B}^0 \left( \bar{p} \right) \,.
\end{equation}
Parity invariance demands that the reflection at the left boundary results in the same reflection phase \cite{Hofman:2007xp}. Therefore,
\begin{equation*}
\label{right reflection singlet-invariance condition-2}
R_{A}^L (-p) S^{AB} (-p, \bar{p}) R_{B}^L (-\bar{p}) R_{A}^R (p) S^{AB} (p, -\bar{p}) R_{B}^R (\bar{p}) |1_{AB} \rangle (p, \bar{p}) = r(p)^2 |1_{AB} \rangle (p, \bar{p}) \,.
\end{equation*}
By demanding \eqref{right reflection singlet-invariance condition} we get
\begin{equation}
\label{crossing eq-1}
r(p)^2=1 \,.
\end{equation}
Between the solutions $r(p)=1$ and $r(p)=-1$, we will shortly see that only the latter is 
compatible with weak-coupling results. Consequently,
\begin{equation}
\label{crossing eq-2}
R_{A}^0 (p) R_{B}^0 \left( \bar{p} \right) = - \frac{\frac{1}{x^+}+x^+}{\frac{1}{x^-}+x^-}  \frac{1}{\sigma \left( p, -\bar{p} \right)} \,.
\end{equation}
Finally, we have to impose also the unitarity constraints \cite{Ghoshal:1993tm}
\begin{equation}
\label{unitarity constraints}
\begin{aligned}
R_{A}^0(-p) R_{A}^0(p) &=1 \,, \\
R_{B}^0(-p) R_{B}^0(p) &=1 \,.
\end{aligned}
\end{equation}

\subsubsection*{Weak-coupling analysis}
\label{sec: weak coupling}

In order to get insight towards the construction of an all-loop solution of the crossing equation, let us focus now on the weak-coupling expansion of the dressing phases $R_{A}^0$ and $R_{B}^0$. 

Let us begin by studying an $SU(2)$ scalar sub-sector for the odd sites of the chain, where type $A$ impurities can be allocated. We will consider states of the form
\begin{equation}
\label{A impurity states}
| C_{I_0} , C_{I_1},\cdots C_{I_{\ell}}\rangle := \sqrt{2} \left( \frac{2k}{N} \right)^{\ell+\frac{1}{2}} \left( \begin{array}{cc} 0 & C_{I_0}\bar{C}^2C_{I_1}\bar{C}^2\cdots \bar{C}^2C_{I_{\ell}}
\\ 
0 & 0 \end{array} \right) , 
\end{equation}
where $k$ is the Chern-Simons level, $N$ is the number of colors and the $ C_{I_n}$ fields can be either $C_1$ or $C_3$. The overall constant in \eqref{A impurity states} has been included to get a trivial normalization in the tree-level contribution to the two-point functions.

We shall now turn to the computation of the Hamiltonian ${\bf H}^{A}$, which governs the quantum 
dynamics of the states $| C_{I_0} , C_{I_1},\cdots C_{I_{\ell}}\rangle$. Let us recall that ${\bf H}^{A}$ is given by the perturbative mixing matrix of anomalous dimensions of the corresponding operators, which can be computed from the correlator between an operator \eqref{A impurity states} inserted at $\tau_2$ and a conjugate operator inserted at $\tau_1$.
It is useful to distinguish between the two types of Feynman diagrams contributing to the mixing matrix of anomalous dimensions: those in which the contribution of the Wilson line is trivial and those including  propagators from the Wilson line. The former give rise to the bulk Hamiltonian ${\bf H}_{\rm bulk}^{A}$. The latter, in contrast,
specify the boundary Hamiltonian ${\bf H}_{\rm bdry}^{A}$.

Since for the moment we are just interested in the computation of the reflection matrix, we can ignore the right boundary and focus on the left one. 
Therefore, we will take the $\ell \to \infty$ limit and we will deal with a semi-infinite chain whose only boundary is at the left.
The bulk Hamiltonian ${\bf H}_{\rm bulk}^{A}$ should not be different from the periodic spin-chain Hamiltonian, and so at two-loops we have 
\cite{Minahan:2008hf}
\begin{equation}
\label{bulk hamiltonian}
{\bf H}_{\rm bulk}^{A} = \lambda^2 \sum_{n=0}^{\infty} ({\bf 1}-{\bf P}_{n,n+1}) \,,
\end{equation}
where ${\bf P}_{n,n+1}$ is the permutation operator between fields at the sites $n$ and $n+1$. 

\begin{figure}
\centering
\includegraphics[scale=0.7]{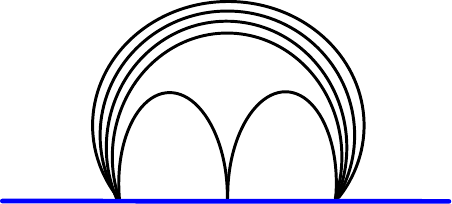}
\hspace{0.5cm}
\includegraphics[scale=0.7]{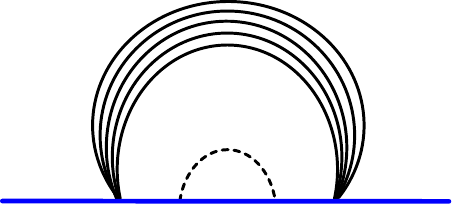}
\caption{Feynman diagram contributing to ${\bf H}_{\rm bdry}^{A}$ at 1-loop. Solid and dashed lines represent scalar and fermionic  propagators respectively. The blue line is the Wilson loop.}
\label{calzonfermion}
\end{figure}

Therefore, we just have to compute the diagrams that give ${\bf H}_{\rm bdry}^{A}$. As customary when evaluating Feynman diagrams in $D = 3-2\epsilon$, the anomalous dimensions can be read from the residue in the $1/\epsilon$ divergences. The only diagrams that contribute at the one-loop order are the ones depicted in Fig. \ref{calzonfermion}. As anticipated, there is no mixing between type $A$ and $B$ impurities and the action of ${\bf H}_{\rm bdry}^{A}$ is diagonal. When computing the divergence of these two one-loop diagrams one gets
\begin{equation}
\label{divergence alpha}
\frac{\lambda}{(\tau_2-\tau_1)^{2\ell+1}} \frac{1}{\epsilon}
\frac{1}{2}\left(M^{J_0}_{I_0} -\delta^{J_0}_{I_0} \right)
\delta^{J_1}_{I_1} \cdots \delta^{J_{\ell}}_{I_{\ell}}
+ \mathcal{O}(\epsilon) \,.
\end{equation}
From \eqref{divergence alpha} we see that the diagram for the boundary scalar interaction depends on the flavor of left-most field, and the total contribution is non-vanishing when the first site is occupied by a $C_3$. 
With these results we get that the action  of ${\bf H}_{\rm bdry}^{A}$ is of the form
\begin{equation}
\label{bdry hamiltonian}
{\bf H}_{\rm bdry}^{A} =  ( \lambda + \lambda^2 \, \beta_0) \,  {\bf V}_0 \,,
\end{equation}
where
\begin{equation}
\label{Vi operators}
{\bf V}_n | C_{I_0} , C_{I_1},\cdots C_{I_{\ell}}\rangle := \delta^3_{I_n} \, | C_{I_0} , C_{I_1},\cdots C_{I_{\ell}}\rangle \,. \end{equation}
Diagrams contributing to $\beta_0$ are more involved, as some of them include integrals over gluon vertices. Their computation is beyond the scope of our analysis, as we are only interested in the reflection factor at the leading weak-coupling order.

Collecting the expressions \eqref{bulk hamiltonian} and \eqref{bdry hamiltonian} we arrive at
\begin{equation}
\label{semi infinite chain hamitonian}
{\bf H}^{A}= ( \lambda + \lambda^2 \, \beta_0 ) {\bf V}_0 + \lambda^2 \sum_{n=0}^{\infty} ({\bf 1}-{\bf P}_{n,n+1}) \,.
\end{equation}
The above Hamiltonian can be diagonalized by the standard perturbative methods, taking
\begin{equation}
 {\bf H}^0 =  \lambda  {\bf V}_0\,,
 \qquad
 \delta {\bf H} =  \lambda^2 \beta_0  {\bf V}_0
 + \lambda^2 \sum_{n=0}^{\infty} ({\bf 1}-{\bf P}_{n,n+1}) \,.
\end{equation}
Let us consider single impurity states $|n \rangle$, where $n$ indicates the position of the excitation. For these states the unperturbed energies are simply
$E_n^0 = \lambda \delta_{n0}$. However, as the unperturbed spectrum is degenerate, a good basis $| \psi^0_n \rangle$ to apply the methods of perturbation theory should satisfy
\be 
\langle \psi^0_n | \delta H | \psi^0_m \rangle = 0\,,
\qquad\text{for}\quad n\neq m \quad \& \quad 
E^0_n = E^0_m\,.
\label{condi}
\ee 
Thus, we will instead consider the basis
\begin{equation}
|\psi^0_0\rangle := |0\rangle\,, \qquad
|\psi^0(p) \rangle := \sum_{n=1}^{\infty} \left[ e^{-ipn} + R(p) \, e^{ipn} \right] |n \rangle\,, 
\end{equation}
where, in order to satisfy the condition \eqref{condi}, one needs to impose
\begin{equation}
    R(p) = -1\,.
\end{equation}
For a magnon state with momentum $p$ 
the perturbative solution is
\begin{equation}
\label{magnon eigenstates}
|\psi(p) \rangle = \sum_{n=1}^{\infty} \left( e^{-ipn} -  e^{ipn} \right) \, |n \rangle +  \lambda \, (e^{-ip}-e^{ip}) \, |0\rangle + \mathcal{O} (\lambda^2) \,,
\end{equation}
whose energy is
\begin{equation}
\label{magnon energy-perturbative computation}
E(p) = 4 \lambda^2 \sin^2 \frac{p}{2}+ \mathcal{O} (\lambda^3) \,,
\end{equation}
as expected from \eqref{magnon energy}. In addition to the magnon states with momentum $p$ we have a state in which the impurity remains close to the boundary, i.e. a boundary bound state 
\begin{equation}
    \label{boundary bound-states}
    |{\rm B} \rangle = |0 \rangle - \lambda  |1 \rangle + \mathcal{O} (\lambda^2) \,,
\end{equation}
whose energy is
\begin{equation}
\label{bound state energy-perturbative computation}
E_{\rm B} (p, \lambda) = \lambda +  (1+ \beta_0) \, \lambda^2 + \mathcal{O} (\lambda^3) \,.
\end{equation}
One could compute the coefficient $\beta_0$ to know the two-loop correction to the boundary bound-state energy given in \eqref{bound state energy-perturbative computation}, but such computation is not needed to determine the leading weak-coupling correction to the type $A$ magnon dressing phase. Therefore, we conclude that
\begin{equation}
\label{perturbative dressing A}
R_{A}^0(p)=-1 + \mathcal{O} (\lambda^2) \,.
\end{equation}

Finally, let us turn to a weak-coupling analysis 
of the other dressing factor $R_B^0$. We will now consider a $SU(2)$ sub-sector for the even sites of the chain, where the type $B$ impurities can propagate. Therefore, we will work with the states
\begin{equation}
\label{B impurity states}
|\bar C^{I_0} , \bar C^{I_1},\cdots \bar C^{I_{\ell}}\rangle := \sqrt{2} \left( \frac{2k}{N} \right)^{\ell+\frac{1}{2}} \left( \begin{array}{cc} 0 & C_1\bar{C}^{I_0}C_1\bar{C}^{I_1}\cdots \bar{C}^{I_{\ell}}C_{1}
\\ 
0 & 0 \end{array} \right) , 
\end{equation}
where the $\bar C^{I_n}$ are taken to be either $\bar C^{2}$ or $\bar C^{3}$.

A quick diagrammatic analysis shows that the first non-trivial terms in the Hamiltonian ${\bf H}^{B}$ appear at the two-loop order. Diagrams that could potentially contribute to a one-loop order, like the ones depicted in Fig. \ref{calzonfermion}, cancel between each other as we have a $C_1$ field at the left-most site. At two-loop order the boundary term in the Hamiltonian acts as the identity operator or it comes with a matrix $M^{I_0}_{J_0}$. 
As the latter does not distinguish between  $\bar C^{2}$ and $\bar C^{3}$, and since we know that $|\bar C^{2} , \bar C^{2},\cdots \bar C^{I_{2}}\rangle$ has vanishing anomalous dimension, the Hamiltonian in this case must be
\begin{equation}
\label{type B hamiltonian}
{\bf H}^{B}= \lambda^2 \sum_{n=0}^{\infty} ({\bf 1}-{\bf P}_{n,n+1})\,.
\end{equation}
For a single magnon impurity we can diagonalize this Hamiltonian with a usual Bethe Ansatz wave-function of the form
\begin{equation}
    |\psi(p) \rangle = \sum_{n=0}^{\infty} \left[ e^{-ipn} + R(p) \, e^{ipn} \right] |n \rangle\,,
\end{equation}
if we fix $R(p)= e^{ip}$. Therefore, we conclude that the type $B$ right-boundary dressing phase is, in the weak coupling limit,
\begin{equation}
\label{weak coupling type B dressing}
R_{B}^0(p)= e^{-ip} + \mathcal{O}(\lambda^2)\,.
\end{equation}

\subsubsection*{All-loop proposal}

We will now make an all-loop proposal for the boundary dressing phases, such that they simultaneously solve the crossing equation \eqref{crossing eq-2} and reproduce the results of the last section when considered in the weak-coupling limit. 

As we have seen, there exists a type $A$ excited state with energy of order $\lambda$. The fact that the dispersion relation \eqref{magnon energy} is expressed as an expansion in even powers of $\lambda$ indicates that such state is not an ordinary magnon, but rather it is a boundary bound state. Let us consider the factor
\begin{equation}
\label{crossing eq factor}
\frac{\frac{1}{x^+}+x^+}{\frac{1}{x^-}+x^-} \,,
\end{equation}
which appears in the r.h.s. of the crossing equation \eqref{crossing eq-2}. Interestingly, it has precisely a pole for $x^- \to -i$ whose energy is\footnote{In all the weak-coupling expansions we are using that $h(\lambda) = \lambda+{\cal O}(\lambda^3)$.}
\begin{equation}
\label{energy pole}
E_{\rm pole} = \lambda + \mathcal{O} (\lambda^2) \,,
\end{equation}
in accordance with \eqref{bound state energy-perturbative computation}.  This suggests that the boundary bound state \eqref{boundary bound-states} could arise from a pole in the reflection matrix if the factor \eqref{crossing eq factor} is included in $R_{A}^0$. Moreover, such factor should be absent in $R_{B}^0$, as we have not  observed boundary bound states associated to type $B$ particles. It is also useful to note that, for real fixed momentum,
\begin{equation}
\frac{\frac{1}{x^+}+x^+}{\frac{1}{x^-}+x^-}
= e^{ip} + {\cal O}(\lambda^2)\,,
\end{equation}
which could serve to explain the relative factor between \eqref{perturbative dressing A} and \eqref{weak coupling type B dressing}.

Taking all these considerations into account, we propose that the all-loop dressing factors are
\begin{equation}
\label{crossing ansatz}
\begin{aligned}
R_{A}^0(p) &= - \frac{1}{R_0 (p)} \left( \frac{\frac{1}{x^+}+x^+}{\frac{1}{x^-}+x^-} \right) \left( \frac{x^-}{x^+} \right)\,, \\
R_{B}^0(p) &= \frac{1}{R_0 (p)} \, \left( \frac{x^-}{x^+} \right)\,,  \\
\end{aligned}
\end{equation}
with
\begin{equation}
\label{crossing eq R0}
R_{0} (p) R_{0} \left( \bar{p} \right) =  \sigma \left( p, -\bar{p} \right) \,,
\end{equation}
and
\begin{equation}
\label{unitarity R0}
R_{0}(-p) R_{0}(p) =1 \,.
\end{equation}
Equations \eqref{crossing eq R0} and \eqref{unitarity R0} can be solved if we take $R_0(p)$ to be the square root of the dressing phase proposed by \cite{Correa:2012hh,Drukker:2012de} for the ${\cal N}=4$ sYM case,
 replacing $g=\frac{\sqrt{\lambda_{YM}}}{4\pi}$ by $h(\lambda)$,
i.e.
\begin{equation}
\label{solution R0}
R_0 (p) = \left[ \frac{1}{\sigma_B (p) \sigma (p,-p)} \left( \frac{1+\frac{1}{(x^-)^2}}{1+\frac{1}{(x^+)^2}} \right) \right]^{\frac{1}{2}} \,,
\end{equation}
with
\begin{equation}
\label{sigma B}
\begin{aligned}
\sigma_B (p) &= e^{i \chi(x^+)-i \chi(x^-)} \,, \\
i \chi (x) &= \left \{ \begin{array}{ccc}
 i \Phi(x)= \oint_{|z|=1} \frac{dz}{2\pi i} \frac{1}{x-z} \log \left\{ \frac{\sinh[2\pi h(z + \frac{1}{z})]}{2\pi h \left( z+\frac{1}{z} \right) } \right\} & {\rm if} & |x|>1 \\
 i \Phi(x)+ \log \left\{ \frac{\sinh[2\pi h(x + \frac{1}{x})]}{2\pi h \left( x+\frac{1}{x} \right) } \right\} & {\rm if} & |x|<1 
\end{array} \right.
\end{aligned}
\end{equation}
Using the results of \cite{Correa:2012hh,Drukker:2012de} we have
\begin{equation}
\label{boundary sigma expansion}
R_0(p)=1+\mathcal{O}(\lambda^2) \,,
\end{equation}
and therefore we recover the expressions \eqref{perturbative dressing A} and \eqref{weak coupling type B dressing} for the weak coupling expansions of the dressing phases $R_{A}^0$ and $R_{B}^0$. Let us note that in order to reproduce \eqref{perturbative dressing A} and \eqref{weak coupling type B dressing} from \eqref{crossing ansatz} we have chosen $r(p)=-1$. 

The fact that $R_0(p)$ is the square root of the dressing factor proposed for the ${\cal N}=4$ sYM case can be naturally understood as follows. In the strong-coupling limit, dressing phases are computed from the scattering of excitations propagating on the worldsheet of strings carrying large angular momentum. These strings propagate in an $AdS_2\times S^2$ sub-space of the geometry (see Appendix \ref{app: dual string vacuum state}) and, after a Pohlmeyer reduction, the propagation of excitations in the worldsheet is described in terms of sine/sinh Gordon solitons \cite{Hofman:2006xt}. Being the open string restricted to an $AdS_2\times S^2$ sub-space of the geometry, its classical dynamics is identical to that of the open string propagating in $AdS_2\times S^2 \subset AdS_5\times S^5$ \cite{Drukker:2006xg}. Thus, 
the reflection of worldsheet excitations is described in exactly the same way as done in \cite{Correa:2012hh,Drukker:2012de}
\begin{equation}
 R_A^0(p) \simeq R_B^0(p) 
 \simeq e^{-i2T\cos\tfrac{p}{2}\left[\log\left(\frac{1-\sin\frac{p}2}{1+\sin\frac{p}2}\right)+2\log\cos\tfrac{p}{2}\right]}\,,
 \label{stronR0}
\end{equation}
where $T$ is the effective string tension. What changes between one case and the other is how $T$ is related to the 't Hooft coupling. In the  $AdS_4\times {\mathbb{CP}^3}$ background
$2T = \sqrt{2\lambda} \simeq 2 h(\lambda)$ for large $\lambda$. 
Thus, the result \eqref{stronR0}
is in agreement with the strong-coupling limit of our proposal \eqref{crossing ansatz}. Also note that \eqref{stronR0} gives the reflection phase for the two types of magnons. The relative factor in our proposal is an order 1 quantity in the strong-coupling limit and therefore it is not observed in the semiclassical computation.

As said before, the result \eqref{stronR0} also holds for the $AdS_5\times S^5$ background.  However, the relation between the effective string tension and 't Hooft coupling is  $2T = \frac{\sqrt{\lambda_{YM}}}{\pi} = 4 g$ in this case, which explains that the same boundary dressing phase appears squared in the ${\cal N}=4$ sYM case. The fact that bulk S-matrices come with $\sigma(x_1,x_2)$ in one case and with $\sigma(x_1,x_2)^2$ in the other is also explained by the same argument.

Let us note
that type $A$ and type $B$ magnons reflect \textit{differently} at the boundaries. This implies a striking contrast to what is observed for the bulk scattering properties. At weak coupling one can see this as a straightforward consequence of the fact that type $A$ particles are in closer interaction with the boundaries than type $B$ particles. The range of the interactions between the impurities and the boundary is of order $\lambda$ for type $A$ magnons and of order $\lambda^2$ for type $B$ magnons.  As an important implication of this, one type of particle can form a bound state with the boundary while the other can not. 
Let us also mention that this is a distinctive property of the open boundary set by Wilson loops. Extending the ideas of \cite{Berenstein:2005vf}, open spin chains for the anomalous dimensions of determinant operators in ABJM were studied in \cite{Chen:2018sbp,Chen:2019igg}, and no distinction was observed for the reflection of type $A$ and $B$ magnons in those cases.

To conclude this section, let us comment on the possibility of inserting operators with a non-trivial lower off-diagonal block. A possible supersymmetric lower off-diagonal block vacuum, over which magnon excitations can propagate, is the hermitian conjugate of the upper off-diagonal block
\be 
{\cal V}^{\dagger}_\ell = 
\left(
\begin{array}{cc}
 0 & 0 
\\
(\bar C^1 C_2)^\ell \bar C^1 &  0
\end{array} \right)\,.
\label{susyinsertion0 conjugate}
\ee
The spin-chain vacuum states ${\cal V}_\ell$ and ${\cal V}^{\dagger}_\ell$ are interchanged under charge conjugation.
Consider for example the $U(1)_{\rm extra}$ symmetry of the ABJM spin chain \cite{Klose:2010ki}, under which the fields $\bar{C}^1,\bar{C}^2,C_3$ and $C_4$ have charge +1 and the fields $C_1,C_2,\bar{C}^3$ and $\bar{C}^4$ have charge -1: while ${\cal V}_\ell$ has charge $-1$, ${\cal V}^{\dagger}_\ell$ has charge $+1$. Similarly, type $A$ magnons in the lower off-diagonal block are the charge conjugates of type $B$ magnons in the upper off-diagonal block, and vice-versa. When considering their reflection from the boundaries, type $A$ and $B$ magnons in the lower block behave as type $B$ and $A$ impurities in the upper block, respectively. Therefore, their dressing phases are interchanged. This is manifest in the weak-coupling regime, as type $B$ magnons in the lower block 
interact with the boundaries at order $\lambda$, while the range of the  interaction with the boundaries is of order $\lambda^2$ for type $A$ magnons in the same block.

\section{Y-system for the cusped Wilson line}
\label{sec: Y-system}

The main goal of this chapter is to compute the cusp anomalous dimension $\Gamma_{\rm cusp}$ of ABJM using an integrability approach. In the previous sections we have studied the spin-chain description of the anomalous dimensions of operators inserted at the 1/2 BPS Wilson line of ABJM. In order to compute $\Gamma_{\rm cusp}$ we shall consider instead a cusped Wilson loop, which is correspondingly described by an open spin chain with an appropriate twist in one of its boundaries. We will consider the insertion of the vacuum ${\cal V}_{\ell}$ at the position of the cusp, and we will study the corresponding anomalous dimension as a function of $\ell$. This anomalous dimension is due to finite-size effects, which can be computed using a Boundary Thermodynamic Bethe Ansatz (BTBA). Eventually, $\Gamma_{\rm cusp}$ will be obtained in the limit in which no operator is inserted at the cusp. As discussed at the beginning of this chapter, the integral BTBA equations that give the finite-size corrections usually can be rewritten as a set of functional equations, known as $Y$-system. In this section we will propose a set of $Y$-system equations for the cusped Wilson loop of ABJM, which we will later use to compute the  one-loop cusp anomalous dimension from a leading-order finite-size correction. 

\subsubsection*{The $Y$-system}

Interestingly, there are many examples, in particular within the framework of the AdS/CFT correspondence, in which the introduction of integrable boundary conditions in a system modifies the analytical and asymptotic properties of the $Y$-functions without changing the $Y$-system \cite{Correa:2012hh,Drukker:2012de,Behrend:1995zj,OttoChui:2001xx,Gromov:2010dy,Ahn:2010ws,Ahn:2011xq,vanTongeren:2013gva,Bajnok:2012xc}. This could be related to the fact that either for periodic or open boundary conditions in the physical theory, in the mirror theory one deals with exactly the same system of mirror excitations. We will follow this insight  and we will assume that, as it was the case for the ${\cal N}=4$ sYM Wilson loop \cite{Correa:2012hh,Drukker:2012de}, the same $Y$-system that describes the ABJM spectrum with periodic boundary conditions can be used to describe the spectrum with open boundary conditions set by the ABJM Wilson line.
More specifically, we propose that the $Y$-system of the cusped line of ABJM is \cite{Gromov:2009at,Bombardelli:2009xz,Cavaglia:2013lgg}
\begin{align}
\label{y system ABJM-1}
Y_{a,s}^+Y_{a,s}^- &= \frac{(1+Y_{a,s+1})(1+Y_{a,s-1})}{(1+1/Y_{a+1,s})(1+1/Y_{a-1,s})} \, , \qquad s>1, \, (a,s) \neq (2,2) \,, \\
\label{y system ABJM-2}
Y_{a,1}^+Y_{a,1}^- &= \frac{(1+Y_{a,2})(1+Y_{a,0}^I)(1+Y_{a,0}^{II})}{(1+1/Y_{a+1,1})(1+1/Y_{a-1,1})} \,, \\
\label{y system eq momentum carrying nodes-1}
Y_{a,0}^{\alpha,+}  Y_{a,0}^{\beta,-} &= \frac{(1+Y_{a,1})}{(1+1/Y_{a+1,0}^{\beta})(1+1/Y_{a-1,0}^{\alpha})} \,, \qquad a>1 , \, \alpha \neq \beta \,,  \\
\label{y system eq momentum carrying nodes-2}
Y_{1,0}^{\alpha,+}  Y_{1,0}^{\beta,-} &= \frac{(1+Y_{1,1})}{(1+1/Y_{2,0}^{\beta})} \,, \qquad \qquad \qquad \qquad \quad \alpha \neq \beta \,,
\end{align}
where the set of non-vanishing nor infinite $Y$-functions is given by
\begin{equation}
\label{abjm y functions}
\begin{aligned}
& Y_{a,0}^{\alpha}\,, \, \, a \geq 1, \, \alpha \in \{ I,II \} \,;  \quad  Y_{a,s}\,, (a,s) \in \mathbb{N}^+ \times \mathbb{N}^+, \, a \leq 1 \, \, \text{or} \, \, s \leq 1 \,; \quad \text{and} \quad  Y_{2,2} \,.
\end{aligned}
\end{equation}
The $Y$-functions in \eqref{y system ABJM-1}-\eqref{y system eq momentum carrying nodes-2} are again written as functions of the spectral parameter $u$, defined by
\begin{equation}
    \label{spectral parameter}
    x(u)+\frac{1}{x(u)}= \frac{u}{h(\lambda)}\,, \qquad \frac{x^+}{x^-}=e^{ip} \,.
\end{equation}
Moreover, considering the set of $T$-functions given by
\begin{equation}
\label{abjm t functions}
\begin{aligned}
T_{a,s}^{\alpha} \,, (a,s) \in \mathbb{N} \times \{ -1,0 \}, \,  \alpha \in \{ I, II \} \,; \qquad  T_{a,s}\,, (a,s) \in \mathbb{N} \times \mathbb{N}^+, \, a \leq 2 \, \, \text{or} \, \, s \leq 2 \,,\\
\end{aligned}
\end{equation}
and making the change of variables 
\begin{align}
\label{T to Y-1}
Y_{a,s} &= \frac{T_{a,s+1}T_{a,s-1}}{T_{a+1,s}T_{a-1,s}}\,, \, \quad s \geq 2 \, \qquad a \geq 1 \,, \\
\label{T to Y-2}
Y_{a,1} &= \frac{T_{a,2}T_{a,0}^I T_{a,0}^{II}}{T_{a+1,1}T_{a-1,1}} \,, \, \quad a \geq 1 \,, \\
\label{abjm momentum carrying nodes from t functions}
Y_{a,0}^{\alpha} &= \frac{T_{a,1}T_{a,-1}^{\beta}}{T_{a+1,0}^{\alpha}T_{a-1,0}^{\beta}} \,, \, \quad a \geq 1 \, , \quad \alpha, \beta \in \{ I, II \} \, , \quad \alpha \neq \beta \,,
\end{align}
one can rewrite the $Y$-system equations \eqref{y system ABJM-1}-\eqref{y system eq momentum carrying nodes-2} in terms of the $T$-system \cite{Cavaglia:2013lgg} 
\begin{align}
\label{T system-1}
T^+_{a,s} T^-_{a,s} &= (1-\delta_{a,0}) \, T_{a+1,s}T_{a-1,s}+T_{a,s+1}T_{a,s-1}\, , \quad s \geq 2 \,,\\
T^+_{a,1} T^-_{a,1} &= (1-\delta_{a,0}) \, T_{a+1,1}T_{a-1,1}+T_{a,2} T_{a,0}^I T_{a,0}^{II} \,, \\
T_{a,0}^{\alpha,+}  T_{a,0}^{\beta,-} &= (1-\delta_{a,0}) \, T_{a+1,0}^{\beta} T_{a-1,0}^{\alpha} +T_{a,-1}^{\alpha} T_{a,1} \, , \quad \alpha, \beta \in \{ I, II \} \, , \quad \alpha \neq \beta \,,\\
\label{T system-4}
 T_{a,-1}^{\alpha,+}   T_{a,-1}^{\beta,-}  &=  T_{a+1,-1}^{\beta} T_{a-1,-1}^{\alpha} \, , \quad a \neq 0 \, , \quad \alpha, \beta \in \{ I, II \} \, , \quad \alpha \neq \beta \,.
\end{align}
In this case, the BTBA formula for the vacuum energy becomes\footnote{Let us note the extra 1/2 factor in the r.h.s. of this equation. This can be explained by taking into account that the dispersion relation of magnons in the ABJM picture has an overall 1/2 factor with respect to the similar dispersion relation of ${\cal N}=4$ sYM \cite{Gromov:2009at}.}
\begin{equation}
\label{TBA ABJM}
E_0(L)-E_0(\infty)= - \frac{1}{4\pi} \sum_{a=1}^{\infty} \int_0^{\infty} dq \, \log[1+Y^I_{a,0}(q)]- \frac{1}{4\pi} \sum_{a=1}^{\infty} \int_0^{\infty} dq \, \log[1+Y^{II}_{a,0}(q)] \,.
\end{equation}

\subsubsection*{Asymptotic solution of the $Y$-system}
\label{asymptotic solution}

In order to obtain the leading-order contribution to $\Gamma_{\rm cusp}$ from \eqref{TBA ABJM} we will discuss now the asymptotic large-volume solution that is obtained from the $Y$-system of eqs. \eqref{y system ABJM-1}-\eqref{y system eq momentum carrying nodes-2}. Following \cite{Gromov:2009at}, in the asymptotic limit one gets
\begin{align}
\label{asymptotic solution 1}
Y^I_{a,0} &\sim \left( \frac{z^{[-a]}}{z^{[+a]}} \right)^{2L} \, T_{a,1} \, \prod_{n=-\frac{a-1}{2}}^{\frac{a-1}{2}} \phi_{I}^{\zeta(n,a)} \left( u + i n \right) \phi_{II}^{1-\zeta(n,a)} \left( u + i n \right) \,,\\
\label{asymptotic solution 2}
Y^{II}_{a,0} &\sim \left( \frac{z^{[-a]}}{z^{[+a]}} \right)^{2L} \, T_{a,1} \, \prod_{n=-\frac{a-1}{2}}^{\frac{a-1}{2}} \phi_{II}^{\zeta(n,a)} \left( u + i n \right) \phi_{I}^{1-\zeta(n,a)} \left( u + i n \right) \,,
\end{align}
where the two functions $\phi_I$ and $\phi_{II}$ will be fixed later by comparison with L\"uscher corrections, and with
\begin{equation*}
\label{zeta}
\zeta(n,a) = \left\{ 
\begin{array}{cc}
1     & \text{if }  n+\frac{a-1}{2} \text{ is even} \\
0     & \text{if }  n+\frac{a-1}{2} \text{ is odd}
\end{array} \right.
\end{equation*}
In \eqref{asymptotic solution 1} and \eqref{asymptotic solution 2} we are using the notation $z^{\pm}$ for the spectral variables in the mirror theory. In the particular case in which 
\begin{equation}
\label{phi equality0}
\phi_I(u)=\phi_{II}(u)= \pm \frac{\varphi \left( u- \frac{i}{2} \right)}{\varphi \left( u+ \frac{i}{2} \right)} \,,
\end{equation}
for some function $\varphi$, one gets
\begin{equation}
\label{phi equality}
Y^I_{a,0}=Y^{II}_{a,0} \sim (\pm 1)^a \left( \frac{z^{[-a]}}{z^{[+a]}} \right)^{2L} \, \frac{\varphi \left( u- \frac{ia}{2} \right)}{\varphi \left( u+ \frac{ia}{2} \right)} \, T_{a,1} \,.
\end{equation}

Let us start by discussing the $T_{a,1}$ functions that appear in \eqref{asymptotic solution 1} and \eqref{asymptotic solution 2}, which can be obtained as a solution of the $T$-system of eqs. \eqref{T system-1}-\eqref{T system-4}.  In principle, one could determine the asymptotic $T_{a,1}$ functions with a computation of the double-row transfer matrix for a bound state of $a$ magnons. Instead of directly computing such a double-row transfer matrix we will follow the ideas of \cite{Bajnok:2012xc}, and we will argue that the $T_{a,1}$ functions of a system with $SU(1|2)$ symmetry can be obtained from the corresponding $T$-functions of a system with $SU(2|1)$ symmetry from the identification
\begin{equation}
T_{a,1}^{SU(1|2)} \equiv \left( T_{1,a}^{SU(2|1)} \right)^*\,.
\end{equation}
The $T$-system of the case with $SU(2|1)$ symmetry has already been studied in \cite{Bajnok:2013wsa}. Following their results, we obtain
\begin{equation}
\label{asymptotic T functions}
T_{a,1}^{SU(1|2)}= 2 (-1)^a \left[ b_{0,a} \left( 1+ \frac{u^{[-a]}}{u^{[a]}} \right)  + 2 \sum_{k=1}^{a-1} \frac{b_{k,a} u^{[-a]}}{u^{[-a+2k]}} \right] \, , a \geq 1  \,, \\
\end{equation}
with
$$
\begin{aligned}
b_{0,s}&= \sin^2 \tfrac{\theta}{2} \, P_{s-1}^{(0,1)} \left( 1-2\cos^2 \tfrac{\theta}{2} \right) \,, \\
b_{l,s}&=b_{s-l,s}=b_{0,l}b_{0,s-l} \,,\\
b_{0,0}&=1 \,,
\end{aligned}
$$
and where $P_{s-1}^{(0,1)}$ stands for Jacobi polynomial. For simplicity, from now on we will simply write $T_{a,1}^{SU(1|2)} \equiv T_{a,1}$. The asymptotic solution to the $T$-system proposed in \eqref{T system-1}-\eqref{T system-4} is completed by\footnote{This solution receives finite-size corrections away from the asymptotic limit. In particular, following \eqref{abjm momentum carrying nodes from t functions} we see that the correction to $T_{a,-1}^{\alpha}$ with $\alpha=I,II$ and $a \geq 1$ gives the leading-order contribution to $Y_{a,0}^{\alpha}$, which is presented in \eqref{asymptotic solution 1} and \eqref{asymptotic solution 2}.}
\begin{equation}
\label{cusped WL t system ABJM}
\begin{aligned}
T_{1,s}	&= (-1)^s \frac{4su}{u^{[s]}} \sin^2 \frac{\theta}{2} \, , \, \, s \geq 1 \,; \qquad 
T_{2,s}=T_{s,2}= \frac{16 u^{[s]} u^{[-s]}}{u^{[s-1]u^{[s+1]}}} \sin^4 \frac{\theta}{2} \, , \, \, s \geq 2\,; \\
T_{0,s} &= 1 \, , \, \, s > 0 \,; \qquad \qquad \quad \qquad \quad \; \,
T_{a,0}^I =1 \, , \, \, a \geq 0 \,; \\
T_{a,0}^{II} &=1 \, , \, \, a \geq 0 \,; \qquad \qquad \quad \qquad \; \; \;
T_{0,-1}^I =1 \,;  \\
T_{0,-1}^{II} &=1 \,; \qquad \qquad \quad \qquad \quad \qquad \quad \,
T_{a,-1}^I =0 \, , \, \, a \geq 1 \,; \\
T_{a,-1}^{II} &=0 \, , \, \, a \geq 1 \,.
\end{aligned}
\end{equation}

Having discussed the asymptotic $T_{a,1}$ functions associated with the cusped Wilson line, let us focus now on the $\phi_I$ and $\phi_{II}$ functions that appear in the asymptotic solutions \eqref{asymptotic solution 1} and \eqref{asymptotic solution 2}. 
We will obtain an expression for these functions from the study of the leading-order L\"uscher correction to the vacuum energy $E_0$, which for the ABJM cusped Wilson line we propose to be given as
\begin{align}
\label{Luscher ABJM-1}
Y_{a,0}^I (q) &\sim e^{-2L \, \epsilon_a(q)} \, {\rm Tr} \left[ R_{A}^{\it up}(q) \, C \, R_{B, \theta}^{\it down} \, (-\bar{q}) \, C^{-1} \right] \,, \\
\label{Luscher ABJM-2}
Y_{a,0}^{II} (q) &\sim e^{-2L \, \epsilon_a(q)} \, {\rm Tr} \left[ R_{B}^{\it up}(q) \, C \, R_{A,\theta}^{\it down} \, (-\bar{q}) \, C^{-1} \right] \,.
\end{align}
Let us comment some details about the above formulas. First, the function $\epsilon_a(q)$ gives the energy of an $a$-magnon of mirror-momentum $q$, and
\begin{equation}
\label{charge conjugation}
C= \left( \begin{array}{cc}
-i \epsilon_{ab} & 0 \\
0 & \epsilon_{\alpha \beta}
\end{array}
\right),
\end{equation}
is the charge-conjugation matrix. Moreover, $R_{A/B}$ is the reflection matrix derived in the previous sections for a Wilson line along the $x_0$ direction, while $R_{A/B,\theta}$ is the corresponding reflection matrix for a Wilson loop rotated by an angle $\theta$, 
\begin{equation}
\label{rotated reflection}
R_{A/B,\theta} (q) = {\cal O} (\theta) R_{A/B} (q) {\cal O}^{-1} (\theta) \,,
\end{equation}
where the rotation matrix ${\cal O} (\theta)$ is given by
\begin{equation}
\label{rotation matrix}
{\cal O} (\theta)= \left( 
\begin{array}{cccc}
1 & 0 & 0 & 0 \\
0 & 1 & 0 & 0 \\
0 & 0 & \cos \frac{\theta}{2} & \sin \frac{\theta}{2} \\
0 & 0 & -\sin \frac{\theta}{2} & \cos \frac{\theta}{2} \\
\end{array}
\right) \,.
\end{equation}
Note that we are only considering the case of a Wilson line with a geometric cusp $\theta$, i.e. we are not including an internal cusp angle $\varphi$\footnote{The internal cusp $\varphi$ should account for an R-symmetry rotation in a plane that includes the $I=1$ direction. However, such a rotated line would not share any supersymmetries with the vacuum state written in \eqref{susyinsertion0}, which would notoriously difficult the computation of the corresponding rotation matrix.}. The words ${\it up}$ and ${\it down}$
in \eqref{Luscher ABJM-1} and \eqref{Luscher ABJM-2}
refer to the reflection matrices of impurities in the upper or lower off-diagonal blocks
of the corresponding supermatrix, respectively.
It is crucial to recall that the charge conjugate of a type $A$ (or $B$) magnon 
in the upper off-diagonal block spin chain is a type $B$ (or $A$) magnon in the lower off-diagonal block spin chain. Then, taking into account that
\begin{equation}
\begin{aligned}
R_{A}^{\it down}(q) &= R_{B}^{\it up}(q) \,, \\
R_{B}^{\it down}(q) &= R_{A}^{\it up}(q) \,, \\
\end{aligned}
\end{equation}
we arrive at
\begin{align}
\label{Luscher ABJM-3}
Y_{a,0}^I (q) &\sim e^{-2L \, \epsilon_a(q)} \, {\rm Tr} \left[ R_{A}^{\it up}(q) \, C \, R_{A, \theta}^{\it up} \, (-\bar{q}) \, C^{-1} \right] \,, \\
\label{Luscher ABJM-4}
Y_{a,0}^{II} (q) &\sim e^{-2L \, \epsilon_a(q)} \, {\rm Tr} \left[ R_{B}^{\it up}(q) \, C \, R_{B,\theta}^{\it up} \, (-\bar{q}) \, C^{-1} \right] \,.
\end{align}
Otherwise stated, from now on we will return to work only with reflection matrices of magnons in the upper off-diagonal block. Moreover, to simplify the notation we will again refer to them simply as $R_{A}$ and $R_{B}$.

Let us turn now to the explicit computation of \eqref{Luscher ABJM-3} and \eqref{Luscher ABJM-4}, and let us focus first on the computation of $Y_{1,0}^I$ and $Y_{1,0}^{II}$. Using the results of Sections \ref{sec: open spin chain} and \ref{sec: crossing} we arrive at
\begin{align}
\label{rotated transfer matrix-trace}
Y_{1,0}^I(z^+,z^-) &= e^{-2L \, \epsilon_1} \, \sin^2 \frac{\theta}{2}\, \frac{(z^+ + z^-)^2}{z^+ z^-} \, R_{A}^0 \left( z^+ , z^- \right) R_{A}^0 \left( -\tfrac{1}{z^-} , -\tfrac{1}{z^+} \right) \,, \\
\label{rotated transfer matrix-trace-2}
Y_{1,0}^{II}(z^+,z^-) &= e^{-2L \, \epsilon_1} \, \sin^2 \frac{\theta}{2} \, \frac{(z^+ + z^-)^2}{z^+ z^-} \, R_{B}^0 \left( z^+ , z^- \right) R_{B}^0 \left( -\tfrac{1}{z^-} , -\tfrac{1}{z^+} \right) \,.
\end{align}
Moreover, from \eqref{crossing ansatz} we get
\begin{equation}
\label{a magnon-a=1-2}
\begin{aligned}
R_{A}^0 \left( z^+ , z^- \right) R_{A}^0 \left(- \tfrac{1}{z^-} , -\tfrac{1}{z^+} \right)=R_{B}^0 \left( z^+ , z^- \right) R_{B}^0 \left(- \tfrac{1}{z^-} , -\tfrac{1}{z^+} \right) \,, \\
\end{aligned}
\end{equation}
which implies
\begin{equation}
\label{a magnon-a=1-3}
\begin{aligned}
Y_{1,0}^I(z^+,z^-)=Y_{1,0}^{II}(z^+,z^-)= e^{-2L \, \epsilon_1} \, \sin^2 \frac{\theta}{2} \, \frac{(z^+ + z^-)^2}{z^+ z^-} \, R_{A}^0 \left( z^+ , z^- \right) R_{A}^0 \left( -\tfrac{1}{z^-} , -\tfrac{1}{z^+} \right) \,.
\end{aligned}
\end{equation}
Suggestively, one can rewrite \eqref{rotated transfer matrix-trace} and \eqref{rotated transfer matrix-trace-2} as
\begin{equation}
\label{rotated transfer matrix-trace-a1-1}
Y_{1,0}^I= Y_{1,0}^{II}= - e^{-2L \, \epsilon_1} \frac{(z^+ + z^-)^2}{2 \, z^+ z^- \left( 1+\frac{z^-+\frac{1}{z^-}}{z^+ +\frac{1}{z^+}}\right)}  \, R_{A}^0 \left( z^+ , z^- \right) R_{A}^0 \left( -\tfrac{1}{z^-} , -\tfrac{1}{z^+} \right) \, T_{1,1} \,,
\end{equation}
where $T_{1,1}$ is given in \eqref{asymptotic T functions}. 

Let us focus for the moment on the product of dressing phases that appears in \eqref{rotated transfer matrix-trace-a1-1}. Using \eqref{crossing ansatz} and \eqref{solution R0} we have
\begin{equation}
\label{product dressing phases}
\begin{aligned}
R_{A}^0 \left( z^+ , z^- \right) R_{A}^0 \left( -\tfrac{1}{z^-} , -\tfrac{1}{z^+} \right)  &= \sigma_B^{1/2}(q) \, \sigma_B^{1/2}(-\bar{q}) \, \sigma^{1/2} (q,-q) \, \sigma^{1/2} (-\bar{q},\bar{q}) \, \left( \frac{z^-}{z^+} \right)^3 \,.
\end{aligned}
\end{equation}
The product of bulk dressing phases that appear in \eqref{product dressing phases} can be evaluated using the identity \cite{Correa:2009mz}
\begin{equation}
\sigma(q_i,-q_j) \sigma(-\bar{q}_i,\bar{q}_j)=  \frac{z_i^+}{z_i^-} \frac{z_j^+}{z_j^-}  \frac{f(q_j,-q_i)}{f(q_j,-\bar{q}_i)} \,,    
\end{equation}
with
\begin{equation}
f(z_1,z_2):=\frac{z_1^- -z_2^+}{z_1^- -z_2^-} \frac{1-1/z_1^+ z_2^+}{1-1/z_1^+ z_2^-} \,,
\end{equation}
which gives
\begin{equation}
\sigma^{1/2}(q,-q) \sigma^{1/2}(-\bar{q},\bar{q})= 2 \,  \frac{z^+}{z^-} \left( \frac{1+z^+ z^-}{z^+ + z^-} \right)\frac{1}{\sqrt{\left(z^+ + \frac{1}{z^+} \right) \left(z^- + \frac{1}{z^-} \right)}}  \,.    
\end{equation}
Therefore,
\begin{equation}
\label{product dressing phases-2}
\begin{aligned}
R_{A}^0 \left( z^+ , z^- \right) R_{A}^0 \left( -\tfrac{1}{z^-} , -\tfrac{1}{z^+} \right)  &= 2 \, \sigma_B^{1/2}(q) \, \sigma_B^{1/2}(-\bar{q}) \, \left( \frac{z^-}{z^+} \right)^2 \left( \frac{1+z^+ z^-}{z^+ + z^-} \right)\, \\
& \quad \, \times  \frac{1}{\sqrt{\left(z^+ + \frac{1}{z^+} \right) \left(z^- + \frac{1}{z^-} \right)}}  \,,
\end{aligned}
\end{equation}
and
\begin{equation}
\label{Y1}
Y_{1,0}^I= Y_{1,0}^{II}= - e^{-(2L+2) \, \epsilon_1} \, \sigma_B^{1/2}(q) \, \sigma_B^{1/2}(-\bar{q}) \, \left( \frac{z^+ + \tfrac{1}{z^+}}{z^- + \tfrac{1}{z^-}}\right)^{1/2}   T_{1,1} \,.
\end{equation}
From \eqref{Y1} we can read the expressions for the $\phi_I$ and $\phi_{II}$ functions introduced in \eqref{asymptotic solution 1} and \eqref{asymptotic solution 2}, which we see that behave as in \eqref{phi equality0}. Consequently, following \eqref{phi equality} we can propose
\begin{equation}
\label{Y functions-1}
Y_{a,0}^I= Y_{a,0}^{II}= (-1)^a e^{-(2L+2) \, \epsilon_a} \, \sigma_B^{1/2}(q) \, \sigma_B^{1/2}(-\bar{q}) \, \left( \frac{z^{[+a]} + \tfrac{1}{z^{[+a]}}}{z^{[-a]} + \tfrac{1}{z^{[-a]}}}\right)^{1/2}    T_{a,1} \,,
\end{equation}
for general $a$. Taking into account the discussion given at the beginning of this section, we get that the expressions \eqref{Y functions-1} constitute an asymptotic solution to the Y-system presented in  \eqref{y system ABJM-1}-\eqref{y system eq momentum carrying nodes-2}.

\section{Cusp anomalous dimension from the BTBA formula}
\label{sec: gamma cusp from BTBA}

Let us focus now on the computation of the leading-order finite-size contribution to te vacuum energy $E_0(L)$, that comes from inserting the asymptotic solutions given in \eqref{Y functions-1} into the BTBA formula presented in \eqref{TBA ABJM}.
We will compute the finite-size correction at leading weak-coupling order. As we will see next, the explicit evaluation of \eqref{Y functions-1} shows that
\begin{equation}
Y^{I}_{a,0} \sim Y^{II}_{a,0} \sim 
{\cal O}(h^{4L+4})
\,,
\end{equation}
For the leading-order contribution to \eqref{TBA ABJM} there are two distinct possibilities, depending on whether each $Y$-function has a double pole or not as $q\to 0$ \cite{Correa:2009mz}. On the one hand, for $Y$-functions that are regular as $q\to 0$ one gets
\begin{equation}
\log \left( 1+ Y^{\alpha}_{a,0} \right) \sim Y^{\alpha}_{a,0} \sim 
{\cal O}(h^{4L+4}) \,, 
\end{equation}
which specifies the order of the finite-size correction. This leading asymptotic contribution is mediated by the exchange of a two-particle state in the mirror theory. On the other hand, if a $Y$-function has a double pole in $q$, i.e.
\begin{equation}
Y^{\alpha}_{a,0} = \frac{C_{\alpha,a}^2}{q^2} + \mathcal{O} (1) \,,
\end{equation}
the integral of its contribution to the finite-size correction provides a term proportional to
\begin{equation}
\label{approx TBA double pole}
 C_{\alpha,a} 
 \sim  {\cal O}(h^{2L+2}) \,.
\end{equation}
In this case,  the correction is mediated by the exchange of a one-particle state in the mirror picture.

Let us therefore study the behaviour of the $Y$-functions given in \eqref{Y functions-1} for $q \to 0$ and in the leading weak-coupling limit. From the results of \cite{Correa:2012hh,Drukker:2012de}, the factor that includes the boundary phase $\sigma_B$ behaves as
\begin{equation}
\sigma_B^{1/2} (q) \sigma_B^{1/2} (-\bar{q}) = \left(\frac{a}{q} + {\cal O}(q^0)\right) + {\cal O}(h^2)\,.
\label{polosigmaB}
\end{equation}
Furthermore, we have
\begin{equation}
e^{-(2 L+2) \, \epsilon_{a}(q)} = \left( \frac{4h^2}{a^2+q^2} \right)^{2L+2} +
{\cal O}(h^{4L+6})\,.
\label{exponopole}
\end{equation}
 It remains to evaluate the $T$-functions for $q \to 0$ to leading weak-coupling order.  For odd values of $a=2n+1$, the $T_{a,1}$ vanish linearly in $q$. Thus, altogether with the other factors \eqref{polosigmaB} and \eqref{exponopole}, we have regular $Y$-functions in the limit $q\to 0$,
\begin{equation}
\label{approx Y odd}
 Y^{I}_{2n+1,0} = Y^{II}_{2n+1,0} =\mathcal{O}(q^0) \,.
\end{equation}

On the other hand, for even values of $a$ the $T$-functions present a simple pole
\begin{equation}
\label{even a T functions expansion}
\left( \frac{z^{[+2n]} + \tfrac{1}{z^{[+2n]}}}{z^{[-2n]} + \tfrac{1}{z^{[-2n]}}}\right)^{1/2} T_{2n,1} = \left( \frac{8n b_{0,n}^2}{q}  + {\cal O}(q^{0}) \right) + {\cal O}(h^2) \,,
\end{equation}
Therefore, for even values of $a$,
\begin{equation}
\label{approx Y even}
Y^{I}_{2n,0} = Y^{II}_{2n,0} = \left( \frac{16 \, b_{0,n}^2 h^2}{q^2} \left( \frac{h}{n} \right)^{4L+2} + {\cal O}(q^0) \right) + {\cal O}(h^{4L+6}) \,.
\end{equation}
 Plugging these asymptotic expressions in \eqref{Y functions-1} we obtain
\begin{align}
E_0 (L) - E_0 (\infty) &= - 2h^{2L+2} \sum_{n=1}^{\infty}\frac{b_{0,n}}{n^{2L+1}}  + \mathcal{O}(h^{2L+3})\nonumber
\\ 
&=  - 2h^{2L+2} \sin^2\tfrac{\theta}{2} \sum_{k=0}^{\infty}
\frac{P_k^{(0,1)}(-\cos\theta)}{(k+1)^{2L+1}}  + \mathcal{O}(h^{2L+3}) \,.
\label{TBA result-L}
\end{align}

Let us comment about the sign of \eqref{TBA result-L}. The result of the integrals over $q$ that appear in \eqref{TBA ABJM} are given by the square root of the coefficient in front of the double poles in \eqref{approx Y even}, which entails  an ambiguity in the sign choice of the result. The correct sign can be determined if we consider the limit $\theta\to\pi$, at which the Wilson loop configuration can be related to a quark antiquark pair and  each term should contribute negatively to the energy \cite{Correa:2012hh}. The Jacobi polynomials are normalized such that
$P_k^{(0,1)}(1) = 1$, and this fixes the sign  to be the one in \eqref{TBA result-L}.

Eq. \eqref{TBA result-L} gives the finite-size correction to the energy  of a vacuum state \eqref{susyinsertion0} inserted at the position of the cusp. Its evaluation for arbitrary $L$ is rather complicated. If we identify the TBA-length $L$ with $\ell$, 
the vacuum state insertion is made of $2L+1$ fields. In order to associate the finite-size correction to this vacuum energy with $\Gamma_{\rm cusp}$, we need to consider an insertion with a vanishing number of fields, which
requires to analytically continue  the length $L$ such that
\begin{equation}
2L+1=0 
\qquad
\Rightarrow
\qquad
L = -1/2
\,.
\end{equation}
Therefore, we are interested in computing
\begin{equation}
\label{TBA result-2}
E_0 (-1/2) - E_0 (\infty) = - 2h \, \sin^2 \frac{\theta}{2} \sum_{k=0}^{\infty} P_{k}^{(0,1)} (-\cos\theta)  + \mathcal{O}(h^2) \,.
\end{equation}
Using the generating function of Jacobi Polynomials 
\begin{equation}
\label{Jacobi sum identity}
\sum_{k=0}^{\infty} P_{k}^{(0,1)} \left( x \right) t^k = \frac{2}{(1+t+ \sqrt{1-2x t + t^2})\sqrt{1-2x t + t^2}} \,,
\end{equation}
it is straightforward to compute the sum that appears in \eqref{TBA result-2}, which gives
\begin{equation}
\label{TBA result-3}
E_0 (-1/2) - E_0 (\infty) = -  h \left( \frac{1}{\cos \frac{\theta}{2}}-1 \right) + \mathcal{O}(h^2) \,.
\end{equation}

As shown in \cite{Minahan:2009aq,Leoni:2010tb}, in the weak-coupling limit the interpolating function behaves as $
h(\lambda) = \lambda + \mathcal{O}(\lambda^2) $. Inserting this expansion in \eqref{TBA result-3} we obtain that the leading weak-coupling value of $E_0 (-1/2) - E_0 (\infty) $ completely agrees with the one-loop cusp anomalous dimension of the ABJM theory computed in \cite{Griguolo:2012iq} from an expansion in  Feynman diagrams. This match provides support to our proposal for the $Y$-system of the cusped Wilson line, presented in \eqref{y system ABJM-1}-\eqref{y system eq momentum carrying nodes-2}.

\section{BTBA equations for the cusped Wilson line}
\label{sec: btba eqs}

Finally, let us write the BTBA equations for the cusped Wilson line of ABJM. To that aim, we should recall that in the previous sections we used that the $Y$-system for the cusped Wilson line of ABJM is the same as the one that describes the corresponding periodic system (the only changes are in the analytical and asymptotic properties of the solutions).  The same was observed in the ${\cal N}=4$ sYM theory, and the BTBA equations for the cusped Wilson line in that case were found to be almost the same as the TBA equations for the periodic spin chain. To be more precise, one can obtain the BTBA equations by taking the TBA equations of the periodic system and then subtracting the result of evaluating them in the leading-order finite-size solution  \cite{Correa:2012hh}. 
Assuming that the same holds in the ABJM case and using the TBA equations presented in \cite{Gromov:2009at,Bombardelli:2009xz} for the periodic spin chain, we propose that the BTBA equations for the cusped Wilson line of ABJM are
\begin{align}
\log \left( \frac{Y_{1,1}}{\mathbf{Y}_{1,1}} \right) &= K_{m-1} \star \log \left( \frac{1+ \bar{Y}_{1,m}}{1+Y_{m,1}} \frac{1+\mathbf{Y}_{m,1}}{1+ \mathbf{\bar{Y}}_{1,m}} \right) + {\cal R}^{(01)}_{1m} \star \log (1+Y_{m,0}^I) \nonumber \\
&\quad + {\cal R}^{(01)}_{1m} \star \log (1+Y_{m,0}^{II}) \,, \nonumber \\
\log \left( \frac{\bar{Y}_{2,2}}{\mathbf{\bar{Y}}_{2,2}} \right) &= K_{m-1} \star \log \left( \frac{1+ \bar{Y}_{1,m}}{1+Y_{m,1}} \frac{1+\mathbf{Y}_{m,1}}{1+ \mathbf{\bar{Y}}_{1,m}} \right) + {\cal B}^{(01)}_{1m} \star \log (1+Y_{m,0}^I) \nonumber \\
&\quad + {\cal B}^{(01)}_{1m} \star \log (1+Y_{m,0}^{II}) \,, \nonumber \\
\log \left( \frac{\bar{Y}_{1,n}}{\mathbf{\bar{Y}}_{1,n}} \right) &= - K_{n-1,m-1} \star \log \left( \frac{1+\bar{Y}_{1,m}}{1+\mathbf{\bar{Y}}_{1,m}} \right) - K_{n-1} \circledast \log \left( \frac{1+Y_{1,1}}{1+\mathbf{Y}_{1,1}} \right) \,, \nonumber \\
\log \left( \frac{Y_{n,1}}{\mathbf{Y}_{n,1}} \right) &= - K_{n-1,m-1} \star \log \left( \frac{1+Y_{m,1}}{1+\mathbf{Y}_{m,1}} \right) -  K_{n-1} \circledast \log \left( \frac{1+Y_{1,1}}{1+\mathbf{Y}_{1,1}} \right)  + \nonumber \\
&\quad + \left( {\cal R}^{(01)}_{nm} +{\cal B}^{(01)}_{n-2,m}  \right) \star \log (1+Y_{m,0}^I) + \left( {\cal R}^{(01)}_{nm} +{\cal B}^{(01)}_{n-2,m}  \right) \star \log (1+Y_{m,0}^{II}) \,, \nonumber \\
\log \left( \frac{Y_{n,0}^I}{\mathbf{Y}_{n,0}^I} \right) &= {\cal T}^{\parallel}_{nm} \star \log (1+ Y_{m,0}^I) + {\cal T}^{\perp}_{nm} \star \log (1+ Y_{m,0}^{II}) \, + \nonumber \\
& \quad + {\cal R}^{(10)}_{n1} \circledast \log \left( \frac{1+Y_{1,1}}{1+\mathbf{Y}_{1,1}} \right) + \left( {\cal R}^{(10)}_{nm} +{\cal B}^{(10)}_{n,m-2}  \right) \star \log \left( \frac{1+Y_{m,1}}{1+\mathbf{Y}_{m,1}} \right) \,, \nonumber 
\end{align}
\begin{align}
\log \left( \frac{Y_{n,0}^{II}}{\mathbf{Y}_{n,0}^{II}} \right) &= {\cal T}^{\parallel}_{nm} \star \log (1+ Y_{m,0}^{II}) + {\cal T}^{\perp}_{nm} \star \log (1+ Y_{m,0}^{I}) \, + \nonumber \\
& \quad + {\cal R}^{(10)}_{n1} \circledast \log \left( \frac{1+Y_{1,1}}{1+\mathbf{Y}_{1,1}} \right) + \left( {\cal R}^{(10)}_{nm} +{\cal B}^{(10)}_{n,m-2}  \right) \star \log \left( \frac{1+Y_{m,1}}{1+\mathbf{Y}_{m,1}} \right) \,. \nonumber
\end{align}
Above we are following the conventions of \cite{Gromov:2009at} for the integral kernels and convolutions, and we are using the notation $\bar{Y}_{a,s}=1/Y_{a,s}$. Moreover, the bold face $\mathbf{Y}$'s are used to represent the leading-order finite-size solutions, whose explicit expressions can be obtained from \eqref{T to Y-1}, \eqref{T to Y-2}, \eqref{asymptotic T functions}, \eqref{cusped WL t system ABJM} and \eqref{Y functions-1}.  Let us note that, working as in \cite{Gromov:2009bc,Bombardelli:2009xz,Gromov:2009at}, one can see that the BTBA equations that we have proposed in this section consistently lead to the $Y$-system written in \eqref{y system ABJM-1}-\eqref{y system eq momentum carrying nodes-2}.

%% file: introamplitudes.tex
As discussed in Chapter \ref{ch:intro}, improving our understanding of the properties of scattering amplitudes is a key challenge towards enhancing the predictive power of Quantum Field Theory. In this regard, the AdS/CFT correspondence has proven to be a valuable framework to address the study of these quantities. As an example, scattering amplitudes in the ${\cal N}=4$ super Yang-Mills theory have been found to satisfy remarkable properties, such as their invariance under dual conformal symmetry \cite{Drummond:2008vq} and the all-loop BFCW recursion relation \cite{Arkani-Hamed:2010zjl}. It is precisely in this theory were a positive geometry description of scattering amplitudes was first proposed \cite{Arkani-Hamed:2013jha}, providing a perturbative description of these quantities that is independent from the traditional Feynman diagram expansion.

In Chapter \ref{ch: neg geoms ABJM} we will be interested in the study of infrared-finite functions that naturally arise from the geometric description of scattering amplitudes in the ABJM theory. In view of this, in the present chapter we will briefly review the main properties of scattering amplitudes in the AdS/CFT framework, with an emphasis in their geometrical description in terms of positive geometries. We will begin with a short introduction to the main properties of scattering amplitudes in ${\cal N}=4$ super Yang-Mills. We will later discuss the geometric description of amplitudes in ${\cal N}=4$ sYM in terms of the Amplituhedron. Finally, we will end with a brief introduction to the analysis of scattering amplitudes in the ABJM theory. We refer to Appendix \ref{ch: massless kinematics} for a brief review on the kinematics of massless particles in three and four spacetime dimensions.

\section{Scattering amplitudes in ${\cal N}=4$ super Yang-Mills}
\label{sec: amplitudes N=4}

As discussed in Chapter \ref{ch:adscft}, the particle content of the ${\cal N}=4$ super Yang-Mills theory is given by a gluon of helicity +1, four gluinos of helicity +1/2, six scalars of vanishing helicity, four gluinos of helicity -1/2, and a gluon of helicity -1. The corresponding annihilation operators are $a(p)$, $a^A(p)$, $a^{[AB]}(p)$, $a^{[ABC]}(p)$ and $a^{[ABCD]}(p)$, where $A,B,C,D=1, \dots,4$ are indices in the fundamental representation of $SU(4)$. Note that we are omitting color indices for simplicity, but it should be reminded that all fields transform in the adjoint representation of the gauge group. Crucially, all the particles of this theory transform within a same supermultiplet, i.e. all the corresponding annihilation operators mix between each other under supersymmetry. It is therefore convenient to consider the superfield
\begin{equation}
\label{superfield N=4}
\Phi(p,\eta)=a(p) + \eta_A \, a^A(p) + \eta_{A} \, \eta_B \, a^{AB}(p)+\eta_A \, \eta_B \, \eta_C \, a^{ABC}(p)+ \eta_{A} \, \eta_{B} \, \eta_{C} \, \eta_{D} \, a^{ABCD}(p) \,,
\end{equation}
where the $\eta_A$ are Grassmann-odd variables. As customary, in the following we will focus on $n$-particle scattering processes where all the particles are outgoing\footnote{One may worry about the definition of a scattering process in a conformal field theory such as ${\cal N}=4$ super Yang-Mills, where the asymptotic past and future are ill-defined given the absence of a scale. A possible way out of this problem is to consider scattering amplitudes in the Coulomb branch of the theory (where the conformal invariance is broken), and then define the amplitudes at the origin of the Coulomb branch by taking the limit of vanishing vacuum expectation values.}, i.e. cases in which the amplitude is built only out of annihilation operators. In view of the above discussion, it is useful then to work with a \textit{superamplitude} ${\cal A}_n[\Phi_1(p_1,\eta_1), \dots , \Phi_n(p_n,\eta_n)]$, whose $\eta$-expansion gives all the component amplitudes.

The ${\cal A}_n[\Phi_1(p_1,\eta_1), \dots , \Phi_n(p_n,\eta_n)]$ superamplitudes are invariant under the $PSU(2,2|4)$ supersymmetry of the ${\cal N}=4$ sYM theory, i.e. are annihilated by the generators of the corresponding superalgebra. In the superspace notation introduced above, the momentum and supercharges are written as
\begin{equation}
\label{superspace generators}
p^{\dot{a}a}=\sum_{i=1}^n \tilde{\lambda}_i^{\dot{a}} \lambda_i^a \,, \qquad q^{aA}= \sum_{i=1}^n  \lambda^a_i \eta^A_i \,,\qquad \tilde{q}^{\dot{a}}_A= \sum_{i=1}^n \tilde{\lambda}_i^{\dot{a}}  \frac{\partial}{\partial \eta_i^A} \,,
\end{equation}
where the $\lambda_i^{a}$ and $\tilde{\lambda}_i^{\dot a}$ are the spinor-helicity variables introduced in Appendix \ref{ch: massless kinematics}. Eq. \eqref{superspace generators} implies that the action of $p^{\dot{a}a}$ and $q^{aA}$ is multiplicative. Therefore, the invariance of the superamplitude under their action gives\footnote{The $\frac{1}{\langle 12\rangle \langle 23\rangle \dots \langle n1\rangle}$ normalization is arbitrary.}
\begin{equation}
\label{superamplitude structure}
{\cal A}_n[\Phi_1(p_1,\eta_1), \dots , \Phi_n(p_n,\eta_n)]  = \frac{\delta^{(4)}(p^{\dot{a}a}) \, \delta^{(8)}(q^{aA}) }{\langle 12\rangle \langle 23\rangle \dots \langle n1\rangle}\, {\cal M}_n (\lambda_i^a,\tilde{\lambda}_i^{\dot{a}},\eta_i^A) \,,
\end{equation}
for some function ${\cal M}_n (\lambda_i^a,\tilde{\lambda}_i^{\dot{a}},\eta_i^A)$ and where
\begin{equation}
\label{supercharge conservation}
\delta^{(8)}(q^{aA})= \prod_{a=1}^2 \prod_{A=1}^4 \left( \sum_{i=1}^4  \lambda^a_i \eta^A_i \right) \,.
\end{equation}
It is interesting to note that \eqref{superamplitude structure} and \eqref{supercharge conservation} imply that 
\begin{equation}
{\cal A}_n[\Phi_1(p_1,\eta_1), \dots , \Phi_n(p_n,\eta_n)] \sim \mathcal{O}(\eta^8)\,.
\label{eta order superamplitude}
\end{equation}
Taking into account the definition given in eq. \eqref{superfield N=4}, we see then that eq. \eqref{eta order superamplitude} implies the vanishing of $n$-gluon amplitudes in which all gluons have the same helicity. The same result holds for amplitudes with at most one gluon with an opposite helicity. These results are true at all loop orders in the ${\cal N}=4$ super Yang-Mills theory. Moreover, they extend to pure Yang-Mills theories at tree level, given the equality between pure gluon tree-level amplitudes in pure Yang-Mills and ${\cal N}=4$ sYM. The first non-vanishing gluon amplitude has two gluons of opposite helicity, and it is known as the \textit{maximally helicity violating} (MHV) amplitude. Similarly, amplitudes with $k+2$ gluons of opposite helicity are referred to as N$^k$MHV. 

The R-symmetry invariance of the superamplitude implies that there can not be free $SU(4)$ indices in the ${\cal M}_n$ function introduced in \eqref{superamplitude structure}, i.e. all R-symmetry indices must be contracted with Levi-Civita tensors. Therefore, the ${\cal M}_n$ function admits an expansion of the type
\begin{equation}
\label{NkMHV sectors}
    {\cal M}_n (\lambda_i^a,\tilde{\lambda}_i^{\dot{a}},\eta_i^A) = {\cal M}_{n,0} (\lambda_i^a,\tilde{\lambda}_i^{\dot{a}},\eta_i^A) + {\cal M}_{n,1} (\lambda_i^a,\tilde{\lambda}_i^{\dot{a}},\eta_i^A) + \dots + {\cal M}_{n,n-2}(\lambda_i^a,\tilde{\lambda}_i^{\dot{a}},\eta_i^A) \,,
\end{equation}
where ${\cal M}_{n,k} \sim \mathcal{O}(\eta^{4k})$. Let us note that the N$^k$MHV gluon amplitude is contained in ${\cal M}_{n,k}$. Furthermore, ${\cal M}_{n,k}$ describes every amplitude that is related to the N$^k$MHV gluon amplitude through supersymmetry transformations, i.e. ${\cal M}_{n,k}$ accounts for what is known as the N$^k$MHV \textit{sector}.

Let us consider the $n$-point N$^k$MHV sector ${\mathcal M}_{n,k} [(p_1,h_1,a_1), \dots, (p_n,h_n,a_n) ]$ of a scattering process between fields with adjoint indices $a_1, a_2 , \dots, a_n$ and helicities $h_1, h_2 , \dots, h_n$. As it is well-known for Yang-Mills theories, one can factor out the color dependence of the amplitude by means of several $SU(N)$ identities. More precisely, one gets
\begin{align}
\label{color ordered ampl}
{\mathcal M}_{n,k} [(p_1,h_1,a_1), \dots, (p_n,h_n,a_n) ] &= \sum_{\sigma \in S_n/{\mathbb Z}_n} {\rm Tr}(T^{a_{\sigma_1}} \dots T^{a_{\sigma_n}}) \, \nonumber \\
& \quad \times M_{n,k} [(p_{\sigma_1}h_{\sigma_1}), \dots, (p_{\sigma_n},h_{\sigma_n}) ] \nonumber \\
&  \quad + \dots \,,
\end{align}
where $S_n/{\mathbb Z}_n$ is the set of non-cyclic permutations of $n$ elements, $M_{n,k} [(p_{\sigma_1}h_{\sigma_1}), \dots ]$ is a \textit{color-ordered} component of ${\mathcal M}_{n,k}$ and the $+\dots$ indicate subleading corrections in the large $N$ expansion. In the following we will restrict our analysis to the planar limit of the theory (see Chapter \ref{ch:adscft}), and therefore we will ignore the subleading large $N$ corrections in \eqref{color ordered ampl}. We will sometimes refer to $M_{n,k}$ as the $n$-particle N$^k$MHV amplitude, for simplicity.

\subsubsection*{Duality between scattering amplitudes and Wilson loops}

As previously stated, the analysis presented in this thesis is focused on Wilson lines and scattering amplitudes in the AdS/CFT framework. Interestingly, these objects are closely related in such a context. Let us consider an $n$-point MHV color-ordered gluon amplitude $M_{n,0}$, characterized by a set of $n$ points $x_i^{\mu}$ in dual space (see Appendix \ref{ch: massless kinematics}). Moreover, let us take a Wilson loop $W_n$ whose contour $\mathcal{C}$ describes a polygon with vertices at the $x_i^{\mu}$ points, i.e. with light-like edges. As first proposed by Alday and Maldacena \cite{Alday:2007hr, Alday:2007he} and thoroughly studied later (see e.g. \cite{Drummond:2007aua,Brandhuber:2007yx,Drummond:2007cf,Alday:2010vh}), there is a conjectured duality between the latter objects. More precisely, it is proposed that
\begin{equation}
    \label{WL SA duality}
    \log M_{n,0} \sim \log \langle W_n \rangle \,,
\end{equation}
at the level of the integrands\footnote{To be precise one should specify a prescription to compute the integrands. At the amplitude's side of the duality the integrand is fixed by using dual-space coordinates and requiring the correct pole structure, while at the Wilson loop's side one should use the method of the lagrangian insertions to build the integrand \cite{Eden:2010zz}.}. 

The correspondence between scattering amplitudes and polygonal Wilson loops was first motivated at strong coupling from an holographic analysis \cite{Alday:2007hr, Alday:2007he}. A key point in that argument relies on the self-duality of the $AdS_5 \times S^5$ space under T-duality, which allows to map the worldsheet associated to the scattering process to the worldsheet of the string which is dual to the Wilson loop. The conjecture was later extended to arbitrary N$^k$MHV sectors via the construction of a dual Wilson loop in superspace \cite{Mason:2010yk,Caron-Huot:2010ryg,Belitsky:2011zm}.

An important consequence of the duality is the \textit{dual conformal invariance} of scattering amplitudes in ${\cal N}=4$ super Yang-Mills. This conformal symmetry is naturally formulated in dual space, and exists on top of the usual conformal symmetry of amplitudes in spinor-helicity space. In the context of the Wilson loop/scattering amplitude correspondence, dual conformal invariance can be interpreted as the conformal symmetry of the dual Wilson line in configuration space. This symmetry has proven to be a very powerful tool for constructing loop integrands for scattering amplitudes. As an example, it was shown that dual conformal invariance can be used in combination with dihedral symmetry and certain soft-limit conditions to completely fix the four-particle integrand up to seven loops \cite{Bourjaily:2011hi}. Interestingly, the commutation between the ordinary and dual conformal generators generates the infinite-dimensional \textit{Yangian} algebra $Y[\mathfrak{psu(2,2|4)}]$ \cite{Drummond:2009fd}, which is a signature of the integrability of the theory.

\section{The Amplituhedron}

The search for structures that could account for the underlying simplicity of scattering amplitudes in the ${\cal N}=4$ sYM theory ultimately led to the proposal of the \textit{Amplituhedron} \cite{Arkani-Hamed:2013jha,Arkani-Hamed:2013kca} by Arkani-Hamed and Trnka. This was the first example of a \textit{positive geometry} \cite{Arkani-Hamed:2017tmz} being used to describe the scattering amplitudes of a theory. As we will shortly discuss, positive geometries are characterized by an unique differential form, known as the \textit{canonical form}, that has logarithmic singularities at the boundaries of the geometry. The Amplituhedron proposal states a direct connection between tree-level amplitudes and loop integrands and the canonical forms of the corresponding Amplituhedron geometries. Given that it will be instructive for the discussions that we will have in Chapter \ref{ch: neg geoms ABJM}, throughout this section we will briefly introduce the main properties of the geometric description of amplitudes in the ${\cal N}=4$ super Yang-Mills theory. For useful reviews on this topic see \cite{Ferro:2020ygk,Herrmann:2022nkh}. Later, in Section \ref{sec: amplitudes ABJM}, we will discuss the extension of these ideas to the ABJM theory.

\subsubsection*{Positive geometries}

We will begin by defining the notion of a positive geometry. To that aim, let us consider an ambient space $X$ with a slice $X_{\geq 0}$ of real dimension $D$\footnote{To be more precise, we should take $X$ to be a complex projective algebraic variety of complex dimension $D$, i.e. the solution set in complex projective space of a set of homogeneous algebraic equations. Moreover, $X_{\geq 0} \subset X(\mathbb{R})$ should be taken to be an oriented manifold of real dimension $D$, where $X(\mathbb{R})$ is the real part of $X$ (i.e. the solution set in real projective space of the set of algebraic equations that defines $X$).}. A positive geometry is a pair $(X,X_{\geq 0})$ that admits a unique non-zero differential $D$-form $\Omega(X,X_{\geq 0})$ satisfying
\begin{enumerate}
    \item If $D=0$, then $X=X_{\geq 0}$ is just a single point and $\Omega(X,X_{\geq 0})= \pm 1$ (the sign depends on the orientation).
    \item For $D > 0$ each boundary $(C, C_{\geq 0})$ of $(X,X_{\geq 0})$ is again a positive geometry, and its canonical form is obtained by taking the residue of $\Omega(X,X_{\geq 0})$ at the boundary, i.e.
    \begin{equation}
    \label{canonical form bdry}
        \Omega(C,C_{\geq 0})= {\rm Res}_{C} \, \Omega(X,X_{\geq 0}) \,.
    \end{equation}
\end{enumerate}
The form $\Omega(X,X_{\geq 0})$ is known as the \textit{canonical form} of the positive geometry. To understand the residue operation in \eqref{canonical form bdry}, let us consider a parametrization of $X$ such that the boundary $C$ is located at $z=0$ and
\begin{equation}
\label{asymptotic bdry expansion}
    \Omega(X,X_{\geq 0})= \omega(u) \wedge \frac{dz}{z} + \dots \,,
\end{equation}
for $z \to 0$ (the dots in \eqref{asymptotic bdry expansion} indicate higher orders in the $z$ expansion), where $u$ stands for the remaining coordinates of $X$. Then,
\begin{equation}
    {\rm Res}_{C} \, \Omega(X,X_{\geq 0})= \omega(u)\,.
\end{equation}

As an example, let us consider the case of the line segment $[a,b] \subset \mathbb{R}$, which is the simplest non-trivial example of a positive geometry. In such a case the geometry has two boundaries, located respectively at $x=a$ and $x=b$. The corresponding canonical form is
\begin{equation}
    \Omega([a,b])= \frac{dx}{x-a}- \frac{dx}{x-b} \,.
\end{equation}
One can easily verify that the only singularities of $\Omega([a,b])$ are at the $x=a$ and $x=b$ boundaries, and the corresponding residues are
\begin{equation}
    {\rm Res}_{x=a} \, \Omega([a,b])=1 \,, \qquad {\rm Res}_{x=b} \, \Omega([a,b])=-1 \,.
\end{equation}
Similarly, one can consider the example of a triangle with boundaries at $x=0$, $y=0$ and $x+y=1$. It is straightforward to check that the canonical form
\begin{equation}
    \Omega_{\rm triangle}=\frac{dx \wedge dy}{x y (1-x-y)} \,,
\end{equation}
satisfies all the axioms that define a positive geometry.

\subsubsection*{Tree-level Amplituhedron}

When discussing the Amplituhedron it will prove useful to work with \textit{momentum-twistor} variables, which are introduced in Appendix \ref{ch: massless kinematics}. For a formulation of the Amplituhedron in spinor-helicity space see \cite{Damgaard:2019ztj,Ferro:2022abq}. As stated in Appendix \ref{ch: massless kinematics}, momentum twistors are defined as points
\begin{equation}
\label{momentum twistors def-main text}
    z^I_i:=(\lambda_a, \mu^{\dot a}) \,,
\end{equation}
in $\mathbb{CP}^3$ that satisfy
\begin{equation}
\label{incidence relations-main text}
    \mu^{\dot{a}}=\lambda_a \, x^{\dot{a}a} \,,
\end{equation}
where $x^{\dot{a}a}$ is a point in dual space (see \eqref{dual space def} in Appendix \ref{ch: massless kinematics}). Eq. \eqref{incidence relations-main text} is known as the \textit{incidence relation}. A scattering process involving $n$ massless particles can be described in terms of a set of $n$ momentum twistors $z_i^I=(\lambda_{ia}, \mu_i^{\dot a}) , \; i=1, \dots, n$, for which the incidence relation \eqref{incidence relations-main text} implies
\begin{equation}
\label{incidence relations 2-main text}
    \mu^{\dot{a}}_i=\lambda_a^i \, x_i^{\dot{a}a}=\lambda^i_a \, x_{i+1}^{\dot{a}a} \,.
\end{equation}
where $x_i$ and $x_{i+1}$ are the points in dual space respectively associated to the $(z_{i}^I,z_{i-1}^I)$ and $(z_{i+1}^I,z_{i}^I)$ lines in momentum-twistor space. Moreover, in order to formulate a geometric description of N$^k$MHV amplitudes it turns out useful to define \textit{bosonized} momentum twistors as
\begin{equation}
\label{bosonized momentum twistors}
Z_i=\left( z_i, \, \phi_{1A} \eta^A_i, 
\, \phi_{2A} \eta^A_i , \dots , \, \phi_{kA} \eta^A_i  \right) \,,
\end{equation}
where $\eta^A_i$ (with $A=1 \dots, 4$) are the superspace variables introduced in \eqref{superfield N=4} and $\phi_{1A}\,, \phi_{2A}\,, \dots \phi_{kA}$ are auxiliary Grassmann-odd variables.

As we will shortly see, a key obtect in the Amplituhedron construction turns out to be the \textit{Grassmannian} $G(k,n)$, which is defined as the space of $k$ planes in $n$ dimensions. Each element in $G(k,n)$ can be expressed in terms of a $k \times n$ matrix $C_{\alpha i}$ (with $\alpha=1, \dots, k$ and $i=1, \dots,n$) modulo a $GL(k)$ invariance. Furthermore, one can introduce the \textit{positive Grassmannian} $G_+(k,n)$ as the subspace of $G(k,n)$ where all elements are such that all their ordered $k \times k$ minors are positive.

We are now in position to give the original definition of the tree-level $n$-particle N$^k$MHV Amplituhedron $\mathbb{A}^{(0)}_{n,k}$, as presented by Arkani-Hamed and Trnka in \cite{Arkani-Hamed:2013jha,Arkani-Hamed:2013kca}. It is defined as the image of the map $\Phi: G_+(k,n) \rightarrow G(k,k+4)$ given as
\begin{equation}
\label{tree level amplituhedron}
\Phi: \quad Y^I_{\alpha}= C_{\alpha i} \,  Z^I_i \,, \quad \alpha=1\, \dots\, k\,, \quad I=1\,,\dots \,,k+4 \,, \quad i=1\, \dots \,n \,,
\end{equation}
where $C \in G_+(k,n)$ and $Z =(Z_1 \, Z_2 \dots Z_n) \in G_+(k+4,n)$ is the matrix of bosonized momentum-twistors that parametrizes the external kinematics. The corresponding canonical form $\hat{\omega}^{(0)}_{n,k}$ can generically be written as
\begin{equation}
\label{canonical form tree level}
\hat{\omega}^{(0)}_{n,k} (Y,Z)= \left( \prod_{i=1}^k \langle Y_1  \dots Y_k \, d^{4} Y_a \rangle \right) \omega^{(0)}_{n,k} (Y,Z) \,,
\end{equation}
As proposed in \cite{Arkani-Hamed:2013jha,Arkani-Hamed:2013kca}, one can obtain the tree-level $n$-particle N$^k$MHV color-ordered amplitude $M^{(0)}_{n,k}$ as
\begin{equation}
\label{tree level amplitude from Amplituhedron}
M^{(0)}_{n,k} = n^{(0)}_{n,k}\, \int d^4 \phi_1 \dots d^4 \phi_k \,\, \omega^{(0)}_{n,k} (Y^*,Z) \,,
\end{equation}
with
\begin{equation}
\label{Y*}
Y^*= (0_{k \times 4} | 1_{k \times k})^T \,.
\end{equation}
and where $n^{(0)}_{n,k}$ is a normalization that can be fixed by requiring the correct behavior of the amplitude in certain limits and by comparison with the literature (see Appendix \ref{ch: normalizations} for a detailed computation of these normalizations at loop level in the ABJM case).

The geometric description of amplitudes in ${\cal N}=4$ sYM admits an alternative topological formulation, as proposed in \cite{Arkani-Hamed:2017vfh}. There it was suggested that the tree-level N$^k$MHV Amplituhedron ${\mathbb A}^{(0)}_{n,k} $ can alternatively be taken as the space of bosonic momentum twistors $z_i$ that satisfy
\begin{equation}
\label{top tree level cond 1}
    \langle i \, i+1 \, j \, j+1 \rangle >0 \,,
\end{equation}
and are such that the sequence
\begin{equation}
\label{top tree level cond 2}
\{ \langle 1234 \rangle ,\, \langle 1235 \rangle ,\, \dots ,\, \langle 123 n \rangle  \} \,,
\end{equation}
has $k$ sign flips. For a definition of the $\langle ijkl \rangle$ brackets see eq. \eqref{momentum twistor brackets} in Appendix \ref{ch: massless kinematics}. The amplitude $M^{(0)}_{n,k}$ is then computed from the corresponding canonical form $\Omega^{(0)}_{n,k}$ by replacing the differencials $dz_i$ by the superspace Grassmann variables $\eta_i$, i.e.
\begin{equation}
M^{(0)}_{n,k}=n^{(0)}_{n,k} \, \Omega^{(0)}_{n,k}|_{dz_i \rightarrow \eta_i} \,.
\end{equation}

\subsubsection*{Loop Amplituhedron}

The Amplituhedron construction can also be extended to loop orders, where it can be used to compute the $L$-loop integrand of the scattering amplitude. For simplicity, we will just present the topological formulation of the $n$-particle N$^k$MHV $L$-loop Amplituhedron $\mathbb{A}^{(L)}_{n,k}$ that was given in \cite{Arkani-Hamed:2017vfh}. Details of the original formulation can be found in \cite{Arkani-Hamed:2013jha,Arkani-Hamed:2013kca}. The $\mathbb{A}^{(L)}_{n,k}$ geometry is defined by taking a set of bosonic momentum twistors $z_i^I$ satisfying the tree-level Amplituhedron conditions \eqref{top tree level cond 1} and \eqref{top tree level cond 2} and a set of loop lines $(AB)_{\ell}^{IJ}, \, \ell=1\,, \dots \, L$, such that 
\begin{align}
    &\langle  (AB)_{\ell} \, i \, i+1\rangle >0 \,, \nonumber \\
    &\{ \langle(AB)_{\ell} \, 12 \rangle ,\,  \langle (AB)_{\ell} \, 13 \rangle ,\, \dots ,\, \langle (AB)_{\ell} \, 1n \rangle \} \, \text{ has } k+2 \text{ sign flips,}
    \label{loop ampl constraints}
\end{align}
for all $\ell$ and
\begin{equation}
\label{positivity constraint}
\langle  (AB)_{\ell_1} \, (AB)_{\ell_2} \rangle >0  \quad \forall \ell_1,\, \ell_2 \,.
\end{equation}
The $L$-loop integrand\footnote{Throughout this thesis we will use the convention $\int \frac{d^D x}{i\pi^{D/2}}$ for the $D$-dimensional loop-measure in dual coordinate space.} ${\bf I}^{(L)}_{n,k}$ of the $L$-loop amplitude $M^{(L)}_{n,k}$ is then obtained from the canonical form $\Omega^{(L)}_{n,k}$ in the $\{ z_i, (AB)_{\ell} \}$ space after replacing $dz_i \rightarrow \eta_i$, i.e.
\begin{equation}
\label{integrand from canonical form chapter amplt}
{\bf I}^{(L)}_{n,k}=n^{(L)}_{n,k} \, \Omega^{(L)}_{n,k}|_{dz_i \rightarrow \eta_i} \,.
\end{equation}
where again we have included a relative normalization $n^{(L)}_{n,k}$.

In Chapter \ref{ch: neg geoms ABJM} we will be interested in four-particle MHV scattering processes. In such a case the $L$-loop Amplituhedron constraints presented in \eqref{top tree level cond 1}, \eqref{top tree level cond 2} and \eqref{loop ampl constraints} become
\begin{equation}
    \label{4-pt amplituhedron constraint 1}
    \langle 1234 \rangle >0 \,,
\end{equation}
and
\begin{align}
    \label{4-pt amplituhedron constraint 2}
    \langle (AB)_{\ell} 12 \rangle >0 ,\, \quad \langle (AB)_{\ell} 23 \rangle &>0 ,\, \quad \langle (AB)_{\ell} 34 \rangle >0 ,\, \quad \langle (AB)_{\ell} 14 \rangle >0 \,, \\
    \label{4-pt amplituhedron constraint 3}
    \langle(AB)_{\ell} 13 \rangle &<0 ,\, \quad \langle (AB)_{\ell} 24 \rangle <0 \,.
\end{align}
Moreover, one has to additionally impose the positivity condition \eqref{positivity constraint}. As an example, the one-loop MHV four-particle canonical form $\Omega^{(1)}_{4,0}$ is \cite{Arkani-Hamed:2013jha}
\begin{equation}
    \label{one loop form N=4}
    \Omega^{(1)}_{4,0}= \frac{\langle 1234 \rangle^2 \langle ABd^2A\rangle \langle ABd^2B\rangle}{\langle AB12 \rangle \langle AB23 \rangle \langle AB34 \rangle \langle AB14 \rangle} \,,
\end{equation}
while at two loops one gets \cite{Arkani-Hamed:2013kca}
\begin{equation}
    \label{two loop form N=4}
    \Omega^{(2)}_{4,0}= \frac{\langle 1234 \rangle^3 {\cal N}^{(2)}_{4,0} \langle ABd^2A\rangle \langle ABd^2B\rangle \langle CDd^2C\rangle \langle CDd^2D\rangle}{\langle AB12 \rangle \langle AB23 \rangle \langle AB34 \rangle \langle AB14 \rangle \langle ABCD \rangle \langle CD12 \rangle \langle CD23 \rangle \langle CD34 \rangle \langle CD14 \rangle} \,,
\end{equation}
with
\begin{equation}
{\cal N}^{(2)}_{4,0}= \langle AB12 \rangle \langle CD34 \rangle+ \langle AB23 \rangle \langle CD14 \rangle + (AB \leftrightarrow CD) \,.
\end{equation}

\section{Scattering amplitudes in ABJM}
\label{sec: amplitudes ABJM}

As reviewed in Chapter \ref{ch:adscft}, the particle excitations of the ABJM theory are given by eight massless scalars and eight massless fermions. Note that the gauge fields of the theory do not generate external states in a scattering process, given that their equations of motion do not allow for particle excitations. To be precise, the set of annihilation operators upon which amplitudes are built comprises four massless scalars $(a^I)^i_{\hat{i}}(p)$ and four massless fermions $(\alpha_I)^i_{\hat{i}}(p)$ in the fundamental representation of the $SU(4)$ R-symmetry and the $(\mathbf{N},\bar{\mathbf{N}})$ bi-fundamental representation of the $U(N)\times U(N)$ gauge group. Moreover, there are also four massless scalars $(b_I)^{\hat{i}}_i(p)$ and four massless fermions $(\beta^I)^{\hat{i}}_i(p)$ in the fundamental representation of $SU(4)$ and the $(\bar{\mathbf{N}},\mathbf{N})$ bi-fundamental representation of $U(N)\times U(N)$. As in Chapter \ref{ch:adscft} we are using $i=1, \dots, N$ and $\hat{i}=1, \dots, N$ indices for the left and right $U(N)$ factors of the gauge group, respectively. 

When writing a superfield upon which amplitudes could be expressed, a key difference arises in the ABJM theory with respect to its four-dimensional ${\cal N}=4$ super Yang-Mills cousin. In ABJM all the excitations transform in the fundamental representation of the $SU(4)$ R-symmetry group, which seems to prevent us from constructing a superfield where the full $SU(4)$ R-symmetry is manifest. This was not the case for the ${\cal N}=4$ sYM theory, where all excitations could be accommodated in an $SU(4)$ invariant way at different orders of the expansion in the Grassmann-odd superspace variable (see eq. \eqref{superfield N=4}). Instead, in the ABJM theory the superfield is covariant under an $SU(3) \subset SU(4)$ subgroup of the R-symmetry. This is realized by the introduction of a Grassmann-odd variable $\eta^A$ ($A=1,2,3$) in the fundamental representation of $SU(3)$, which allows to organize the particle excitations into two superfields as \cite{Bargheer:2010hn}
\begin{align}
\Phi^i_{\hat i}(p,\eta)&= (a^4)^i_{\hat i}(p)+ \eta^A (\alpha_A)^i_{\hat i}(p)+\frac{1}{2} \epsilon_{ABC} \eta^A \eta^B (a^C)^i_{\hat i}(p) + \frac{1}{3!} \epsilon_{ABC} \eta^A \eta^B \eta^C  (\alpha_4)^i_{\hat i}(p) \,, \nonumber \\
\Psi_i^{\hat i}(p,\eta)&= (\beta^4)_i^{\hat i}(p)+ \eta^A (b_A)_i^{\hat i}(p)+\frac{1}{2} \epsilon_{ABC} \eta^A \eta^B (\beta^C)_i^{\hat i}(p) + \frac{1}{3!} \epsilon_{ABC} \eta^A \eta^B \eta^C  (b_4)_i^{\hat i}(p) \,.
\label{abjm superfield 2} 
\end{align}
Note that, as opposed to the ${\cal N}=4$ sYM case, in the ABJM theory one has to introduce two superfields (one of which is bosonic, while the other is fermonic), not just one.

As expected, one can use the superspace variables $\lambda^a$ and $\eta^A$ (note that in three dimensions one has only one spinor-helicity variable, see Appendix \ref{ch: massless kinematics}) to construct a representation of the $\mathfrak{osp}(6|4)$ superalgebra of the theory. In this representation we obtain
\begin{equation}
    p^{ab}=\lambda^a \lambda^b \,, \qquad q^{Aa}= \lambda^a \eta^A\,, \qquad \bar{q}^a_A= \lambda^a \frac{\partial}{\partial \eta^A} \,,
\end{equation}
for the momentum and the supercharges. Moreover, the $\mathfrak{su}(4)$ R-symmetry algebra can be expressed in terms of $\mathfrak{su}(3) \oplus \mathfrak{u}(1)$ generators \cite{Bargheer:2010hn}. There is an interesting corollary that can be obtained from the structure of the latter $\mathfrak{u}(1)$ charge, which takes the form
\begin{equation}
\label{u1 gen}
j= \eta^A \frac{\partial}{\partial \eta^A} - \frac{3}{2} \,.
\end{equation}
Let us consider an $n$-particle superamplitude ${\cal A}_n(\lambda_1,\eta_1, \dots,\lambda_n,\eta_n)$. Requiring its invariance under \eqref{u1 gen} imposes
\begin{equation}
\label{n constraint}
    \sum_{i=1}^n \eta^A_i \frac{\partial}{\partial \eta^A_i} {\cal A}_n = \frac{3}{2}n \, {\cal A}_n \,.
\end{equation}
The left-hand-side of eq. \eqref{n constraint} counts the $\eta$-degree of the superamplitude ${\cal A}_n$, which can only take integer values. Therefore, this enforces $n$ in the right-hand side to take \textit{even} values, i.e. scattering amplitudes with an odd number of fields are vanishing in the ABJM theory. Furthermore, we see that the $\eta$-degree of a superamplitude ${\cal A}_n$ is tightened to the number of particles, i.e. the N$^k$MHV level $k$ of ${\cal A}_n$ depends on $n$. As an example, we obtain that four-particle amplitudes have degree 6 in $\eta$ and are MHV, while six-point amplitudes have degree 9 and are NMHV.

The color structure of the amplitude can be stripped out by taking into account the bi-fundamental behaviour of the excitations, which implies \cite{Bargheer:2010hn}
\begin{align}
{\cal A}_n [\Phi^{i_1}_{{\hat i}_1} \Psi^{{\hat j}_1}_{j_1} \Phi^{i_2}_{{\hat i}_2} \dots  \Psi^{{\hat j}_{n/2}}_{j_{n/2}}] &= \sum_{(\sigma,\bar{\sigma}) \in  S_{n/2}/{\mathbb{Z}_{n/2}} \times S_{n/2}/{\mathbb{Z}_{n/2}}} A_n(\lambda_{\sigma_1}, \eta_{\sigma_1}, \lambda_{\bar{\sigma}_1}, \eta_{\bar{\sigma}_1} \dots,\lambda_{\bar{\sigma}_{n/2}}, \eta_{\bar{\sigma}_{n/2}} ) \,  \nonumber \\
& \qquad \qquad \qquad \qquad \qquad \; \; \, \times \delta^{i_{\sigma_1}}_{j_{\bar{\sigma}_1}} \delta^{\hat{j}_{\bar{\sigma}_1}}_{\hat{i}_{\sigma_2}} \dots \delta^{\hat{j}_{\bar{\sigma}_{n/2}}}_{\hat{i}_{\sigma_1}} \,,
\end{align}
where the sum runs over all permutations of even and odd sites among themselves, modulo cyclic permutations.

As for the ${\cal N}=4$ sYM case, one can use the supersymmetries of the theory to constrain the structure of the corresponding superamplitudes. For the four-point case one obtains \cite{Bargheer:2010hn}
\begin{equation}
\label{4-pt amplitude ABJM structure}
A_4(\lambda_1, \eta_1, \dots, \lambda_4, \eta_4)= \frac{\delta^{(3)}(p^{ab})\delta^{(6)}(q^{aA})}{\langle 12 \rangle \langle 4 1\rangle} \, M_4 (\lambda_1, \dots, \lambda_4) \,,
\end{equation}
where $M_4 (\lambda_1, \dots, \lambda_4)$ is a Lorentz-invariant dimensionless function of the spinor-helicity variables. Let us note that, as expected from equation \eqref{n constraint}, the four-point amplitude has degree 6 in $\eta$. The four-particle scattering amplitude is known up to three-loops \cite{Bargheer:2010hn,Chen:2011vv,Bargheer:2012cp,Bianchi:2014iia}, and there is a BDS-like conjecture for the all-loop result \cite{Bianchi:2011aa,Bianchi:2014iia}. Moreover, even non-planar corrections have been computed \cite{Bianchi:2013iha,Bianchi:2013pfa}. Similarly, the six point amplitude can be expressed as
\cite{Bargheer:2010hn}
\begin{align}
A_6(\lambda_1, \eta_1, \dots, \lambda_6, \eta_6)= \delta^{(3)}(p^{ab})\delta^{(6)}(q^{aA}) (  & M_6^+ (\lambda_1, \dots, \lambda_6) \, \delta^{(3)}(\alpha^A)  \nonumber \\
&  +M_6^- (\lambda_1, \dots, \lambda_6) \, \delta^{(3)}(\beta^A) ) \, \,,
\label{6-pt amplitude ABJM structure}
\end{align}
where $M_6^+$ and $M_6^-$ are Lorentz invariant functions of weight -6 in the spinor-helicity variables, and with
\begin{equation}
    \alpha^A := \sum_{i=1}^6 \, x^+_i \eta_i^A \,, \qquad 
    \beta^A := \sum_{i=1}^6 \, x^-_i \eta_i^A \,,
\end{equation}
where
\begin{align}
x_{i}^{\pm} &:= \frac{1}{2\sqrt{2}} \epsilon_{ijk} \frac{\langle jk \rangle}{\sqrt{\langle 13 \rangle^2+ \langle 35 \rangle^2+ \langle 51 \rangle^2}} \qquad \text{for} \qquad i,j,k \; \text{odd} \,, \\
x_{i}^{\pm} &:= \frac{\pm i}{2\sqrt{2}} \epsilon_{ijk} \frac{\langle jk \rangle}{\sqrt{\langle 24 \rangle^2+ \langle 46 \rangle^2+ \langle 62 \rangle^2}} \qquad \text{for} \qquad i,j,k \;  \text{even} \,.
\end{align}
The six-particle amplitude is currently known up to two-loops \cite{Caron-Huot:2012sos,He:2022lfz}. This is also the case for the eight-particle amplitude \cite{He:2022lfz}, and there are one-loop results for scattering processes with arbitrary number of particles \cite{Bianchi:2012cq}.

\subsubsection*{Duality with Wilson loops and dual conformal symmetry}

As opposed to the ${\cal N}=4$ super Yang-Mills case, in the ABJM theory the duality between scattering amplitudes and polygonal Wilson loops (see Section \ref{sec: amplitudes N=4}) holds only for the four-particle case, and it has been proven to fail when considering scattering processes with a higher number of particles \cite{Henn:2010ps,Bianchi:2011dg,Chen:2011vv,Bianchi:2012cq,Bianchi:2013pva}. This breaking of the duality for $n>4$ particles is believed to be related to the fact that the $\eta$-degree of the $A_n$ superamplitude depends on $n$, which contrasts what happens in ${\cal N}=4$ sYM. As seen in eq. \eqref{4-pt amplitude ABJM structure}, the four-particle amplitude $A_4$ of ABJM has $\eta$-degree six and mimics the behaviour of an MHV amplitude in ${\cal N}=4$ sYM. On the other hand, in the six-particle case the amplitude $A_6$ has degree nine (see eq. \eqref{6-pt amplitude ABJM structure}) and behaves as an NMHV amplitude. Considering that in the ${\cal N}=4$ sYM theory the N$^k$MHV amplitudes with arbitrary $k$ are proposed to be dual to a Wilson loop in superspace \cite{Mason:2010yk,Caron-Huot:2010ryg,Belitsky:2011zm}, it has been suggested  \cite{Bianchi:2012cq} that in the ABJM theory scattering amplitudes with arbitrary number of particles should similarly be dual to a Wilson loop in superspace. 

It is worth mentioning that, despite a Wilson loop/scattering amplitude duality has not been found yet beyond four points, there is evidence of dual superconformal \cite{Huang:2010qy,Gang:2010gy,Chen:2011vv,Bianchi:2011fc} and Yangian symmetry \cite{Bargheer:2010hn,Lee:2010du} for higher-point amplitudes.

\subsubsection*{ABJM Amplituhedron}

The geometric formulation of amplitudes in terms of positive geometries was first extended at tree level to the ABJM theory in \cite{Huang:2021jlh,He:2021llb}. Recently, the authors of \cite{He:2022cup,He:2023rou} proposed an all-loop \textit{ABJM Amplituhedron} by imposing a symplectic condition on the amplituhedron of ${\cal N}=4$ sYM. This conjecture has been used to compute loop integrands up to $L=5$ loops for the four-particle case \cite{He:2022cup}. Moreover, two-loop integrands have been computed up to eight points, while one-loop integrands have been obtained for scattering amplitudes with up to ten points \cite{He:2023rou}.

A crucial point in the construction of the ABJM Amplituhedron is the symplectic projection that can be used to obtain three-dimensional momentum twistors from their four-dimensional counterparts. As discussed in more detail in Appendix \ref{ch: massless kinematics}, the kinematics of a scattering process in three dimensions can be described with a set of $n$ momentum twistors $z_1^I, \dots, z_n^I \in \mathbb{CP}^3$ subjected to the symplectic constraint
\begin{equation}
\label{symplectic constraint-main text}
    \Sigma_{IJ} \, z^I_i z^J_{i-1}=0 \,,
\end{equation}
where
\begin{equation}
\label{Sigma-main text}
\Sigma= \left( 
\begin{array}{cc} 
0 & -\mathbb{1}_{2 \times 2} \\
\mathbb{1}_{2 \times 2} & 0
\end{array}
 \right) \,.
\end{equation}
The above projection implies that the momentum twistors $z_i^I=(\lambda_{i,a}, \mu_i^b)$ satisfy the incidence relations
\begin{equation}
\label{3D incidence relations}
\mu_i^b=x_i^{ab} \lambda_{i,a}=x_{i+1}^{ab} \lambda_{i,a} \,,
\end{equation}
where $x_i^{ab}, \; i=1, \dots,n$ are points in the three-dimensional dual space.

It is precisely the projection \eqref{symplectic constraint-main text}, together with a change of signs, that allows to obtain the tree-level ABJM Amplituhedron from its ${\cal N}=4$ super-Yang Mills analogue. More precisely, the tree-level ABJM Amplituhedron $\mathbb{A}_{n}^{(0)}$ is proposed to be the region described by $n$ momentum twistors $z_1^I, \dots, z_n^I \in \mathbb{CP}^3$ such that \cite{Huang:2021jlh,He:2021llb,He:2023rou}
\begin{align}
\label{tree ABJM Amplt 1}
    \langle i \, i+1 \, j \, j+1 \rangle &<0 \,, \\
    \Sigma_{IJ} \, z^I_i z^J_{i+1}&=0 \,,
\label{tree ABJM Amplt 2}
\end{align}
for all $i=1, \dots,n$, and where the sequence
\begin{equation}
\label{tree ABJM Amplt 3}
\{ \langle 1234 \rangle ,\, \langle 1235 \rangle ,\, \dots ,\, \langle 123 n \rangle  \} \,,
\end{equation}
has $k=\frac{n}{2}-2$ sign flips. Note the relation $k=\frac{n}{2}-2$ between the number of sign flips and the number of particles, which is precisely the correspondence between the N$^k$MHV order $k$ and the number of particles $n$ in ABJM amplitudes. We should also emphasize the change of sign in \eqref{tree ABJM Amplt 1} with respect to \eqref{top tree level cond 1}. As discussed in \cite{He:2023rou}, there is a key corollary that can be deduced from the simultaneous application of the change of sign in \eqref{tree ABJM Amplt 1} and the symplectic projection \eqref{tree ABJM Amplt 2}: only even values of $n$ with $k=\frac{n}{2}-2$ give a non-empty $\mathbb{A}_{n}^{(0)}$ region, which is precisely the case in which the ABJM amplitudes are non-vanishing.

Similarly, the $L$-loop ABJM Amplituhedron $\mathbb{A}_{n}^{(0)}$ is defined as the region in the $ \{ z_i^I, (AB)^{IJ}_{\ell} \}$ space (with $i=1, \dots,n$ and $\ell=1, \dots, L$) such that the external kinematics $z_i^I$ satisfies the constraints \eqref{tree ABJM Amplt 1}-\eqref{tree ABJM Amplt 3} and where the loop lines $(AB)^{IJ}_{\ell}$ are subjected to
\begin{align}
\label{ABJM loop Amplt 1}
    &\langle  (AB)_{\ell} \, i \, i+1\rangle <0 \,,  \\
    &\{ \langle (AB)_{\ell} \, 12 \rangle ,\,  \langle (AB)_{\ell} \, 13 \rangle ,\, \dots ,\, \langle (AB)_{\ell} \, 1n \rangle \} \, \text{ has } k+2=\frac{n}{2} \text{ sign flips,}
   \label{ABJM loop Amplt 2} \\
   &\Sigma_{IJ} \, A^I_{\ell} B^J_{\ell} =0 \,,
   \label{ABJM loop Amplt 3}
   \end{align}
for all ${\ell}$, with
\begin{align}
\langle  (AB)_{\ell_1} \, (AB)_{\ell_2} \rangle &<0  \quad \forall \ell_1,\, \ell_2 \,.
\label{ABJM loop Amplt 4}
\end{align}
As in the tree-level case, it should be noted that there is a change of sign in \eqref{ABJM loop Amplt 1} and \eqref{ABJM loop Amplt 4} with respect to the ${\cal N}=4$ sYM counterpart presented in \eqref{loop ampl constraints} and \eqref{positivity constraint}. 

In the rest of this thesis we will focus the analysis on the four-particle case. As in the ${\cal N}=4$ super Yang-Mills theory, we can compute the $L$-loop integrands ${\bf I}^{(L)}_4$ from the corresponding canonical forms $\Omega_4^{(L)}$ as
\begin{equation}
    {\bf I}^{(L)}_4= n^{(L)}_4 \, \Omega_4^{(L)} \,.
\end{equation}
The relative normalizations $n^{(L)}_4$ are computed in Appendix \ref{ch: normalizations}. As an example, at one-loop one obtains \cite{He:2022cup}
\begin{equation}
\label{one loop 4-pt ABJM}
\Omega_4^{(1)}= \frac{\langle 1234 \rangle^{3/2} \sqrt{\langle AB13 \rangle \langle AB24 \rangle } \, \delta(\Sigma_{IJ} \, A^I B^J)\langle AB d^2A \rangle \langle AB d^2B \rangle}{\langle AB12 \rangle \langle AB23 \rangle \langle AB34 \rangle \langle AB14 \rangle } \,.
\end{equation}

%% file: integrated_neg_geoms.tex
In the previous chapter we saw how geometric structures known as positive geometries can be used to organize the perturbative expansion of scattering amplitudes in ${\cal N}=4$ super Yang-Mills and ABJM. Interestingly, the Amplituhedrons of the ${\cal N}=4$ super Yang-Mills (sYM) and ABJM theories can be alternatively reformulated as sums over \textit{negative} geometries \cite{Arkani-Hamed:2021iya,He:2022cup,He:2023rou}, which are characterized by a change of sign in the constraints that relate different loop lines in momentum-twistor space. As shown in \cite{Arkani-Hamed:2021iya}, negative geometries are natural objects in terms of which the logarithm of the amplitude can be expressed. Quite remarkably, there is an infrared-finite observable that can be defined as a consequence of the negative geometry expansion of the logarithm of the amplitude: performing $L-1$ loop integrals over its $L$-loop integrand gives a divergence-free result. From now on we will refer to these quantities as \textit{integrated negative geometries}.

Integrated negative geometries have been shown to posses notable properties in the ${\cal N}=4$ sYM theory. One can show that the integrated result can be expressed in terms of a function of $3n-11$ dual conformal cross-ratios, which is the same number of kinematic variables as for QCD $n$-point amplitudes. In which follows we will focus on the four-particle case, which in the ${\cal N}=4$ sYM case depends on a function ${\cal F}(z)$ of a single cross-ratio $z$. An exciting outcome of the study of $\mathcal{F}(z)$ comes when taking into account the duality between scattering amplitudes and Wilson loops in ${\cal N}=4$ sYM \cite{Alday:2007hr,Drummond:2007aua,Brandhuber:2007yx,Drummond:2007cf}. Interestingly, this duality allows to recover the $L$-loop contribution to the light-like cusp anomalous dimension $\Gamma_{\rm cusp}^{\infty}$ (which regulates the IR divergences of massless amplitudes and the UV divergences of Wilson lines with light-like cusps) from the $(L-1)$-loop term in the perturbative expansion of $\mathcal{F}(z)$ \cite{Arkani-Hamed:2021iya,Alday:2013ip,Henn:2019swt}. In other words, via integrated negative geometries one can access the $L$-th loop order of $\Gamma_{\rm cusp}^{\infty}$ by just performing $L-1$ loop integrations. This prescription has been used to compute the full four-loop contribution to $\Gamma_{\rm cusp}^{\infty}$ both in ${\cal N}=4$ sYM and in QCD, including the first non-planar corrections \cite{Henn:2019swt}. 

As shown by \cite{Arkani-Hamed:2021iya}, in ${\cal N}=4$ sYM one can perform a non-perturbative sum over a particular subset of integrated negative geometries, opening the door for a full all-loop computation of $\mathcal{F}(z)$. Such results would allow a comparison with the non-perturbative derivation of $\Gamma_{\rm cusp}^{\infty}$ coming from integrability \cite{Beisert:2006ez,Correa:2012hh}. Also surprisingly, the leading singularities of these integrated negative geometries enjoy a (hidden) conformal symmetry \cite{Chicherin:2022bov,Chicherin:2022zxo}. Furthermore, identities relating $\mathcal{F}(z)$ to all-plus amplitudes in pure Yang-Mills theory have been found \cite{Chicherin:2022bov,Chicherin:2022zxo}. Finally, one can also note that the perturbative expansion of $\mathcal{F}(z)$ has uniform transcendentality \cite{Henn:2019swt} and respects an alternating sign pattern \cite{Arkani-Hamed:2021iya}.

Higher-point integrated results have also been studied for the ${\cal N}=4$ sYM case, up to $L=3$ for five \cite{Chicherin:2022zxo,Chicherin:2024hes} and six particles \cite{Carrolo:2025pue} and up to $L=2$ for arbitrary number of particles \cite{Chicherin:2022bov}.  Interesting positivity properties have been studied for the five-particle case in \cite{Chicherin:2022zxo, Chicherin:2024hes}. Strong-coupling results have been obtained in the ${\cal N}=4$ sYM case for the four- and six-particle cases \cite{Alday:2011ga,Hernandez:2013kb}. Moreover, all-loop leading singularities of $n$-particle integrated negative geometries have been studied in \cite{Brown:2025plq}.

Taking into account the previous considerations, it seems natural to pose the question of how the above results generalize to the ABJM case. Negative geometries have been studied in this theory in \cite{He:2022cup,He:2023rou}, where for the four-particle case the canonical forms have been computed up to five loops. Crucially, when comparing to the ${\cal N}=4$ sYM case, in the ABJM theory a smaller number of negative geometries contribute to the integrand of the logarithm of the amplitude. More precisely, only \textit{bipartite} negative geometries contribute, allowing for a significant simplification in the perturbative expansion of the integrand. We should note that the study of integrated negative geometries in ABJM is interesting towards an all-loop computation of the ABJM cusp anomalous dimension \cite{Gromov:2008qe,Bianchi:2014ada,Bianchi:2013pfa}. Non-perturbative results would clear the way for the all-loop computation of the interpolating function $h(\lambda)$ of ABJM (which we have discussed in Chapter \ref{ch: integrable line defects in ABJM}), whose knowledge is crucial to exploit the results coming from integrability.

In this chapter we focus again on the ABJM theory, and we explicitly perform the $(L-1)$-loop integrations of the four-particle negative geometries for the $L \leq 4$ cases, showing that the integrated results are given by finite and uniform-transcendental polylogarithmic functions (with transcendental weight $L-1$ for odd values of $L$ and with weight $L-2$ for even values of $L$). Analogously to the five-particle case of ${\cal N}=4$ sYM \cite{Chicherin:2022zxo}, we find it convenient to organize the integrated results in parity-even and parity-odd terms,  which are described by two functions $\mathcal{F}(z)$ and $\mathcal{G}(z)$, respectively.  As we will see, it is straightforward to show that only the former contributes to the cusp anomalous dimension $\Gamma_{\rm cusp}^{\infty}$ after the last loop integration. We will use the integrated negative geometries up to $L=4$ to compute the cusp anomalous dimension of the theory up to four loops. This result is the first explicit four-loop computation of $\Gamma_{\rm cusp}^{\infty}$ in the ABJM theory, and it agrees with the all-loop integrability-based proposal \cite{Gromov:2008qe}. We will also show that at the integrated level the results follow an alternating sign pattern within the Euclidean region $z>0$. Finally, we will see that the leading singularities are restricted to the set $\{ s\sqrt{t}, \, t\sqrt{s}, \, \sqrt{st(s+t)} \}$ in the limit at which the unintegrated loop variable goes to infinity. These results could provide useful insight for higher-loop bootstrap computations of the integrated negative geometries.

The chapter is organized as follows. In Section \ref{sec: review neg geoms} we review the construction of negative geometries and integrated negative geometries in the ${\cal N}=4$ super Yang-Mills theory, while in Section \ref{sec: neg geoms ABJM} we review the structure of negative geometries in ABJM. Section \ref{sec: L up to 3} is devoted to the integration of the ABJM negative geometries up to $L=3$, while in Section \ref{sec: L=4} we focus on the $L=4$ case via the method of differential equations. In Section \ref{sec: properties of integrated results} we discuss some properties of the integrated results, and in Section \ref{sec: gamma cusp from int neg geoms} we use the integrated negative geometries of ABJM to compute the four-loop cusp anomalous dimension of the theory.

This chapter is based on \cite{Henn:2023pkc,Lagares:2024epo}, which is work perform by the author of this thesis in collaboration with J. M. Henn and S. Q. Zhang.

\section{Finite observables from geometry}
\label{sec: review neg geoms}

To begin with, let us review the construction of infrared-finite observables via the integration of negative geometries in the ${\cal N}=4$ super Yang-Mills theory. In the following sections we will extend this construction to the ABJM case.

\subsubsection*{Negative geometries and the logarithm of the amplitude}

From now on we will focus only on the four-particle case. As discussed in Chapter \ref{ch: scatt ampl}, $L$-loop integrands for MHV scattering amplitudes can be expressed in the ${\cal N}=4$ sYM theory via the $L$-loop MHV Amplituhedron, which in the four-particle case is defined as the set of points in momentum-twistor space that are subjected to the constraints given in \eqref{positivity constraint} and \eqref{4-pt amplituhedron constraint 1}-\eqref{4-pt amplituhedron constraint 3}. Following \eqref{integrand from canonical form chapter amplt}, the ${\bf I}^{(L)}_{4,0}$ integrand for the $L$-loop four-particle MHV scattering amplitude is given as 
\begin{equation}
\label{normalization I_L}
{\bf I}^{(L)}_{4,0}=n^{(L)}_{4,0} \, \Omega^{(L)}_{4,0} \,,
\end{equation}
where $n^{(L)}_{4,0}$ is a normalization factor and $\Omega^{(L)}_{4,0}$ is the canonical form of the $L$-loop four-particle MHV Amplituhedron. To simplify the notation, hereafter we will refer to ${\bf I}^{(L)}_{4,0}$, $n^{(L)}_{4,0}$ and $\Omega^{(L)}_{4,0}$ simply as ${\bf I}_L$, $n_L$ and $\Omega_L$, respectively. Moreover, we will introduce the notation
\begin{equation}
    \Omega := \sum_{L=1}^{\infty} \lambda^L \, \Omega_L \,,
\end{equation}
where $\lambda$ is the 't Hooft coupling (we are working on the planar limit). See Appendix \ref{ch: normalizations} for a discussion on the normalization factors $n_L$ in the case of the ABJM theory.

In the following it will prove useful to take into account the pictorial representation introduced in \cite{Arkani-Hamed:2021iya} to describe positive geometries. We will use a node to indicate a one-loop amplituhedron associated to a certain loop variable, i.e. a geometry satisfying the constraints \eqref{4-pt amplituhedron constraint 1}-\eqref{4-pt amplituhedron constraint 3}, and a dashed light-blue line to describe a mutual positivity condition between a pair of loop variables (see eq. \eqref{positivity constraint}). As an example, the four-loop amplituhedron will be drawn as
\begin{equation}
\includegraphics[scale=0.10]{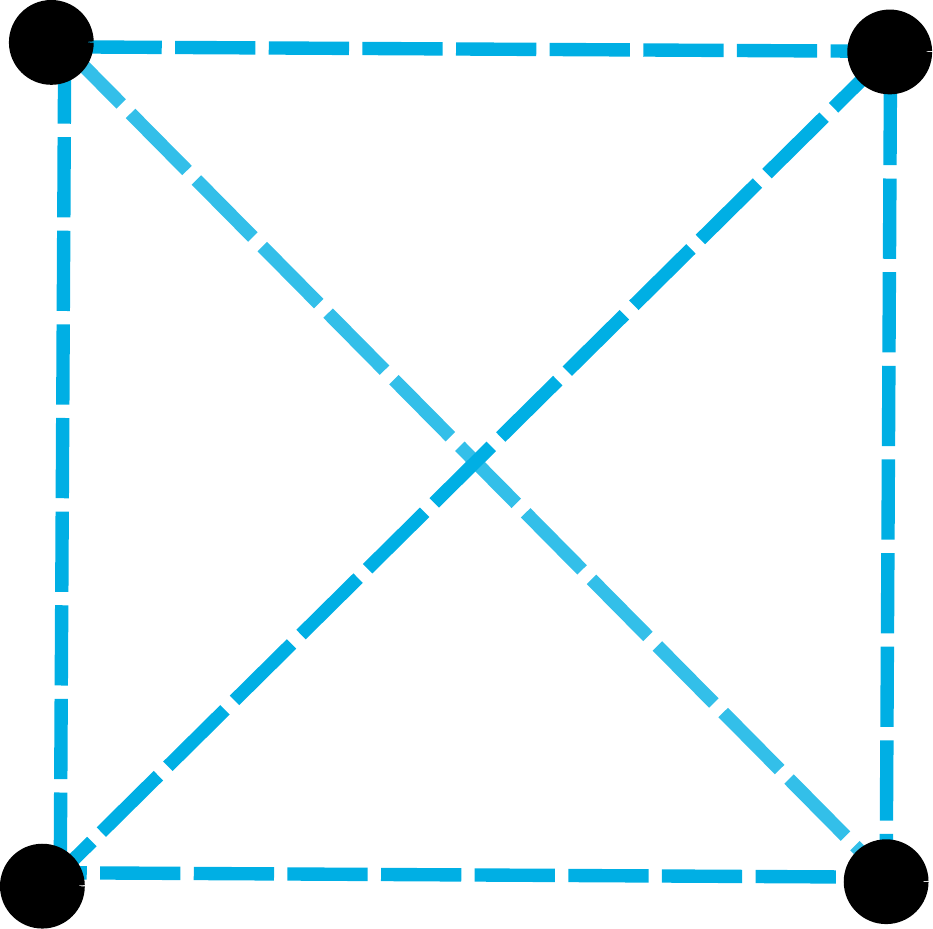} 
\label{four-loop positive geometry}
\end{equation}
As pointed out in \cite{Arkani-Hamed:2021iya}, it turns out to be very convenient to consider also mutual negativity conditions between loop variables. That is, constraints given by
\begin{equation}
    \label{mutual negativity}
    \langle (AB)_{\ell_i} (AB)_{\ell_j} \rangle <0 \,,
\end{equation}
for two loop lines $(AB)_{\ell_i}$ and $(AB)_{\ell_j}$. We will use thick red lines to graphically represent these negativity constraints, e.g. 
\begin{equation}
\includegraphics[scale=0.10]{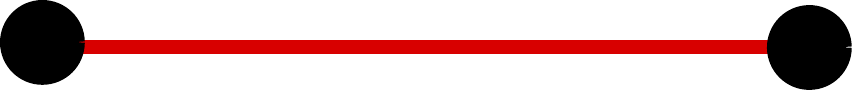} \label{example_neg_geom}
\end{equation}
and we will refer to geometries including them as \textit{negative geometries}. In order to understand the advantages of using negative geometries for the computation of integrands, let us introduce the notation 
\begin{equation}
    \label{log of positive geometries}
    \tilde{\Omega}:= \log \Omega \,,
\end{equation}
with 
\begin{equation}
    \tilde{\Omega} = \sum_{L=1}^{\infty} \lambda^L \, \tilde{\Omega}_L \,.
\end{equation}
Then, as described in \cite{Arkani-Hamed:2021iya}, one can expand $\tilde{\Omega}$ in terms of \textit{connected} negative geometries. More precisely, 
\begin{equation}
\includegraphics[scale=0.23]{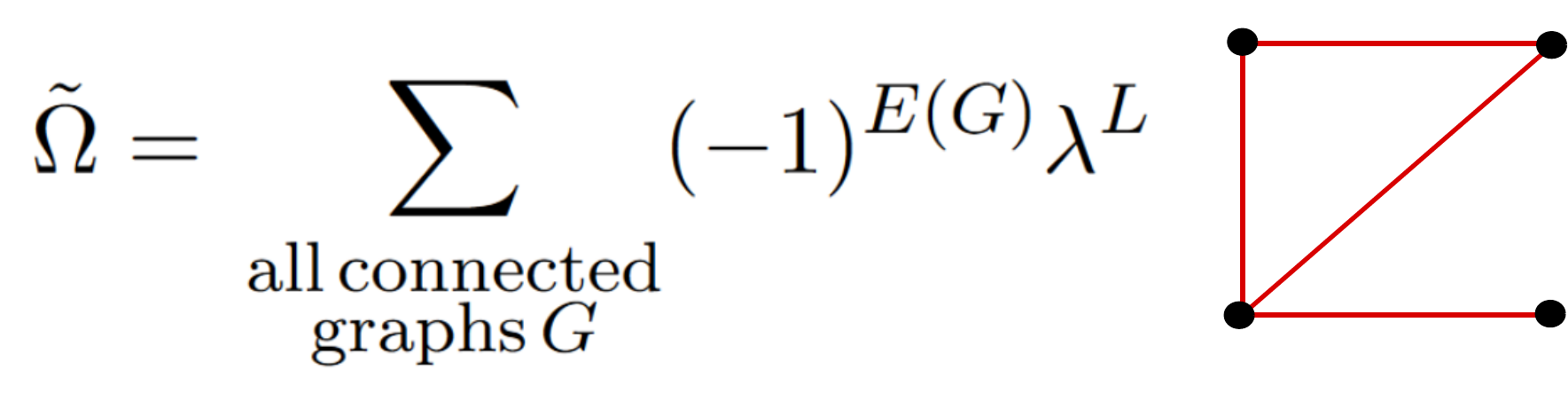} \label{expansion connected neg geom}
\end{equation}
where $E(G)$ is the number of edges of a graph $G$ and $L$ is the corresponding number of vertices. Therefore,
\begin{equation} 
\includegraphics[scale=0.3]{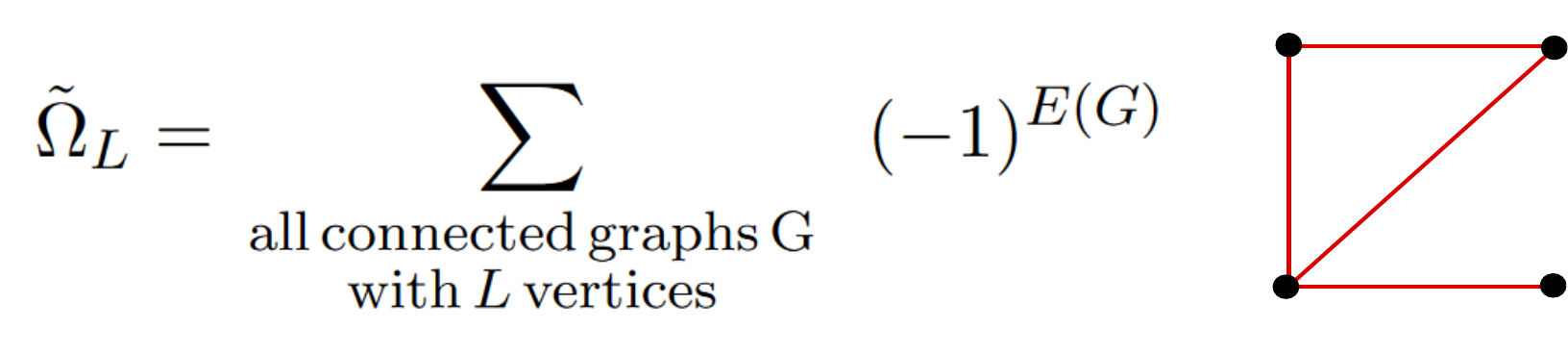} 
\label{expansion tilde Omega-connected graphs}
\end{equation}

Let us define the $L$-loop integrand ${\bf L}_L$ for the logarithm of the scattering amplitude as
\begin{equation}
\label{log of amplitude conventions}
\log M \,  \bigg|_{\rm L \, loops} := \left( \prod_{j=5}^{4+L} \int \frac{d^D x_j}{i\pi^{D/2}} \right) \, {\bf L}_L \,.
\end{equation}
where $M$ is the four-particle color-ordered MHV scattering amplitude, normalized by the corresponding tree-level amplitude. The integrand ${\bf L}_L$ can be obtained from $\tilde{\Omega}_L$ as
\begin{equation}
\label{normalization L_L}
{\bf L}_L= \tilde{n}_L \, \tilde{\Omega}_L \,.
\end{equation}
We refer again to the Appendix \ref{ch: normalizations} for the discussion on the computation of the relative normalizations $\tilde{n}_L$ in the case of the ABJM theory.

\subsubsection*{Infrared-finite observable}

As discussed in \cite{Arkani-Hamed:2021iya}, an interesting observable naturally arises from the expansion of the logarithm of the amplitude in terms of connected negative geometry diagrams. Indeed, such a decomposition implies that the loop variables can not access the collinear region separately, i.e. the only configuration in which loop lines can probe the collinear region is if all of them access such a region simultaneously. Therefore, all divergences concentrate in the last loop integration. One can then define an infrared-finite quantity by performing all but one of the loop integrations over the $L$-loop integrand for the logarithm of the amplitude.

Let us focus on the four-particle case, although the results can be extended to higher-particle scenarios. In four dimensions, the dual conformal invariance of the logarithm of the amplitude implies that there exists a function ${\cal F}_{L-1}$ such that \cite{Alday:2011ga}
\begin{equation}
\label{n=4 general expression}
\left( \prod_{j=6}^{4+L} \int \frac{d^4 x_j}{i\pi^{2}} \right) \, {\bf L}_L = \frac{x_{13}^2 x_{24}^2}{x_{15}^2 x_{25}^2 x_{35}^2 x_{45}^2} \, \frac{{\cal F}_{L-1} \left( z \right)}{\pi^2} \,,
\end{equation}
where the cross-ratio $z$, defined as
\begin{equation}
z= \frac{x_{15}^2 x_{35}^2 x_{24}^2}{x_{25}^2 x_{45}^2 x_{13}^2} \,,
\end{equation}
is the only dual conformally invariant cross-ratio that can be built using the external kinematic data and the unintegrated loop variable $x_5$. The function ${\cal F}_{L-1}(z)$ has been computed in the literature up to $L=4$ \cite{Alday:2011ga,Alday:2012hy,Alday:2013ip,Henn:2019swt}, where it has been found to have uniform transcendental weight $2L$. Moreover, certain subsets of negative geometry diagrams have been integrated to all loop orders \cite{Arkani-Hamed:2021iya}. It was conjectured in \cite{Arkani-Hamed:2021iya} that 
\begin{align}
    {\cal F}_{L-1}(z)&<0 \qquad \text{for even $L$} \,, \nonumber \\
    {\cal F}_{L-1}(z)&>0 \qquad \text{for odd $L$} \,. 
\end{align}
for $z>0$. 

When using dimensional regularization, the logarithm of the scattering amplitude diverges as $1/\epsilon^2$, where $\epsilon$ is the dimensional regularization parameter. The coefficient which governs that leading divergence is the cusp anomalous dimension $\Gamma^{\infty}_{\rm cusp}$, which regulates the UV divergences of Wilson lines with light-like cusps as well as the IR divergences of massless scattering amplitudes. It can be obtained from the angle-dependent cusp anomalous dimension discussed in Chapter \ref{ch: integrable line defects in ABJM} using that
\begin{equation}
    \Gamma_{\rm cusp}(\theta,\varphi) \bigg|_{\varphi=0}= \frac{1}{2} \theta \, \Gamma^{\infty}_{\rm cusp} + \mathcal{O}(\theta ^0) \,,
\end{equation}
for $\theta \to \infty$ \cite{Correa:2012nk}. Remarkably, one can compute $\Gamma^{\infty}_{\rm cusp}$ from ${\cal F}_{L-1}$ without the need to perform the last loop integration, which would give a divergent result. As discussed in \cite{Alday:2013ip}, one can extract the leading divergence of the last loop integration by just applying a functional on ${\cal F}_{L-1}$. More precisely, one obtains
\begin{equation}
g \, \partial_g \Gamma^{\infty}_{\rm cusp} (g) =-2 \, {\cal I} [{\cal F}(z,g)] \,,
\end{equation}
with
\begin{equation}
    {\cal I} [z^p]= \frac{\sin \pi p}{\pi p} \,,
\end{equation}
and where we are using the notations $g^2=\frac{\lambda}{16 \pi^2}$ and ${\cal F}(z,g)=g^2 \sum_{L=1}^{\infty} g^{2L}{\cal F}_{L-1} (z)$. This prescription has been used in \cite{Henn:2019swt} to derive the four-loop cusp anomalous dimension of ${\cal N}=4$ sYM and of QCD, even at the level of the first non-planar corrections.

\section{Negative geometries in ABJM}
\label{sec: neg geoms ABJM}

The negative geometry formulation of the Amplituhedron has also been extended to the ABJM theory \cite{He:2022cup,He:2023rou}. One major simplification occurs in ABJM when considering the expansion of $\tilde{\Omega}_L$ into negative geometries, namely that only \textit{bipartite} connected graphs are required \cite{He:2022cup}. The latter are defined as those graphs where, after assigning an orientation to each edge, each node is either a sink or a source. Examples of bipartite graphs are
\begin{center}
\includegraphics[scale=0.3]{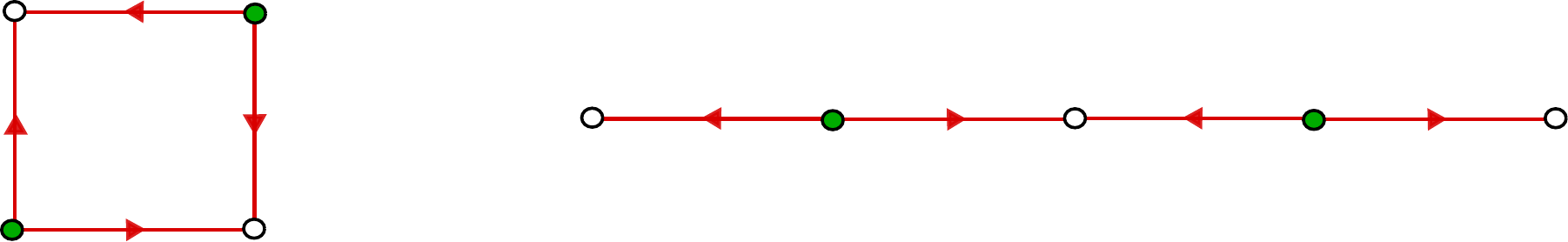}
\end{center}
where green nodes represent sources and white nodes correspond to sinks.
Taking into account the above simplification, the expansion \eqref{expansion tilde Omega-connected graphs} now becomes
\begin{equation}
\includegraphics[scale=0.3]{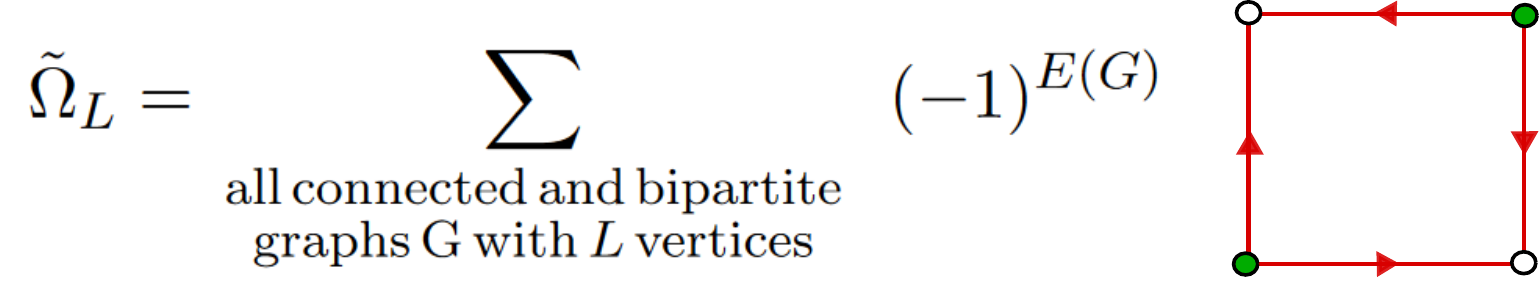} \label{bipartite expansion}
\end{equation}
Using \eqref{bipartite expansion}, the canonical forms $\tilde{\Omega}_L$ were computed in \cite{He:2022cup} up to $L=5$. In particular, for the first three loop orders one gets
\begin{equation}
    \label{amplituhedron canonical forms}
\includegraphics[scale=0.27]{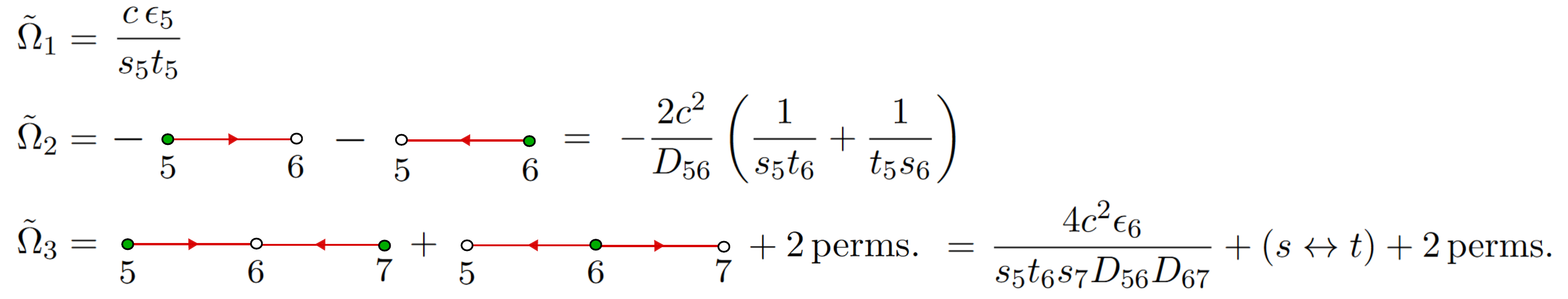}
\end{equation}
with
\begin{align}
\label{s,t,c,D,epsilon definitions}
s_i &:= \langle (AB)_{\ell_i} 12 \rangle \langle (AB)_{\ell_i} 34 \rangle \,, 
&t_i &:= \langle (AB)_{\ell_i} 23 \rangle \langle (AB)_{\ell_i} 14 \rangle \,, \nonumber \\ 
D_{ij} &:= -\langle (AB)_{\ell_i} (AB)_{\ell_j} \rangle \,, 
&c &:= \langle 1234 \rangle \,, \nonumber \\
\epsilon_i &:= \sqrt{\langle (AB)_{\ell_i} 13 \rangle \langle (AB)_{\ell_i} 24 \rangle \langle 1234 \rangle} \,, &&
\end{align}
and where the permutations are over all nonequivalent configurations of the loop variables. Let us note that for simplicity of notation we are omitting the $d^3(AB)_{\ell_i}$ factors in the differential forms.

In the four-loop case there are only three type of bipartite graphs that contribute to the $\tilde{\Omega}_4$ canonical form \cite{He:2022cup}. On the one hand, one has to consider the \textit{ladder} diagrams
$$
\includegraphics[scale=0.27]{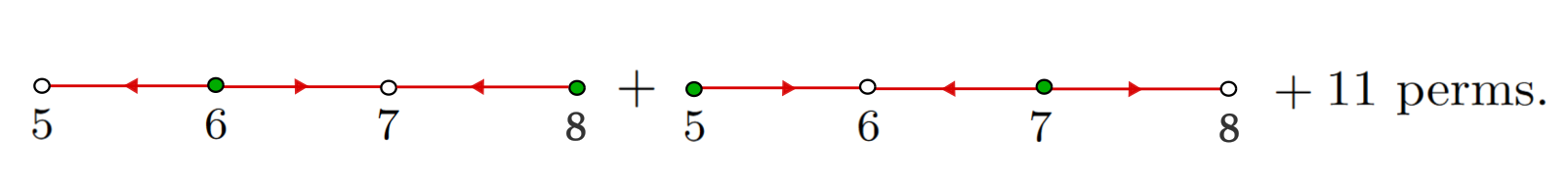}
$$
whose canonical form is
\begin{equation}
\label{ladder canonical form}
\tilde{\Omega}_{4}^{\rm Ladd} = 8 c^2 \frac{\epsilon_6\epsilon_7}{s_5 t_6 s_7 t_8 D_{56}D_{67}D_{78}} + (s \leftrightarrow t) + {\rm perms}.
\end{equation}
On the other hand, there is a contribution from the \textit{star} diagrams
\vspace{0.15cm}
$$
\includegraphics[scale=0.27]{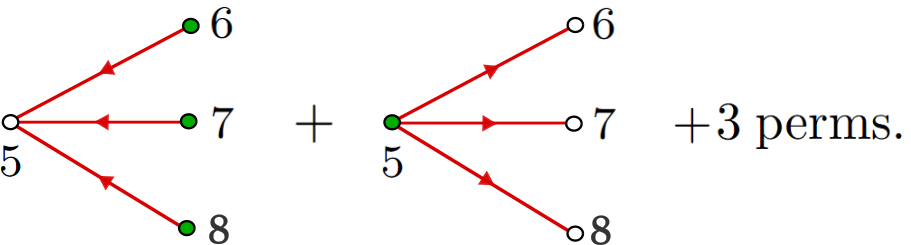}
$$
with
\begin{equation}
\label{star canonical form}
\tilde{\Omega}_{4}^{\rm Star} = 8 c^3 \frac{t_5}{s_5 t_6 t_7 t_8 D_{56}D_{57}D_{58}} + (s \leftrightarrow t) + {\rm perms}.
\end{equation}
Finally, one has to take into account also the \textit{box} diagrams
\vspace{0.15cm}
$$
\includegraphics[scale=0.27]{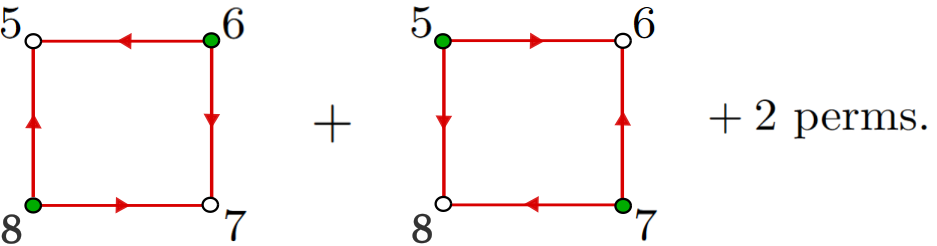}
$$
that have
\begin{equation}
\label{box canonical form}
\tilde{\Omega}_{4}^{\rm Box} =4 \frac{4 \epsilon_5 \epsilon_6 \epsilon_7 \epsilon_8-c( \epsilon_5 \epsilon_7 N^t_{68}+\epsilon_6 \epsilon_8 N^s_{57})-c^2 N^{\rm cyc}_{5,6,7,8}}{s_5 t_6 s_7 t_8 D_{56}D_{67}D_{78}D_{58}} + (s \leftrightarrow t) + {\rm perms}.
\end{equation}
and where
\begin{align}
\label{box twistor def-1}
N^s_{57} &= \langle l_5 12 \rangle \langle l_7 34 \rangle + \langle l_7 12 \rangle \langle l_5 34 \rangle \,, \\
N^t_{68} &= \langle l_6 14 \rangle \langle l_8 23 \rangle + \langle l_8 14 \rangle \langle l_6 23 \rangle \,, \\
\label{box twistor def-3}
N^{\rm cyc}_{ijkl} &= \langle l_i 12 \rangle \langle l_j 34 \rangle \langle l_k 12 \rangle \langle l_l 34 \rangle + \langle l_i 23 \rangle \langle l_j 14 \rangle \langle l_k 23 \rangle \langle l_l 14 \rangle  \,.
\end{align}
The canonical form $\tilde{\Omega}_4$ is then given as
\begin{equation}
    \label{four-loop canonical form}
    \tilde{\Omega}_4 = -\tilde{\Omega}_{4}^{\rm Ladd}-\tilde{\Omega}_{4}^{\rm Star}+\tilde{\Omega}_{4}^{\rm Box} \,.
\end{equation}

\section{Integrated negative geometries up to $L=3$}
\label{sec: L up to 3}

Having introduced the canonical forms $\tilde{\Omega}_L$ for the ABJM negative geometries with $L \leq 4$, we will now turn to their integration. 

\subsubsection*{Constraints from dual conformal invariance}

Interestingly, in the three-dimensional case the expression \eqref{n=4 general expression} for the integrated negative geometries is incomplete. To see this, it is convenient to use five-dimensional notation to describe the coordinates of the dual space (see for example \cite{Chen:2011vv,Caron-Huot:2012sos} and Appendix \ref{app: five-dimensional notation}).
One can then see that the most general expression that one can construct in order to generalize \eqref{n=4 general expression} to the three-dimensional case is
\begin{equation}
\label{n=4 general expression D=3}
\left( \prod_{j=6}^{4+L} \int \frac{d^3 X_j}{i\pi^{3/2}} \right) \, {\bf L}_L = \sqrt{\pi} \, \left( \frac{X_{13}^2 X_{24}^2}{X_{15}^2 X_{25}^2 X_{35}^2 X_{45}^2} \right)^{\frac{3}{4}} {\cal F}_{L-1}\left( z \right)  + \frac{i \, \epsilon\left( 1,2,3,4,5 \right)}{X_{15}^2 X_{25}^2 X_{35}^2 X_{45}^2} \, \frac{{\cal G}_{L-1} \left( z \right)}{\sqrt{\pi}} \,,
\end{equation}
where capital letters refer to five-dimensional coordinates and 
\begin{equation}
\epsilon(1,2,3,4,5):=\epsilon_{\mu \nu \rho \sigma \eta} X_1^{\mu} X_2^{\nu} X_3^{\rho} X_4^{\sigma} X_5^{\eta} \,.
\end{equation}
Therefore, when going to three dimensions we have to include in \eqref{n=4 general expression} an additional parity-odd term given by a function ${\cal G}_{L-1}(z)$. This is analogous to what was found in ${\cal N}=4$ sYM for the five-particle case \cite{Chicherin:2022zxo}.

\subsubsection*{$L=1$ case}

We will turn now to the explicit $L-1$ loop integrations of the negative geometries obtained from the ABJM Amplituhedron \cite{He:2022cup,He:2023rou}, seeking for the perturbative expansion of the ${\cal F}$ and ${\cal G}$ functions defined in \eqref{n=4 general expression D=3}.

Let us begin by computing the tree-level values of ${\cal F}(z)$ and ${\cal G}(z)$. From \eqref{amplituhedron canonical forms} we have
\begin{equation}
\label{Omega tilde 1}
\tilde{\Omega}_1
= 
\frac{\langle 1234 \rangle^{3/2} \sqrt{\langle (AB)_{\ell_5} 13 \rangle \langle (AB)_{\ell_5} 24 \rangle }}{\langle (AB)_{\ell_5} 12 \rangle \langle (AB)_{\ell_5} 23 \rangle \langle (AB)_{\ell_5} 34 \rangle \langle (AB)_{\ell_5} 14 \rangle} .
\end{equation}
Using the Schouten identity 
\begin{equation}
\label{13 24 twistor identity}
\langle (AB)_{\ell_5} 13 \rangle \langle (AB)_{\ell_5} 24 \rangle= \langle (AB)_{\ell_5} 12 \rangle \langle (AB)_{\ell_5} 34\rangle + \langle (AB)_{\ell_5} 23 \rangle \langle (AB)_{\ell_5} 14\rangle \,.
\end{equation}
we can rewrite \eqref{Omega tilde 1} in terms of five-dimensional dual coordinates as
\begin{align}
\label{Omega1new}
\tilde{\Omega}_1 = \frac{\sqrt{ X_{13}^2 X_{24}^2 \, (X_{24}^2 X_{15}^2 X_{35}^2+X_{13}^2 X_{25}^2 X_{45}^2)}}{X_{15}^2 X_{25}^2 X_{35}^2 X_{45}^2}
\,.
\end{align}
In order to compute ${\cal F}_0(z)$ and ${\cal G}_0(z)$ it is important to note that 
the integration is over the three-dimensional Minkowski space.
However, \eqref{Omega1new} was derived 
within
the Amplituhedron region,
and therefore we need to extend its definition.
Indeed, one can see that 
naively
integrating \eqref{Omega1new} over the whole kinematic space 
gives a non-zero result,
in contradiction with 
what is expected for the one-loop four-particle amplitude of ABJM \cite{Agarwal:2008pu,Bianchi:2011fc}. 
This 
issue can be resolved by taking into account the identity\footnote{Any possible overall sign ambiguity that could arise when using \eqref{epsilon identity} should be absorbed in the sign of the overall normalization of the amplitude.}
\begin{equation}
\label{epsilon identity}
 \epsilon(1,2,3,4,5)= -\frac{1}{4} \sqrt{X_{13}^2 X_{24}^2 \, (X_{24}^2 X_{15}^2 X_{35}^2+X_{13}^2 X_{25}^2 X_{45}^2)} \,,
\end{equation}
to analytically continue to Minkowski space. Then, taking into account the normalization $\tilde{n}_1$ discussed in Appendix \ref{ch: normalizations} (see \eqref{tilde_n_1}) and using \eqref{epsilon identity} to rewrite \eqref{Omega1new} in terms of the five-dimensional Levi-Civita tensor we arrive at
\begin{equation}
\label{L1}
{\bf L}_1  = - \frac{2i}{\sqrt{\pi}} \, \frac{\epsilon(1,2,3,4,5)}{X_{15}^2 X_{25}^2 X_{35}^2 X_{45}^2} \, .
\end{equation}
Therefore, comparing with the definition presented in \eqref{n=4 general expression D=3} we conclude
\begin{equation}
\label{F0, G0}
{\cal F}_0(z) = 0\,, \quad \text{and} \quad {\cal G}_0(z) = -2 \,.
\end{equation}

\subsubsection*{$L=2$ case}

As discussed in previous sections, the integrand ${\bf L}_2$ that is obtained from the canonical form $\tilde{\Omega}_2$ given in \eqref{amplituhedron canonical forms} reads 
\begin{equation}
\label{L2}
\begin{aligned}
{\bf L}_2
&= \frac{c^2}{4 \pi \, D_{56}} \left( \frac{1}{s_5 t_6} + \frac{1}{t_5 s_6} \right) = \\
 & = -\frac{X_{13}^2 X_{24}^2}{4 \pi \, X_{56}^2} \left( \frac{1}{X_{15}^2 X_{26}^2 X_{35}^2  X_{46}^2} + \frac{1}{X_{16}^2 X_{25}^2 X_{36}^2  X_{45}^2} \right) \,.
\end{aligned} 
\end{equation}
To obtain ${\cal F}_1(z)$ and ${\cal G}_1(z)$ we should now perform one of the loop integrations, which we choose to be the one over $X_6$ (i.e. we take $X_5$ to be the frozen loop variable). This integral turns out to be a triangle integral with three massive legs. Consequently, using the results of Appendix \ref{app: useful integrals}, we arrive at
\begin{equation}
\label{two loop form amplituhedron-v2-2}
\begin{aligned}
\int \frac{d^3 X_6}{i \pi^{3/2}} \, {\bf L}_2 &= -\frac{X_{13}^2 X_{24}^2}{4 \pi} \, \int \frac{d^3 X_6}{i \pi^{3/2}} \, \frac{1}{ X_{56}^2} \left( \frac{1}{X_{15}^2 X_{26}^2 X_{35}^2  X_{46}^2} + \frac{1}{X_{16}^2 X_{25}^2 X_{36}^2  X_{45}^2} \right) \\
&= - \frac{\sqrt{\pi}}{4} \left( \frac{X_{13}^2 X_{24}^2}{X_{15}^2 X_{25}^2 X_{35}^2 X_{45}^2} \right)^{3/4} \left( z^{1/4} + \frac{1}{z^{1/4}} \right)\,.
\end{aligned}
\end{equation}
Therefore, from \eqref{n=4 general expression D=3} we deduce
\begin{equation}
\label{F1, G1}
{\cal F}_1 (z)= -\frac{1}{4} \left( z^{1/4} + \frac{1}{z^{1/4}} \right) \,, \quad\ \text{and} \;\;\, {\cal G}_1 (z)=0 \,.
\end{equation}

\subsubsection*{L=3 case}

The integrand ${\bf L}_3$, which can be obtained from \eqref{amplituhedron canonical forms} and \eqref{normalization L_L}, is
\begin{equation}
\label{L3}
{\bf L}_3 
= - \frac{i}{12 \pi^{3/2}} \, \frac{c^2 \epsilon_6}{s_5 t_6 s_7 D_{56} D_{67}} + (s \leftrightarrow t) + 2 \, \text{perms}. 
\end{equation}
In order to compute ${\cal F}_2$ and ${\cal G}_2$ we will perform two of the loop integrations. We will choose to integrate over $X_6$ and $X_7$, and again we will keep $X_5$ frozen. Let us begin by considering the first term of the r.h.s in \eqref{L3}, 
which is explicitly given by
\begin{equation}
\label{first term two loops}
\begin{aligned}
{\bf L}_3^{(1)} 
:= \frac{i}{3\pi^{3/2}} \, \frac{X_{13}^2 \, \epsilon(1,2,3,4,6)}{X_{15}^2 X_{35}^2 X_{26}^2 X_{46}^2 X_{17}^2 X_{37}^2 X_{56}^2 X_{67}^2} \,.
\end{aligned}
\end{equation}
First, the integral over $X_7$ is again a triangle integral, and therefore we get
\begin{equation}
\label{first term two loops-2}
\int \frac{d^3X_7}{i \pi^{3/2}} \, {\bf L}_3^{(1)} = \frac{i X_{13} \, \epsilon(1,2,3,4,6)}{3 X_{15}^2 X_{35}^2 X_{26}^2 X_{46}^2 X_{56}^2 X_{16} X_{36}} \,.
\end{equation}
We will turn now to the integration over $X_6$. To compute this integral we will make use of the results derived in Appendix \ref{app: useful integrals}. In particular, we have 
\begin{equation}
\label{epsilon num integral}
\int \frac{d^3 X_6}{i\pi^{3/2}} \, \frac{\epsilon(1,2,3,4,6)}{X_{26}^2 X_{46}^2  X_{56}^2 X_{16} X_{36}}  = \frac{\epsilon(1,2,3,4,5)}{\left( X_{15}^2 X_{25}^2 X_{35}^2 X_{45}^2 X_{24}^2 \right)^{1/2}} \, \sqrt{\frac{z}{\pi}} {\cal H}(z)\,,
\end{equation}
where the weight-two function ${\cal H}(z)$ takes the form
\begin{equation}
\label{H}
\begin{aligned}
{\cal H}(z) &= \frac{1}{\sqrt{1+z}} \left[ -\text{Li}_2\left(\frac{2 \left(\sqrt{z+1}-1\right)}{z}\right)+\text{Li}_2\left(-\frac{2
   \left(\sqrt{z+1}+1\right)}{z}\right) \right.\\
   & \left. \quad \, +2 \log \left(\frac{4}{z}\right) \log
   \left(\frac{\sqrt{z+1}+1}{\sqrt{z}}\right) + \pi^2 \right] \,.
\end{aligned}
\end{equation}
Consequently,
\begin{equation}
\label{first term two loops-3}
\int \frac{d^3X_6}{i\pi^{3/2}} \int \frac{d^3X_7}{i\pi^{3/2}} \, {\bf L}_3^{(1)} = \frac{i}{3\sqrt{\pi}} \, \frac{\epsilon(1,2,3,4,5)}{X_{15}^{2} X_{25}^2 X_{35}^{2} X_{45}^2} \, {\cal H}(z) \,.
\end{equation}
For the $X_6 \leftrightarrow X_7$ permutation, i.e. for 
\begin{equation}
\label{second term two loops}
\begin{aligned}
{\bf L}_3^{(2)} := \frac{i}{3\pi^{3/2}} \, \frac{X_{13}^2 \, \epsilon(1,2,3,4,7)}{X_{15}^2 X_{35}^2 X_{27}^2 X_{47}^2 X_{16}^2 X_{36}^2 X_{57}^2 X_{67}^2} \, ,
\end{aligned}
\end{equation}
we similarly get
\begin{equation}
\label{second term two loops-2}
\int \frac{d^3X_6}{i\pi^{3/2}} \int \frac{d^3X_7}{i\pi^{3/2}} \, {\bf L}_3^{(2)} = \frac{i}{3\sqrt{\pi}} \, \frac{\epsilon(1,2,3,4,5)}{X_{15}^{2} X_{25}^2 X_{35}^{2} X_{45}^2} \, {\cal H}(z) \,.
\end{equation}
Finally, let us consider the $X_5 \leftrightarrow X_6$ permutation. That is, let us take
\begin{equation}
\label{third term two loops}
\begin{aligned}
{\bf L}_3^{(3)} := \frac{i}{3\pi^{3/2}} \, \frac{X_{13}^2 \, \epsilon(1,2,3,4,5)}{X_{16}^2 X_{36}^2 X_{25}^2 X_{45}^2 X_{17}^2 X_{37}^2 X_{56}^2 X_{67}^2} \,.
\end{aligned}
\end{equation}
We can see that the integrals of ${\bf L}_3^{(3)}$ over $X_6$ and $X_7$ are simply two triangle integrals, and therefore
\begin{equation}
\label{third term two loops-2}
\begin{aligned}
\int \frac{d^3X_6}{i\pi^{3/2}} \int \frac{d^3X_7}{i\pi^{3/2}} \, {\bf L}_3^{(3)} &=  \frac{i\pi^{3/2}}{3} \, \frac{ \epsilon(1,2,3,4,5)}{X_{15}^2 X_{26}^2 X_{35}^2 X_{45}^2} \,.\\
\end{aligned}
\end{equation}
Finally, adding the corresponding $(s \leftrightarrow t)$ terms we get
\begin{align}
\label{F2, G2}
{\cal F}_2(z)&=0 \,, \nonumber\\
{\cal G}_2(z)&=\frac{2}{3} \left[ \frac{\hat{{\cal H}}(z) + \pi^2}{\sqrt{1+z}} \,  +  \frac{\pi^2}{2} + \left( z \to \frac{1}{z} \right)  \right] \,,
\end{align}
where we have introduced the notation
\begin{equation}
\label{hat H}
    \hat{{\cal H}} (z)= \sqrt{1+z} \, ({\cal H}(z)-\pi^2) \,.
\end{equation}

\section{The $L=4$ case}
\label{sec: L=4}

In this section we will turn to the three-loop integration of the $L=4$ negative geometry of the ABJM theory, which we will perform by the method of differential equations. We will show how to decompose the integrals into a basis of master integrals, which can be computed from a set of first order differential equations. We will discuss how to take those differential equations into a canonical form, which highly simplifies their analysis. Finally, we will solve the canonical differential equations and we will use their solution to compute the $L=4$ integrated negative geometries. 

\subsubsection*{Star diagram}

Before turning to the computation of the integrals over $\tilde{\Omega}_4^{\rm Ladd}$ and $\tilde{\Omega}_4^{\rm Box}$ with the method of differential equations, let us discuss the integration over $\tilde{\Omega}_{4}^{\rm Star}$, which can be easily done by direct integration.

We should recall that in order to compute the ${\cal F}_3$ and ${\cal G}_3$ functions that were defined in \eqref{n=4 general expression D=3} one should freeze one of the loop integrations over ${\bf L}_4$ and then perform the remaining three-loop integrations. With the freedom of arbitrarily choosing a frozen node, we will
always leave the $X_5$ variable unintegrated. In order to diagrammatically represent this integration process we will use a squared node to denote the loop variable that is left unintegrated. For example, for the star diagrams we have two possibilities: on the one hand, one can freeze the node that is at the center of the graph, i.e.
\vspace{0.15cm}
\begin{equation}
\includegraphics[scale=0.27]{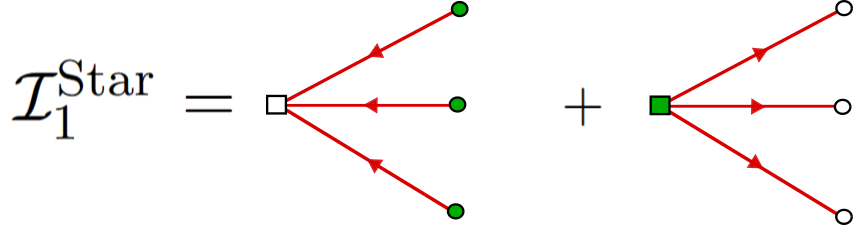}
\end{equation}
On the other hand, the unintegrated node can be in one of the legs of the star, i.e.
\vspace{0.15cm}
\begin{equation}
\includegraphics[scale=0.27]{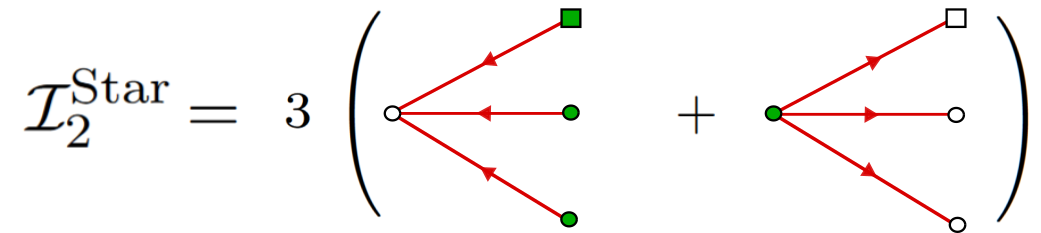}
\end{equation}
where the factor of 3 comes from the permutation of the loop variables that are being integrated. Both the ${\cal I}_1^{\rm Star}$ and the ${\cal I}_2^{\rm Star}$ integrals can be straightforwardly solved by taking into account the triangle integral \eqref{triangle integral-1} that is discussed in Appendix \ref{app: useful integrals}. One gets
\begin{equation}
\label{I1 Star}
\begin{aligned}
{\cal I}_1^{\rm Star} &= 8\int \frac{d^3l_6}{i \pi^{3/2}} \int \frac{d^3l_7}{i \pi^{3/2}} \int \frac{d^3l_8}{i \pi^{3/2}} \, \frac{c^3 \, t_5}{s_5 t_6 t_7 t_8 D_{56}D_{57}D_{58}} + (s \leftrightarrow t)  \\
&= -8 \pi^{9/2} \left( \frac{X^2_{13}X^2_{24}}{X^2_{15}X^2_{25}X^2_{35}X^2_{45}} \right)^{3/4} \, \left( z^{1/4}+ \frac{1}{z^{1/4}} \right) \,,
\end{aligned}
\end{equation}
and
\begin{equation}
\label{I2 Star}
\begin{aligned}
{\cal I}_2^{\rm Star} &= 24 \int \frac{d^3l_6}{i \pi^{3/2}} \int \frac{d^3l_7}{i \pi^{3/2}} \int \frac{d^3l_8}{i \pi^{3/2}} \, \frac{c^3 \, t_6}{s_6 t_5 t_7 t_8 D_{56}D_{67}D_{68}} + (s \leftrightarrow t)  \\
&= -24 \pi^{9/2} \left( \frac{X^2_{13}X^2_{24}}{X^2_{15}X^2_{25}X^2_{35}X^2_{45}} \right)^{3/4} \, \left( z^{1/4}+ \frac{1}{z^{1/4}} \right) \,.
\end{aligned}
\end{equation}
Therefore, summing up \eqref{I1 Star} and \eqref{I2 Star} and taking into account the normalization $\tilde{n}_4$ discussed in Appendix \ref{ch: normalizations} we arrive at
\begin{align}
\label{F and G Star}
{\cal F}_3^{\rm Star} (z) &= -\frac{\pi^2}{12} \left( z^{1/4} + \frac{1}{z^{1/4}} \right) \,, \; &{\cal G}_3^{\rm Star} (z) &=0 \,.
\end{align}

\subsubsection*{Ladder and box diagrams}

Let us now discuss the integration of the ladder and box diagrams. For the case of the ladders we can either integrate a node at the end of the chain, i.e.
\vspace{0.15cm}
\begin{equation}
\includegraphics[scale=0.27]{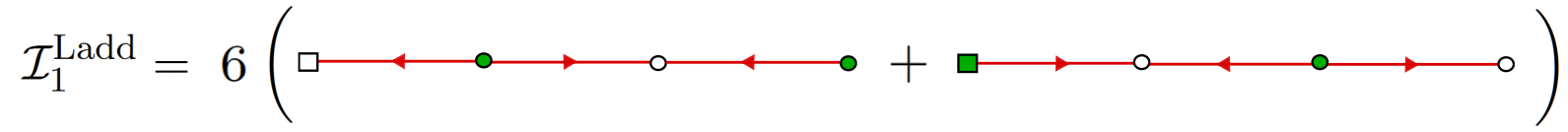}
\end{equation}
or a node in the middle of the chain, i.e.
\vspace{0.15cm}
\begin{equation}
\includegraphics[scale=0.27]{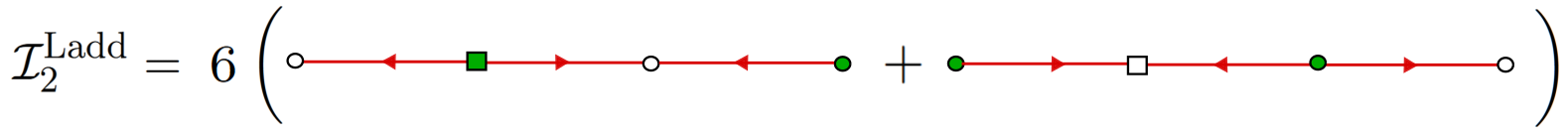}
\end{equation}
On the other hand, for the integration of the box diagram we have only one contribution, given by
\vspace{0.15cm}
\begin{equation}
\includegraphics[scale=0.27]{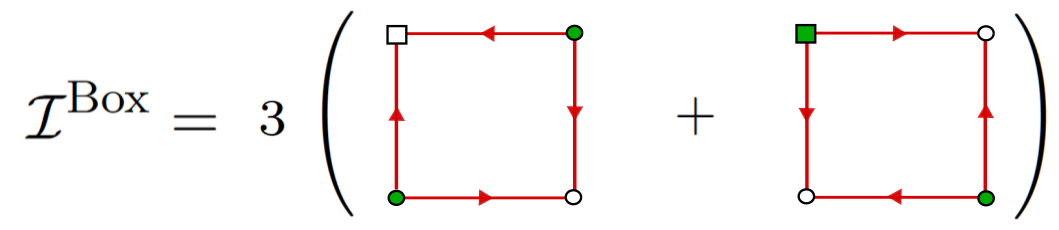}
\end{equation}
To be more specific, the integrals that we seek to compute are
\begin{align}
\label{I1 ladder}
{\cal I}_1^{\rm Ladd}&= 48 \int \frac{d^3l_6}{i \pi^{3/2}} \int \frac{d^3l_7}{i \pi^{3/2}} \int \frac{d^3l_8}{i \pi^{3/2}} \, \frac{c^2 \epsilon_6\epsilon_7}{s_5 t_6 s_7 t_8 D_{56}D_{67}D_{78}} + (s \leftrightarrow t) \,, \\
\label{I2 ladder}
{\cal I}_2^{\rm Ladd}&= 48 \int \frac{d^3l_6}{i \pi^{3/2}} \int \frac{d^3l_7}{i \pi^{3/2}} \int \frac{d^3l_8}{i \pi^{3/2}} \, \frac{c^2 \epsilon_5\epsilon_7}{t_6 s_5 t_7 s_8 D_{56}D_{57}D_{78}} + (s \leftrightarrow t) \,, \\
\label{I box}
{\cal I}^{\rm Box}&= 12 \int \frac{d^3l_6}{i \pi^{3/2}} \int \frac{d^3l_7}{i \pi^{3/2}} \int \frac{d^3l_8}{i \pi^{3/2}} \, \frac{4 \epsilon_5 \epsilon_6 \epsilon_7 \epsilon_8-c( \epsilon_5 \epsilon_7 N^t_{68}+\epsilon_6 \epsilon_8 N^s_{57})-c^2 N^{\rm cyc}_{5,6,7,8}}{s_5 t_6 s_7 t_8 D_{56}D_{67}D_{78}D_{58}} \nonumber \\
& \quad \, + (s \leftrightarrow t) \,.
\end{align}

\subsubsection*{Differential equations}

In the rest of this section we will discuss the computation of \eqref{I1 ladder}-\eqref{I box}
by the method of differential equations \cite{Henn:2013pwa}.
As the first step towards that goal, let us begin by defining the family of integrals given as
\begin{equation}
\label{three loop family}
G_{a_1, \, a_2, \, a_3, \, \dots , \, a_{15}}= \int \frac{d^D l_1}{i\pi^{3/2}} \, \int \frac{d^D l_2}{i\pi^{3/2}} \, \int \frac{d^D l_3}{i\pi^{3/2}} \, \prod_{j=1}^{15} \frac{1}{P_j^{a_j}} \,,
\end{equation}
with $D=3-2\epsilon$ and
\begin{center}
\begin{tabular}{l l l}
$P_1 = -(l_1+p_1)^2 \,,$ & $P_6 = -(l_3-p_4)^2 \,,$ & $P_{11} = -(l_2+p_1)^2 \,,$ \\
$P_2 = -(l_1-p_4)^2 \,,$ & $P_7 = -(l_1-l_2)^2 \,,$ & $P_{12} = -(l_2-p_4)^2 \,,$\\
$P_3 = -l_2^2 \,,$ & $P_8 = -(l_2-l_3)^2 \,,$ & $P_{13} = -l_3^2 \,,$ \\
$P_4 = -(l_2+p_1+p_2)^2 \,,$ & $P_9 = -l_1^2 \,,$ & $P_{14} = -(l_3+p_1+p_2)^2 \,,$\\
$P_5 = -(l_3+p_1)^2 \,,$ & $P_{10} = -(l_1+p_1+p_2)^2 \,,$ & $P_{15} = -(l_1-l_3)^2 \,,$
\end{tabular}
\end{center}
and where the momenta $p_i,\, i=1 , \, \dots , \, 4$ are all massless. In order to express \eqref{I1 ladder}-\eqref{I box} as a linear combination of integrals of the family \eqref{three loop family} we can use the identity \eqref{5D epsilon product}, which allows us to rewrite \eqref{I1 ladder}-\eqref{I box} in terms of integrals that only depend on $X_{ij}^2$ distances. However, let us note that these integrals still do not belong to the family defined in \eqref{three loop family}, given that there is still a non-trivial dependence on $x_5$. One can get rid of that dependence by taking into account the dual conformal covariance of the integrals \eqref{I1 ladder}-\eqref{I box}, which allows one to go to the $x_5 \to \infty$ frame. Therefore, we define
\begin{align}
\label{I1 ladder infty}
\hat{\cal I}_1^{\rm Ladd} &= \lim_{x_5 \to \infty} (x_5^2)^3 \, {\cal I}_1^{\rm Ladd} \,, \\
\label{I2 ladder infty}
\hat{\cal I}_2^{\rm Ladd} &= \lim_{x_5 \to \infty} (x_5^2)^3 \, {\cal I}_2^{\rm Ladd} \,, \\
\label{I box infty}
\hat{\cal I}^{\rm Box} &= \lim_{x_5 \to \infty} (x_5^2)^3 \, {\cal I}^{\rm Box} \,.
\end{align}
The advantage of taking the $x_5 \to \infty$ limit relies in the fact that it allows us to express the integrals \eqref{I1 ladder infty}-\eqref{I box infty} solely in terms of four-particle kinematic variables, i.e. in that limit one can rewrite \eqref{I1 ladder infty}-\eqref{I box infty} as a linear combination of integrals of the type defined in \eqref{three loop family}.
Moreover, we can further decompose \eqref{I1 ladder infty}-\eqref{I box infty} in terms of a set of master integrals for the family \eqref{three loop family}. In order to look for that basis we have used the \texttt{FIRE} \cite{Smirnov:2023yhb} and \texttt{LiteRed} \cite{Lee:2013mka} algorithms, that have allowed us to find the following basis of 19 master integrals:
\begin{align}
    M_1 &= s^{-1/2 + 3 \epsilon} \, \hat{G}_{0, 1, 0, 0, 1, 0, 1, 1} \,, &M_2 &= s^{3/2 + 3 \epsilon} \, \hat{G}_{0, 1, 0, 1, 1, 1, 1, 1}\,, \nonumber \\
    M_3 &= s^{3/2 + 3 \epsilon}\, \hat{G}_{0, 1, 1, 1, 0, 1, 1, 1}\,,  &M_4 &= s^{3/2 + 3 \epsilon} \, \hat{G}_{0, 1, 1, 1, 1, 0, 1, 1} \,, \nonumber\\
    M_5 &= s^{5/2 + 3 \epsilon}\, \hat{G}_{0, 1, 1, 1, 1, 0, 1, 2}\,,  &M_6 &= s^{3/2 + 3 \epsilon} \, \hat{G}_{0, 1, 1, 1, 1, 1, 1, 0}\,, \nonumber\\
    M_7 &= s^{5/2 + 3 \epsilon}\, \hat{G}_{0, 1, 1, 1, 1, 1, 1, 1} \,, &M_8 &= s^{7/2 + 3 \epsilon} \, \hat{G}_{0, 1, 1, 1, 1, 1, 1, 2}\,, \nonumber\\
    \label{non can MI}
    M_9 &= s^{7/2 + 3 \epsilon}\, \hat{G}_{0, 1, 1, 1, 1, 1, 2, 1}\,,  &M_{10} &= s^{3/2 + 3 \epsilon} \, \hat{G}_{1, 1, 0, 0, 1, 1, 1, 1} \,, \\
    M_{11} &= s^{3/2 + 3 \epsilon}\, \hat{G}_{1, 1, 0, 1, 1, 1, 0, 1} \,, &M_{12} &= s^{5/2 + 3 \epsilon} \, \hat{G}_{1, 1, 0, 1, 1, 1, 1, 1}\,,\nonumber \\
    M_{13} &= s^{3/2 + 3 \epsilon}\, \hat{G}_{1, 1, 1, 1, 1, 1, 0, 0} \,, &M_{14} &= s^{5/2 + 3 \epsilon} \, \hat{G}_{1, 1, 1, 1, 1, 1, 0, 1}\,, \nonumber\\
    M_{15} &= s^{7/2 + 3 \epsilon}\, \hat{G}_{1, 1, 1, 1, 1, 1, 0, 2}\,,  &M_{16} &= s^{7/2 + 3 \epsilon} \, \hat{G}_{1, 1, 1, 1, 1, 1, 1, 1}\,,\nonumber \\
    M_{17} &= s^{9/2 + 3 \epsilon}\, \hat{G}_{1, 1, 1, 1, 1, 1, 1, 2}  \,,&M_{18} &= s^{7/2 + 3 \epsilon} \, \hat{G}_{1, 1, 1, 1, 1, 2, 0, 1} \,,\nonumber \\
    M_{19} &= s^{9/2 + 3 \epsilon}\, \hat{G}_{1, 1, 1, 1, 1, 2, 1, 1} \,, &  & \nonumber
\end{align}
where we are using the shorthand notation
\begin{equation}
    \hat{G}_{a_1, a_2, a_3, a_4, a_5, a_6, a_7, a_8} :=G_{a_1, a_2, a_3, a_4, a_5, a_6, a_7, a_8, 0,0,0,0,0,0,0} \,.
\end{equation}
Let us note that we have chosen to normalize each element of the basis with a factor
\begin{equation}
    \label{3L normalization}
    s^{-\frac{9}{2}+3 \epsilon+ \sum_{i=1}^{15} a_i} \,,
\end{equation}
in order to get dimensionless integrals.

As is well known, for a given basis of master integrals one can generically derive a set of first order differential equations that in principle allows for their computation. For the case of interest to us, we have
\begin{equation}
\label{non canonical diff eq}
\partial_{\omega} \vec{M} = {\bf A}(\omega,\epsilon) \, \vec{M} \,,
\end{equation}
where $\vec{M}$ is a vector constructed with the 19 master integrals presented in \eqref{non can MI}, ${\bf A}$ is a $19\times19$ matrix presented in the file {\tt non\_can\_DE.txt} of Ref. \cite{Lagares:2024epo}, $\epsilon$ is the dimensional regularization parameter (defined as $D=3-2\epsilon$), and $\omega$ is defined by\footnote{For each value of $t/s$ there are four possible solutions of \eqref{omega}. Throughout this chapter we will always choose to work with the solution that lies in the $0< \omega <1$ interval.}
\begin{equation}
\label{omega}
\frac{t}{s}=\frac{1}{4} \left( \omega - \frac{1}{\omega} \right)^2 \,,
\end{equation}
where the Mandelstam variables are taken to be $s=(p_1+p_2)^2$ and $t=(p_1+p_4)^2$. The reason for introducing the $\omega$ variable defined in \eqref{omega} will become clear soon. 

As shown in \cite{Henn:2013pwa}, in order to solve \eqref{non canonical diff eq} it is better to look for a change of basis 
\begin{equation}
\label{change of basis}
\vec{M}= {\bf T} \, \vec{C} \,,
\end{equation}
that takes the differential equation \eqref{non canonical diff eq} into a \textit{canonical form}
\begin{equation}
\label{canonical diff eq}
\partial_{\omega} \vec{C} = \epsilon \, {\bf A}_c(\omega) \, \vec{C} \,,
\end{equation}
where the matrix ${\bf A}_c(\omega)$ is given as
\begin{equation}
\label{canonical matrix}
{\bf A}_c(\omega) = \sum_{j=1}^{n} \, {\bf a}_j \, {\rm dlog} [W_j(\omega)] \,,
\end{equation}
with constant ${\bf a}_j$ matrices and where the $W_j$ are a set of algebraic functions of $\omega$. The $W_j$ functions are known as \textit{letters}, and the set of all letters is known as the \textit{alphabet}. Then, expanding the canonical basis as\footnote{In order to get \eqref{C epsilon expansion} one has to properly normalize the canonical basis so that all integrals are finite in $\epsilon \to 0$. }
\begin{equation}
\label{C epsilon expansion}
\vec{C}(\omega, \epsilon)=\sum_{k=0}^{\infty} \epsilon^k \, \vec{C}^{(k)} (\omega) \,,
\end{equation}
one gets
\begin{equation}
\vec{C}^{(0)} (\omega)= {\rm const.} \,,
\end{equation}
and
\begin{equation}
\label{formal solution canonical DE}
\vec{C}^{(k)} (\omega)=\vec{C}^{(k)} (0)+ \int_{0}^{\infty} d\tau \, {A}_c(\tau) \, \vec{C}^{(k-1)} (\tau) \,,
\end{equation}
for $k \geq 1$, where $\vec{C}^{(k)} (0)$ are boundary constants at the boundary point $\omega=0$\footnote{The choice of the boundary point is arbitrary.}. Therefore, from \eqref{canonical matrix} and \eqref{formal solution canonical DE} one can see that the solution to the differential equation \eqref{canonical diff eq} will be a set of integrals with uniform transcendental weight in their $\epsilon$ expansion, provided the $\vec{C}^{(0)}$ vector has a given transcendental weight $w_0$.

In order to take the differential equation \eqref{non canonical diff eq} into the canonical form we have used the algorithm \texttt{CANONICA} \cite{Meyer:2017joq}. We have found that, given the change of basis presented in the file {\tt change\_of\_basis.txt} of Ref. \cite{Lagares:2024epo} (which corresponds to the $\bf{T}$ matrix in \eqref{change of basis}), the alphabet is
\begin{align}
W_1 &= \omega \,, \qquad & W_2 &=1+\omega \,, \\
W_3 &= 1-\omega \,, \qquad & W_4 &=1+\omega^2 \,.
\end{align}
The corresponding ${\bf a}_j$ matrices are presented in the ancillary files ${\tt a1}.{\tt txt}$, ${\tt a2}.{\tt txt}$, ${\tt a3}.{\tt txt}$ and ${\tt a4}.{\tt txt}$ of Ref. \cite{Lagares:2024epo}. At this point we are in the position of arguing the choice of $\omega$ as the kinematic variable to construct the differential equations. Had we used $z=t/s$ instead of $\omega$, we would not have obtained a rational alphabet, and the letters would have included square roots of $z$. The decompositions of \eqref{I1 ladder infty}-\eqref{I box infty} into the canonical basis $\vec{C}$ are presented in the ancillary files ${\tt I1} \_ {\tt ladd} \_ {\tt dec}.{\tt txt}$, ${\tt I2} \_ {\tt ladd} \_ {\tt dec}.{\tt txt}$ and ${\tt I} \_ {\tt box} \_ {\tt dec}.{\tt txt}$ of Ref. \cite{Lagares:2024epo}.

\subsubsection*{Solving the differential equations}

As discussed above, given a set of canonical differential equations, solving them reduces to determining the boundary values $\vec{C}^{(k)}(0)$ for $k \geq 0$. We expect the ${\cal F}_3$ and ${\cal G}_3$ functions to have transcendental weight $2$, and therefore we will only be interested in solving the differential equations up to second order in the $\epsilon$ expansion. Consequently, we only need to compute $\vec{C}^{(0)}(0), \, \vec{C}^{(1)}(0)$ and $\vec{C}^{(2)}(0)$. To that aim we take into consideration that
\begin{enumerate}
    \item Given that we are working with planar integrals, the symbol\footnote{The
 symbol \cite{Goncharov:2010jf} is an important concept related to a transcendental function. Let us consider a transcendental function $f$ of weight $n$ whose total derivative can be written as
\begin{equation}
    \label{symbol def-1}
    df= \sum_i g_i \, d\log \omega_i \,,
\end{equation}
where the $g_i$ are functions of weight $n-1$ and the $\omega_i$ are rational functions called \textit{letters}. The set of all letters of a transcendental function is known as its \textit{alphabet}. Then, the symbol ${\cal S}$ of $f$ is defined recursively as
\begin{equation}
    \label{symbol def-2}
    {\cal S}(f)= \sum_i {\cal S}(g_i) \otimes \omega_i \,.
\end{equation}} of the  $\vec{C}$ integrals should always have $z=\frac{(1-\omega^2)^2}{4 \omega^2}=\frac{t}{s}$ as its first letter. Moreover, there should not be singularities for $u \to 0$, which taking into account the pole structure of the ${\bf A}_c$ matrix implies
\begin{equation}
    \label{two loop sing constraint}
    \lim_{\omega^2 \to -1} {\bf a}_4 \, \vec{C}^{(k)} (\omega)=0 \,.
\end{equation}
for all $k \geq 0$.

    \item Four out of the nineteen integrals in the canonical basis can be computed exactly by Feynman parametrization \cite{Brandhuber:2013gda}. More precisely, we have 
    \begin{align}
    \label{three loop input data 1}
        C_1 (z) &= -z^{-3\epsilon} \, \frac{9 \left(24 \epsilon ^2-10 \epsilon +1\right) \Gamma \left(\frac{1}{2}-\epsilon \right)^4 \Gamma \left(3 \epsilon -\frac{1}{2}\right)}{10 \epsilon ^3 (2 \epsilon +1)^3 \Gamma (2-4 \epsilon )} \,, \\
        C_6 (z) &= z^{-\epsilon} \, \frac{54 \Gamma \left(\frac{1}{2}-\epsilon \right)^4 \Gamma (-2 \epsilon ) \Gamma \left(\epsilon +\frac{1}{2}\right) \Gamma \left(\epsilon +\frac{3}{2}\right) \Gamma (2 \epsilon +2)}{\epsilon ^2 (2 \epsilon +1)^5 \Gamma \left(\frac{1}{2}-3 \epsilon \right) \Gamma (1-2 \epsilon )} \,, \\
            \label{three loop input data 3}
        C_{11} (z) &=z^{-3\epsilon} \, \frac{216 \Gamma \left(\frac{1}{2}-\epsilon \right)^4 \Gamma (-2 \epsilon ) \Gamma \left(\epsilon +\frac{1}{2}\right) \Gamma \left(\epsilon +\frac{3}{2}\right) \Gamma (2 \epsilon +2)}{\epsilon ^2 (2 \epsilon +1)^5 \Gamma \left(\frac{1}{2}-3 \epsilon \right) \Gamma (1-2 \epsilon )} \,, \\
             \label{three loop input data 4}
        C_{13} (z) &=z^{-2\epsilon} \, \frac{108 \Gamma \left(\frac{1}{2}-\epsilon \right)^6 \Gamma \left(\epsilon +\frac{1}{2}\right)^3}{\epsilon  (2 \epsilon +1)^3 \Gamma (1-2 \epsilon )^3} \,.
    \end{align}
Consequently, the solutions obtained from \eqref{canonical diff eq} should agree with \eqref{three loop input data 1}-\eqref{three loop input data 4}. 

\item Within the euclidean region, which in our conventions\footnote{We are taking the metric to have signature $(-,+,+)$ and we are defining $s:=(p_1+p_2)^2$ and $t:=(p_1+p_4)^2$.} is defined by the $s>0$ and $t>0$ constraints (or, equivalently, $0<\omega<1$), all the integrals in the $\vec{C}$ basis should be real. 

\end{enumerate}

We have used the \texttt{PolyLogTools} package \cite{Duhr:2019tlz} to manipulate the iterated integrals that appear when solving the differential equations. All the above considerations allow us to completely fix the zeroth-order and first-order constants $\vec{C}^{(0)}$ and $\vec{C}^{(1)}$. As for the second-order boundary vector $\vec{C}^{(2)}$, the previous constraints leave only two unknown constants. We have fixed them by demanding consistence of the $\vec{C}$ integrals with their numerical evaluations. The $\vec{C}$ canonical basis is presented, up to second order in its $\epsilon$-expansion, in the ancillary file ${\tt DE} \_ {\tt sol}.{\tt txt}$ of Ref. \cite{Lagares:2024epo}.

\subsubsection*{Integrated results}

The solution to the canonical differential equations discussed above allows us to compute the ladder and box integrals defined in \eqref{I1 ladder infty}-\eqref{I box infty}. We get
\begin{align}
\label{I1 ladder infty-result}
\hat{\cal I}_1^{\rm Ladd} &= -24 \, s \sqrt{t} \, \pi^{5/2}   \left[ \left( \log(z)-2 \log(2) \right)^2 + 2\pi^2 \right]   \nonumber \\
    & \quad - 24 \, t \sqrt{s} \, \pi^{5/2}   \left[  \left( \log(z)+2 \log(2) \right)^2 +2 \pi^2 \right] \,, \\
\label{I2 ladder infty-result}
\hat{\cal I}_2^{\rm Ladd} &= -48 \pi^{5/2} \sqrt{s t (s+t)} \left[ \hat{{\cal H}}(z) + \hat{{\cal H}} \left( \frac{1}{z} \right) + 2 \pi^2 \right] \,, \\
\label{I box infty-result}
\hat{\cal I}^{\rm Box} &= -48 \, \pi^{5/2} \,\sqrt{s t (s+t)} \, \left[ \hat{{\cal H}}(z) + \hat{{\cal H}} \left( \frac{1}{z} \right) \right]  \nonumber \\
&\quad -32\pi^{9/2} \,\left( 3\, \sqrt{s t (s+t)}- \, s \sqrt{t}-\, t \sqrt{s}
  \right) \,,
\end{align}
where the $\hat{{\cal H}}$ function was defined in \eqref{hat H}. Let us note that the ${\cal I}_2^{\rm Ladd}$ integral can also be easily obtained from direct integration using the two-loop integrals computed in Appendix \ref{app: useful integrals}, in agreement with the result presented in \eqref{I2 ladder infty-result}. Reversing the $x_5 \to \infty$ limit introduced in \eqref{I1 ladder infty}-\eqref{I box infty} and taking into account the normalizations presented in Appendix \ref{ch: normalizations} we arrive at
\begin{align}
\label{F Ladder}
{\cal F}_3^{\rm Ladd} (z) &= -\frac{z^{-1/4}}{16} \,  \left[ \left( \log(z)-2 \log(2) \right)^2 + 2\pi^2 \right]   \nonumber \\
&\quad - \frac{z^{1/4}}{16} \,   \left[ \left( \log(z)+2 \log(2) \right)^2 + 2\pi^2 \right]  \nonumber \\
&\quad -\frac{z^{-1/4} \sqrt{1+z}}{8} \left[  \hat{{\cal H}}(z)+  \hat{{\cal H}} \left( \frac{1}{z} \right) + 2 \pi^2 \right] \\
\label{F Box}
{\cal F}_3^{\rm Box} (z) &= -\frac{z^{-1/4}\sqrt{1+z}}{8} \, \left[ \hat{{\cal H}} (z) + \hat{{\cal H}} \left( \frac{1}{z} \right) \right]  \nonumber \\
&\quad -\frac{\pi^2}{12} \left( 3 z^{-1/4}\sqrt{1+z}-z^{1/4} - z^{-1/4} \right) \,,
\end{align}
and
\begin{align}
\label{G Ladd and Box}
{\cal G}_3^{\rm Ladd} (z)= {\cal G}_3^{\rm Box} (z)  &= 0 \,.
\end{align}
Finally, adding up the contributions from the different diagrams as in \eqref{four-loop canonical form}, we get
\begin{align}
{\cal F}_3 (z) &= -{\cal F}_3^{\rm Ladd} (z)-{\cal F}_3^{\rm Star} (z)+{\cal F}_3^{\rm Box} (z) \\
\label{F total}
&= \frac{z^{-1/4} }{16} \, \left[ \log(z)-2 \log(2) \right]^2 + \frac{z^{1/4}}{16} \,   \left[ \log(z)+2 \log(2) \right]^2 \nonumber \\
&\quad + \frac{7\pi^2}{24} \left( z^{1/4} + z^{-1/4}  \right)  \,, \\
\label{G total}
{\cal G}_3 (z) &=0 \,.
\end{align}

\section{Properties of the integrated results}
\label{sec: properties of integrated results}

After computing the $L \leq 4$ integrated negative geometries of the ABJM theory, we will turn now to the analysis of the integrated results. We will begin with an analysis of their transcendental weight. Later we will move on to a discussion on their leading singularities, and finally we will comment on the sign patterns that are observed in the integrated results.

\subsubsection*{Transcendental weight}

Let us recall that the {\it{transcendental weight}} (also called \textit{degree of transcendentality}) $T$ of a function $f$ is defined as the number of iterated integrals that are needed to compute $f$ \cite{Henn:2013pwa}, e.g. $T({\rm Li}_n)=n$. Moreover, one can extend the definition of $T$ to transcendental numbers, i.e. numbers that can not be obtained as the solution of a polynomial equation with rational coefficients. For example, $T(\pi)=1$ and $T(\zeta_n)=n$. A series expansion is said to have uniform degree of transcendentality (often abbreviated as UT) when all its terms have the same degree $T$. 
Moreover, when discussing Laurent expansions in the dimensional regularization parameter $\epsilon$, it is natural to assign weight $-1$ to $\epsilon$ (see \cite{Henn:2020omi} for a review).

We note that the $L \leq 4$ formulas computed in Sections \ref{sec: L up to 3} and \ref{sec: L=4} have uniform transcendental weight. More precisely, the $L \leq 4$ results suggest that
\begin{align}
 {\cal F}_{2k+1} (z) \text{ and } {\cal G}_{2k} (z)  \text{ have transcendental degree $2k$}
\end{align}
for $k \geq 0$.

\subsubsection*{Leading singularities}

One can always generically write the integrated negative geometries as
\begin{equation}
\label{eq:leading_singularities_def}
\left( \prod_{j=6}^{4+L} \int \frac{d^3 x_j}{i\pi^{3/2}} \right) \, {\bf L}_L = \sum_{i=1}^k R_{L-1,i}\, T_{L-1,i}\,,
\end{equation}
for some integer $k$, where the $T_{L-1,i}$ are transcendental functions and the $R_{L-1,i}$ are rational functions known as \textit{leading singularities}. Having a good understanding of the leading singularities of the integrated negative geometries is an important step towards a bootstrap computation of these infrared-finite functions. For the sake of more generality, we will separately study the contributions of each individual negative geometry diagram, without considering the cancellations that may occur when adding up all contributions according to \eqref{bipartite expansion}. Taking into account the $L \leq 4$ results presented in Sections \ref{sec: L up to 3} and \ref{sec: L=4}, we see that the only leading singularities that contribute to the integrated results are\footnote{The $R_6$ leading singularity appears when integrating the box and ladder diagrams at $L=4$, but it cancels when summing all contributions as in \eqref{bipartite expansion}.}
\begin{align}
\label{even leading sing}
R_{1} &= \frac{4 \, \epsilon(1,2,3,4,5)}{X_{15}^2X_{25}^2X_{35}^2X_{45}^2} \,, & R_{2} &= \frac{R_{1}}{\sqrt{1+z}} \,, \quad & R_{3} &= \sqrt{\frac{z}{z+1}} \, R_{1} \,, \\
R_{4} &= \left( \frac{X_{13}^2 X_{24}^2}{X_{15}^2 X_{25}^2 X_{35}^2 X_{45}^2} \right)^{3/4} \,  z^{-1/4}\, \,, & R_{5} &= \sqrt{z} \, R_4 \,, & R_6&= \sqrt{1+z} \, R_4 \,.
\end{align}
In order to work only with four-particle kinematic variables it is best to perform the analysis of leading singularities in the frame in which the unintegrated variable goes to infinity. Therefore, defining
\begin{align}
\label{x5 infty LS}
r= \lim_{x_5 \to \infty} (x_5^2)^3 \, R \,,
\end{align}
we see that in the $x_5 \to \infty$ limit the leading singularities reduce to
\begin{align}
r_1 &= \sqrt{s \, t \, (s+t)} \,, \qquad &r_2 &= s \sqrt{t} \,,  \qquad &r_3 &= t \sqrt{s} \,, \\
r_4 &= s \sqrt{t} \,, \qquad &r_5 &= t \sqrt{s} \,, \qquad &r_6 &= \sqrt{s \, t \, (s+t)} \,. \\
\end{align}
That is, we get that up to $L \leq 4$ the leading singularities of each individual integrated negative geometry diagram belong to the set 
\begin{equation}
\{ \sqrt{st(s+t)} , \, s\sqrt{t}, \, t\sqrt{s}\} \,.
\end{equation}

\subsubsection*{Sign patterns}

We will close this section with a positivity analysis of the integrated negative geometries. To that end, let us first recall the sign properties that have been observed for the integrated negative geometries of the ${\cal N}=4$ super Yang-Mills theory \cite{Arkani-Hamed:2021iya}. As discussed in Section \ref{sec: review neg geoms}, perturbative results suggest that 
\begin{align}
{\cal F}_{L-1}(z)&<0  \qquad \text{for odd }L \,, \\
{\cal F}_{L-1}(z)&>0  \qquad \text{for even }L \,,
\end{align}
when restricted to the Euclidean region $z>0$. Moreover, similar results were found for the five-particle case in \cite{Chicherin:2022zxo}. There it was found that
\begin{equation}
\label{five particle signs}
    f_5^{(0)} \big|_{\rm Eucl^+}>0 \,, \qquad f_5^{(1)} \big|_{\rm Eucl^+}<0 \,, \qquad f_5^{(2)} \big|_{\rm Eucl^+}>0 \,,
\end{equation}
where
\begin{equation}
f_5^{(L-1)}=\lim_{x_6 \to \infty}  \, \left( \prod_{j=7}^{5+L} \int \frac{d^4 x_j}{i\pi^{2}} \right) \, {\bf L}_L^{(n=5)} \,,
\end{equation}
with ${\bf L}_L^{(n=5)}$ the five-particle $L$-loop integrand for the logarithm of the amplitude. Let us note that $x_5$ is no longer an unintegrated loop variable in \eqref{five particle signs}, and such a role is played by $x_6$ in that case. Furthermore, the region ${\rm Eucl^+}$ in \eqref{five particle signs} is defined as the one were all adjacent five-particle Mandelstam invariants are negative and
\begin{equation}
4i \epsilon_{\mu \nu \rho \sigma} p_1^{\mu} p_2^{\nu} p_3^{\rho} p_4^{\sigma} >0 \,.
\end{equation}
Interestingly, it was shown in \cite{Chicherin:2024hes} that the individual negative geometry diagrams that contribute to the two-loop five-particle integrated negative geometry also have a definite sign within the ${\rm Eucl^+}$ region.

Let us turn now to the analysis of the ABJM case. For the $L \leq 3$ the results obtained in Section \ref{sec: L up to 3} one can read that in the Euclidean region, i.e. for $z>0$,
\begin{align}
{\cal G}_0(z)\big|_{z>0}<0 \,, \qquad {\cal F}_1(z)\big|_{z>0}<0 \,, \qquad {\cal G}_2(z)\big|_{z>0}>0 \,.
\end{align}
Moreover, from Section \ref{sec: L=4} we see that in the $L=4$ case the individual negative geometry diagrams behave as
\begin{align}
{\cal F}_3^{\rm Ladd}(z)\big|_{z>0}<0 \,, \qquad {\cal F}_3^{\rm Star}(z)\big|_{z>0}<0 \,, \qquad {\cal F}_3^{\rm Box}(z)\big|_{z>0}>0 \,.
\end{align}
Therefore, from \eqref{four-loop canonical form} we get
\begin{align}
{\cal F}_3(z)\big|_{z>0}>0  \,.
\end{align}
Consequently, we conjecture that
\begin{align}
{\cal F}_{2k+1}(z)\big|_{z>0}&<0 \,, \qquad \text{and} \qquad {\cal G}_{2k}(z)\big|_{z>0}<0 \,, \qquad \text{for even }k\,,\\
{\cal F}_{2k+1}(z)\big|_{z>0}&>0 \,, \qquad \text{and} \qquad {\cal G}_{2k}(z)\big|_{z>0}>0 \,, \qquad \text{for odd } k\,,
\end{align}
for $k \geq 0$.

\section{Cusp anomalous dimension}
\label{sec: gamma cusp from int neg geoms}

As discussed in Section \ref{sec: review neg geoms} for the case of the ${\cal N}=4$ super Yang-Mills theory, integrated negative geometries constitute a powerful tool to compute the light-like cusp anomalous dimension $\Gamma_{\rm cusp}^{\infty}$ that governs the IR divergences of massless scattering amplitudes (and the UV divergences of Wilson loops with light-like cusps). In this section we will study how to generalize these ideas to the ABJM theory.

\subsubsection*{All-loop proposal for $\Gamma_{\rm cusp}^{\infty}$}

In ${\cal N}=4$ sYM, an all-loop prediction for $\Gamma_{\rm cusp}^{\infty}$ was obtained following integrability ideas \cite{Beisert:2006ez}. This result was expressed as a function of the  interpolating function $h_{{\cal N}=4} (\lambda)$, which as seen in Chapter \ref{ch: integrable line defects in ABJM} governs the dispersion relation of magnons in the integrability picture. At weak coupling, it was shown that
\begin{equation}
\label{f n=4}
\Gamma_{\rm cusp}^{\infty}(h_{{\cal N}=4})= 4 h_{{\cal N}=4}^2 - \frac{4}{3} \pi^2 h_{{\cal N}=4}^4 + \frac{44}{45} \pi^4 h_{{\cal N}=4}^6 + \dots \,.
\end{equation}
where $h_{{\cal N}=4} (\lambda)=\frac{\sqrt{\lambda}}{4 \pi}$ to all loops, as presented in \eqref{h N=4}. 

Based on a proposal for the all-loop asymptotic Bethe Ansatz that describes the anomalous dimensions of large-charge single-trace operators in the ABJM theory \cite{Gromov:2008qe,Ahn:2008aa}, it was stated in \cite{Gromov:2008qe} that the cusp anomalous dimension of ABJM is related to its ${\cal N}=4$ sYM counterpart by
\begin{equation}
\label{Gamma cusp proposal ABJM}
 \Gamma_{\rm cusp}^{\infty,\rm ABJM}= \frac{1}{4} \Gamma_{\rm cusp}^{\infty,{\cal N}=4}\bigg|_{h_{{\cal N}=4} \to h} \,.
\end{equation}
where $h(\lambda)$ is the interpolating function of the ABJM theory, which has been conjectured to behave as in \eqref{h conjecture ABJM} to all loops \cite{Gromov:2014eha,Cavaglia:2016ide}. Then, combining \eqref{h N=4} with \eqref{h conjecture ABJM}, \eqref{f n=4} and \eqref{Gamma cusp proposal ABJM} we get
\begin{equation}
\label{Gamma cusp conjecture ABJM-weak coupling-2}
\Gamma_{\rm cusp}^{\infty, \rm ABJM} (\lambda)= \lambda^2 - \pi^2 \lambda^4 + \frac{49 \pi^4}{30} \lambda^6 + \dots \,.
\end{equation}
The above proposal is consistent with the leading-order perturbative result computed in \cite{Griguolo:2012iq}. From now on we will write $\Gamma_{\rm cusp}^{\infty}$ when referring to $\Gamma_{\rm cusp}^{\infty, \rm ABJM}$, for simplicity.

\subsubsection*{$\Gamma_{\rm cusp}^{\infty}$ from integrated negative geometries}

Let us now study how we can compute $\Gamma_{\rm cusp}^{\infty}$ from the knowledge of the integrated negative geometries. To that aim it will be crucial to use the Wilson loop/scattering amplitude duality, which was discussed in Chapter \ref{ch: scatt ampl}. 
In the ABJM theory such a duality was observed only in the four-particle case, in which it identifies the $L$-loop integrand ${\bf L}_L$ for the logarithm of the scattering amplitude with the $L$-loop integrand for $\log \langle W_4 \rangle$, where $W_4$ is a polygonal Wilson loop whose vertices locate at the four $x_i$ points that characterize the kinematics of the scattering process in dual space.

We can use the Wilson loop/scattering amplitude duality to relate $\Gamma_{\rm cusp}^{\infty}$ to the integrated negative geometries. To that aim we should recall that the renormalization theory of light-like Wilson loops \cite{Korchemskaya:1992je} implies
\begin{equation}
\label{log WL in terms of Gamma cusp}
\log \langle W_4 \rangle= - 2 \sum_{L=1}^{\infty} \frac{\lambda^L \, \Gamma_{{\rm cusp},L}^{\infty}}{(L \epsilon)^2} + \mathcal{O}(1/\epsilon) \,,
\end{equation}
where $\Gamma_{{\rm cusp},L}^{\infty}$ is the $L$-loop coefficient of the cusp anomalous dimension. Therefore, we get
\begin{align}
\label{Gamma cusp from F and G}
\int \frac{d^{D} X_5}{i\pi^{D/2}} \left[ \sqrt{\pi}\left( \frac{X_{13}^2 X_{24}^2}{X_{15}^2 X_{25}^2 X_{35}^2 X_{45}^2} \right)^{\frac{3}{4}} {\cal F} \left( z \right) + \frac{i \epsilon\left( 1,2,3,4,5 \right)}{X_{15}^2 X_{25}^2 X_{35}^2 X_{45}^2} \, \frac{{\cal G} \left( z \right)}{\sqrt{\pi}} \right] = &- 2 \sum_{L=1}^{\infty} \frac{\lambda^L \, \Gamma_{{\rm cusp},L}^{\infty}}{\epsilon^2} \nonumber \\
& + \mathcal{O}\left( \frac{1}{\epsilon} \right) \,,
\end{align}
with $D=3-2\epsilon$. Then, after defining
\begin{align}
\label{F functional}
I_{\cal F}[{\cal F}_{L-1}] &:= \left[-\frac{\sqrt{\pi}}{2} \int  \frac{d^D X_5}{i\pi^{D/2}} \, \left( \frac{X_{13}^2 X_{24}^2}{X_{15}^2 X_{25}^2 X_{35}^2 X_{45}^2} \right)^{\frac{3}{4}} {\cal F}_{L-1} \left( z \right) \right]_{1/\epsilon^2 \, {\rm term}} \,, \\
\label{G functional}
I_{\cal G}[{\cal G}_{L-1}] &:= \left[ -\frac{1}{2\sqrt{\pi}} \int \frac{d^D X_5}{i\pi^{D/2}} \,  \frac{i \epsilon(1,2,3,4,5)}{X_{15}^2 X_{25}^2 X_{35}^2 X_{45}^2} \, {\cal G}_{L-1} \left( z \right) \right]_{1/\epsilon^2 \, {\rm term}}  \,,
\end{align}
we have
\begin{equation}
\label{Gamma cusp from F and G-functionals}
\Gamma_{{\rm cusp},L}^{\infty} = I_{\cal F}[{\cal F}_{L-1}] + I_{\cal G}[{\cal G}_{L-1}] \,.
\end{equation}
As shown in Appendix \ref{app: useful integrals}, using Feynman parametrization we get
\begin{align}
\label{F functional-zp}
I_{\cal F}[z^p] &= -\frac{2\sqrt{\pi}}{ \Gamma \left( \frac{3}{4} +p \right) \Gamma \left( \frac{3}{4} -p \right)} \,, \\
\label{G functional-zp}
I_{\cal G}[z^p] &= 0 \,.
\end{align}

Therefore, from the $L \leq 3$ integrated results computed in Section \ref{sec: L up to 3} we obtain
\begin{align}
\label{If functional up to L=3}
I_{\cal F}[{\cal F}_0] &=0 \,, \qquad &I_{\cal F}[{\cal F}_1] &=1 \,, \qquad &I_{\cal F}[{\cal F}_2] &=0 \,.
\end{align}
The results of eq. \eqref{If functional up to L=3} give $\Gamma_{\rm cusp}^{\infty} (\lambda)= \lambda^2 + \mathcal{O}(\lambda^4)$, in agreement with \eqref{Gamma cusp conjecture ABJM-weak coupling-2} and the two-loop Feynman diagram computation of \cite{Griguolo:2012iq}.

In order to compute the four-loop contribution $\Gamma_{\rm cusp,4}^{\infty}$ from the $L=4$ integrated negative geometries obtained in Section \ref{sec: L=4} it is useful to notice that, as in the ${\cal N}=4$ sYM case \cite{Henn:2019swt},
\begin{align}
\label{F functional-zp log a}
I_{\cal F}[z^p \log^a(z)] &= \lim_{\zeta \to 0} \frac{\partial^a}{\partial \zeta^a} \, I_{\cal F}[z^{p+\zeta}] \,.
\end{align}
In particular, this allows us to get
\begin{equation}
\label{Gamma H}
    I_{\cal F} \left[ z^{-1/4}\sqrt{1+z} \left( \hat{\cal H}(z) + \hat{\cal H} \left( \frac{1}{z} \right) \right) \right]= \frac{4\pi^2}{3} \,.
\end{equation}
Then, applying \eqref{F functional-zp} and \eqref{F functional-zp log a} to the different $L=4$ integrated negative geometry diagrams we arrive at
\begin{align}
I_{\cal F} \left[ {\cal F}_3^{\rm Ladd} \right]&= \frac{2\pi^2}{3} \,, \\
I_{\cal F} \left[ {\cal F}_3^{\rm Star} \right]&= \frac{\pi^2}{3} \,, \\
I_{\cal F} \left[ {\cal F}_3^{\rm Box} \right]&= 0 \,.
\end{align}
Interestingly, the four-loop box diagram has a vanishing contribution to the cusp anomalous dimension. It would be interesting to see if this result continues to be valid for higher-loop ``loops of loops'' diagrams. Combining all contributions we get
\begin{equation}
I_{\cal F} \left[ {\cal F}_3 \right]=-I_{\cal F} \left[{\cal F}_3^{\rm Ladd} \right]-I_{\cal F} \left[{\cal F}_3^{\rm Star} \right]+I_{\cal F} \left[{\cal F}_3^{\rm Box} \right] \,.
\end{equation}
Therefore, we obtain that the four-loop contribution to the cusp anomalous dimension of ABJM is
\begin{equation}
\label{four loop gamma cusp}
\Gamma_{\rm cusp,4}^{\infty}=-\pi^2 \,.
\end{equation}
This constitutes a direct four-loop computation of the cusp anomalous dimension of the ABJM theory, and shows perfect agreement with the integrability-based proposal \eqref{Gamma cusp conjecture ABJM-weak coupling-2}.

\subsubsection*{Contour integral representation for $I_{\cal F}$}

Finally, let us end this section by discussing a contour integral representation of the $I_{\cal F}$ functional. To that end, it is useful to note that the result \eqref{F functional-zp} can be rewritten as
\begin{align}
\label{F functional integral -1}
I_{\cal F}[z^p] &= - \frac{4}{B\left( \frac{3}{4}+p, \frac{3}{4}-p \right)} \,,
\end{align}
where $B$ is the well-known Euler Beta function. Then, we can use the identity
\begin{equation}
\frac{1}{B(a,b)}= b \int_{c-i \infty}^{c+i \infty}  \frac{dz}{2\pi} \, z^{-a} (1-z)^{-1-b} \,, \qquad 0<c<1 \,, \quad  {\rm Re}(a+b)>0 \,,
\end{equation}
to get
\begin{align}
\label{F functional integral -2}
I_{\cal F}[z^p] &= -4 \left( \frac{3}{4}-p \right) \int_{c-i \infty}^{c+i \infty}  \frac{dz}{2\pi} \, z^{-3/4} (1-z)^{-7/4} \, \left[ \frac{1-z}{z} \right]^p \,.
\end{align}
Therefore,
\begin{align}
\label{F functional integral -3}
I_{\cal F}[{\cal F}] &= - \int_{c-i \infty}^{c+i \infty}  \frac{dz}{2\pi} \, z^{-3/4} (1-z)^{-7/4}\, \left[ 3 \, {\cal F}\left(\frac{1-z}{z}\right) + 4 \, z (1-z) \, \frac{d{\cal F}}{dz}\left(\frac{1-z}{z}\right)  \right] \,,
\end{align}
for $0<c<1$. One of the advantages of this representation of the $I_{\cal F}$ functional is that it is more amenable for numerical computations of the cusp anomalous dimension. We have tested \eqref{F functional integral -3} at $L=4$ loops, finding excellent numerical agreement with \eqref{four loop gamma cusp}.

%% file: conclusions.tex
This thesis focused on the study of line defects and scattering amplitudes in the framework of the AdS/CFT correspondence. The analysis was centered on the application of non-perturbative methods for the description of line defects, and on the study of infrared-finite functions that emerge in the ABJM theory as a consequence of the positive geometry description of scattering amplitudes. More specifically, we explored the use of analytic conformal bootstrap and integrability methods for the description of superconformal line defects in AdS$_3$/CFT$_2$ and AdS$_4$/CFT$_3$ dualities, respectively, and we studied the integrated negative geometries of the ABJM theory.

The results of this thesis represent a further step in our understanding of non-perturbative methods and scattering amplitudes. We have efficiently implemented the analytic conformal bootstrap program to the strong-coupling description of 1/2 BPS line defects in the holographic dual to string theory in $AdS_3 \times S^3 \times T^4$ with mixed Ramond-Ramond and Neveu Schwarz-Neveu Schwarz three-form flux, yielding a successful application of analytic conformal bootstrap ideas to a bi-parametric problem that is invariant under two Poincaré supercharges. Moreover, we have delivered strong evidence about the integrability of the 1/2 BPS line defect of the ABJM theory, and we have given a proposal for the Boundary Thermodynamic Bethe Ansatz equations associated to the cusped Wilson line of the theory. These results set the stage for an all-orders integrability-based derivation of the interpolating function of the theory, that appears in every all-loop result obtained through integrability. Finally, we have computed the integrated negative geometries of the ABJM theory up to three-loops, and we have used them to compute the four-loop light-like cusp anomalous dimension of the theory. Our analysis provides a tool to compute the aforementioned anomalous dimension at $L$ loops by just performing $L-1$ loop integrals. Moreover, we have identified useful properties that could offer insight into a higher-loop bootstrap computation of the integrated negative geometries. Below we provide a detailed discussion of the topics addressed in this thesis and comment on promising future directions.

\subsubsection*{Line defects in AdS$_3$/CFT$_2$}

We have studied line defects in the context of AdS$_3$/CFT$_2$ dualities, with an emphasis on the application of the analytic conformal bootstrap program to their study. Our analysis was performed in the context of AdS$_3$/CFT$_2$ realizations in which the gravity side is given by type IIB string theory in $AdS_3 \times S_+^3 \times S_-^3 \times S^1$ with mixed R-R and NS-NS three-form flux and with a constant dilaton. In particular, the bootstrap results were obtained in the limit in which the dual metric becomes $AdS_3 \times S^3 \times T^4$.

We began our analysis in the general $AdS_3 \times S_+^3 \times S_-^3 \times S^1$ case, adopting a string theory approach that enabled us to identify a rich family of supersymmetric strings describing BPS line defects in the dual CFT$_2$. These strings, that range from 1/2 BPS to 1/8 BPS, are subjected to a wide variety of supersymmetric boundary conditions that describe configurations either localized or delocalized in the compact space. The former case is associated to strings with Dirichlet boundary conditions, while the latter case can either correspond to smeared or Neumann boundary conditions. Such a richness of supersymmmetric boundary conditions is closely tied to the presence of massless fermionic degrees of freedom in the spectrum of fluctuations, as it was observed for the case of the 1/2 BPS line of ABJM \cite{Correa:2019rdk}. Interestingly, we have found cases in which a string smeared over a given manifold ${\cal M}$ is left invariant under a different number of supersymmetries than the string with Neumann boundary conditions over ${\cal M}$. This contrasts the situation that was observed for strings in $AdS_5 \times S^5$ or in  $AdS_4 \times \mathbb{CP}^3$, where smeared and Neumann boundary conditions seem to always preserve the same amount of supersymmetry. We have also found a network of interpolating strings that connect all the aforementioned localized and delocalized strings between each other.

We applied analytic conformal bootstrap techniques to study 1/2 BPS line defects in the limit at which the supergravity metric describes an $AdS_3 \times S^3 \times T^4$ background. In particular, we have studied two-, three- and four-point functions of the line defects excitations contained in the displacement and tilt supermultiplets, performing the analysis up to next-to-leading order in the strong-coupling expansion. We have found that supersymmetry completely fixes the two-point and three-point functions at most up to a single parameter. As for the four-point functions, the analytic bootstrap determines the full result up to two coefficients. From these results we have obtained CFT data associated to the line defect, up to next-to-leading order and expressed in terms of these two parameters that are not fixed by the bootstrap procedure. We have checked that the bootstrap results perfectly match the holographic expectations coming from the Witten diagram expansion of the correlators. This holographic description allowed us to obtain an interpretation of the two coefficients that parametrize the bootstrap result, relating them to the string tension, the AdS radius and the tilt angle of the dual string with respect to the boundary of AdS$_3$. Considering that the defects we have studied are invariant under eight real supercharges and are parametrized by two couplings (the 't Hooft coupling and the $\lambda$ parameter that interpolates between the pure R-R and pure NS-NS backgrounds), we see that our results provide a step towards the development of analytic conformal bootstrap methods in theories with decreasing supersymmetry and an increasing number of coupling constants.

It is a natural question to understand how far we might push the bootstrap program as we reduce the number of supersymmetries. In this regard, we should make the distinction between the number of supersymmetries preserved by the line defect and the number of supersymmetries preserved by the bootstrapped multiplet. Concerning the former, it would be interesting to explore  whether analytic conformal bootstrap techniques can be applied to study the bosonic 1/6 BPS Wilson loop of ABJM \cite{Drukker:2008zx}, which preserves only 2 supercharges. As for the bootstrap of less supersymmetric supermultiplets, one could consider the case of the displacement supermultiplet associated to the 1/2 BPS line defect of the CFT$_2$ which is dual to type IIB string theory in $AdS_3 \times S_+^3 \times S_-^3 \times S^1$. Notably, this multiplet is just 1/4 BPS, an therefore is left invariant by only 1 supercharge. 
The correlators of this less supersymmetric multiplet should depend on the additional parameter $\Omega$ that accounts for the relative size of the two $S^3$. Interestingly, in the limit $S_-^3 \times S^1 \rightarrow T^4$ this supermultiplet splits into two smaller multiplets\footnote{See the supersymmetry transformations presented in \eqref{susy massive fermions-1}-\eqref{susy s3- scalars-1}.}, which are precisely the displacement and tilt supermultiplets that we have studied in Chapter \ref{ch: line defects in AdS3CFT2}. The study of four-point correlators of the displacement supermultiplet in the $AdS_3 \times S_+^3 \times S_-^3 \times S^1$ case would naturally provide the mixed correlators between two displacement and two tilt supermultiplets in the $S_-^3 \times S^1 \rightarrow T^4$ limit.

Furthermore, the bootstrap of the 1/4 BPS displacement multiplet in the $AdS_3 \times S_+^3 \times S_-^3 \times S^1$ case opens another interesting new question. Let us note that in our analysis a crucial step to constrain the superconformal blocks and four-point functions was the existence of a topological sector. To construct it we used the fact that our tilt and displacement multiplets were annihilated by two supercharges. Then, the R-symmetry rotating those supercharges could be combined with time translations to build the topological translations. This does not seem to be possible if the bootstrapped multiplets were annihilated by only 1 supercharge, which is in fact what happens with the 1/4 BPS displacement multiplet in the $AdS_3 \times S_+^3 \times S_-^3 \times S^1$ case. We believe it might be interesting to explore the possibility of applying analytic bootstrap methods without the existence of a topological sector.

Another avenue for further investigation is the extension of the bootstrap analysis performed in Chapter \ref{ch: line defects in AdS3CFT2} beyond the next-to-leading order in the strong coupling expansion. At these orders we start to get loop Witten diagrams, which are extremely challenging for direct integration methods \cite{Carmi:2018qzm}. Therefore, the analytic bootstrap appears as a very appealing method to obtain further subleading corrections to the correlators. 

It would also be interesting to explore the Lagrangian description discussed in \cite{OhlssonSax:2014jtq} for the bulk CFT$_2$ in the limit of pure R-R flux, and to see if field representation could be constructed for the 1/2 BPS line defects we have studied. This could open the door to a supersymmetric localization analysis of these line defects, which might enable the exact determination of the bremsstrahlung function. These results could then be compared with the first strong-coupling orders of the bremsstrahlung function, that could be obtained from the four-point correlators presented here using integral formulas as in \cite{Drukker:2022pxk}. 

Another interesting direction is the bootstrap of multi-point correlators. These types of correlators have recently gathered interest \cite{Barrat:2021tpn,Barrat:2022eim,Bliard:2023zpe,Giombi:2023zte,Artico:2024wut}, and many of the tools used in this paper are applicable when extended to more than four points. In higher-point cases the conformal blocks become generalized Appell functions \cite{Rosenhaus:2018zqn}, the ansatz contains more complicated polylogarithms, and the Ward identities \cite{Bliard:2024und} play an important role. 

\subsubsection*{Integrable line defects in ABJM}

We have studied the integrability properties of line defects in the ABJM theory. In particular, we have focused on the 1/2 BPS straight line and the cusped Wilson line of the theory.

We showed that the anomalous dimensions of operators inserted along the contour of the 1/2 BPS line are described by an open and integrable spin chain. We computed the corresponding all-loop reflection matrices including their over-all dressing phases (the bulk scattering matrix is the same as for the periodic spin chain associated to single-trace operators) . Those phases, that were obtained using a crossing symmetry equation, were shown to reproduce results obtained both at weak and strong coupling. Interestingly, the dressing phases account for boundary bound states, a novelty in comparison with the integrability analysis of Wilson lines in ${\cal N}=4$ super Yang-Mills. 

We have also proposed a $Y$-system of equations for the cusped Wilson line of the theory. Moreover, we constructed an asymptotic solution to those equations and used it to reproduce the one-loop cusp anomalous dimension of the theory via a BTBA formula. This result serves as a test of the $Y$-system proposal, for which we have also written an equivalent set of integral BTBA equations.

The analysis presented in Chapter \ref{ch: integrable line defects in ABJM} suggests several compelling directions for future research. It would be interesting to use the BTBA equations of the cusped Wilson line to derive the bremsstrahlung function of ABJM, as done in \cite{Gromov:2012eu} for the ${\cal N}=4$ super Yang-Mills (sYM) theory. That would require to exactly solve the BTBA system in the small cusp angle limit. A comparison of this result with the localization-based computation of \cite{Bianchi:2017svd,Bianchi:2018scb} would allow for a direct derivation of the interpolating function $h(\lambda)$ of ABJM, in terms of which every all-loop integrability results are expressed. The result of such computation could be then compared to the current conjecture given in \cite{Gromov:2014eha}.

Finite-size corrections to the vacuum energy of the cusped Wilson line of ${\cal N}=4$ sYM were reformulated in terms of the Quantum Spectral Curve (QSC) formalism in \cite{Gromov:2015dfa}. There the authors found that the functional relations of the QSC were the same as for the corresponding periodic system, and only the asymptotic and analytic properties of the solutions had to be modified. This is suggestively similar to what we have found for the $Y$-system   in our setup. It would be interesting to follow this insight in order to propose a Quantum Spectral Curve for  the cusped Wilson line in ABJM.

Furthermore, it would be valuable to investigate the generalization of our results to the ABJ theory \cite{Aharony:2008gk}, i.e. to the case in which the ranks of the gauge group factors are not necessarily equal. In this picture, evidence of integrability was found in \cite{Minahan:2009aq} in the weak-coupling limit, while an all-loop proposal for the interpolating function $h$ was made in \cite{Cavaglia:2016ide}. However, the all-loop integrability of the ABJ theory is not trivially guaranteed from the assumed integrability of the ABJM limit. It is known that the string sigma model dual to the ABJ theory contains a theta-angle term proportional to $\lambda-\hat{\lambda}$ \cite{Aharony:2008gk}, where $\lambda$ and $\hat{\lambda}$ are the 't Hooft couplings of the ABJ theory. Such term implies a violation of parity, which is sometimes related to a breakdown of integrability (see for example the discussion in \cite{Minahan:2009te}). In this regard, it is suggestive to note that the 
generalization of the boundary Hamiltonian ${\bf H}^A_{\rm bdry}$ discussed in Section \ref{sec: weak coupling} 
would have both an order $\lambda$ and an order $\hat{\lambda}$ term when computed in the ABJ theory,  which would also imply a violation of parity.

One crucial ingredient of our results is the all-loop proposal for the boundary dressing phases. It would be interesting to further check the solutions to the crossing-unitarity equations that we have given, for example by a two-loop computation of the energy of the boundary bound states.

Another direction to consider is the extension of the TBA-based computation of $\Gamma_{\rm cusp}$ to the next-to-leading order, as done for the ${\cal N}=4$ sYM case in \cite{Bajnok:2013sya}. This would require an iteration of the BTBA equations, whose  result could eventually be compared with the Feynman-diagram computation done in \cite{Griguolo:2012iq}.
This would constitute a test for the proposed dressing factors as well as for the BTBA equations.

\subsubsection*{Integrated negative geometries in ABJM}

We have studied the IR-finite functions that arise when performing $L-1$ loop integrations over the $L$-loop integrand for the logarithm of the scattering amplitude in the ABJM theory. With a focus on the four-particle case, we have computed the integrated results up to $L=4$. To perform the loop integrations we have used either direct integration or the method of differential equations, therefore providing an application of this method to a three-dimensional theory.

We have found that the integrated negative geometries of order $L$ split into a parity-even term, proportional to a function ${\cal F}_{L-1}(z)$ of a dual conformally invariant cross ratio $z$, and a parity-odd contribution, described by a function ${\cal G}_{L-1}(z)$. Our results show that both the parity-even function ${\cal F}_{2k+1}$ and the parity-odd function ${\cal G}_{2k}$ have transcendentality degree $2k$ for $k\leq 1$. Moreover, we have found that the leading singularities are surprisingly simple in the limit in which the unintegrated variable goes to infinity. More precisely, in this frame the leading singularities are restricted to the set $\{ s\sqrt{t}, \, t\sqrt{s}, \, \sqrt{st(s+t)} \}$, up to the loop order we have studied. Furthermore, we have found that in the Euclidean region $z>0$ the signs of ${\cal F}_{2k+1}(z)$ and ${\cal G}_{2k}(z)$ alternate with $k$ for $k \leq 1$, suggesting a pattern for all $k$. This generalizes the results found for the ${\cal N}=4$ super Yang Mills theory in \cite{Arkani-Hamed:2021iya,Chicherin:2022zxo,Chicherin:2024hes} in the four- and five-particle cases. Finally, we have given a prescription to compute the $L$-loop order $\Gamma_{{\rm cusp},L}^{\infty}$ of the light-like cusp anomalous dimension of the theory in terms of the ${\cal F}_{L-1}(z)$ contribution to the integrated negative geometries. We have used this result to obtain the cusp anomalous dimension up to four loops, providing a direct computation of that quantity that perfectly agrees with the integrability-based prediction that follows from \cite{Gromov:2008qe,Ahn:2008aa}.

There are many exciting questions that arise from the analysis of Chapter \ref{ch: neg geoms ABJM}. As an example, one could further explore the computation of the integrated negative geometries of ABJM to higher loops. While the integration of the $L=5$ order is expected to give a vanishing contribution to the cusp anomalous dimension, reaching the $L=6$ order would serve as the first six-loop test of the all-loop integrability-based proposal for $\Gamma_{\rm cusp}^{\infty}$ \cite{Gromov:2008qe,Ahn:2008aa}. In turn, it would allow for a $\mathcal{O}(\lambda^5)$ test of the current all-loop proposal for the interpolating function $h(\lambda)$ of ABJM \cite{Gromov:2014eha}. It would be interesting to investigate whether a bootstrap approach \cite{Dixon:2011pw,Dixon:2011nj,Dixon:2014voa,Dixon:2015iva,Caron-Huot:2016owq,Drummond:2014ffa,Dixon:2016nkn} could provide access to these higher-loop integrated results, along the lines of the recent bootstrap computation of the two-loop six-point integrated negative geometries in ${\cal N}=4$ super Yang-Mills \cite{Carrolo:2025pue}. To this end, one could draw insight from the properties analyzed in Chapter \ref{ch: neg geoms ABJM} for the integrated negative geometries of ABJM.

Another question to address is the structure of higher-point integrated negative geometries. To that end, the six- and eight-point integrands presented in \cite{He:2023rou} could serve as a starting point. It would be worth exploring if some of the properties found for the four-particle case, such as the sign patterns and the apparent simplicity of the leading singularities, also extend to the higher-multiplicity integrated results. Furthermore, an interesting problem to study would be the description of the leading singularities of the $n$-point integrated negative geometries in terms of a Grassmannian formula \cite{Arkani-Hamed:2009ljj,Arkani-Hamed:2012zlh,Huang:2013owa,Huang:2014xza}, as done in \cite{Chicherin:2022bov} for an arbitrary number of particles and in the context of the ${\cal N}=4$ sYM theory.

Another interesting direction to explore is the computation of the last loop integral that steps in the way between the integrated negative geometries and the logarithm of the corresponding scattering amplitude. Starting from our $L=4$ result, that integration would give the four-loop four-particle scattering amplitude of ABJM, which is currently unknown.

Finally, let us conclude by mentioning an exciting question that merges two of the main topics that have been explored in this thesis, namely the study of integrability methods and the analysis of integrated negative geometries. As discussed for example in \cite{Alday:2010vh,Basso:2013vsa,Basso:2013aha,Basso:2014koa,Basso:2014nra}, the integrability properties of ${\cal N}=4$ super Yang-Mills manifest also at the level of the theory’s scattering amplitudes, providing a decomposition of these quantities in terms of building blocks known as pentagon transitions. A similar analysis was initiated for the ABJM theory in \cite{Basso:2018tif}. It would be interesting to explore whether integrability also emerges at the level of the integrated negative geometries, both in the ${\cal N}=4$ super Yang-Mills and ABJM theories.

%% file: Scalar_and_spinor_fields_in_AdSCFT.tex
Let us briefly discuss the behavior of free scalar and spinor fields in the AdS/CFT context. A free scalar field $\phi$ of mass $m_B^2$ in EAdS$_{D+1}$ satisfies the Klein-Gordon equation
\begin{equation}
\label{kg}
(\Box - m_B^2) \; \phi =0 \,, \qquad \text{with} \qquad \Box= \frac{1}{\sqrt{g}} \partial_{A}(\sqrt{g} g^{AB}\partial_{B}) \,,
\end{equation}
where the notation $A,B=0,1, \dots, D$ is used for the AdS$_{D+1}$ indices. Working in Poincaré coordinates and in units where the radius of AdS is set to one
\begin{equation}
ds^2=\frac{1}{z^2} \left( dt^2+\sum_{i=1}^{D-1} dx_i^2+dz^2 \right) \,,
\end{equation}
one obtains that the solutions to the Klein-Gordon equation asymptotically behave as
\begin{equation}
\label{asympt sol KG}
\phi(z,x^{\mu}) = \alpha(x^{\mu}) \, z^{\Delta_{-}}+ \mathcal{O} (z^{\Delta_-+2}) +\beta(x^{\mu}) \, z^{\Delta_{+}} + \mathcal{O}(z^{\Delta_+ +2}) \,,
\end{equation}
where
\begin{equation}
\label{delta pm scalar}
\Delta_{\pm}:= \frac{D}{2} \pm \nu  \qquad \text{with} \qquad \nu:=\sqrt{\frac{D^2}{4}+m_B^2} \,,
\end{equation}
The $\alpha(x^{\mu})$ and $\beta(x^{\mu})$ functions in \eqref{asympt sol KG} satisfy the relation
\begin{equation}
\label{alpha beta KG}
\beta(k_\mu)= \frac{\Gamma\left( -\nu \right)}{\Gamma\left( \nu \right)} \left( \frac{k}{2} \right)^{2\nu} \alpha(k_{\mu}) \,,
\end{equation}
in Fourier space and depend on the boundary conditions that one imposes on the field. 

The path integral is performed over the space of fields that asymptotically behave as \eqref{asympt sol KG} but do not necessarily satisfy the on-shell constraint \eqref{alpha beta KG}, and it has to be supplemented with a boundary condition for the $\alpha$ and $\beta$ fields. In turn, to define a well-posed variational problem  (i.e. such that $\delta S=0$ for solutions to the Klein-Gordon equation that satisfy the desired boundary conditions) one has to add an appropriate boundary term to the action. Therefore, the on-shell action will be strongly dependent on the choice of boundary conditions for the field, and so will be the dual CFT correlators obtained according to the GKPW prescription (see \eqref{GKPW} and \eqref{GKPW 2} in Chapter \ref{ch:adscft}). As an example, one obtains that a scalar field of mass $m_B^2$ with regular boundary conditions
\begin{equation}
\label{Dirichlet bc scalar field}
\alpha(x^{\mu})=J(x^{\mu}) \,,
\end{equation}
gives
\begin{equation}
\big< \mathcal{O}(x_{1})\mathcal{O}(x_{2}) \big> \propto \frac{1}{\rvert x_1-x_2 \rvert^{2 \Delta_+}} \,,
\end{equation}
Consequently, such a scalar field in the gravity side is dual to a scalar field of scaling dimension $\Delta_+$ in the CFT \cite{Witten:1998qj,Gubser:1998bc}. Similarly, a field with the alternative boundary conditions
\begin{equation}
\label{Neumann bc scalar field}
\beta(x^{\mu})=J(x^{\mu}) \,,
\end{equation}
is dual to a scalar field of dimension $\Delta_-$ \cite{Klebanov:1999tb}, which is possible in the range of masses $-\tfrac{d^2}{4}\leq m_B^2 \leq 1-\tfrac{d^2}{4}$.

As for free spinor fields of mass $m_F$ in EAdS$_{D+1}$, the Dirac equation is
\begin{equation}
\label{diracsignaturegrav}
(\slashed{D}-m_F) \, \psi=0 \quad , \quad \slashed{D}=e^A_a \gamma^a (\partial_A + \frac{1}{2} \omega_A^{bc} \Sigma_{bc} ) \,,
\end{equation}
with $\Sigma_{bc}= \frac{1}{4} [\gamma_b,\gamma_c]$ and where $a,b= 0 ,\dots, D$. Asymptotically, solutions behave as
\begin{equation}
\label{diracasymptotic}
\psi (z,x^{\mu}) = z^{\frac{d}{2} - m_F} \alpha^{\psi} (x^{\mu}) + \mathcal{O}(z^{\frac{d}{2} - m_F+1}) + z^{\frac{d}{2} + m_F} \beta^{\psi} (x^{\mu}) + \mathcal{O}(z^{\frac{d}{2} + m_F+1}) \,,
\end{equation} 
with
\begin{equation}
\label{diracasymptotic 2}
P_{+} \alpha^{\psi}= \alpha^{\psi} \, \qquad \text{and} \qquad  P_{-} \beta^{\psi}= \beta^{\psi} \,,
\end{equation}
where $P_{\pm} :=  \frac{1}{2} (1 \mp \gamma^0)$ and
\begin{equation}
\label{on shell constraint spinors}
\beta^\psi (k_{\mu}) = i \frac{\gamma^{\mu} k_{\mu}}{k} \frac{\Gamma(1/2-m_F)}{\Gamma(1/2+m_F)} \bigg( \frac{k}{2} \bigg)^{2m_F} \; \alpha^\psi (k_{\mu}) \,.
\end{equation}
In this case the path integral is again performed over the space of fields that asymptotically behave as \eqref{diracasymptotic} with \eqref{diracasymptotic 2} but do not necessarily satisfy the on-shell constraint \eqref{on shell constraint spinors}. The usual choice of boundary conditions consists of fixing either $\alpha^{\psi}$ or $\beta^{\psi}$, which translates into correlators of dual spinor fields of scaling dimensions 
\begin{equation}
\label{spinor holographic dict}
\Delta_{+}^{\psi}=\frac{d}{2}+m_F \qquad   \text{or} \qquad \Delta_{-}^{\psi}=\frac{d}{2}-m_F \,,
\end{equation}
respectively \cite{Henneaux:1998ch}.

%% file: ads3cft2_killing_spinors.tex
In this appendix we will focus on the $AdS_3\times S^3_+ \times S^3_- \times S^1$ supergravity background discussed in Chapter \ref{ch: line defects in AdS3CFT2}, and we will provide a detailed computation of its Killing spinors.

We will use the coordinates \eqref{poincare}-\eqref{spherecoor}, for which the vielbeins can be taken to be
\begin{empheq}{alignat=9}
\label{vierbein1}
    e^0 & = \frac{L}{z}dt\,, &\qquad  e^1 & = \frac{L}{z}dx \,, &\qquad & e^2 = \frac{L}{z}dz\,,
    \\
e^3 & = \frac{L}{\sin \Omega}d\beta_+\,, &\qquad  e^4 & = \frac{L}{\sin \Omega}\cos\beta_+d\gamma_+ \,,&\qquad & e^5 = \frac{L}{\sin \Omega}\cos\beta_+\cos\gamma_+d\varphi_+\,,
    \\
 e^6 & = \frac{L}{\cos \Omega}d\beta_-\,, &\qquad  e^7 & = \frac{L}{\cos \Omega}\cos\beta_-d\gamma_- \,,&\qquad & e^8 = \frac{L}{\cos \Omega}\cos\beta_-\cos\gamma_-d\varphi_-\,,
   \\
 e^9 & = l d\theta\,. & & & & & &
 \label{vierbein4}
\end{empheq}
Then, the non-vanishing components of the spin connections turn out to be
\begin{empheq}{alignat=9}
    \omega^{20} & = \frac{1}{z}dt\,, &\qquad  \omega^{21} & = \frac{1}{z}dx \,, &\qquad &     
    \\
\omega^{34} & = \sin\beta_+d\gamma_+\,, &\qquad  \omega^{35} & = \sin\beta_+\cos\gamma_+d\varphi_+ \,,&\qquad & \omega^{45} = \sin\gamma_+d\varphi_+\,,
    \\
    \omega^{67} & = \sin\beta_-d\gamma_-\,, &\qquad  \omega^{68} & = \sin\beta_-\cos\gamma_-d\varphi_- \,,&\qquad & \omega^{78} = \sin\gamma_-d\varphi_-\,.
\end{empheq}

The conditions defining the Killing spinors can be conveniently presented in terms of the complex three-form field strength
\begin{align}
G_{(3)} &= - e^{-\Phi/2} H_{(3)} - i e^{\Phi/2} F_{(3)} \nonumber \\
 &= 2 i L^2 e^{-\Phi/2+i\vartheta}
\left({\rm vol}(AdS_3)+\frac{1}{\sin^2 \Omega} {\rm vol}(S^3_+)+ \frac{1}{\cos^2 \Omega} {\rm vol}(S^3_-) \right)
\,.   
\end{align}
More precisely, for a real representation of Dirac matrices $\Gamma_\mu = e^m_\mu \gamma_m$, a Killing spinor $\epsilon$ satisfies
\begin{align}
\Gamma^{\mu\nu\rho} G_{\mu\nu\rho}   \epsilon = & 0\,,
\label{killingcond1}
\\
D_\mu\epsilon +\frac{e^{\Phi/2}}{96}\left(9\Gamma^{\nu\rho}G_{\mu\nu\rho}  - \Gamma_\mu^{\ \, \nu\rho\sigma}G_{\nu\rho\sigma}\right)\epsilon^{*} = & 0\,,
\label{killingcond2}
\end{align}
where $D_\mu$ stands for the covariant derivative
\begin{equation}
D_\mu = \partial_\mu +\frac{1}{4} \omega^{mn}_\mu
\gamma_{mn}\,.
\end{equation}
Additionally, the Killing spinors are Weyl spinors obeying
\begin{equation}
\gamma_{11}\epsilon = -\epsilon\,,
\end{equation}
where $\gamma_{11} = \gamma^0\gamma^1\cdots\gamma^9$. Defining
\begin{equation}
\gamma_* =   \gamma^0\gamma^1\gamma^2\,,
\qquad
\gamma_*^+ =  i \gamma^3\gamma^4\gamma^5\,,
\qquad
\gamma_*^- =   i \gamma^6\gamma^7\gamma^8\,,
\end{equation}
and
\begin{equation}
    \label{sigmaprojector}
    P^\Sigma_{\pm} :=  \frac12(1 \pm \Sigma) \qquad \text{for}\qquad
     \Sigma := i\left(\sin \Omega \, \gamma_*\gamma_*^++\cos \Omega \, \gamma_*\gamma_*^-\right) \,
\end{equation}
the condition \eqref{killingcond1} becomes
\begin{equation}
   P^\Sigma_{-}\epsilon =0\,.
    \label{killingcond1bis}
\end{equation}

Since the condition \eqref{killingcond2} relates $\epsilon$ with $\epsilon^*$  it is convenient to separate the Killing spinor into
\begin{equation}
\epsilon = \eta + i \xi\,.
\label{etaxi}
\end{equation}

For the different values of $\mu$, and using the condition \eqref{killingcond1bis}, the equation \eqref{killingcond2}  becomes
\begin{empheq}{alignat=9}
\label{killeq1}
   \partial_t  \eta  -  \frac{1}{z} M_x \eta 
   -\frac{1}{z} \sin\vartheta M_t \eta +\frac{1}{z} \cos\vartheta M_t \xi
   & =0\,, \\
   \partial_t  \xi  -  \frac{1}{z} M_x \xi 
   +\frac{1}{z} \sin\vartheta M_t \xi +\frac{1}{z} \cos\vartheta M_t \eta
   & =0\,,     
    \\
    \label{killeq2}
    \partial_x  \eta  -  \frac{1}{z} M_t \eta 
   -\frac{1}{z} \sin\vartheta M_x \eta -\frac{1}{z} \cos\vartheta M_x \xi
   & =0\,,\\
   \partial_x  \xi  -  \frac{1}{z} M_t \xi 
   +\frac{1}{z} \sin\vartheta M_x \xi -\frac{1}{z} \cos\vartheta M_x \eta
   & =0\,,     
   \\
   \partial_z  \eta   
   -\frac{1}{z} \sin\vartheta M_z \eta +\frac{1}{z} \cos\vartheta M_z \xi
   & =0\,, \\
   \partial_z  \xi  
   +\frac{1}{z} \sin\vartheta M_z \xi +\frac{1}{z} \cos\vartheta M_z \eta
   & =0\,,  
   \label{killeq4}
\end{empheq}
and
\begin{empheq}{alignat=9}
   \label{killeq5}
   \partial_{\theta}  \eta   & =0\,, \\ 
   \label{killeq5-2}
   \partial_{\theta}  \xi  & =0\,,    \\
 \label{killeq6}
   \partial_{\beta_\pm}  \eta   
   -\sin\vartheta M_{\beta_\pm} \eta +\cos\vartheta M_{\beta_\pm} \xi
   & =0\,, \\
    \label{killeq6-2}
   \partial_{\beta_\pm}  \xi 
   + \sin\vartheta M_{\beta_\pm} \xi + \cos\vartheta M_{\beta_\pm} \eta
   & =0\,,  \\
   \label{killeq7}
   \partial_{\gamma\pm}  \eta   +\sin\beta_{\pm} M_{\varphi_{\pm}}\eta
   -\sin\vartheta \cos\beta_{\pm} M_{\gamma_\pm} \eta +\cos\vartheta \cos\beta_{\pm} M_{\gamma_\pm} \xi
   & =0\,, 
   \\
   \label{killeq8}
   \partial_{\gamma\pm}   \xi +\sin\beta_{\pm} M_{\varphi_{\pm}}\xi 
 +\sin\vartheta \cos\beta_{\pm} M_{\gamma_\pm} \xi + \cos\vartheta \cos\beta_{\pm} M_{\gamma_\pm} \eta
   & =0\,, 
   \\
   \label{killeq9}
     \partial_{\varphi\pm}  \eta   -\sin\beta_{\pm}\cos\gamma_{\pm} M_{\gamma_{\pm}}\eta
    +\sin\gamma_{\pm} M_{\beta_{\pm}}\eta
   -\sin\vartheta \cos\beta_{\pm}\cos\gamma_{\pm} M_{\varphi_\pm} \eta & \nonumber \\
   +\cos\vartheta \cos\beta_{\pm}\cos\gamma_{\pm} M_{\varphi_\pm} \xi
   & =0\,, 
   \\
   \label{killeq10}
        \partial_{\varphi\pm}  \xi   -\sin\beta_{\pm}\cos\gamma_{\pm} M_{\gamma_{\pm}}\xi
    +\sin\gamma_{\pm} M_{\beta_{\pm}}\xi
   +\sin\vartheta \cos\beta_{\pm}\cos\gamma_{\pm} M_{\varphi_\pm} \xi & \nonumber \\
   +\cos\vartheta \cos\beta_{\pm}\cos\gamma_{\pm} M_{\varphi_\pm} \eta
   & =0\,, 
   \end{empheq}
where the matrices $M$ are the ones defined in \eqref{Mtxz}-\eqref{Mminus}.

By solving the Killing equations \eqref{killeq1}-\eqref{killeq10} one by one, we will construct a factorized expression for the Killing spinors. We can start with equations \eqref{killeq5} and \eqref{killeq5-2}, which simply state that the Killing spinors are independent of $\theta$. Next, we consider the equations \eqref{killeq6} and \eqref{killeq6-2}. They can be disentangled by taking an additional derivative with respect to $\beta_\pm$, which gives
\begin{equation}
\partial_{\beta_\pm}^2 \eta = M_{\beta_\pm}^2 \eta\,,\qquad   
\partial_{\beta_\pm}^2 \xi = M_{\beta_\pm}^2 \xi\,.
\label{kill5derivated}
\end{equation}
Let us first solve for the $\beta_+$ dependence. In order to fulfill \eqref{kill5derivated} we need
\begin{align}
\eta & = e^{\beta_+ M_{\beta_+}} a + e^{-\beta_+ M_{\beta_+}} b \,,
\\
\xi & = e^{\beta_+ M_{\beta_+}} c + e^{-\beta_+ M_{\beta_+}} d 
\,.
\end{align}
Replacing in \eqref{killeq6} and \eqref{killeq6-2} we find a relation between $a$ and $c$ and between $b$ and $d$. Defining $U_2 = e^{\beta_+ M_{\beta_+}}$ and $V_2 = e^{-\beta_+ M_{\beta_+}}$ we obtain
\begin{align}
\eta & = U_2 \epsilon_0^{(2)} + \frac{\cos\vartheta}{\sin\vartheta +1} V_2 \epsilon_1^{(2)} \,,
\\
\xi & =-\frac{\cos\vartheta}{\sin\vartheta +1} U_2 \epsilon_0^{(2)} + V_2 \epsilon_1^{(2)} 
\,,
\end{align}
where $\epsilon_0^{(2)}$ and $\epsilon_1^{(2)}$ are not constant but independent of $\theta$ and $\beta_+$. In the same way we can deal with the $\beta_-$ dependence. Now we define 
$U_3 = e^{\beta_+ M_{\beta_+}}e^{\beta_- M_{\beta_-}}$ and $V_3 = e^{-\beta_+ M_{\beta_+}}e^{-\beta_- M_{\beta_-}}$, and then
\begin{align}
\eta & = U_3 \epsilon_0^{(3)} + \frac{\cos\vartheta}{\sin\vartheta +1} V_3 \epsilon_1^{(3)} \,,
\\
\xi & =-\frac{\cos\vartheta}{\sin\vartheta +1} U_3 \epsilon_0^{(3)} + V_3 \epsilon_1^{(3)} 
\,.
\end{align}
It is straightforward to extend this to account completely for the dependence on the compact space coordinates, obtaining
\begin{empheq}{alignat=9}
\eta & = U_7 \epsilon_0^{(7)}(t,x,z) + \frac{\cos\vartheta}{\sin\vartheta +1} V_7 \epsilon_1^{(7)}(t,x,z) \,,
\\
\xi & =-\frac{\cos\vartheta}{\sin\vartheta +1} U_7 \epsilon_0^{(7)}(t,x,z) + V_7 \epsilon_1^{(7)}(t,x,z) 
\,,
\end{empheq}
where
\begin{align}
 U_7 & = e^{\beta_+ M_{\beta_+}+\beta_- M_{\beta_-}}
 e^{\gamma_+ M_{\gamma_+} +\gamma_- M_{\gamma_-}}
 e^{\varphi_+ M_{\varphi_+}+\varphi_- M_{\varphi_-}}\,, \nonumber \\
 V_7 & = e^{-\beta_+ M_{\beta_+}-\beta_- M_{\beta_-}}
 e^{-\gamma_+ M_{\gamma_+}-\gamma_- M_{\gamma_-}}
 e^{-\varphi_+ M_{\varphi_+}-\varphi_- M_{\varphi_-}}\,.
 \label{UyV}
\end{align}

Finally, one can solve for the remaining eqs. \eqref{killeq1}-\eqref{killeq4} to get
\begin{empheq}{alignat=9}
\label{eps07}
\epsilon_0^{(7)}(t,x,z)  & = e^{\log z M_z}e^{(x+t)(M_t+M_x)}\epsilon_0  \nonumber \\
&= \frac12 \left[\frac{1}{\sqrt{z}}(1-2M_z)\left(1+2(x+t)M_t\right)+\sqrt{z}(1+2M_z)\right]\epsilon_0
\,, 
\\
\label{eps17}
\epsilon_1^{(7)}(t,x,z)  & = e^{-\log z M_z}e^{(x-t)(M_t-M_x)}\epsilon_1 \nonumber \\
&= \frac12 \left[\frac{1}{\sqrt{z}}(1+2M_z)\left(1+2(x-t)M_t\right)+\sqrt{z}(1-2M_z)\right]\epsilon_1
\,.
\end{empheq}

%% file: quadratic_fluctuations.tex
In this appendix we will focus on the fluctuations around the classical string solution presented in eq. \eqref{susyconfiguration}. The discussion provided here complements the results of Chapter \ref{ch: line defects in AdS3CFT2}.

\subsubsection*{Bosonic fluctuations}

The classical string presented in \eqref{susyconfiguration} emerges as a saddle-point solution of the Nambu-Goto
action for the bosonic coordinates of a string coupled to a NS-NS $B$-field
\begin{equation}
\label{bosonic action}
S_B=-\frac{L^2}{2\pi\alpha'}\int d^2 \sigma \; \sqrt{-h} + \frac{L^2}{4\pi\alpha'} \int d^2 \sigma \; \epsilon^{\alpha \beta} B_{\alpha\beta}\,.
\end{equation}
Here $h_{\alpha\beta}$ and $B_{\alpha\beta}$ are the induced metric and the pullback of the NS-NS $B$-field on the worldsheet, respectively. Note that in \eqref{bosonic action} we have factored out the AdS radius $L$ in order to define $g:=\tfrac{L^2}{2\pi\alpha'}$, where $g^2$ is a parameter proportional to the 't Hooft coupling in the dual CFT$_2$.  Thus, we can implement a semiclassical expansion by taking the large $g$ limit and organizing terms according to their inverse powers of $g$. For the $B$-field we will use the gauge
\begin{align}
    \label{b field gauge}
    B= 2 \sin &\vartheta \,\bigg( \frac{t}{z^3} \; dx\wedge dz+ \frac{\varphi_+}{\sin^2 \Omega} \; \cos^2 \beta_+ \; \cos \gamma_+ \; d\beta_+ \wedge d\gamma_+ \nonumber \\
     & \quad \, + \frac{\varphi_-}{\cos^2 \Omega} \; \cos^2 \beta_- \; \cos \gamma_- \; d\beta_- \wedge d\gamma_- \bigg)\,.
\end{align}

Making the ansatz 
\begin{equation}
t = \omega\tau\,,\qquad
x = x(\sigma)\,,\qquad 
z = \sigma\,,\qquad 
\beta_\pm,\gamma_\pm,\varphi_\pm,\theta = \text{const.}
\label{ansatzapp}
\end{equation}
the equations of motions reduce to
\begin{equation}
2\left(1+(x')^2\right)\left(
x' +\sin\vartheta \sqrt{1+(x')^2}\right)
-\sigma x'' = 0\,,
\end{equation}
which is straightforwardly solved by $x' = -\tan\vartheta$.

Let us now turn to the effective action for the bosonic fluctuations. The standard approach would be to perform a series expansion of the action \eqref{bosonic action} in powers of $\delta X^{\mu}:={X}^{\mu}-X^{\mu}_{\sf cl}$, with $X^{\mu}_{\sf cl}$ the classical solution.
However, and because the coefficients of such expansion will not be manifestly covariant under reparametrizations \cite{Drukker:2000ep,Forini:2015mca}, it is more convenient to make the expansion in terms of Riemann normal coordinates. Let $X^{\mu}(q)$ be a geodesic such that $X^{\mu}(0)=X^{\mu}_{\sf cl}$. Using the geodesic equation we get
\begin{equation}
    \label{geodesic eq solution}
    \begin{aligned}
    X^{\mu}(q) &= X^{\mu}_{\sf cl}+ \frac{1}{g} q \zeta^{\mu}-\frac{1}{g^2}\frac{q^2}{2} \Gamma^{\mu}_{\nu \sigma} \zeta^{\nu} \zeta^{\sigma} + \frac{1}{g^3}\frac{q^3}{6} \left( 2 \Gamma^{\mu}_{\nu \kappa} \Gamma^{\kappa}_{\rho \sigma}-\partial_{\nu} \Gamma^{\mu}_{\rho \sigma} \right) \zeta^{\nu} \zeta^{\rho} \zeta^{\sigma} + \\
    &\quad + \frac{1}{g^4}\frac{q^4}{24} \bigg( 2 \Gamma^{\kappa}_{\rho \sigma} \partial_{\alpha} \Gamma^{\mu}_{\beta \kappa} + 2 \Gamma^{\mu}_{\alpha \kappa} \partial_{\beta} \Gamma^{\kappa}_{\rho \sigma}- 2 \Gamma^{\mu}_{\kappa \vartheta} \Gamma^{\kappa}_{\alpha \beta} \Gamma^{\vartheta}_{\rho \sigma} -4 \Gamma^{\mu}_{\alpha \kappa} \Gamma^{\kappa}_{\beta \vartheta} \Gamma^{\vartheta}_{\rho \sigma} - \partial_{\alpha} \partial_{\beta} \Gamma^{\mu}_{\rho \sigma} +  \\
  & \quad + \Gamma^{\kappa}_{\rho \sigma} \partial_{\kappa} \Gamma^{\mu}_{\alpha \beta}  + 2 \Gamma^{\kappa}_{\rho \sigma} \partial_{\alpha} \Gamma^{\mu}_{\kappa \beta} \bigg) \zeta^{\alpha} \zeta^{\beta} \zeta^{\rho} \zeta^{\sigma} + \mathcal{O}(\tfrac{1}{g^5})\,,
    \end{aligned}
\end{equation}
where the Christoffel symbols are to be evaluated in the classical solution $X^{\mu}_{\sf cl}$ and we have defined
\begin{equation}
    \label{rho coord riemann}
   \frac{1}{g} \zeta^{\mu}:=\left.\frac{dX^{\mu}}{dq}\right|_{q=0}\,.
\end{equation}
If we now consider $X^{\mu} = X^{\mu}(1)$, \eqref{geodesic eq solution} serves as an expansion of $X^\mu$ around $X^\mu_{\sf cl}$ in powers of a vector $\zeta^{\mu}$. Moreover, in order to get an expansion in terms of scalar fields, it is useful to define 
\begin{equation}
    \label{scalar phi}
    \phi^m= e^{m}_{\mu} \zeta^{\mu}\,,
\end{equation}
where $e^a_{\mu}$ are the inverse of the background-space vielbeins given in \eqref{vierbein1}-\eqref{vierbein4}. 

The effective action arises from substituting this expansion into \eqref{bosonic action}. Up to some boundary terms which can be discarded by appropriate counterterms, we get
\begin{equation}
\label{bosonic action expansion}
S_B= g S^{(0)}+S^{(2)}+ \mathcal{O}(g^{-1/2})
\end{equation}
where $S^{(0)}$ is the Nambu-Goto action evaluated in the classical solution and where the quadratic action $S^{(2)}$ for the eight transverse modes is
\begin{align}
\label{bosonic action expansion-quadratic}
S^{(2)}&= -\frac{1}{2} \int d^2 \sigma \; \sqrt{-h_{\sf cl}} \left[ \partial_{\alpha} \phi^{\sf tr} \partial^{\alpha} \phi^{\sf tr} + 2 \cos^2 \vartheta (\phi^{\sf tr})^2 + \sum_{j=3}^9 \left(  \partial_{\alpha} \phi^{j} \partial^{\alpha} \phi^{j} \right) \right]\,,
\end{align}
where $(h_{\alpha\beta})_{\sf cl}$ stands for the induced metric evaluated in the classical solution (in what follows we will refer to $(h_{\alpha\beta})_{\sf cl}$ simply as $h_{\alpha\beta}$). Moreover, we have introduced the notation $\phi_{\sf tr}$ for the transverse fluctuation in $AdS_3$, defined as
\begin{equation}
    \label{transverse fluctuation}
    \phi_{\sf tr} (\tau,\sigma):= \cos \vartheta \; \phi^1(\tau,\sigma) +\sin \vartheta\; \phi^2(\tau,\sigma)\,,
\end{equation}
and we are using 
\begin{align}
\label{definition phi su2+}
\phi^{a}_{b}  &:= \phi^{3} \left( \sigma^1 \right)^{a}_{b}  + \phi^{4} \left( \sigma^2 \right)^{a}_{b}  - \phi^{5} \left( \sigma^3 \right)^{a}_{b} \,, \\
\label{definition phi su2-}
\phi^{\dot{a}}_{\dot{b}} &:= \phi^{6} \left( \sigma^1 \right)^{\dot{a}}_{\dot{b}} + \phi^{7} \left( \sigma^2 \right)^{\dot{a}}_{\dot{b}} - \phi^{8} \left( \sigma^3 \right)^{\dot{a}}_{\rm \dot{b}}\,.
\end{align}
for the three-sphere fluctuations. We see from \eqref{bosonic action expansion-quadratic} that the spectrum of bosonic fluctuations consists in one massive (with mass $m_{\sf tr}^2=\tfrac{2}{R^2}$) and seven massless scalar fields.

\subsubsection*{Fermionic fluctuations}

We will now turn to the analysis of fermionic fluctuations, which we will study in the Green-Schwarz formalism. In order to simplify the calculations, as customary for open strings in type IIB backgrounds \cite{Drukker:2000ep} we will fix the $\kappa$-symmetry gauge condition to be $\Theta^1=\Theta^2:=\Theta$, where $\Theta^1$ and $\Theta^2$ are the fermionic fluctuations. Then, up to quadratic order in the fluctuations the action reads \cite{Cvetic:1999zs}
\begin{equation}
    \label{fermionic action}
    S_F= S_{F1}+S_{F2}+S_{F3}\,,
\end{equation}
with 
\begin{align}
    \label{SF1}
    S_{F1} &:=-{2 i} 
\int d^2 \sigma \sqrt{-h} \; h^{\alpha\beta} \bar{\Theta} \Gamma_{\alpha}  {\cal D}_{\beta}  \Theta\,,    \\
      \label{SF2}
    S_{F2} &:=-\frac{i e^{\Phi}}{24}  \int d^2 \sigma \sqrt{-h} \; h^{\alpha\beta}  \; \bar{\Theta} \Gamma_{\alpha}  \Gamma^{\delta \rho \eta} F_{\delta \rho \eta} \Gamma_{\beta} \Theta = 
    i \cos \vartheta \, \int d^2 \sigma \sqrt{-h} \; \bar{\Theta} \gamma_* P_{\Sigma} \Theta
    \,, \\
     \label{SF3}
    S_{F3} &:=-\frac{i}{2} \int d^2 \sigma  \; \epsilon^{\alpha\beta} \partial_{\alpha} X^{\mu} \partial_{\beta} X^{\nu} \; \bar{\Theta} \Gamma_{\mu}\;^{\rho \eta} H_{\nu\rho\eta}  \Theta = 0\,,
\end{align}
where 
\begin{equation}
    \label{string covariant derivative}
\Gamma_{\alpha}:= \partial_{\alpha} X^{\mu} \Gamma_{\mu}\,, \qquad
{\cal D}_{\alpha}:=\partial_{\alpha} + \frac{1}{4} \tilde\omega_{\alpha}^{ab} {\tilde\gamma}_{ab}\,,
\end{equation}
with ${\tilde\gamma}^{{a}}:=\tilde{e}^{{a}}_{\alpha} \Gamma^{\alpha}$, for $\tilde{e}^{{a}}_{\alpha}$ and $\tilde\omega_{\alpha}^{ab}$ the vielbeins and spin connections of $AdS_2$. Further splitting $\Theta$ as
\begin{equation}
    \Theta = \Theta_0 +\Theta_1\,, \qquad\text{for}\qquad
P^\Sigma_-\Theta = \Theta_0\,,\quad P^\Sigma_+\Theta =\Theta_1\,,
\label{Sigmaprojections}
\end{equation}
the total action for the fermionic fluctuations becomes
\begin{equation}
    \label{fermionic action-v2}
    S_F= -2i \int d^2 \sigma \sqrt{-h} \;\bar{\Theta}_0  \Gamma^{\alpha}  {\cal D}_{\alpha}  \Theta_0+ \bar{\Theta}_1 \left( \Gamma^{\alpha}  {\cal D}_{\alpha} - \cos \vartheta \, \gamma_* \right) \Theta_1\,.
\end{equation}

\subsubsection*{Supersymmetry of the fluctuations}

The Green-Schwarz action is invariant under the supersymmetry transformations
\begin{equation}
\delta_{\sf susy} X^\mu = \bar\eta \Gamma^\mu \Theta^{1}
+\bar\xi \Gamma^\mu \Theta^{2}
\,,\qquad
\delta_{\sf susy} \Theta^1 = \eta
\,,\qquad
\delta_{\sf susy} \Theta^2 = \xi\,,
\end{equation}
for $\eta$ and $\xi$ the real and imaginary parts of the Killing spinors \eqref{etaxi}. It is also invariant under the $\kappa$-symmetry transformations
\begin{equation}
\delta_{\kappa} X^\mu = \bar\Theta^1\Gamma^\mu P^\kappa_+\kappa^1
+\bar\Theta^2 \Gamma^\mu P^\kappa_-\kappa^2
\,,\qquad
\delta_{\kappa} \Theta^1 = P^\kappa_+\kappa^1
\,,\qquad
\delta_{\kappa} \Theta^2 = P^\kappa_-\kappa^2\,,
\end{equation}
where the projectors
\begin{equation}
P^\kappa_\pm = \frac12\left(1\pm \tilde\Gamma\right) \,,
\qquad 
\tilde\Gamma = -\frac{\epsilon^{\alpha\beta}\Pi^\mu_\alpha\Pi^\nu_\beta\Gamma_{\mu\nu}}{2\sqrt{-H}}\,,
\end{equation}
are defined in terms of
\begin{equation}
\Pi^\mu_\alpha = \partial_\alpha X^\mu -  \bar\Theta^1 \Gamma^\mu \partial_\alpha\Theta^{1}-  \bar\Theta^2 \Gamma^\mu \partial_\alpha\Theta^{2}\,,
\qquad
H_{\alpha\beta} = g_{\mu\nu} \Pi^\mu_\alpha \Pi^\nu_\beta\,.
\end{equation}
Whenever a combined transformation
$\delta:=\delta{\sf susy}+\delta_\kappa$ leaves a configuration invariant, we say it is supersymmetric. A classical configuration with vanishing fermions is then supersymmetric for $\kappa_1 = -\eta$ and $\kappa_2 = -\xi$ if
\begin{equation}
 P^\kappa_- \eta = 0\,, \qquad P^\kappa_+ \xi = 0\,,
\label{kappapro}
\end{equation}
or, equivalently, if
\begin{equation}
\Gamma\epsilon = \epsilon\,,    
\label{kappapro0}
\end{equation}
with
\begin{equation}
    \Gamma = -\frac{\partial_{\tau} X^\mu \partial_{\sigma} X^\nu\Gamma_{\mu\nu}}{\sqrt{h}}K\,.
\end{equation}
and where $K$ indicates complex conjugation.

As we show in Chapter \ref{ch: line defects in AdS3CFT2}, the string configuration \eqref{susyconfiguration} is supersymmetric. In what follows we shall derive the supersymmetry transformations for the fluctuations around it. The preservation of the $\kappa$-symmetry gauge condition requires $\delta \Theta^1 = \delta \Theta^2$ and from this, labelling with $(0), (1), \ldots $ the successive orders of the expansion in powers of the fluctuations, we get
\begin{equation}
\eta_{(1)} + P_{+(0)}^{\kappa} \kappa^1_{(1)} -  P_{+(1)}^{\kappa} \eta_{(0)}  =
\xi_{(1)} + P_{ -(0)}^{\kappa} \kappa^2_{(1)} -  P_{ -(1)}^{\kappa}  \xi_{(0)}\,.
\end{equation}

For the variation of the bosonic transverse fluctuations we obtain\footnote{ The total transformation must preserve the static gauge choice, which requires $\delta X^0=\delta X^{\sf lg}=0$. In order to achieve this, a diffeomorphism should be added to the total transformation. 
As the transformation should be first order in the fluctuations, $\delta_{\sf diff} X^{\mu}=-\partial_{\alpha} X^{\mu}_{\sf cl} \delta \sigma^{\alpha}$. Then, this additional transformation does not affect the transverse fluctuations and $\delta\sigma^\alpha$ can be chosen so that $\delta X^0=\delta X^{\sf lg}=0$. Similarly, $\delta_{\sf diff} \Theta=-\partial_{\alpha} \Theta_{\sf cl} \, \delta \sigma^{\alpha}=0$ and so the diffeomorphism does not modifies the transformation of the fermionic fluctuations either. }
\begin{align}
\label{susy bosonic fluctuations}
\delta\phi^{\sf tr} &=    -2 \bar\Theta \gamma^{\sf tr}\left(\eta_{(0)}+\xi_{(0)}\right)\,, \\
\delta\phi^a &=    -2 \bar\Theta \gamma^a\left(\eta_{(0)}+\xi_{(0)}\right)\,, \qquad a=3, \dots, 9
\end{align}
while for the fermionic fluctuations we get
\begin{equation}
\delta\Theta = \frac{1}{2}\left(\eta_{(1)}+ \xi_{(1)} \right) -
\frac{1}{2}\tilde\Gamma_{(0)}\left(\eta_{(1)} - \xi_{(1)} \right)-
\frac{1}{2}\tilde\Gamma_{(1)}\tilde\Gamma_{(0)}\left(\eta_{(0)} + \xi_{(0)} \right)\,,
\label{deltaTheta}
\end{equation}
where
\begin{align}
\tilde\Gamma_{(0)} & = -2\sin\vartheta M_z -2 \cos\vartheta M_x := -2 M_{\sf tr}\,, 
\\
\tilde\Gamma_{(1)} & = -2 \cos \vartheta\left[ 
\sigma \partial_\tau \phi^{\sf tr} M_t -\left(\phi^{\sf tr}
+\sigma\partial_\sigma \phi^{\sf tr}\right)M_{\sf lg}
+\frac{\sigma}{2} \sum_{a=3}^9\gamma^a
\left(\partial_\tau\phi^a \gamma^{\sf lg}+\partial_\sigma\phi^a \gamma^{0}\right)
\right]\,,
\label{Gamma1a}
\end{align}
with
\begin{equation}
M_{\sf lg} = \cos\vartheta M_z - \sin\vartheta M_x \,.     
\end{equation}
For getting \eqref{deltaTheta} we have used that
\begin{equation}
\eta_{(0)} - \xi_{(0)} = \tilde\Gamma_{(0)}\left(\eta_{(0)} + \xi_{(0)} \right) \,.
\end{equation}

Using this relation, it is also possible to express all the dependence of $\delta\Theta$ on the Killing spinors through $\eta_{(0)} + \xi_{(0)}$, since
\begin{align}
\eta_{(1)} + \xi_{(1)} =  &  -\bigg[\bigg( \phi^{\sf tr}M_{\sf tr}+
\sin \Omega\sum_{a=3}^5\phi^a M_a+\cos \Omega\sum_{a=6}^8\phi^a M_a
\bigg)(\cos\vartheta-\sin\vartheta\tilde\Gamma_{(0)}) \nonumber \\
& \qquad \quad \, -\cos\vartheta \phi^{\sf tr} M_t
\bigg](\eta_{(0)} + \xi_{(0)})\,, \nonumber
\\
\eta_{(1)} - \xi_{(1)} =  &
 \bigg[\bigg(\phi^{\sf tr}M_{\sf tr}+
\sin \Omega\sum_{a=3}^5\phi^a M_a+\cos \Omega\sum_{a=6}^8\phi^a M_a
\bigg)(\sin\vartheta+\cos\vartheta\tilde\Gamma_{(0)}) \nonumber \\
& \quad \quad  +\cos\vartheta \phi^{\sf tr} M_t \tilde{\Gamma}_{(0)}
\bigg](\eta_{(0)} + \xi_{(0)})\,, \nonumber
\end{align}
Thus, splitting into the variation of massless and massive fermions, taking into account that $\Theta_0 =P^\Sigma_-\Theta$,  $\Theta_1 =P^\Sigma_+\Theta$ and that $P^\Sigma_-(\eta_{(0)} + \xi_{(0)}) = 0$, we have
\begin{align}
\delta\Theta_0 = &
\left(-\cos \Omega\sum_{a=3}^5\Gamma^\alpha \partial_\alpha\phi^a M_a+\sin \Omega\sum_{a=6}^8 \Gamma^\alpha \partial_\alpha\phi^a M_a
+\tfrac{1}{2}\Gamma^\alpha \partial_\alpha\phi^9\right)\gamma^9(\eta_{(0)} + \xi_{(0)})\,,
\\
\delta\Theta_1 = & \bigg(\tfrac{1}{2}\Gamma^\alpha\partial_\alpha\phi^{\sf tr}\gamma^{\sf tr}
-\cos \vartheta \, \phi^{\sf tr}M_{\sf tr}
+
\sin \Omega\sum_{a=3}^5\left(\Gamma^\alpha \partial_\alpha\phi^a\gamma_*-\cos \vartheta \, \phi^a \right)M_a \\
& \quad +\cos \Omega\sum_{a=6}^8\left(\Gamma^\alpha \partial_\alpha\phi^a\gamma_*-\cos \vartheta \, \phi^a \right)M_a
\bigg)(\eta_{(0)} + \xi_{(0)}). \nonumber
\end{align}

\subsubsection*{Fluctuations as AdS$_2$ fields}

In the previous subsection, fermionic fluctuations and Killing spinors were treated as 32-component spinors. We can alternatively describe the quadratic fluctuations and their supersymmetry transformations in terms of two-component $AdS_2$ spinors. In order to do that, it is necessary to identify longitudinal and transverse directions (to the worldsheet) and decompose the Dirac matrices into products are $SO(1,1)$ and $SO(8)$ matrices:
\begin{align}
\gamma^0    & = \uptau^0\otimes \mathbb{1}_{16 \times 16}\,,
\\
\gamma^{\sf lg} & = -\sin\vartheta \gamma^1+\cos\vartheta \gamma^2 =
\uptau^1\otimes \mathbb{1}_{16 \times 16}\,,
\\
\gamma^{\sf tr}  & =  \cos\vartheta \gamma^1+\sin\vartheta \gamma^2 =
\uptau_3\otimes \uprho^8\,,
\\
\gamma^{m}  & = \uptau_3\otimes \uprho^{m-2}\,,\qquad\text{for } m\geq 3\,,
\end{align}
where $\uptau^a$ and $\uprho^A$ are $SO(1,1)$ and $SO(8)$ Dirac matrices. Then, the 32-component spinors can be written as $\Theta = \psi(\tau,\sigma)\otimes\varphi$, where $\psi(\tau,\sigma)$ are two-component worldsheet spinors and $\varphi$ are constant 16-component spinors. All these spinors can be accommodated into representations of the $SU(2)$ rotations that leave invariant the $\Sigma$ projections \eqref{Sigmaprojections}. Thus,
\begin{equation}
\Theta_0 = \psi_{+}^{ a\dot{a}}\otimes  \varphi_{{ a\dot{a}}}^{++}+\psi_{-}^{a\dot{a}}\otimes  \varphi_{{ a\dot{a}}}^{--}\,,\qquad
\Theta_1 = \chi_{+}^{ a\dot{a}}\otimes  \varphi_{{ a\dot{a}}}^{+-} +\chi_{-}^{ a\dot{a}}\otimes  \varphi_{{ a\dot{a}}}^{-+}\,,\qquad
\end{equation}
where ${ a}$ and $\dot{ a}$ are fundamental $SU(2)$ indices and
\begin{align}
\uptau_3\psi_\pm = \pm \psi_\pm\,,\qquad
    \uprho_9\varphi^{\pm s} = \pm \varphi^{\pm s}\,,\qquad 
    \upsigma\varphi^{s\pm} = \pm \varphi^{s\pm}\,,
\end{align}
for
\begin{equation}
\uptau_3 = \, \uptau^0\uptau^1\,,\qquad 
\uprho_9 = \uprho^1\cdots\uprho^8\,,\qquad
\upsigma = \sin \Omega \ \uprho^8\uprho^1\uprho^2\uprho^3+\cos \Omega \ \uprho^8\uprho^4\uprho^5\uprho^6\,.
\end{equation}

It is easy and convenient to use a basis
\footnote{In the need of an explicit basis we shall use
\begin{align}
    \uptau^0 = i\sigma^2\,,\qquad \uptau^1 = \sigma^1 \,,
    \nonumber
\end{align}
and
\begin{empheq}{alignat=9}
    \uprho^1 &= \sigma^2\otimes\sigma^2\otimes\sigma^1\otimes \mathbb{1}_{2 \times 2}\,, &\qquad 
    \uprho^2 &= \sigma^2\otimes\sigma^2\otimes\sigma^3\otimes \mathbb{1}_{2 \times 2}\,, &\qquad
    \uprho^3 &= \sigma^2\otimes\sigma^1\otimes \mathbb{1}_{2 \times 2} \otimes\sigma^2\,, \nonumber\\
    \uprho^4 &= \sigma^2\otimes\sigma^3\otimes \mathbb{1}_{2 \times 2} \otimes\sigma^2\,, &\qquad
    \uprho^5 &= \sigma^2\otimes \mathbb{1}_{2 \times 2}\otimes\sigma^2\otimes\sigma^1\,,  &\qquad 
    \uprho^6 &= \sigma^2\otimes \mathbb{1}_{2 \times 2}\otimes\sigma^2\otimes\sigma^3\,,  \nonumber\\
    \uprho^7 &= \sigma^1\otimes\mathbb{1}_{2 \times 2}\otimes\mathbb{1}_{2 \times 2}\otimes \mathbb{1}_{2 \times 2}\,,  &\qquad 
    \uprho^8 &= -\sigma^3\otimes\mathbb{1}_{2 \times 2}\otimes\mathbb{1}_{2 \times 2}\otimes \mathbb{1}_{2 \times 2}\,.
    \nonumber
\end{empheq}}
 in which the constant spinors satisfy
\begin{align}
    (\varphi^\dagger)^{{ a\dot{a}}}_{\pm\pm}\cdot\varphi_{{ b\dot{b}}}^{\pm\pm} = \frac14\delta^{ a}_{ b}
    \delta^{ \dot a}_{\dot b}\,,
    \qquad
(\varphi^\dagger)^{{ a\dot{a}}}_{\pm\pm}\cdot\uprho^8\cdot\varphi_{{ b\dot{b}}}^{\mp\mp} = \frac14\delta^{ a}_{ b}
    \delta^{ \dot a}_{ \dot b}\,, \\
    (\varphi^\dagger)^{{ a\dot{a}}}_{\pm\mp}\cdot\varphi_{{ b\dot{b}}}^{\pm\mp} = \frac14\delta^{ a}_{ b}
    \delta^{ \dot a}_{ \dot b}\,,
    \qquad
(\varphi^\dagger)^{{ a\dot{a}}}_{\pm\mp}\cdot\uprho^8\cdot\varphi_{{ b\dot{b}}}^{\mp\pm} = \frac14\delta^{ a}_{ b}
    \delta^{ \dot a}_{ \dot b}\,,
\end{align}
and 0 otherwise.

Substituting in \eqref{fermionic action-v2}, the action for the fermionic fluctuations  becomes
\begin{equation}
    \label{fermionic action-v2d2}
     S_F= -\frac{i}2\! \int\! d^2 \sigma \sqrt{-h} \left(\bar{\psi}_{{ a}\dot{ a}} \ \slash\!\!\!\!{\cal D}^{(2)}  \psi^{{ a}\dot{ a}}  +
    \bar{\chi}_{{ a}\dot{ a}} \left( \slash\!\!\!\!{\cal D}^{(2)}
    -\cos \vartheta \right)\chi^{{ a}\dot{ a}}\right)\,,
\end{equation}
where
\begin{equation}
   \psi^{{ a}\dot{ a}} = \psi_{+}^{{ a}\dot{ a}}+ \psi_{-}^{{ a}\dot{ a}}\,,
   \qquad
   \chi^{{ a}\dot{a}} = \chi_{+}^{{ a}\dot{ a}}+ \chi_{-}^{{ a}\dot{ a}}\,,
\end{equation}
and $^{(2)}$ indicates that Dirac matrices and covariant derivatives are now those of $AdS_2$. 
This is the action of four massless and four massive (with mass $\tfrac{1}{R}$) worldsheet fermions. 

We would like to also have expressions for the supersymmetry transformations in terms of $AdS_2$ spinors. For this we would need the decomposition of $\eta_{(0)} + \xi_{(0)}$ as well.
 In terms of the matrices $U$ and $V$ \eqref{etadef}-\eqref{xidef} we can write
\begin{equation}
\eta_{(0)} + \xi_{(0)} = \left(1-p(\vartheta)\right) U_{(0)}\epsilon_0
+(1+p(\vartheta)) V_{(0)}\epsilon_1\,,
\qquad{\rm for}\  p(\vartheta) = \frac{\cos\vartheta}{\sin\vartheta + 1}\,.
\end{equation}
Evaluating \eqref{eps07}-\eqref{eps17} on the classical solutions, using the relation \eqref{susy constraint dirichlet}  and defining
\begin{equation}
    \varepsilon(\tau) = \frac{1}{2(\sin\vartheta+1)}\left(1+2M_{\sf lg}\right)\left(1+2M_{\sf tr}\right)(1+2M_z)\left[1+2(x_0-\tau\sec\vartheta )M_t\right]\epsilon_1\,,
\end{equation}
we get
\begin{equation}
 \eta_{(0)} + \xi_{(0)} =  \sigma^{-1/2} \varepsilon
 +\sigma^{1/2} 2M_t \dot\varepsilon\,.
\end{equation}
It is useful to note that
\begin{equation}
\label{properties varepsilon}
2M_{\sf lg} \varepsilon = \varepsilon \quad \text{and} \quad \ddot{\varepsilon}=0 \, .
\end{equation}
Then, writing $\eta_{(0)} + \xi_{(0)}$ and $\varepsilon$ as
\begin{equation}
\eta_{(0)} + \xi_{(0)}=\kappa_{+}^{ a\dot{a}}\otimes  \varphi_{{ a\dot{a}}}^{+-} +\kappa_{-}^{ a\dot{a}}\otimes  \varphi_{{ a\dot{a}}}^{-+} \qquad \text{and} \qquad \varepsilon=\varepsilon_{+}^{ a\dot{a}}\otimes  \varphi_{{ a\dot{a}}}^{+-} +\varepsilon_{-}^{ a\dot{a}}\otimes  \varphi_{{ a\dot{a}}}^{-+}
\end{equation}
we obtain
\begin{equation}
 \kappa^{ a\dot{a}} =  \sigma^{-1/2} \varepsilon^{ a\dot{a}} 
 +\sigma^{1/2} \uptau^0 \dot{\varepsilon}^{ a\dot{a}} 
\end{equation}
where
\begin{equation}
   \kappa^{{ a}\dot{ a}} = \kappa_{+}^{{ a}\dot{a}}+ \kappa_{-}^{{ a}\dot{ a}} 
   \qquad \text{and} \qquad
  \varepsilon^{{ a}\dot{a}} = \varepsilon_{+}^{{ a}\dot{ a}}+ \varepsilon_{-}^{{ a}\dot{ a}} \, .
\end{equation}
The properties \eqref{properties varepsilon} can now be expressed as
\begin{equation}
\label{properties varepsilon2}
\uptau^1 \varepsilon^{{ a}\dot{ a}} =- \varepsilon^{{ a}\dot{ a}} \quad \text{and} \quad \ddot{\varepsilon}^{{ a}\dot{ a}}=0 \, .
\end{equation}

Choosing a representation of $\uptau$ and $\uprho$ matrices such that the $\varphi$ spinors satisfy
\begin{align}
    &(\varphi^\dagger)^{{ a\dot{a}}}_{\pm\mp}\cdot \uprho^4 \cdot \uprho^5 \cdot \varphi_{{ b\dot{b}}}^{\pm\mp} = -\frac{i}{4} \delta^{\dot{a}}_{\dot{b}} \left( \sigma^1 \right)^{a}_{b}  \,,
    \quad
 \quad \quad \; (\varphi^\dagger)^{{ a\dot{a}}}_{\pm\mp}\cdot \uprho^8 \cdot \uprho^4 \cdot \uprho^5 \cdot \varphi_{{ b\dot{b}}}^{\mp\pm} = -\frac{i}{4} \delta^{\dot{a}}_{\dot{b}} \left( \sigma^1 \right)^{a}_{b}  \,, \nonumber \\
  &(\varphi^\dagger)^{{ a\dot{a}}}_{\pm\mp}\cdot \uprho^3 \cdot \uprho^5 \cdot \varphi_{{ b\dot{b}}}^{\pm\mp} = \frac{i}{4} \delta^{\dot{a}}_{\dot{b}} \left( \sigma^2 \right)^{a}_{b}  \,,
    \quad
  \quad \quad \; \; \; (\varphi^\dagger)^{{ a\dot{a}}}_{\pm\mp}\cdot \uprho^8 \cdot \uprho^3 \cdot \uprho^5 \cdot \varphi_{{ b\dot{b}}}^{\mp\pm} = \frac{i}{4} \delta^{\dot{a}}_{\dot{b}} \left( \sigma^2 \right)^{a}_{b}  \,, \nonumber \\
  &(\varphi^\dagger)^{{ a\dot{a}}}_{\pm\mp}\cdot \uprho^3 \cdot \uprho^4 \cdot \varphi_{{ b\dot{b}}}^{\pm\mp} = \frac{i}{4} \delta^{\dot{a}}_{\dot{b}} \left( \sigma^3 \right)^{a}_{b}  \,,
    \quad
 \quad \quad \; \; \; (\varphi^\dagger)^{{ a\dot{a}}}_{\pm\mp}\cdot \uprho^8 \cdot \uprho^3 \cdot \uprho^4 \cdot \varphi_{{ b\dot{b}}}^{\mp\pm} = \frac{i}{4} \delta^{\dot{a}}_{\dot{b}} \left( \sigma^3 \right)^{a}_{b}  \,, \nonumber \\
     &(\varphi^\dagger)^{{ a\dot{a}}}_{\pm\mp}\cdot \uprho^7 \cdot \uprho^8 \cdot \varphi_{{ b\dot{b}}}^{\pm\mp} = -\frac{i}{4} \delta^{a}_{b}  \left( \sigma^1 \right)^{\dot{a}}_{\dot{b}} \,,
    \quad
 \quad \quad \; (\varphi^\dagger)^{{ a\dot{a}}}_{\pm\mp}\cdot \uprho^8 \cdot \uprho^7 \cdot \uprho^8 \cdot \varphi_{{ b\dot{b}}}^{\mp\pm} = -\frac{i}{4} \delta^{a}_{b}  \left( \sigma^1 \right)^{\dot{a}}_{\dot{b}} \,,\nonumber  \\
  &(\varphi^\dagger)^{{ a\dot{a}}}_{\pm\mp}\cdot \uprho^6 \cdot \uprho^8 \cdot \varphi_{{ b\dot{b}}}^{\pm\mp} = \frac{i}{4} \delta^{a}_{b}  \left( \sigma^2 \right)^{\dot{a}}_{\dot{b}} \,,
    \quad
 \quad \quad \; \;\; (\varphi^\dagger)^{{ a\dot{a}}}_{\pm\mp}\cdot \uprho^8 \cdot \uprho^6 \cdot \uprho^8 \cdot \varphi_{{ b\dot{b}}}^{\mp\pm} = \frac{i}{4} \delta^{a}_{b}  \left( \sigma^2 \right)^{\dot{a}}_{\dot{b}} \,, \nonumber \\
  &(\varphi^\dagger)^{{ a\dot{a}}}_{\pm\mp}\cdot \uprho^6 \cdot \uprho^7 \cdot \varphi_{{ b\dot{b}}}^{\pm\mp} = \frac{i}{4} \delta^{a}_{b}  \left( \sigma^3 \right)^{\dot{a}}_{\dot{b}} \,,
    \quad
 \quad \quad \; \; \; (\varphi^\dagger)^{{ a\dot{a}}}_{\pm\mp}\cdot \uprho^8 \cdot \uprho^6 \cdot \uprho^7 \cdot \varphi_{{ b\dot{b}}}^{\mp\pm} = \frac{i}{4} \delta^{a}_{b}  \left( \sigma^3 \right)^{\dot{a}}_{\dot{b}} \,, \nonumber \\
 &(\varphi^\dagger)^{{ a\dot{a}}}_{\pm\pm}\cdot \uprho^9 \cdot \uprho^4 \cdot \uprho^5 \cdot \varphi_{{ b\dot{b}}}^{\pm\mp} = \pm \frac{i}{4} \delta^{\dot{a}}_{\dot{b}} \left( \sigma^1 \right)^{a}_{b}  \,, 
    \quad
 \; \; (\varphi^\dagger)^{{ a\dot{a}}}_{\pm\pm}\cdot \uprho^9 \cdot \uprho^7 \cdot \uprho^8 \cdot \varphi_{{ b\dot{b}}}^{\pm\mp} = \pm \frac{i}{4} \delta^{a}_{b}  \left( \sigma^1 \right)^{\dot{a}}_{\dot{b}} \,, \nonumber \\
 &(\varphi^\dagger)^{{ a\dot{a}}}_{\pm\pm}\cdot \uprho^9 \cdot \uprho^3 \cdot \uprho^5 \cdot \varphi_{{ b\dot{b}}}^{\pm\mp} = \mp \frac{i}{4} \delta^{\dot{a}}_{\dot{b}} \left( \sigma^2 \right)^{a}_{b}  \,, 
    \quad
 \; \; (\varphi^\dagger)^{{ a\dot{a}}}_{\pm\pm}\cdot \uprho^9 \cdot \uprho^6 \cdot \uprho^8 \cdot \varphi_{{ b\dot{b}}}^{\pm\mp} = \mp \frac{i}{4} \delta^{a}_{b}  \left( \sigma^2 \right)^{\dot{a}}_{\dot{b}} \,, \nonumber \\
 &(\varphi^\dagger)^{{ a\dot{a}}}_{\pm\pm}\cdot \uprho^9 \cdot \uprho^3 \cdot \uprho^4 \cdot \varphi_{{ b\dot{b}}}^{\pm\mp} = \mp \frac{i}{4} \delta^{\dot{a}}_{\dot{b}} \left( \sigma^3 \right)^{a}_{b}  \,, 
    \quad
 \; \; (\varphi^\dagger)^{{ a\dot{a}}}_{\pm\pm}\cdot \uprho^9 \cdot \uprho^6 \cdot \uprho^7 \cdot \varphi_{{ b\dot{b}}}^{\pm\mp} = \mp \frac{i}{4} \delta^{a}_{b}  \left( \sigma^3 \right)^{\dot{a}}_{\dot{b}} \,, \nonumber \\
&(\varphi^\dagger)^{{ a\dot{a}}}_{\pm\pm}\cdot \uprho^9 \cdot \varphi_{{ b\dot{b}}}^{\pm\mp} = \mp \frac{1}{4}\delta^{a}_{b} 
    \delta^{\rm \dot a}_{\rm \dot b}\,,
\end{align}
and 0 otherwise and defining
\begin{align}
\label{definition phi su2+}
\phi^{a}_{b}  &:= \phi^{\beta_+} \left( \sigma^1 \right)^{a}_{b}  + \phi^{\gamma_+} \left( \sigma^2 \right)^{a}_{b}  - \phi^{\varphi_+} \left( \sigma^3 \right)^{a}_{b} \,, \\
\label{definition phi su2-}
\phi^{\dot{a}}_{\dot{b}} &:= \phi^{\beta_-} \left( \sigma^1 \right)^{\dot{a}}_{\dot{b}} + \phi^{\gamma_-} \left( \sigma^2 \right)^{\dot{a}}_{\dot{b}} - \phi^{\varphi_-} \left( \sigma^3 \right)^{\dot{a}}_{\rm \dot{b}} \,,
\end{align}
we get that the supersymmetry transformations of the $AdS_2$ fields can be expressed as
\begin{align}
\label{susy massive fermions}
\delta \chi^{a \dot{a}} &= \frac{1}{2} \left( \slashed{\partial} \phi^{\sf tr} + \cos \vartheta \, \phi^{\sf tr}  \right) \uptau^3 \kappa^{a\dot{a}} \nonumber \\
&\quad -\frac{i}{2} \left[ \sin \Omega \left( \slashed{\partial} \phi^{a}_{b} - \cos \vartheta \, \phi^{a}_{b} \right) \kappa^{b\dot{a}} + \cos \Omega \left( \slashed{\partial} \phi^{\dot{a}}_{\dot{b}} - \cos \vartheta \, \phi^{\rm \dot{a}}_{\dot{b}} \right) \kappa^{a\dot{b}} \right] \,, \\
\label{susy massless fermions}
\delta \psi^{ a \dot{a}} &= \frac{1}{2} \slashed{\partial} \phi^{9} \kappa^{ a \dot{a}}+\frac{i}{2} \left( \cos \Omega \; \slashed{\partial} \phi^{a}_{b}  \kappa^{b \dot{a}}-\sin \Omega \; \slashed{\partial} \phi^{\dot{a}}_{\dot{b}} \kappa^{ a \dot{b}} \right) \,, \\
\label{susy s1 scalar}
\delta \phi^{9} &=  -\frac{1}{2} \; \overline{ \psi}_{ a \dot{a}}  \kappa^{ a \dot{a}} \,, \\
\label{susy tr scalar}
\delta \phi^{\sf tr} &=  -\frac{1}{2} \; \overline{ \chi}_{ a \dot{a}} \uptau^3 \kappa^{ a \dot{a}} \,, \\
\label{susy s3+ scalars}
\delta \phi^{a}_{b}  &=-\frac{i}{2} \left[ \cos \Omega \left( 2 \; \overline{\psi}_{b\dot{c}} \kappa^{a\dot{c}} - \delta^{a}_{b}  \overline{\psi}_{c\dot{c}} \kappa^{c\dot{c}} \right) -\sin \Omega \left(2 \; \overline{\chi}_{ b\dot{c}} \kappa^{a\dot{c}} - \delta^{a}_{b}  \overline{\chi}_{c\dot{c}} \kappa^{c\dot{c}} \right) \right] \,, \\
\label{susy s3- scalars}
\delta \phi^{\dot{a}}_{\dot{b}} &= \frac{i}{2} \left[ \sin \Omega \left( 2 \; \overline{\psi}_{c\dot{b}} \kappa^{c\dot{a}} - \delta^{\dot{a}}_{\dot{b}} \overline{\psi}_{ c\dot{c}} \kappa^{c\dot{c}} \right) +\cos \Omega \left(2 \; \overline{\chi}_{c\dot{b}} \kappa^{c\dot{a}} - \delta^{\dot{a}}_{\dot{b}} \overline{\chi}_{c\dot{c}} \kappa^{c\dot{c}} \right) \right] \,.
\end{align}

%% file: defectalgebra.tex
In Chapter \ref{ch: line defects in AdS3CFT2} we study a line defect with $\mathfrak{psu}(1,1|2) \times \mathfrak{su}(2)_A$ invariance from the point of view of the analytic conformal bootstrap program. In order to do so it is crucial to have a good understanding of the symmetry algebra of the dCFT$_1$ and of its representation theory, which we discuss in this appendix. Moreover, we will show that correlators of some short supermultiplets exhibit a topological sector, which will be of use when studying the superconformal block decomposition of four-point functions.

\subsubsection*{Symmetry algebra}

We are interested in the 1/2 BPS line defect which is dual the 1/2 BPS Dirichlet string discussed in Chapter \ref{ch: line defects in AdS3CFT2}. In particular, we will analyze the case in which the supergravity background is described by an $AdS_3 \times S^3 \times T^4$ metric. The symmetries of the such a background form a $\mathfrak{psu}(1,1|2)^2\times \mathfrak{so}(4)_A \times \mathfrak{u}(1)_A$ algebra \cite{Borsato:2013qpa}, where the extra $\mathfrak{so}(4)_A\times \mathfrak{u}(1)_A$ acts as an automorphism. For the 1/2 BPS line defect the corresponding symmetry algebra is $\mathfrak{psu}(1,1|2)\times \mathfrak{su}(2)_A$. The bosonic $\mathfrak{su}(1,1) \times \mathfrak{su}(2)_R$ subalgebra of the $\mathfrak{psu}(1,1|2)$ describes the conformal and R-symmetries of the defect, while the extra $\mathfrak{su}(2)_A$ is again an automorphism. Let us note that in the literature it is common to preserve only the $\mathfrak{u}(1)$ generated by the Cartan of the $\mathfrak{su}(2)_A$ automorphism\footnote{An interesting discussion on this subject is given in Section 2.2 of \cite{Beisert:2006qh}.}. The non-vanishing commutators of this algebra are \cite{Borsato:2013qpa} \small{
\begin{align}
[D, P] &= P, 
& [D, K] &= -K,  & [P, K] &= -2 D, 
\nonumber \\
[R^a_b, R^c_d] &= \delta^c_b R^a_d-\delta^a_d R^c_b,
&[R^a_b, Q_{c\dot{c}}]  &= \delta^a_c Q_{b \dot{c}}-\frac{1}{2} \delta^a_{b} Q_{c \dot{c}}, & [R^a_b, S_{c\dot{c}}]  &= \delta^a_c S_{b \dot{c}}-\frac{1}{2} \delta^a_{b} S_{c \dot{c}},  \nonumber \\
[L^{\dot{a}}_{\dot{b}}, L^{\dot{c}}_{\dot{d}}] &= \delta^{\dot{c}}_{\dot{b}} L^{\dot{a}}_{\dot{d}}-\delta^{\dot{a}}_{\dot{d}} L^{\dot{c}}_{\dot{b}}, 
&[L^{\dot{a}}_{\dot{b}}, Q_{c\dot{c}}]  &= \delta^{\dot{a}}_{\dot{c}} Q_{c \dot{b}}-\frac{1}{2} \delta^{\dot{a}}_{\dot{b}} Q_{c \dot{c}}, 
&[L^{\dot{a}}_{\dot{b}}, S_{c\dot{c}}]  &= \delta^{\dot{a}}_{\dot{c}} S_{c \dot{b}}-\frac{1}{2} \delta^{\dot{a}}_{\dot{b}} S_{c \dot{c}}, 
\nonumber \\
[D, Q_{a\dot{a}}] &= \frac{1}{2} Q_{a\dot{a}},  & [D, S_{a\dot{a}}] &= -\frac{1}{2} S_{a\dot{a}},
& \{ Q_{a\dot{a}}, Q_{b\dot{b}}\} 
&=-\epsilon_{ab} \epsilon_{\dot{a}\dot{b}} P, \nonumber \\
[P,  S_{a\dot{a}}] &= -Q_{a\dot{a}}, & [K,  Q_{a\dot{a}}] &= S_{a\dot{a}}, & \{ S_{a\dot{a}}, S_{b\dot{b}}\} &=-\epsilon_{ab} \epsilon_{\dot{a}\dot{b}} K, \nonumber 
\\
& & \{ Q_{a\dot{a}}, S_{b\dot{b}}\} &=-\epsilon_{ab} \epsilon_{\dot{a}\dot{b}} D  -  \epsilon_{\dot{a}\dot{b}} R_{ab},    & & 
\label{algebra}
\end{align}
}\normalsize{where $P,K$ and $D$ are the generators of the conformal group, $R^a_b$ are the generators of the $\mathfrak{su}(2)_R$ R-symmetry, $L^{\dot{a}}_{\dot{b}}$ are the generators of the $\mathfrak{su}(2)_A$ automorphism, and $Q_{a\dot{a}}$ and $S_{a\dot{a}}$ are the supercharges\footnote{To uplift the $\mathfrak{psu}(1,1|2)\times \mathfrak{su}(2)_A$ algebra to a $\mathfrak{d}(2,1;\sin^2{\Omega})$ algebra (that describes 1/2 BPS line operators in the CFT$_2$ dual to type IIB string theory on AdS$_3\times$S$_+^3\times$S$_-^3\times$S$^1$) we just have to modify the $\{ Q_{a\dot{a}}, S_{b\dot{b}}\}$ anticommutator in order for it to be
\begin{equation}
 \{ Q_{a\dot{a}}, S_{b\dot{b}}\} =-\epsilon_{ab} \epsilon_{\dot{a}\dot{b}} D - \sin^2\Omega \,\epsilon_{\dot{a}\dot{b}} R_{ab} - \cos^2\Omega  \,\epsilon_{ab} L_{\dot{a}\dot{b}} \,.
 \label{d21a}
\end{equation}}. Note that the dotted $\mathfrak{su}(2)_A$ and undotted $\mathfrak{su}(2)_R$ indices take values in $\dot{\pm}$ and $\pm$, respectively, and that we are using $\epsilon_{+-}=\epsilon_{\dot{+}\dot{-}}=1$ and $R_{ab}=\epsilon_{bc}R_a^c$.}

\subsubsection*{Representation theory}

In order to bootstrap four-point functions of operators inserted along the defect one has to be exhaustive about all the possible representations that can appear in the corresponding OPE expansions. Therefore, let us study the representations of the $\mathfrak{psu}(1,1|2) \times \mathfrak{su}(2)_A$ symmetry algebra of the defect. To describe the states in those representations we will use the notation $|\Delta,r,\ell \rangle$, with
\begin{align}
   D |\Delta,r,\ell \rangle &= \Delta \,|\Delta,r,\ell \rangle \,, \nonumber \\
   R_0 |\Delta,r,\ell \rangle &= \frac{r}{2} \, |\Delta,r,\ell \rangle \,, \nonumber \\
   L_0 |\Delta,r,\ell \rangle &= \frac{\ell}{2} \, |\Delta,r,\ell \rangle \,,  
   \label{state notation}
\end{align}
and where $R_0:=R^+_+=-R^-_-$ and $L_0:=L^+_+=-L^-_-$ are the corresponding Cartan generators of the R-symmetry and automorphism groups.

On the one hand, we have $\textit{long representations}$ ${\cal L}^{\Delta}_{r,\ell}$, which are characterized by the three quantum numbers $\Delta, r$ and $\ell$ of their corresponding superprimary state. These representations are given by 16 conformal primaries, that organize as
\footnote{The multiplicity of the different states in the decomposition can be seen from a character analysis.}
\begin{align}
    [\Delta,&r,\ell]\nonumber \\
    [\Delta+\frac 12,r+1,\ell+1] \quad [\Delta+\frac 12,r+1,\ell-1] &\; \; [\Delta+\frac 12,r-1,\ell+1] \quad [\Delta+\frac 12,r-1,\ell-1] \nonumber\\
    [\Delta+1,r\pm2,\ell]\qquad 2 \, [\Delta+1,&r,\ell]\qquad[\Delta+1,r,\ell\pm2]\\
   [\Delta+\frac 32,r+1,\ell+1] \quad [\Delta+\frac 32,r+1,\ell-1] &\; \; [\Delta+\frac 32,r-1,\ell+1] \quad [\Delta+\frac 32,r-1,\ell-1] \nonumber\\
    [\Delta+2,&r,\ell] \nonumber
\end{align}
These satisfy a unitarity bound
\begin{align}
    \Delta \geq \frac r 2.
\end{align}
On the other hand, we have \textit{short representations} that preserve half of the supersymmetries, i.e. that are 1/2 BPS. These multiplets are subjected to the shortening condition\footnote{Naively one could also derive $\Delta=-\frac{r}{2}$ from the algebra. However, this condition is not consistent with the $\Delta \geq 0$ constraint that comes from unitarity.}
\begin{equation}
    \Delta=\frac{r}{2} \,,
\end{equation}
that can be derived from the anti-commutation relations
\begin{align}
\label{qs 1}
    \{Q_{\pm \dot{\pm}}, S_{\mp\dot{\mp}} \} &= -D \pm R_0 \,, \\
\label{qs 2}
    \{Q_{\pm \dot{\mp}}, S_{\mp \dot{\pm}}\} &= D \mp R_0 \,.
\end{align}
Therefore, there are only two independent quantum numbers that characterize the superprimary state of these supermultiplets, which we will choose to be $\Delta$ and $\ell$. Consequently, we will use the notation ${\cal A}^{\Delta}_{\ell}$ and ${\cal B}^{\Delta}_{\ell}$ to denote these representations. The structure of the short multiplets is
\begin{align}
\label{short multiplets} 
    \mc{A}^{\Delta}_{\ell}&:\qquad [\Delta,2\Delta,\ell]\rightarrow [\Delta+\frac 12,2\Delta+1,\ell\pm 1]\rightarrow[\Delta+ 1,2\Delta+2,\ell] \nonumber \\
    \mc{B}^{\Delta}_{\ell}&:\qquad [\Delta,2\Delta,\ell]\rightarrow [\Delta+\frac 12,2\Delta-1,\ell\pm 1]\rightarrow[\Delta+ 1,2\Delta-2,\ell]
\end{align}
For example, we have:
\begin{align}
\label{short multiplets 2}
    &\mc{B}^{\frac 12 }_{1}: [\frac 12,1,1]\rightarrow [1,0,2]\oplus [1,0,0] \nonumber\\
    &\mc{B}^{1}_{0}: [1,2,0]\rightarrow [\frac 32,1,1]\rightarrow[2,0,0] \nonumber\\
    &\mc{B}^{1}_{2}: [1,2,2]\rightarrow [\frac 32,1,1]\oplus [\frac 32,1,3]\rightarrow[2,0,2]
\end{align}

Finally, let us comment about the absence of other short representations. More precisely, we will argue that the symmetry algebra of the defect forbids the existence of 1/4 BPS representations\footnote{This is in contrast with the analysis of \cite{Agmon:2020pde}. The main difference is that in \cite{Agmon:2020pde} the authors consider one-dimensional defects embedded in a three-dimensional bulk space, while we study a defect embedded in a two-dimensional space. Therefore, in the former case there is an extra $\mathfrak{u}(1)$ symmetry (describing rotations around the defect) that is absent in our case.}. In order to show this, let us suppose that we have a superprimary state $|\Delta, r, \ell \rangle $ that is annihilated by the supercharge $Q_{+\dot{+}}$, i.e.
\begin{equation}
Q_{+\dot{+}} |\Delta, r, \ell \rangle=0 \,,
\end{equation}
Then, from \eqref{qs 1} we get
\begin{equation}
(D-R_0) |\Delta, r, \ell \rangle=0 \,,
\end{equation}
and therefore, using \eqref{qs 2} we arrive at
\begin{equation}
 \{Q_{+ \dot{-}}, S_{-\dot{+}}\} |\Delta, r, \ell \rangle= S_{-\dot{+}} Q_{+ \dot{-}} |\Delta, r, \ell \rangle=0 \,,
\end{equation}
were in the last step we have used that $|\Delta, r, \ell \rangle$ is a superprimary. Then,
\begin{equation}
|| Q_{+ \dot{-}} |\Delta, r, \ell \rangle ||^2= \langle \Delta, r, \ell|S_{-\dot{+}}Q_{+ \dot{-}} |\Delta, r, \ell \rangle=0 \,,
\end{equation}
and consequently
\begin{equation}
     Q_{+ \dot{-}} |\Delta, r, \ell \rangle=0 \,. \label{Eq:All shorts are 1/2 BPS}
\end{equation}
Therefore, we have shown that each superprimary that is annihilated by the $Q_{+\dot{+}}$ charge is also annihilated by the $Q_{+ \dot{-}}$ charge (and one can similarly prove the reciprocal). Then, each short representation is 1/2 BPS\footnote{Let us note that this statement is not true for the more general $\mathfrak{d}(2,1;\sin^2{\Omega})$ supergroup. Indeed, the displacement supermultiplet is 1/4 BPS for the case of 1/2 BPS line defects in the holographic dual to string theory in $AdS_3 \times S^3_+ \times S^3_- \times S^1$, where the symmetry group of the defect is $\mathfrak{d}(2,1;\sin^2{\Omega})$.}.

\subsubsection*{Topological sector}

In this section we will show the existence of a topological sector for correlators of primaries of the ${\cal B}^{\Delta}_{\ell}$ short multiplets. This proves useful when studying the OPE decomposition of such correlators, given that it allows to derive the structure of the superconformal blocks associated to that OPE expansion. 

Let us begin the analysis by noticing that it is possible to define \textit{twisted} conformal generators $\hat{P},\hat{K}$ and $\hat{D}$ as
\begin{equation}
\hat{P}=P +  R^+_- \,, \qquad \hat{K}=K-  R^-_+ \,, \qquad \text{and} \qquad \hat{D}=D-R_0 \,.
\end{equation}

Crucially, these operators are $\mathbb{Q}$-exact, which can be seen from
\begin{align}
	\hat{P} &= -\{ \mathbb{Q}_{\dot{+}} ,Q_{-\dot{-}} \} =
 \{ \mathbb{Q}_{\dot{-}} ,  Q_{-\dot{+}} \} \,,
\nonumber \\
	\hat{K} &= \{ \mathbb{Q}_{\dot{+}} ,S_{+\dot{-}} \} =-\{ \mathbb{Q}_{\dot{-}} ,S_{+\dot{+}} \} \,, \nonumber \\
 \hat{D} &= -\frac{1}{2}\{ \mathbb{Q}_{\dot{+}} ,\mathbb{Q}_{\dot{+}}^{\dagger} \} =-\frac{1}{2} \{ \mathbb{Q}_{\dot{-}} ,\mathbb{Q}_{\dot{-}}^{\dagger} \} \,,
\end{align}
with
\begin{equation}
\mathbb{Q}_{\dot{a}}=Q_{+\dot{a}}+S_{-\dot{a}} \,,
\end{equation}
and where we are using that
\begin{equation}
Q_{a\dot{a}}^{\dagger}=S^{a\dot{a}} \,.
\end{equation}

Let us consider the superprimary operator ${\cal O}^{\Delta}_{\ell}(t,Y,W)$ of a ${\cal B}^{\Delta}_{\ell}$ short multiplet, where $t$ is the time coordinate, and $Y^{a}$ and $W^{\dot{a}}$ are auxiliary variables that are respectively introduced to contract every possible R-symmetry and automorphism index in the superprimary field. Then, we have that
\begin{equation}
\label{superprimary annihilation}
    [\mathbb{Q}_{\dot{a}}, {\cal O}^{\Delta}_{\ell} (0,0,W)]=0 \,.
\end{equation}
At this point it is useful to define
\begin{equation}
\label{timeev}
\hat{{\cal O}}^{\Delta}_{\ell} (t, W) := e^{-t \hat{P}} {\cal O}^{\Delta}_{\ell} (0,0,W)e^{t \hat{P}} \,.
\end{equation}
Given that $\hat{P}$ is $\mathbb{Q}$-exact, it commutes with $\mathbb{Q}_{\dot{a}}$. Therefore,
\begin{equation}
\label{superprimary annihilation 2}
    [\mathbb{Q}_{\dot{a}}, {\cal O}^{\Delta}_{\ell} (t,W)]=0 \,.
\end{equation}
Interestingly, when restricting to the frame\footnote{We can always do this given that correlators are homogeneous functions of the $Y$ variables \cite{Dolan:2004mu}. In this frame we get $Y_{ij}=y_i-y_j$.}
\begin{equation}
\label{top frame}
    Y(y)=(y,1) \,,
\end{equation}
we obtain
\begin{equation}
\hat{{\cal O}}^{\Delta}_{\ell} (t, W)= {\cal O}^{\Delta}_{\ell} (t,Y(t),W) \,.
\end{equation}
Then, we see that the $\hat{{\cal O}}^{\Delta}_{\ell}$ operators satisfy
\begin{align}
\partial_{t_1} \langle \hat{{\cal O}}^{\Delta}_{\ell} (t_1,W_1) \dots \hat{{\cal O}}^{\Delta}_{\ell} (t_n,W_n) \rangle &= -\langle [\hat{P}, \hat{{\cal O}}^{\Delta}_{\ell} (t_1,W_1)] \dots \hat{{\cal O}}^{\Delta}_{\ell} (t_n,W_n) \rangle  \nonumber \\
&=\langle [\{ \mathbb{Q}_{\dot{+}}, Q_{-\dot{-}} \}, \hat{{\cal O}}^{\Delta}_{\ell} (t_1,W_1)] \dots \hat{{\cal O}}^{\Delta}_{\ell} (t_n,W_n) \rangle \nonumber \\
&=0 \,,
\end{align}
where in the last step we have used \eqref{superprimary annihilation 2} and  $\mathbb{Q}_{\dot{+}} |0\rangle= \mathbb{Q}_{\dot{+}}^\dagger |0\rangle=0 $.
Therefore, we can conclude that the correlation functions of superprimaries of 1/2 BPS multiplets are topological when taking $Y_i(t_i)=(t_i,1)$, i.e.
\begin{align}
\label{top sector}
\partial_{t_i} \langle {\cal O}^{\Delta}_{\ell} (t_1,Y_1(t_1),W_1) \dots {\cal O}^{\Delta}_{\ell} (t_n, Y_n(t_n),W_n) \rangle &= 0 \,.
\end{align}

For a four-point function this expression becomes particularly useful. Due to conformal and R-symmetry, the tensorial structure and time dependence of four-point correlators is known up to functions of the cross ratios $\chi$, $\rho_1$ and $\rho_2$  associated to the coordinates $t$, $Y$ and $W$, respectively, which we define as
\begin{equation}
\label{cross ratio defs}
    \chi= \frac{t_{12}t_{34}}{t_{13}t_{24}} \,, \qquad \rho_1= \frac{Y_{12}Y_{34}}{Y_{13}Y_{24}} \,, \qquad \rho_2= \frac{W_{12}W_{34}}{W_{13}W_{24}}\,,
\end{equation}
with $t_{ij}=t_i-t_j$, $Y_{ij}=\epsilon_{ab} Y^a Y^b$ and $W_{ij}=\epsilon_{\dot{a} \dot{b}} W^{\dot{a}} W^{\dot{b}}$. For example,
\begin{align}
\label{top4pt}
\langle {\cal O}^{\Delta}_{\ell} (t_1,Y_1,W_1) {\cal O}^{\Delta}_{\ell} (t_2,Y_2,W_2) {\cal O}^{\Delta}_{\ell} (t_3,Y_3,W_3) & {\cal O}^{\Delta}_{\ell} (t_4, Y_4,W_4) \rangle \nonumber \\
   =&\left(\frac{Y_{12}Y_{34}}{t_{12}t_{34}}\right)^\Delta (W_{12}W_{34})^{\ell} f(\chi,\rho_1,\rho_2) 
\end{align}
The form of the function $f(\chi,\rho_1,\rho_2)$ is constrained by the existence of a topological sector. This can be seen by evaluating the four-point function  \eqref{top4pt} in the frame \eqref{top frame}, as then the condition \eqref{top sector} implies
\begin{align}
\label{topfin}
  \partial_\chi f(\chi,\chi,\rho_2)=0 \,.
\end{align}

%% file: D_integrals.tex
In this appendix we review the conventions used in Chapter \ref{ch: line defects in AdS3CFT2} (see Section \ref{sec: holographic 4 pts}) for the AdS$_2$ propagators and the resulting $D$-integrals, which appear in the computation of certain tree-level Witten diagrams \cite{Freedman:1998tz,DHoker:1999kzh,Fitzpatrick:2011ia,Penedones:2010ue}. For a scalar field bulk-to-boundary propagator, we 
consider
\begin{equation}
K_\Delta(z,t,t') := {\cal C}_\Delta \bar K_\Delta(z,t,t') = {\cal C}_\Delta \left(\frac{z}{z^2+(t-t')^2}\right)^\Delta\,,
\end{equation}
where
\begin{equation}
{\cal C}_\Delta = \frac{\Gamma(\Delta)}{2\sqrt{\pi}\Gamma(\Delta+\tfrac{1}{2})}\,.
\end{equation}
For the scalar field bulk-to-bulk propagator, we have
\begin{equation}
G_\Delta(z,z',t,t') =
{\cal C}_\Delta (2u)^{-1}{}  _2F_1(\Delta,\Delta,2\Delta,2 u^{-1})\,,
\qquad
u = \frac{(z-z')^2+(t-t')^2}{2 z z'}\,.
\end{equation}

Tree-level Witten diagrams with a single vertex are computed using the so-called $D$-integrals. For quartic vertices, one defines
\begin{equation}
D_{\Delta_1,\Delta_2,\Delta_3,\Delta_4}(t_1,t_2,t_3,t_4)=
\int\frac{d^2\sigma}{z^2}
\prod_{i=1}^4 \bar K_{\Delta_i}(z,t,t_i)\,.
\end{equation}
When dealing with integrals that come from vertices involving derivatives it becomes useful to consider the identity
\begin{align}
z^2\delta^{\alpha\beta}&\partial_{\alpha}\bar K_{\Delta_1}(z,t,t_1)\partial_{\beta}\bar K_{\Delta_2}(z,t,t_2)
\label{dKdK}
\\
&=\Delta_1\Delta_1\left(
\bar K_{\Delta_1}(z,t,t_1)\bar K_{\Delta_2}(z,t,t_2)-2(t_2-t_1)^2
\bar K_{\Delta_1+1}(z,t,t_1)\bar K_{\Delta_2+1}(z,t,t_2)
\right)\,.\nonumber
\end{align}

We can specify the $D$-integrals in terms of the functions of the cross-ratio associated to them,
\begin{equation}
D_{\Delta_1,\Delta_2,\Delta_3,\Delta_4}
=\frac{\pi^{d/2}\Gamma\left(\Sigma-\tfrac{d}{2}\right)}{2\Gamma(\Delta_1)\Gamma(\Delta_2)\Gamma(\Delta_3)\Gamma(\Delta_4)}\frac{t_{14}^{2\Sigma-2\Delta_1-2\Delta_4}t_{34}^{2\Sigma-2\Delta_3-2\Delta_4}}{t_{13}^{2\Sigma-2\Delta_4}t_{24}^{2\Delta_2}}
\bar D_{\Delta_1,\Delta_2,\Delta_3,\Delta_4}(\chi) \,,
\end{equation}
where $\Sigma = \tfrac{1}{2}(\Delta_1+\Delta_2+\Delta_3+\Delta_4)$. 

The $\bar{D}$-integrals needed for our computations are
\begin{align}
\bar D_{1,1,1,1}(\chi) = &\ -2 \frac{\log\chi}{1-\chi}-2 \frac{\log(1-\chi)}{\chi}\,,
\\
\bar D_{1,1,2,2}(\chi) = &\ \frac{1}{3(1-\chi)} + \frac{\chi^2\log\chi}{3(1-\chi)^2} - \frac{(2+\chi)\log(1-\chi)}{3\chi}
\\
\bar D_{1,2,1,2}(\chi) = &\ - \frac{1}{3\chi(1-\chi)} - \frac{(3-2\chi)\log\chi}{3(1-\chi)^2} - \frac{(1+2\chi)\log(1-\chi)}{3\chi^2}\,,
\\
\bar D_{2,2,2,2}(\chi) = &\
- \frac{2(  1- \chi + \chi^2)}{15\chi^2(1 - \chi)^2 } - \frac{(5 - 5 \chi + 2 \chi^2) \log(\chi)}{15(1-\chi)^3} - \frac{(2 + \chi+ 2 \chi^2) \log(1 - \chi)}{15 \chi^3}\,,
\\
\bar D_{2,2,3,3}(\chi) = &\   -\frac{24 - 48\chi - 5\chi^2 + 29\chi^3 - 18\chi^4}{210 \chi^2(1 - \chi)^3} 
- \frac{(12 + 6\chi + 8\chi^2 + 9\chi^3) \log(1 - \chi)}{105 \chi^3}\nonumber\\
 &\ + \frac{\chi^2(28 - 28\chi + 9\chi^2) \log(\chi)}{105 (1 - \chi)^4}\,,
\\
\bar D_{2,3,2,3}(\chi) = &\ -\frac{18 - 29\chi + 5\chi^2 + 48\chi^3 - 24\chi^4}{210\chi^3(1 - \chi)^3 } - \frac{(9 + 8\chi + 6\chi^2 + 12\chi^3) \log(1 - \chi)}{105 \chi^4} \nonumber \\
&\ + \frac{(-35 + 56\chi - 42\chi^2 + 12\chi^3) \log(\chi)}{105 (1 - \chi)^4}\,,
\\
\bar D_{3,3,3,3}(\chi) = &\ \frac{-24 + 72\chi - 74\chi^2 + 28\chi^3 - 74\chi^4 + 72\chi^5 - 24\chi^6}{315(1 - \chi)^4 \chi^4}\\
&\hspace{-1.4cm} - \frac{4(2 + \chi + \chi^2 + \chi^3 + 2\chi^4) \log(1 - \chi)}{105 \chi^5}
 - \frac{4(7 - 14\chi + 16\chi^2 - 9\chi^3 + 2\chi^4) \log(\chi)}{105 (1 - \chi)^5}\,,\nonumber
\end{align}
along with some others obtained from the following identities \cite{Dolan:2000ut}
\begin{align}
\bar D_{\Delta+1,\Delta+1,\Delta,\Delta}(\chi) = &\ \tfrac{1}{\chi^2}\bar D_{\Delta,\Delta,\Delta+1,\Delta+1}(\chi)\,, 
\\
\bar D_{\Delta,\Delta+1,\Delta+1,\Delta}(\chi) = &\ \tfrac{1}{(1-\chi)^2}\bar D_{\Delta,\Delta,\Delta+1,\Delta+1}(1-\chi)\,,
\\
\bar D_{\Delta+1,\Delta,\Delta,\Delta+1}(\chi) = &\ \bar D_{\Delta,\Delta,\Delta+1,\Delta+1}(1-\chi)\,,
\\
\bar D_{\Delta+1,1,\Delta+1,1}(\chi) = &\ \bar D_{1,\Delta+1,1,\Delta+1}(\chi)\,.
\end{align}

%% file: vacuumapp.tex
In this appendix we will discuss a string dual to the $1/6$ BPS vacuum state described in \eqref{susyinsertion0}. We acknowledge J. Aguilera-Damia for obtaining these results.

\subsubsection*{Conventions for the $AdS_4\times {\mathbb{CP}^3}$ supergravity background}

As discussed in Chapter \ref{ch:adscft}, the ABJM theory is conjectured to be dual to type IIA string theory on $AdS_4\times {\mathbb{CP}^3}$ \cite{Aharony:2008ug}. This background is characterized by the metric\footnote{We refer to eq. \eqref{ads4cft3 other background fields} for the other non-vanishing fields that describe the supergravity solution.}
\be
ds^2 =\frac{R^3}{4k}\left(ds^2_{AdS_4} + 4ds^2_{\mathbb{CP}^3}\right)\,,
\ee
We will take the coordinates to be such that
\begin{alignat}{2}
ds^2_{AdS_4} & = -\cosh^2\rho \, dt^2 + d\rho^2 + \sinh^2\rho \, (d\theta^2 + \sin^2\theta d\psi^2)\,,
\\
ds^2_{\mathbb{CP}^3} &=  dz^a d\bar{z}_a -  z^a \bar{z}_b dz^b d\bar{z}_a \,,
\end{alignat}
where $a=1,\dots,4$ and with $\bar{z}_a z^a=1$. Let us also recall that the $AdS_4\times {\mathbb{CP}^3}$ space described above can be obtained by taking the large $k$ limit in the $AdS_4\times S^7/\mathbb{Z}_k$ background whose metric is
\begin{equation}
\label{S7 orbifold}
ds^2_{S^7/\mathbb{Z}_k}=ds^2_{\mathbb{CP}^3}+\left(\frac{d\xi}{4}+ A\right)^2 \,,
\end{equation}
with $\xi \equiv \xi + \frac{8\pi}{k}$ and
\begin{equation}
 A=\frac{1}{2i} (\bar{z}_a dz^a-z^a d\bar{z}_a) \,.
\end{equation}

It will prove convenient to write the four complex projective coordinates $z^a$ in terms of angles\footnote{The ranges of the angular variables are the following: $0\leq\alpha,\vartheta_1,\vartheta_2\leq\pi$, $0\leq\varphi_1,\varphi_2\leq2\pi$ and $0\leq\chi\leq4\pi$.}
 $\alpha$,$\vartheta_1$,$\varphi_1$,$\vartheta_2$,$\varphi_2$,$\chi$ and $\xi$,
according to
\be
\begin{aligned}
z^1& =\cos{\frac{\alpha}{2}}\cos{\frac{\vartheta_1}{2}}e^{\frac{i}{4}\left(2\varphi_1+\chi+\xi\right)}\,, &
z^2& =\cos{\frac{\alpha}{2}}\sin{\frac{\vartheta_1}{2}}e^{\frac{i}{4}\left(-2\varphi_1+\chi+\xi\right)}\,,\\
z^3& =\sin{\frac{\alpha}{2}}\cos{\frac{\vartheta_2}{2}}e^{\frac{i}{4}\left(2\varphi_2-\chi+\xi\right)}\,,&
z^4& =\sin{\frac{\alpha}{2}}\sin{\frac{\vartheta_2}{2}}e^{\frac{i}{4}\left(-2\varphi_2-\chi+\xi\right)}\,.
\end{aligned}
\nn
\ee

The transverse scalar directions $(C_1,C_2,C_3,C_4)$ should be identified with the complex coordinates $(z^1,z^2,z^3,z^4)$. For example, the ${\rm Tr}[(C_1 \bar{C}^2)^\ell]$ vacuum is the dual operator to a string moving along the null-geodesic defined by $\rho=0$, $\alpha=0$ and
$\vartheta_1= \tfrac{\pi}{2}$.

\subsubsection*{String with large angular momentun in $AdS_4\times \mathbb{CP}^3$ }

Let us turn now to the construction of the string solution dual to the ${\cal V}_{\ell}$ vacuum  presented in \eqref{susyinsertion0}. To that aim, one should recall that for the 1/2 BPS Wilson line (\ref{general WL}) the dual open string worldsheet is an $AdS_2\subset AdS_4$ located at a fixed point in the ${\mathbb{CP}^3}$ space \cite{Drukker:2008zx}. For the choice (\ref{WLchoice}), that singles out $I=1$, one should take a point with $|z^1| = 1$, which corresponds to placing the string at $\alpha =0$ and $\vartheta_1 = 0$ in ${\mathbb{CP}^3}$. When considering the insertion of the ${\cal V}_{\ell}$ vacuum one can generalize the ideas of \cite{Drukker:2006xg}. That is, one can consider an open string carrying a large amount of angular momentum $\ell$ in the plane 12. This configuration is a folded string, whose folding point follows the null-geodesic defined by $\rho=0$, $\alpha=0$ and
$\vartheta_1= \tfrac{\pi}{2}$. In that regard, one can then take the ansatz
\begin{alignat}{4}
t &= \tau\,, & \qquad \quad &  & \alpha & = 0 \,, \\
\rho &= \rho(\sigma)\,, & \qquad \quad & & \vartheta_1 &=  \vartheta_1(\sigma)\,,
\\
\theta & = 0 \,, & \qquad \quad &  & \varphi_1&=  \omega\tau\,.
\end{alignat}
with all the other coordinates being zero. This ansatz fits within a $AdS_2 \times S^2 \subset AdS_4\times {\mathbb{CP}^3}$ geometry, and therefore the classical motion
is the same as the one described in \cite{Drukker:2006xg}. More precisely, in terms of the worldsheet spatial parameter  $\sigma \in (-\infty,\infty)$
 one has
\begin{align}
 \rho &= {\rm arccosh}\left(\tfrac{1}{\tanh\sigma}\right)\,,
\label{xin1}
\\
 \vartheta_1 &= \arccos\left(\tfrac{1}{\cosh \sigma}\right)\,,
\label{xin2}
\\
 \varphi_1 &= \tau\,.
\label{xin3}
\end{align}

\subsubsection*{Supersymmetry of the folded string}

We will now discuss the supersymmetry invariance of the dual string solution proposed in the last section. In order to do so one should
first identify the Killing spinors of $AdS_4\times {\mathbb{CP}^3}$, which can be given in terms of the Killing spinors of $AdS_4\times S^7/\mathbb{Z}_k$. These can be written as \cite{Drukker:2008zx}
\be
\epsilon=\mathcal{M}\epsilon_0 \,,
\label{killesp}
\ee
with
\begin{align}
\mathcal{M} =
&
e^{\frac{\alpha}{4}(\hat{\gamma}\gamma_4-\gamma_9\gamma_{*})}
e^{\frac{\vartheta_1}{4}(\hat{\gamma}\gamma_5-\gamma_7\gamma_{*})}
e^{\frac{\vartheta_2}{4}(\gamma_{98}+\gamma_{46})}
e^{-\frac{\xi_1}{2}\hat{\gamma}\gamma_{*}}e^{-\frac{\xi_2}{2}\gamma_{57}}
e^{-\frac{\xi_3}{2}\gamma_{49}}
e^{-\frac{\xi_4}{2}\gamma_{68}}
\nn\\
&
\times
e^{\frac{\rho}{2}\hat{\gamma}\gamma_1}
e^{\frac{t}{2}\hat{\gamma}\gamma_0}
e^{\frac{\theta}{2}\gamma_{12}}e^{\frac{\psi}{2}\gamma_{23}}\,.
\end{align}
Above we have used $\xi_i$ for the phases appearing in the complex coordinates,
\be
\xi_1=\frac{2\varphi_1+\chi+\xi}{4}, \quad
\xi_2=\frac{-2\varphi_1+\chi+\xi}{4}, \quad
\xi_3=\frac{2\varphi_2-\chi+\xi}{4}, \quad
\xi_4=\frac{-2\varphi_2-\chi+\xi}{4},
\ee
and we are using the notation $\gamma_i$, $i=0, \dots ,9$, for the ten-dimensional Dirac matrices, with $\hat\gamma=\gamma_0\gamma_1\gamma_2\gamma_3$ and $\gamma_*=\prod_{i=0}^9 \gamma_i$.

When restricting to the Killing spinors of $AdS_4\times{\mathbb{CP}^3}$, one should consider only those spinors given in \eqref{killesp} that are invariant
under translations of the variable $\xi$. Under the translation $\xi\rightarrow\xi+\delta\xi$ one gets
\be
\epsilon^{\prime}=\mathcal{M}e^{i\frac{\delta\xi}{2}(i\hat{\gamma}\gamma_{*}+i\gamma_{57}+i\gamma_{49}+i\gamma_{68})}\epsilon_0\,.
\label{xitrasl}
\ee
For $\epsilon'$ to be equal to $\epsilon$, one should take $\epsilon_0$ to be an eigenstate of the matrices $\{ i\hat{\gamma}\gamma_{*}$, $i\gamma_{57}$, $i\gamma_{49}$, $i\gamma_{68}\}$ with eigenvalues $\{s_1,s_2,s_3,s_4\}$. Furthermore, these eigenvalues can only be $+1$ or $-1$, and they must satisfy the constraint
\be
s_1+s_2+s_3+s_4=0\,.
\label{cond}
\ee
Since in $\{s_1,s_2,s_3,s_4\}$ one can only have even numbers of $+1$ and $-1$, there are 8 combinations of eigenvalues in total.
However, the condition (\ref{cond}) rules out $\{+1,+1,+1,+1\}$ and $\{-1,-1,-1,-1\}$, and one is therefore left with
3/4 of the 32 supersymmetries,  i.e. there are 24 supersymmetries in $AdS_4\times{\mathbb{CP}^3}$.
~

As is well known, in type IIA string theory a given string configuration is supersymmetric if
\be
(1-\Gamma){\cal M}\epsilon_0 = 0\,,
\label{kappacon}
\ee
where the projector $\Gamma$ is defined as
\be
\Gamma=i\frac{\partial_\tau X^m\partial_\sigma X^n}{\sqrt{-h}}\Gamma_{mn}\gamma_{*} \,,
\label{proj}
\ee
where $h$ is the determinant of the induced metric on the worldsheet. For the family of solutions presented in (\ref{xin1})-(\ref{xin3}) one has
\be
\epsilon=e^{\frac{\vartheta_1}{4}(\hat{\gamma}\gamma_5-\gamma_7\gamma_{*})}
e^{-\frac{\tau}{4}(\hat{\gamma}\gamma_{*}-\gamma_{57})}
e^{\frac{\rho}{2}\hat{\gamma}\gamma_1}
e^{\frac{\tau}{2}\hat{\gamma}\gamma_0}\epsilon_0 \,.
 \label{killespsol}
 \ee
Moreover, the corresponding $\Gamma$ projector is
\be
\Gamma=\frac{i\left(\rho'\cosh{\rho} \,\gamma_{01} + \vartheta_1' \cosh{\rho} \, \gamma_{05}
-\rho'\sin{\vartheta_1} \, \gamma_{17} - \vartheta_1'\sin{\vartheta_1} \, \gamma_{57}\right) \, \gamma_{*}}{\sinh^2{\rho}+\cos^2{\vartheta_1}} \,,
\label{projsol}
\ee
where we have used that the solution \eqref{xin1}-\eqref{xin3} implies
\begin{equation}
\label{useful properties}
\sqrt{-h}=\sqrt{[(\rho')^2+(\vartheta_1')^2](\cosh^2 \rho-\sin^2 \vartheta_1)}=\sinh^2 \rho+ \cos^2 \vartheta_1 \,.
\end{equation}
The projector equation (\ref{kappacon}) must hold for any value of $\sigma$ and $\tau$. In particular, the only dependence on $\tau$ comes from the Killing spinor (\ref{killespsol}). Since $\hat{\gamma}\gamma_{11}$ and $\gamma_{57}$ commute with
$\hat{\gamma}\gamma_{1}$, one can reshuffle factors in (\ref{killespsol}) to have
\be
\epsilon=e^{\frac{\vartheta_1}{4}(\hat{\gamma}\gamma_5-\gamma_7\gamma_{*})}
e^{\frac{\rho}{2}\hat{\gamma}\gamma_1}
e^{-\frac{\tau}{4}(\hat{\gamma}\gamma_{*}-\gamma_{57}-2\hat{\gamma}\gamma_0)}\epsilon_0\,.
\label{killespsol2}
\ee
To eliminate the $\tau$-dependence one imposes the following projection condition over the constant spinor,
\be
(-i s_1 + i s_2-2\hat{\gamma}\gamma_{0})\epsilon_0 = 0\,.
\label{proj2}
\ee
Since $\hat{\gamma}\gamma_0$ does not have any zero eigenvalue, the condition (\ref{proj2}) is only satisfied if $s_1 = -s_2$. Furthermore, the constraint (\ref{proj2}) is equivalent to
\be
(1+\gamma_0\gamma_{*})\epsilon_0 =0\,, \qquad
(1-\hat{\gamma}\gamma_0\gamma_{57})\epsilon_0 =0\,.
\label{cond2}
\ee
Also, using \eqref{proj2} the equation (\ref{killespsol2}) can be rewritten as
\begin{align}
\epsilon &=
 e^{\frac{\rho}{2}\hat{\gamma}\gamma_1 +\frac{\vartheta_1}{2}\hat{\gamma}\gamma_5} \epsilon_0\,.
\end{align}
At this point one still needs to impose the kappa symmetry projection (\ref{kappacon}). Moving $e^{\frac{\rho}{2}\hat{\gamma}\gamma_1+\frac{\vartheta_1}{2}\hat{\gamma}\gamma_5}$ to the left and using \eqref{xin1}-\eqref{xin2} one gets
\begin{align}
\Gamma{\cal M}\epsilon_0
= \frac{i e^{\frac{\rho}{2}\hat{\gamma}\gamma_1+\frac{\vartheta_1}{2}\hat{\gamma}\gamma_5}}{\sinh^2{\rho}+\cos^2{\vartheta_1}}
&\left[ e^{-\vartheta_1\hat{\gamma}\gamma_5}
\left(  -\sinh{\rho}\cosh{\rho}\, \gamma_{01}+\cos\vartheta_1\cosh{\rho} \,\gamma_{05}\right)+\right.\nn\\
&+\left.
e^{-\rho\hat{\gamma}\gamma_1}
\left(\sinh\rho\sin\vartheta_1\gamma_{17}-\cos\vartheta_1\sin\vartheta_1\gamma_{57}\right)\right]
\gamma_{*} \, \epsilon_0 \,,
\end{align}
and the projection (\ref{kappacon}) becomes
\begin{align}
(\sinh^2{\rho}+\cos^2{\vartheta_1}) \, \epsilon_0
=&\left[ e^{-\vartheta_1\hat{\gamma}\gamma_5}
\left(  -\sinh{\rho}\cosh{\rho} \, \gamma_{01}+\cos\vartheta_1\cosh{\rho} \, \gamma_{05}\right)+\right.\nn\\
&+\left.
e^{-\rho\hat{\gamma}\gamma_1}
\left(\sinh\rho\sin\vartheta_1\gamma_{17}-\cos\vartheta_1\sin\vartheta_1\gamma_{57}\right)\right]
\gamma_{*} \, \epsilon_0\,.
\end{align}
Expanding the exponentials and moving $\gamma_{*}$ to the left one arrives at
\begin{align}
(\sinh^2{\rho}+\cos^2{\vartheta_1}) \, \epsilon_0
=& \, \gamma_{*}\left[-\cosh{\rho}\sinh{\rho}\cos{\vartheta_1}\gamma_{01}
 -\cosh{\rho}\sinh{\rho}\sin{\vartheta_1}\hat{\gamma}\gamma_5\gamma_{01}+\right.\nn\\
 &+\left.\cosh{\rho}\cos^2{\vartheta_1}\gamma_{05}
 -\cosh{\rho} \cos{\vartheta_1} \sin{\vartheta_1} \hat{\gamma}\gamma_{0}+\right.\nn\\
&+\left.\cosh{\rho}\sinh\rho\sin\vartheta_1\gamma_{17}
+\sinh^2\rho\sin\vartheta_1\hat{\gamma}\gamma_{7}-\right.\nn\\
&-\left.\cosh\rho\cos\vartheta_1\sin\vartheta_1\gamma_{57}
-\sinh\rho\cos\vartheta_1\sin\vartheta_1\hat{\gamma}\gamma_1\gamma_{57}\right]
\epsilon_0\,,
\end{align}
which reduces, when replacing with (\ref{xin1})-(\ref{xin3}) and using (\ref{cond2}), to
\be
(1+\gamma_1)\epsilon_0=0 \,.
\ee
Therefore, taking into account \eqref{cond2} and noticing that $\gamma_1$ commutes with $\gamma_0 \gamma_*$ and $\hat{\gamma} \gamma_0 \gamma_{57}$ one concludes that the configuration preserves 4 supersymmetries, i.e. corresponds to a $1/6$ BPS solution.

%% file: kinematics.tex
In this appendix we will briefly review different ways to express the kinematics of a scattering process involving $n$ massless particles\footnote{For literature on this topic see \cite{Elvang:2013cua,Badger:2023eqz}.}. We will focus on the three- and four-dimensional cases.

\subsubsection*{Spinor-helicity variables in four dimensions}

Let us consider a particle with momentum $p^{\mu}$ in four dimensions. It is useful to write the momentum as a bi-spinor $p_{\dot{a}a}$ given by
\begin{equation}
    p_{\dot{a}a}= (\sigma_{\mu})_{\dot{a}a} \, p^{\mu}= \left( \begin{array}{cc}
p^0+p^3 & p^1-i p^2 \\
p^1+ip^2 & p^0-p^3
\end{array}
\right)\,,
\end{equation}
with $\sigma_{\mu}=(\mathbb{1},\sigma^i)$, where $\mathbb{1}$ is the $2\times 2$ identity matrix and $\sigma^i$ are the usual Pauli matrices. Then, one has
\begin{equation}
    \det \left( p_{\dot{a}a} \right)=-p^2 \,.
\end{equation}
For generic values of $p^2$ one can write
\begin{equation}
    p_{\dot{a}a}= \tilde{\lambda}_{\dot{a}} \lambda_a + \tilde{\zeta}_{\dot{a}} \zeta_a   \,,
\end{equation}
where $\lambda, \tilde{\lambda}, \zeta$ and $\tilde{\zeta}$ are Grassmann-even Weyl spinors known as \textit{spinor-helicity variables}. The above decomposition greatly simplifies in the massless case, where the momentum can be written in terms of just two spinors as
\begin{equation}
\label{massless spinor helicity}
    p_{\dot{a}a}= \tilde{\lambda}_{\dot{a}} \lambda_a  \,.
\end{equation}
We would like to stress that spinor-helicity variables make the massless on-shell condition manifest. Moreover, from \eqref{massless spinor helicity} one can read that the rescaling
\begin{equation}
    \lambda_a \to t \, \lambda_a \,, \qquad \tilde{\lambda}_{\dot a} \to t^{-1} \tilde{\lambda}_{\dot a} \,,
\end{equation}
leaves the momentum $p_{\dot{a}a}$ invariant, and it is precisely the $U(1)$ little group invariance associated to massless particles.

Let us focus on an $n$-particle scattering process. Lorentz invariants can then be written as 
\begin{align}
\label{spinor helicity brackets}
    \langle i j \rangle &:= \epsilon^{ab} \lambda^i_a \lambda^j_b \,, \\
    [ i j ] &:= -\epsilon^{\dot{a} \dot{b}} \tilde{\lambda}^i_{\dot{a}} \tilde{\lambda}^j_{\dot{b}} \,,
\end{align}
which satisfy
\begin{equation}
\langle i j \rangle \,[ j i ]= -2 p_i.p_j=-s_{ij} \,,
\end{equation}
where $s_{ij}:=(p_i+p_j)^2$. 

In spinor-helicity notation the helicity generator $h$ becomes
\begin{equation}
    h=-\frac{1}{2} \sum_{i=1}^n \lambda_a^i \frac{\partial}{\partial \lambda^i_a}+ \tilde{\lambda}_{\dot a}^i \frac{\partial}{\partial \tilde{\lambda}^i_{\dot a}} \,,
\end{equation}
from which one can read that $\lambda^a$ and $\tilde{\lambda}^{\dot a}$ have $-1/2$ and $+1/2$ helicity contributions, respectively.

\subsubsection*{Dual space coordinates}

When studying the scattering of $n$ massless particles it is usually convenient to work with \textit{dual space coordinates}, which are defined as
\begin{equation}
\label{dual space def}
    p_i^{\mu}=x_{i}^{\mu}-x_{i+1}^{\mu} \,.
\end{equation}
with $x_{n+1}^{\mu}=x_1^{\mu}$. Note that the definition \eqref{dual space def} is such that momentum conservation $\sum_{i=1}^n p_i^{\mu}=0$ is manifest when using dual space coordinates. In these coordinates the massless on-shell condition is written as $(x_{i}^{\mu}-x_{i+1}^{\mu})^2=0$ for all $i$, which implies that the $x_i^{\mu}$ coordinates form a polygon with light-like edges.

\subsubsection*{Momentum twistors in four dimensions}

It is usually of interest to consider conformal transformations in dual coordinate space, which are usually known as \textit{dual conformal transformations}. So far we have introduced spinor-helicity variables and dual space coordinates, in terms of which the dual conformal generators are quadratic operators. Interestingly, one can linearize the action of those generators by introducing a set of variables known as \textit{momentum twistors} \cite{Hodges:2009hk}. These are defined as points
\begin{equation}
\label{momentum twistors def}
    z^I_i:=(\lambda_a, \mu^{\dot a}) \,,
\end{equation}
in $\mathbb{CP}^3$ that satisfy
\begin{equation}
\label{incidence relations}
    \mu^{\dot{a}}=\lambda_a \, x^{\dot{a}a} \,,
\end{equation}
where $x^{\dot{a}a}$ is a point in dual space. The constraints introduced in \eqref{incidence relations} are known as \textit{incidence relations}, and imply that a given point in dual space is associated to a line in momentum twistor space. Indeed, given two points $z_A^I=(\lambda_{Aa}, \mu^{\dot a}_A)$ and $z_B^I=(\lambda_{Ba}, \mu^{\dot a}_B)$ in momentum-twistor space one can always define a point
\begin{equation}
x_{AB}^{\dot{a}a}=\frac{\lambda_B^a \mu_A^{\dot{a}}-\lambda_A^a \mu_B^{\dot{a}}}{\langle \lambda_A \lambda_B \rangle} \,,
\end{equation}
in dual space such that $\mu^{\dot{a}}_A=\lambda_{Aa} \, x^{\dot{a}a}_{AB}$ and $\mu^{\dot{a}}_B=\lambda_{Ba} \, x^{\dot{a}a}_{AB}$.

Interestingly, a set of $n$ momentum twistors $z_1^I, \dots, z_n^I$ naturally define the kinematics of scattering process between $n$ massless particles. Indeed, the set of $n$ \textit{intersecting} lines $(z_i^I,z_{i-1}^I)$ introduce $n$ points $x_i^{\dot{a}a}$ in dual coordinate space that are subjected to light-like separations $(x_i-x_{i+1})^2=0$. 
To see this last point, let us consider the pair of lines $(z_i^I,z_{i-1}^I)$ and $(z_{i+1}^I,z_{i}^I)$, which intersect at $z_i^I$. As seen before, these lines define the points $x_{i}^{\dot{a}a}$ and $x_{i+1}^{\dot{a}a}$ in dual space, respectively. Using the incidence relations \eqref{incidence relations} and the fact that the lines intersect at $z_i^I$ we can write
\begin{equation}
\label{incidence relations 2}
    \mu^{\dot{a}}_i=\lambda_a^i \, x_i^{\dot{a}a}=\lambda^i_a \, x_{i+1}^{\dot{a}a} \,,
\end{equation}
which implies
\begin{equation}
\label{two dual points one momentum twistor}
\lambda_a^i \, (x_i-x_{i+1})^{\dot{a}a}=0 \,.
\end{equation}
Eq. \eqref{two dual points one momentum twistor} implies that $(x_i-x_{i+1})^{\dot{a}a}$ has a vanishing eigenvalue, and therefore $\det ((x_i-x_{i+1})^{\dot{a}a})=-(x_i-x_{i+1})^2=0$. One should also note that momentum conservation is guaranteed by the fact that the $z_i$'s define a closed contour. 

As previously stated, one immediate advantage of momentum-twistor variables is that they linearize the action of the $SU(2,2)$ group of dual conformal transformations. Indeed, the corresponding generators are written in momentum-twistor notation as
\begin{equation}
    G^I_J= \sum_{i=1}^n z^I_i \frac{\partial}{\partial z_i^J} \,,
\end{equation}
and the algebra becomes
\begin{equation}
    [G^I_J,G^K_L]=\delta^K_J \, G^I_L- \delta^I_L \, G^K_J \,.
\end{equation}
Therefore, dual conformal invariants can simply be written as
\begin{equation}
\label{momentum twistor brackets}
    \langle ijkl\rangle:=\epsilon_{IJKL} \, z^I_i z^J_j z^K_k z^L_l \,.
\end{equation}
Given a pair of lines $(z_i,z_j)$ and $(z_k,z_l)$ in momentum-twistor space and their corresponding points $x$ and $y$ in dual space one can derive the relation
\begin{equation}
    (x-y)^2=\frac{\langle ijkl \rangle}{\langle i j \rangle \langle kl \rangle} \,.
\end{equation}
where $\langle i j \rangle$ and $\langle kl \rangle$ are the spinor-helicity brackets defined in \eqref{spinor helicity brackets}.

\subsubsection*{Three-dimensional kinematics}

The momentum of a particle in three spacetime dimensions can be written in terms of a symmetric matrix as
\begin{equation}
    p_{ab}=p_{\mu} (\tilde{\sigma}^{\mu})_{ab} = \left( \begin{array}{cc}
p^0-p^1 & p^2 \\
p^2 & p^0+p^1
\end{array}
\right)\,,
\end{equation}
with
\begin{equation}
    \det (p_{ab})=-p^2 \,.
\end{equation}
The above construction allows to describe the momentum of massless particles as
\begin{equation}
    p_{ab}= \lambda_a \lambda_b \,,
\end{equation}
where $\lambda^a$ is a three-dimensional spinor-helicity variable. Note that, in contrast to the four-dimensional case, in three-dimensions only one spinor is needed to describe the momentum of a particle. Lorentz invariants can be written as
\begin{equation}
\label{sij spinor helicity}
s_{ij} := (p_i + p_j)^2= -\langle ij \rangle^2\,.
\end{equation}

Three-dimensional dual space coordinates are defined in the same way as in four dimensions, i.e. $p_{i}^{ab}=(x_{i}-x_{i+1})^{ab}$. On the other hand, three-dimensional momentum twistors can be obtained from a projection of the corresponding four-dimensional variables \cite{Elvang:2014fja}. More precisely, one can show that a point $x_{i}^{ab}$ in the three-dimensional dual space can be associated to a line $(z_i^I,z_{i-1}^I)$ in the space of four-dimensional momentum twistors if one imposes the symplectic constraint
\begin{equation}
\label{symplectic constraint}
    \Sigma_{IJ} \, z^I_i z^J_{i-1}=0 \,,
\end{equation}
with
\begin{equation}
\label{Sigma}
\Sigma= \left( 
\begin{array}{cc} 
0 & -\mathbb{1}_{2 \times 2} \\
\mathbb{1}_{2 \times 2} & 0
\end{array}
 \right) \,.
\end{equation}
Indeed, eq. \eqref{symplectic constraint} implies
\begin{equation}
\epsilon_{ab} \, \lambda_i^a \mu_{i-1}^b -\epsilon_{ab} \, \lambda_{i-1}^a \mu_{i}^b=0 \,,
\end{equation}
for $z_i^I=(\lambda_{i,a},\mu_{i}^b)$ and $z_{i-1}^I=(\lambda_{i-1,a},\mu_{i-1}^b)$, which is solved if the latter satisfy the incidence relations
\begin{align}
\mu_{i}^b &= x_{i}^{ab} \lambda_{i,a} \,, \\
\mu_{i-1}^b &= x_{i}^{ab} \lambda_{i-1,a} \,.
\end{align}
Note that in the above argument it is crucial to use that the $x_{i}^{ab}$ matrix is symmetric in its indices, which is a characteristic of the three-dimensional kinematics.

%% file: app_normalizations.tex
As shown in \eqref{normalization I_L} and \eqref{normalization L_L}, there are relative normalizations $n_L$ and $\tilde{n}_L$ between the integrands ${\bf I}_L$ and ${\bf L}_L$ and the canonical forms $\Omega_L$ and $\tilde{\Omega}_L$. We will discuss their computation in this appendix.

As a first step, we should note that the definitions of ${\bf I}_L$ and ${\bf L}_L$ imply
\begin{align}
\label{IL in terms of LL-1}
{\bf I}_1 &= {\bf L}_1 \,, \\
\label{IL in terms of LL-2}
{\bf I}_2 &= {\bf L}_2 + \frac{1}{2} {\bf L}_1^2 \,, \\
\label{IL in terms of LL-3}
{\bf I}_3 &= {\bf L}_3 + {\bf L}_2 \, {\bf L}_1+ \frac{1}{6} {\bf L}_1^3 \,.
\end{align}
On the other hand, from \eqref{expansion tilde Omega-connected graphs} we get
\begin{align}
\label{Omega_L in terms of Tilde_Omega_L-1}
\Omega_1 &= \tilde{\Omega}_1 \,, \\
\label{Omega_L in terms of Tilde_Omega_L-2}
\Omega_2 &= \tilde{\Omega}_2 + \tilde{\Omega}_1^2 \,, \\
\label{Omega_L in terms of Tilde_Omega_L-3}
\Omega_3 &= \tilde{\Omega}_3 +3 \, \tilde{\Omega}_2 \tilde{\Omega}_1 + \tilde{\Omega}_1^3 \,.
\end{align}
Therefore, using the definitions \eqref{normalization I_L} and \eqref{normalization L_L} and the expansions \eqref{IL in terms of LL-1}-\eqref{Omega_L in terms of Tilde_Omega_L-3} we have
\begin{align}
\label{L in terms of omega tilde-1}
{\bf L}_1 &= n_1 \, \tilde{\Omega}_1 \,, \\
\label{L in terms of omega tilde-2}
{\bf L}_2 &=  n_2 \tilde{\Omega}_2 + \left( n_2 - \frac{n_1^2}{2} \right) \tilde{\Omega}_1^2 \,, \\
\label{L in terms of omega tilde-3}
{\bf L}_3 &= n_3 \tilde{\Omega}_3 + \tilde{\Omega}_1 \, \tilde{\Omega}_2 \left( 3n_3 -n_2 n_1 \right)+ \tilde{\Omega}_1^3 \left( n_3 -n_2 n_1 + \frac{n_1^3}{3} \right) \,.
\end{align}

To fix the values of the $n_L$ coefficients we will follow the ideas of \cite{Bourjaily:2011hi}. These authors used the fact that the integrands ${\bf L}_L$ in planar ${\cal N}=4$ sYM should behave as ${\cal O}(1/\delta)$ in the limit 
\begin{equation}
\label{normalization limit}
\langle (AB)_{\ell} 12 \rangle \sim \delta \,, \qquad \quad \text{and} \qquad \langle (AB)_{\ell} 23 \rangle \sim \delta \,,
\end{equation}
for some $\ell$, while all other brackets remain non-vanishing. This property makes ensures that infrared divergences exponentiate after integration. 
A similar analysis can be done in the ABJM case. Noticing that the $\tilde{\Omega}_L$ forms behave as\footnote{This property can be verified in the expressions for $\tilde{\Omega}_L$ presented in Chapter \ref{ch: neg geoms ABJM}.}
\begin{equation}
    \label{tilde omegas in normalization limit} 
 \tilde{\Omega}_L \sim {\cal O} (1/\delta)  \,,
\end{equation}
in the limit \eqref{normalization limit}, and demanding the same behavior for the l.h.s. of \eqref{L in terms of omega tilde-1}-\eqref{L in terms of omega tilde-3}, we get
\begin{equation}
\label{relation between normalization coefficients-2}
n_1=\tilde{n}_1 \,, \qquad n_2=\tilde{n}_2=\frac{\tilde{n}_1^2}{2!} \,, \qquad n_3=\tilde{n}_3=\frac{\tilde{n}_1^3}{3!} \,.
\end{equation}
Similarly, at higher-loop orders we obtain
\begin{equation}
    n_L=\tilde{n}_L= \frac{\tilde{n}_1^L}{L!} \,.
\end{equation}

Finally, comparing the explicit formulas for ${\bf I}_1$ and ${\bf I}_2$ given in \cite{Bianchi:2011dg,Chen:2011vv} to the expressions for $\Omega_1$ and $\Omega_2$ obtained from the results of \cite{He:2022cup} we get that in the ABJM theory
\begin{equation}
\label{tilde_n_1}
\tilde{n}_1= \frac{i}{2\sqrt{\pi}} \,.
\end{equation}

%% file: app_five_dim_not.tex
When working with three-dimensional dual-coordinates it turns out useful to consider the embedding of the three-dimensional Minkowski space into the five-dimensional projective light-cone. One of the main advantages of this parametrization lies in the fact it allows to write three-dimensional dual-conformal invariants simply as five-dimensional expressions that respect Lorentz and scale invariance.

Let us consider a five-dimensional Minkowski space with $(-,-,+,+,+)$ signature and with coordinates $(X^1,X^2,X^3,X^4,X^5)$. Then, the light-cone is defined by the constraint
\begin{equation}
\label{5D light-cone-1}
-(X^1)^2-(X^2)^2+(X^3)^2+(X^4)^2+(X^5)^2=0 \,.
\end{equation}
Let us note that the constraint \eqref{5D light-cone-1} is invariant under a rescaling of the coordinates, and therefore defines a projective space with 3 degrees of freedom, as expected. It is useful to switch to light-cone coordinates $(X^+,X^-,X^2,X^4,X^5)$, with $X^+$ and $X^-$ given by
\begin{equation}
\label{light-cone coordinates}
X^+=\frac{X^1+X^3}{\sqrt{2}} \,, \qquad \text{and} \qquad X^-=\frac{X^1-X^3}{\sqrt{2}} \,,
\end{equation}
so that \eqref{5D light-cone-1} becomes
\begin{equation}
\label{5D light-cone-2}
-2 \, X^+ \, X^- -(X^2)^2+(X^4)^2+(X^5)^2=0 \,.
\end{equation}
The embedding of the three-dimensional Minkowski space\footnote{We are using the $(-,+,+)$ signature for the three-dimensional Minkowski space.} with coordinates $(x^0,x^1,x^2)$ into the five-dimensional space can be defined as
\begin{equation}
\label{embedding}
(X^+,X^-,X^2,X^4,X^5)=\left( \frac{x^{\mu} x_{\mu}}{2},1,x^0,x^1,x^2 \right) \,.
\end{equation}
It is straightforward then to check that \eqref{embedding} satisfies \eqref{5D light-cone-2}. Moreover, under this parametrization we have
\begin{equation}
\label{3D and 5D invariants}
(X_i-X_j)^2=-2 X_i.X_j=(x_i-x_j)^2 \,,
\end{equation}
where $x_i$ and $x_j$ are points in the three-dimensional space and $X_i$ and $X_j$ are their corresponding images under the mapping \eqref{embedding}.

In order to simplify notation, we will write the contraction of the dual coordinates with the five-dimensional Levi-Civita tensor as
\begin{equation}
\label{5D epsilon}
\epsilon(1,2,3,4,5):=\epsilon_{\mu \nu \rho \sigma \eta} X_1^{\mu} X_2^{\nu} X_3^{\rho} X_4^{\sigma} X_5^{\eta} \,.
\end{equation}
Let us discuss some properties of \eqref{5D epsilon}. In the first place, one can rewrite \eqref{5D epsilon} in terms of three-dimensional dual-coordinates as \cite{Chen:2011vv}
\begin{equation}
\label{5D epsilon-2}
\epsilon(1,2,3,4,5)= \frac{1}{2} \left( x_{51}^2 \epsilon_{\mu\nu\rho} x_{21}^{\mu} x_{31}^{\nu} x_{41}^{\rho} + x_{31}^2 \epsilon_{\mu\nu\rho} x_{51}^{\mu} x_{21}^{\nu} x_{41}^{\rho} \right) \,.
\end{equation}
Also, the product of two contractions is given by
\begin{equation}
\label{5D epsilon product}
\epsilon(1,2,3,4,5) \, \epsilon(1,2,3,4,6)= \frac{X^4_{13} X^4_{24}}{32} \left( \frac{X^2_{15} X^2_{36}+X^2_{16} X^2_{35}}{X^2_{13}} + \frac{X^2_{25} X^2_{46}+X^2_{26} X^2_{45}}{X^2_{24}} - X^2_{56}\right) \,.
\end{equation}
In particular, we have
\begin{equation}
\label{5D epsilon^2}
\epsilon(1,2,3,4,5)^2= \frac{X_{13}^2 X_{24}^2}{16} (X_{24}^2 X_{15}^2 X_{35}^2+X_{13}^2 X_{25}^2 X_{45}^2) \,.
\end{equation}
Finally, following \cite{Caron-Huot:2012sos} we can define a measure on the five-dimensional light-cone as
\begin{equation}
\label{measure light-cone}
d^3 X := \int \frac{d^5 X}{\text{Vol}[\text{GL(1)}]} \, \delta(X^2) \,,
\end{equation}
where the factor $\delta(X^2)$ is included to satisfy the constraint given in \eqref{5D light-cone-1}, while the denominator $\text{Vol}[\text{GL(1)}]$ eliminates the redundancy coming from the projective invariance of the light-cone. Therefore, we get
\begin{equation}
\label{relation measures}
\int d^3 X \equiv \int d^3 x \,.
\end{equation}

%% file: app_useful_integrals.tex
We present here several integrals that are useful for obtaining the results of Chapter \ref{ch: neg geoms ABJM}.

\subsubsection*{Triangle integral}
Let us begin with a triangle integral in three dimensions and with three massive legs. This integral first appears in the one-loop analysis in \eqref{two loop form amplituhedron-v2-2}, and it is explicitly given by
\begin{equation}
\label{triangle integral-1}
{\cal T} := \int \frac{d^3X_6}{i \pi^{3/2}} \, \frac{1}{X_{26}^2 X_{46}^2 X_{56}^2} \,.
\end{equation}
Using the standard Feynman parametrization one gets 
\begin{equation}
\label{triangle integral-2}
{\cal T} = \frac{\pi^{3/2}}{\sqrt{X_{25}^2 X_{45}^2 X_{24}^2}} \,.
\end{equation}
It is interesting to note that the functional form of the result (\ref{triangle integral-2}) can also be obtained from noticing that the integral \eqref{triangle integral-1} has dual conformal invariance.

\subsubsection*{Five-leg integral with an epsilon numerator}

Let us consider now the integral
\begin{equation}
\label{integral 2 loops-epsilon numerator-1}
{\cal E} := \int \frac{d^3 X_6}{i\pi^{3/2}} \, \frac{\epsilon(1,2,3,4,6)}{X_{26}^2 X_{46}^2  X_{56}^2 X_{16} X_{36}} \,.
\end{equation}
which shows up at the two-loop computation in \eqref{epsilon num integral}. Introducing Feynman parameters, we have
\begin{equation}
\label{integral 2 loops-epsilon numerator-2}
\begin{aligned}
{\cal E} &=\frac{\epsilon_{\mu \nu \rho \sigma \eta} X_1^{\mu} X_2^{\nu} X_3^{\rho} X_4^{\sigma}}{\pi \, \text{Vol}[\text{GL(1)}]}  \left( \prod_{i=1}^5 \int_{0}^{\infty} d\alpha_i \, \right) \, (\alpha_1 \alpha_3)^{-1/2} \, \partial^{\eta}_Y \left[ \int \frac{d^3 X_6}{i\pi^{3/2}} \, \frac{1}{\left( -2Y.X_6 \right)^3} \right] \,,
\end{aligned}
\end{equation}
where we have defined
\begin{equation}
\label{Y 5D}
Y:= \sum_{i=1}^5 \alpha_i \, X_i \,,
\end{equation}
and we have used \eqref{3D and 5D invariants}. Then, performing the space-time integral we have
\begin{equation}
\label{integral 2 loops-epsilon numerator-3}
\begin{aligned}
{\cal E} &=  \frac{3}{4 \sqrt{\pi} \, \text{Vol}[\text{GL(1)}]} \left( \prod_{i=1}^5 \int_{0}^{\infty} d\alpha_i \, \right) \, \frac{(\alpha_1 \alpha_3)^{-1/2} \, \epsilon(1,2,3,4,Y)}{\left(  -Y^2 \right)^{\frac{5}{2}}} \,. \\
\end{aligned}
\end{equation}
At this point is useful to define
\begin{equation}
\label{beta variables}
\beta_i := \alpha_i X_{i5}^2 \quad \text{for} \quad i=1,\dots ,4 \,,
\end{equation}
and to mod out the ${\rm GL}(1)$ invariance by setting
\begin{equation}
\label{constraint betas}
\sum_{i=1}^4 \beta_i=1 \,.
\end{equation}
Then, performing the integral over $\alpha_5$ we get
\begin{equation}
\label{integral 2 loops-epsilon numerator-4}
\begin{aligned}
{\cal E} &= \frac{ \epsilon(1,2,3,4,5)}{\sqrt{\pi} \, \left( X_{15}^2 X_{35}^2 \right)^{1/2} X_{25}^2 X_{45}^2}{} \left( \prod_{i=1}^4 \int_{0}^{\infty} d\beta_i \, \right) \delta \left( \sum_{i=1}^4 \beta_i -1 \right) \, \frac{(\beta_1 \beta_3)^{-1/2}}{\left( \beta_1 \beta_3 \frac{X_{13}^2}{X_{15}^2 X_{35}^2} + \beta_2 \beta_4 \frac{X_{24}^2}{X_{25}^2 X_{45}^2} \right)^{\frac{1}{2}}} \,.
\end{aligned}
\end{equation}
The number of remaining integrals can be further simplified by defining
\begin{equation}
\label{gamma variables}
\begin{aligned}
\beta_1 &:= \gamma_1 \gamma_2 \,, \qquad \qquad \, \, \, \beta_2 := \gamma_1 (1-\gamma_2) \,, \\
\beta_3 &:= (1-\gamma_1) \gamma_3 \,, \qquad \beta_4 := (1-\gamma_1) (1-\gamma_3) \,.
\end{aligned}
\end{equation}
Let us note that the constraint \eqref{constraint betas} is trivially satisfied by the $\gamma$'s. In terms of these variables we get
\begin{equation}
\label{integral 2 loops-epsilon numerator-5}
\begin{aligned}
{\cal E} &= \frac{\epsilon(1,2,3,4,5)}{\left( X_{15}^2 X_{25}^2 X_{35}^2 X_{45}^2 X_{24}^2 \right)^{1/2}} \,  \sqrt{\frac{z}{\pi}} \, {\cal H}(z) \,,
\end{aligned}
\end{equation}
with
\begin{equation}
\label{H-1}
{\cal H} (z) :=  \int_{0}^{1} d\gamma_2 \, \int_{0}^{1} d\gamma_3 \, \frac{(\gamma_2 \gamma_3)^{-1/2}}{\left[ \gamma_2 \gamma_3 z+ z \, (1-\gamma_2) (1-\gamma_3) \right]^{\frac{1}{2}}} \,.
\end{equation}
Let us focus on the integral \eqref{H-1}, which as we shall see can be solved by iterating Feynman parametrizations. Making the change of variables
\begin{equation}
\label{first change of variables}
\gamma_2 \to \frac{1}{1+\gamma_2} \, , \qquad \gamma_3 \to \frac{\gamma_3}{1+\gamma_3} \,,
\end{equation}
and introducing Feynman parameters one gets
\begin{equation}
\label{H-2}
{\cal H}(z) = \frac{\sqrt{z}}{\pi} \, \left( \prod_{i=1}^3 \int_0^{\infty} d\eta_i \right) \, \frac{1}{\sqrt{\eta_3}  \, (\eta_1 + \eta_2) (\eta_1+1)  (z \eta_2 + z \eta_3 + 1)} \,.
\end{equation}
Moreover, taking
\begin{equation}
\label{second change of variables}
\eta_3 \to \eta_3^2 \,,
\end{equation}
and making a further Feynman parametrization we have
\begin{equation}
\label{H-3}
{\cal H}(z) =  \int_0^{\infty} d\nu_1 \, \int_0^{\infty} d\nu_2 \,  \frac{1}{(\nu_1 + \nu_2) (\nu_1 + 1) \sqrt{z \nu_2+ 1}} \,.
\end{equation}
Finally, defining 
\begin{equation}
\label{theta}
\theta:= \sqrt{z \nu_2 +1} \,,
\end{equation}
and integrating over $\theta$ we arrive at
\begin{equation}
\begin{aligned}
{\cal H}(z) &= \frac{1}{\sqrt{1+z}} \left[ -\text{Li}_2\left(\frac{2 \left(\sqrt{z+1}-1\right)}{z}\right)+\text{Li}_2\left(-\frac{2
   \left(\sqrt{z+1}+1\right)}{z}\right) \right.\\
   & \left. \quad \, +2 \log \left(\frac{4}{z}\right) \log
   \left(\frac{\sqrt{z+1}+1}{\sqrt{z}}\right) + \pi^2 \right] \,.
   \label{H-4}
\end{aligned}
\end{equation}

\subsubsection*{$I_{\cal F}$ functional}

Let us consider the integral that appears in the definition \eqref{F functional} of the $I_{\cal F}$ functional, i.e.
\begin{equation}
\label{F functional-zp-1}
I_{\cal F}[z^p] \sim -\frac{\sqrt{\pi}}{2} \int  \frac{d^D X_5}{i\pi^{D/2}} \, \left( \frac{X_{13}^2 X_{24}^2}{X_{15}^2 X_{25}^2 X_{35}^2 X_{45}^2} \right)^{\frac{3}{4}} z^p \,,
\end{equation}
with $p \in \mathbb{Z}$, $D=3-2\epsilon$, and where to simplify notation we have chosen to use the symbol $\sim$ to indicate that we are only retaining the leading $1/\epsilon^2$ divergence. Using Feynman parametrization we get
\begin{align*}
\label{F functional-zp-2}
I_{\cal F}[z^p]  \sim &-\frac{\sqrt{\pi} \, X_{13}^{3/2-2p} X_{24}^{3/2+2p}}{\Gamma \left( \frac{3}{4}+p \right)^2 \Gamma \left( \frac{3}{4}-p \right)^2} \, \frac{1}{{\rm Vol}[{\rm GL}(1)]}  \\
& \times \left( \prod_{i=1}^4 \int_0^{\infty} d\alpha_i \right) \, \int \frac{d^D X_5}{i\pi^{D/2}} \,  \frac{(\alpha_1 \alpha_3)^{-\frac{1}{4}-p} (\alpha_2 \alpha_4)^{-\frac{1}{4}+p} }{\left( -2X_5.W \right)^3} \,,
\end{align*}
with 
\begin{equation}
\label{W}
W:= \sum_{i=1}^4 \alpha_i X_i  \,.
\end{equation}
Working as with the ${\cal E}$ integral discussed in the previous section we arrive at
\begin{align*}
\label{F functional-zp-3}
I_{\cal F}[z^p]  \sim &-\frac{\sqrt{\pi} \, \Gamma \left( \frac{3}{2}+\epsilon \right) \Gamma(-\epsilon)^2}{2 \, \Gamma \left( \frac{3}{4}+p \right)^2 \Gamma \left( \frac{3}{4}-p \right)^2 \Gamma(-2\epsilon)} \, \\
& \times \int_0^{1} d\gamma_2  \, \int_0^{1} d\gamma_3 \, \frac{(\gamma_2 \gamma_3)^{-\frac{1}{4}-p} [(1-\gamma_2) (1-\gamma_3)]^{-\frac{1}{4}+p} }{\left[ \gamma_2 \gamma_3 + (1-\gamma_2) (1-\gamma_3) \right]^{3/2+\epsilon}} \,.
\end{align*}
At this point is useful to use the Mellin-Barnes formula, which allows us to write
\begin{equation}
\label{F functional-zp-4}
\begin{aligned}
I_{\cal F}[z^p]  &\sim -\frac{\sqrt{\pi}}{2 \, \Gamma \left( \frac{3}{4}+p \right)^2 \Gamma \left( \frac{3}{4}-p \right)^2 \Gamma(-2\epsilon)} \, \times \\
& \qquad \int_{\zeta-i \infty}^{\zeta+i \infty} \frac{dz}{2\pi i} \, \Gamma \left( z+ \frac{3}{2}+\epsilon \right) \Gamma(-z) \Gamma^2 \left( \frac{3}{4} -p +z \right) \Gamma^2 \left( -\frac{3}{4} +p -z - \epsilon \right) \,,
\end{aligned}
\end{equation}
with
\begin{equation}
\label{eta constraint}
-\frac{3}{4}+p < \zeta < -\frac{3}{4}+p - \epsilon \,.
\end{equation}
To compute the leading divergence of \eqref{F functional-zp-4} we have chosen to follow the method described in \cite{Tausk:1999vh}, which was later automatized in a \textit{Mathematica} package by \cite{Czakon:2005rk}. Then, we finally get
\begin{equation}
\label{F functional-zp-final}
I_{\cal F}[z^p] = -\frac{2 \sqrt{\pi}}{ \Gamma \left( \frac{3}{4} +p \right) \Gamma \left( \frac{3}{4} -p \right)} \,.
\end{equation}

\subsubsection*{$I_{\cal G}$ functional}

Finally, let us focus now on the integral that defines the $I_{\cal G}$ functional in \eqref{G functional}. That is, we will consider
\begin{equation}
\label{G functional-zp-1}
I_{\cal G}[z^p] \sim -\frac{1}{2 \sqrt{\pi}} \int \frac{d^D X_5}{i\pi^{D/2}} \,  \frac{i\epsilon(1,2,3,4,5)}{X_{15}^2 X_{25}^2 X_{35}^2 X_{45}^2} \, z^p \,.
\end{equation}
where again we are using the symbol $\sim$ to indicate that we are only keeping the $1/\epsilon^2$ contribution.
Introducing Feynman parameters we get
\begin{align*}
\label{G functional-zp-2}
I_{\cal G}[z^p] \sim &-\frac{i X_{13}^{-2p} X_{24}^{2p} \, \epsilon_{\mu \nu \rho \sigma \eta} X_1^{\mu} X_2^{\nu} X_3^{\rho} X_4^{\sigma}}{2 \sqrt{\pi} \, \Gamma^2(1+p) \, \Gamma^2(1-p) \, \text{Vol}[\text{GL(1)}] }  \\
& \times \left( \prod_{i=1}^4 \int_{0}^{\infty} d\alpha_i \, \right) \, \left( \frac{\alpha_2 \alpha_4}{\alpha_1 \alpha_3} \right)^{p}\, \partial^{\eta}_W \left[ \int \frac{d^D X_5}{i\pi^{D/2}} \,  \frac{1}{(-2 X_5.W)^3} \right] \,,
\end{align*}
where $W$ was defined in \eqref{W}. The integral 
\begin{equation}
\label{G functional-epsilon integral}
\epsilon_{\mu \nu \rho \sigma \eta} X_1^{\mu} X_2^{\nu} X_3^{\rho} X_4^{\sigma} \, \partial^{\eta}_W \left[ \int \frac{d^D X_5}{i\pi^{D/2}} \,  \frac{1}{(-2 X_5.W)^3} \right] \,,
\end{equation}
was solved in \cite{Chen:2011vv,Caron-Huot:2012sos} using a regularization scheme that allows one to dimensionally regularize the integral without losing the projective invariance that comes from the constraint \eqref{5D light-cone-1}, getting as a result that \eqref{G functional-epsilon integral} has an $\mathcal{O}(\epsilon)$ behavior. Therefore,
\begin{equation}
\label{G functional-zp-final}
I_{\cal G}[z^p] =0 \,.
\end{equation}